\documentclass[dsc]{on}     
\usepackage{graphicx}
\usepackage{txfonts}
\usepackage{natbib}
\usepackage{float}
\usepackage{orcidlink}
\usepackage{xspace}
\usepackage{xcolor}
\usepackage{lscape} 
\usepackage{ragged2e}
\usepackage[symbol]{footmisc}
\usepackage{caption}        
\usepackage{float}          
\usepackage[export]{adjustbox} 

\usepackage[export]{adjustbox}
\usepackage{sidecap}
\usepackage{mcaption}
\usepackage{placeins}
\sidecaptionvpos{figure}{c}

\usepackage{tabularx}
\usepackage{graphicx}
\usepackage{adjustbox}
\usepackage{longtable}
\usepackage{rotating}
\usepackage{longtable,lscape}
\usepackage{rotating}
\usepackage{multirow,multicol}
\usepackage{gensymb}
\usepackage{ragged2e}
\usepackage{sectsty}
\usepackage{caption}
\usepackage{numprint}
\usepackage{coffeestains}
\usepackage{icomma}
\usepackage{epigraph}
\usepackage{graphicx}	
\usepackage{amsmath}	
\usepackage{orcidlink}
\usepackage{atbegshi}
\AtBeginDocument{\AtBeginShipoutNext{\AtBeginShipoutDiscard}}

\newcommand{\MJup}{\ensuremath{M_{\mathrm{Jup}}}\xspace}
\newcommand{\as}{\hbox{$^{\prime\prime}$}\xspace}

\def\ts     {\thinspace} 
\def\arcsec  {$''\ts$}

\definecolor{theme}{rgb}{0.0, 0.06, 0.54} 
\definecolor{arsenic}{rgb}{0.23, 0.27, 0.29}
\definecolor{ashgrey}{rgb}{0.7, 0.75, 0.71}
\definecolor{davysgrey}{rgb}{0.33, 0.33, 0.33}
\definecolor{darkgray}{rgb}{0.66, 0.66, 0.66}

\definecolor{burgundy}{rgb}{0.5, 0.0, 0.13}
\definecolor{burntumber}{rgb}{0.54, 0.2, 0.14}
\definecolor{pantone_222_c}{RGB}{108, 29, 69}

\usepackage[pagestyles]{titlesec}
\titleformat{\chapter}[display]
{\bfseries\large}
{
    \filright
    \textsc{\textcolor{davysgrey}{
        \MakeUppercase{\chaptertitlename}
        \Huge
        \thechapter
    }}
}
{-0.1cm}
{\Large \color{theme}\MakeUppercase} 
[\vspace{-0.05cm}\color{black}\titlerule]

\chapterfont{ \color{theme} }
\sectionfont{ \color{theme}}  
\subsectionfont{ \color{theme}}
\usepackage{fancyhdr}

\titleformat{name=\chapter,numberless}[display]
  {\normalfont\huge}
  {}
  {-150pt}
  {\huge \color{black}\MakeUppercase}
[\vspace{-0.05cm}\color{black}\titlerule]

\titleformat{name=\section,numberless}[display]
  {\normalfont\huge}
  {}
  {-150pt}
  {\huge \color{black}\MakeUppercase}
[\vspace{-0.05cm}\color{black}\titlerule]

\vspace*{-0.8cm}

\begin{document}
\begin{titlepage} 
\pagenumbering{roman}
\setcounter{page}{1}
\begin{center}
\includegraphics[scale=.4]{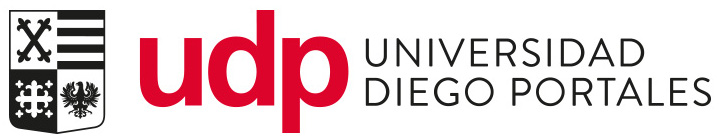}
\\[15mm]

\hrule  height 1.3pt
\vspace{0.3cm}
\textcolor{black}{ \Large \MakeUppercase { \textbf{A MM AND NEAR-IR STUDY OF YSOS:\\ FROM OUTBURSTING PROTOSTARS
TO SATELLITES}}} \\[0.01cm]
\vspace{0.3cm}
\hrule  height 1.3pt
\vspace{1.2cm}

{\textbf{{\large{A THESIS SUBMITTED BY\\[1.3cm]}}}} 
         
{\color{red}{\fontsize{16    }{15}{\textbf{{PEDRO HENRIQUE SOARES DA SILVA DE PINHO NOGUEIRA\\[1.3cm]}}}}} 

{\textbf{{\large{TO THE\\[.3cm]}}}} 

{\textbf{{\large{FACULTAD DE INGENIERÍA Y CIENCIAS\\}}}} 
{\textbf{{\large{INSTITUTO DE ESTUDIOS ASTROFÍSICOS\\[1.7cm]}}}} 

{\textbf{\large{In partial fulfillment of the requirements for the  degree of Doctor of Philosophy in Astrophysics.\\[1.3cm]}}} 

{\textbf{\large{Supervisor: Alice Zurlo\\}}}          
{\textbf{\large{Co-Supervisor: Lucas Cieza\\[1.5cm]}}} 

{\textbf{\fontsize{12}{12}\selectfont{UNIVERSIDAD DIEGO PORTALES\\ Santiago, Chile\\ 
2024\\}}}  
\end{center}
\end{titlepage}


\pagenumbering{roman}
\newgeometry{inner=1.25in,outer=1.0in, top=2.in, bottom=1in}

\section*{} 
\vspace{-5pt}
\addcontentsline{toc}{section}{Author copyrights}

        \copyright 2024, Pedro Henrique Soares da Silva de Pinho Nogueira
                      
        \begin{flushleft}
        
            All rights reserved
        \end{flushleft}
    \begingroup
  \begin{figure}[hb!]
    \centering
    \includegraphics[width=0.9\textwidth]{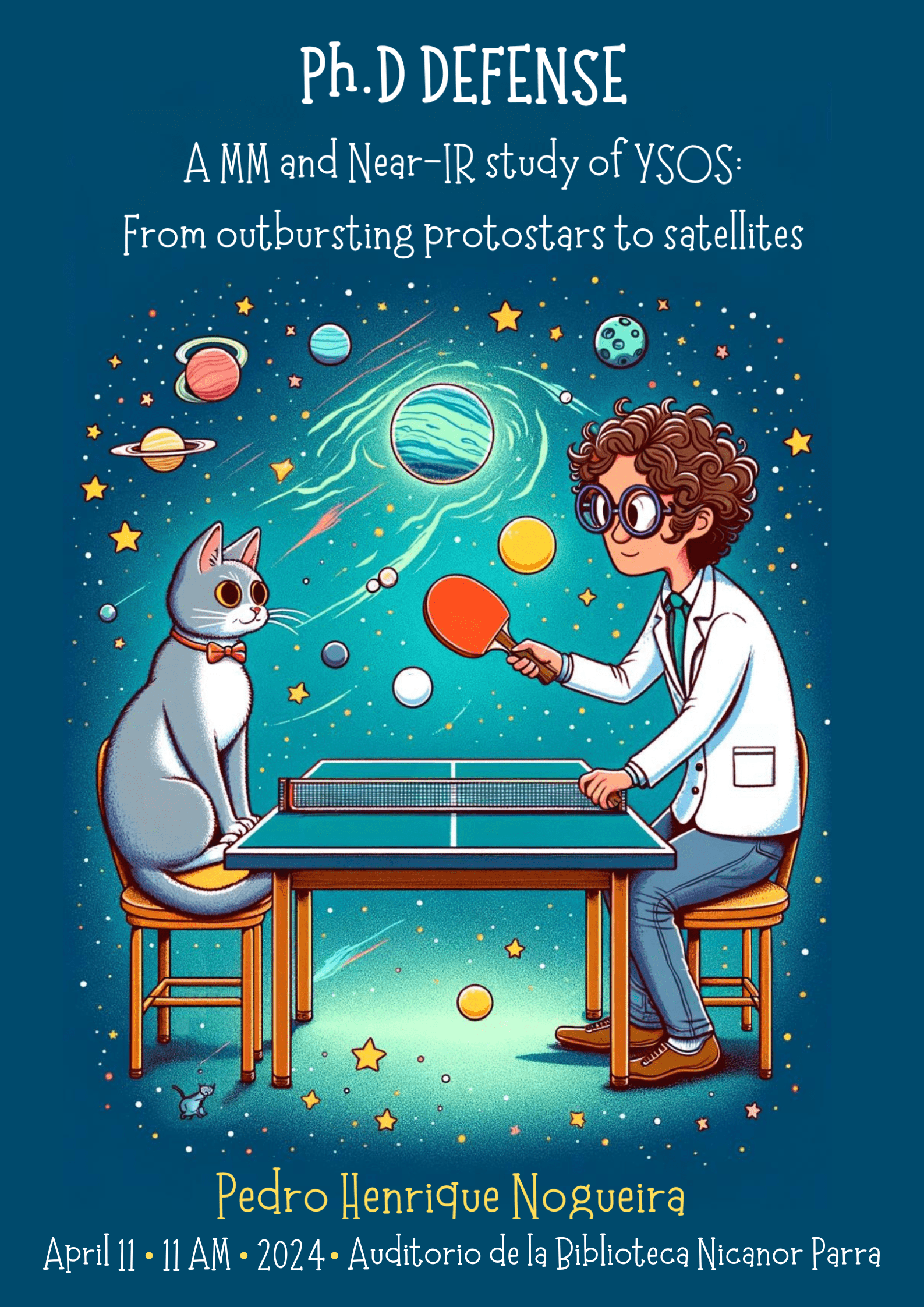}
    \caption{Poster secretly made by UDP friends to advertise others regarding my defense.}
    \label{fig:enter-label}
\end{figure}
\endgroup

\begingroup
  \pagestyle{empty}
  \null
  \newpage
\endgroup

\chapter*{Approval report} 
\addcontentsline{toc}{section}{Approval Report - Doctoral thesis}

A copy of the approval report signed by the Thesis committee is displayed: 

\begin{figure}[!h]
    \centering    \includegraphics[width=0.9\textwidth]{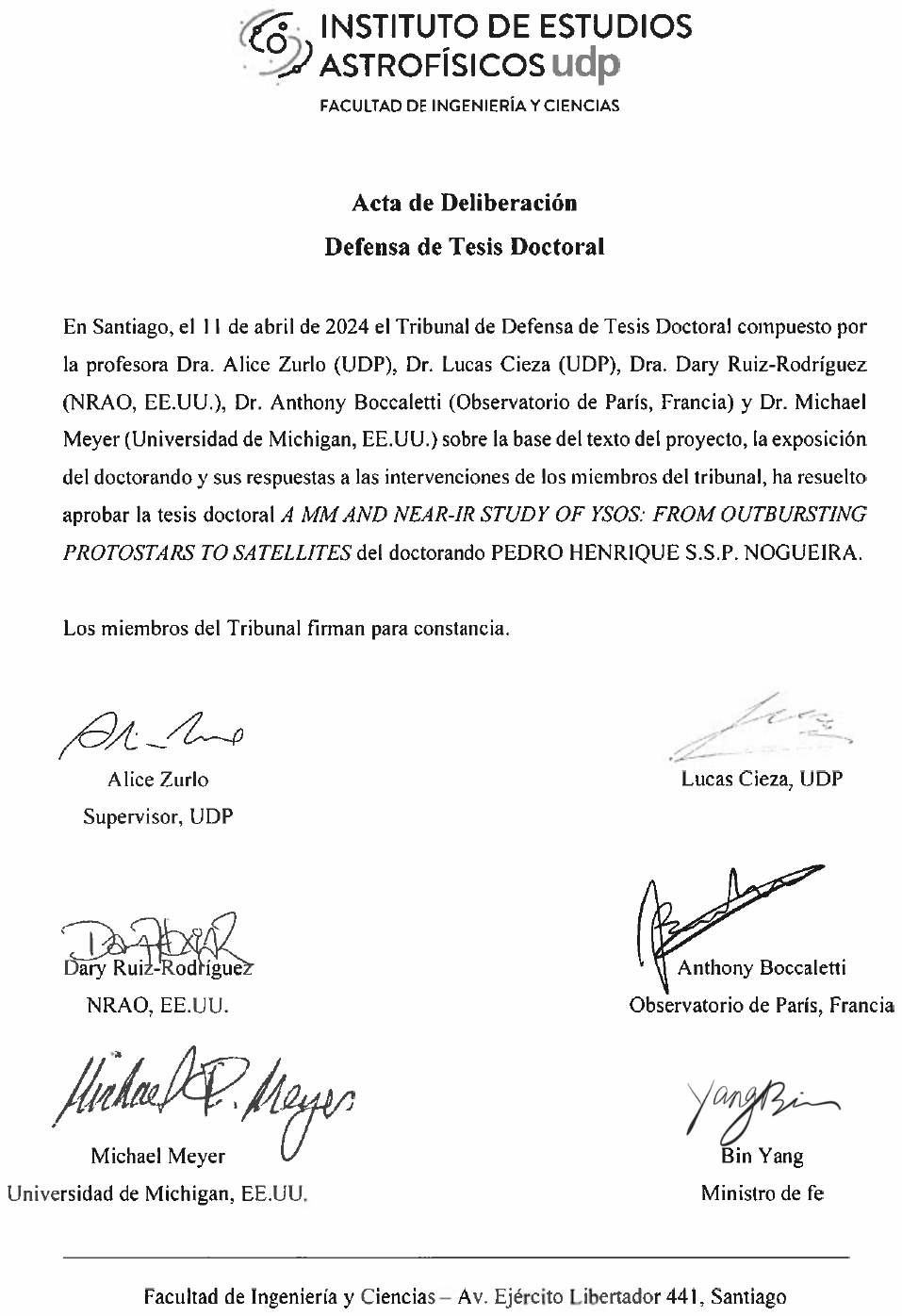}
\end{figure}
%
%
%
%


\clearpage 

\dedication{To and for clearer skies ahead}
\coffeestainA{0.9}{0.85}{-25}{5cm}{1.3cm}  
\chapter*{ACKNOWLEDGEMENTS}

Acknowledging so many great people who have supported me on this journey is a challenging task, but I will do my best.

First of all, I would like to express my gratitude to my scientific mentors and advisors, Alice Zurlo and Lucas Cieza. Your guidance and support were fundamental in making this achievement possible. I also extend my thanks to the committee members for their help during the development of my thesis and the defense itself. Special thanks go to Camilo, Cecilia, and Trisha. You helped me tremendously behind the scenes, showing immense patience and dedication while teaching and supporting me through the toughest moments.
 

Thanks especially to Trisha, my extra advisor and super duper friend. You helped me a lot with everything—in work, and in life. Sometimes, just your presence in the office was enough to give me strength and courage to advance more, do more, and be more. You are engraved in my heart, Trishita!


Thanks to the UDP community. I was pleasantly surprised when I first arrived at UDP. The Astronomy Institute is truly an oasis of gentle people. Everyone there is so nice! I really wish all academic environments were like the one in which I had the opportunity to see and live.

Eu queria especialmente agradecer, do fundo do meu coração, meus pais, Vitor e Roseli, e minha irmã, Ana Clara, os quais sempre me apoiaram com esse sonho doido, de entender e alcançar as estrelas, apesar de me levar ao ponto de ficar mais velhinho e ainda viver de forma instável hahaha. É o preço de sonhadores nesse mundo louco em que hoje vivemos, mas posso dizer, ainda hoje, que não me arrependo da trilha que percorri. Agradeço muito todo o apoio que tive para concluir esses suados 12 anos de estudo e dedicação. Para vocês, o meu mais sincero OBRIGADO e AMO VOCÊS!

Agradeço aos amigos que fiz nessa minha longa jornada, alguns que ficaram comigo até hoje e outros que deixaram suas marcas em uma página que já folheei, porém foram fundamentais para mim. A mencionar alguns deles: grupinho do rock de Amparo - a gente se fala e se encontra um dia a cada 1 ou 2 anos mas sempre é muito prazerosa a companhia de vocês. É muito legal acompanhar a nossa trajetória desde o período de crianças/adolescentes até jovens adultos. Desejo o melhor para vocês! Também menciono meu grande amigo Victor. Companheiro desde início/meio da graduação, foi e continua sendo uma ótima pessoa, de grande coração e humor duvidoso (porém similar ao meu). Te agradeço pela companhia e por tudo até hoje. Te desejo o melhor! Agradeço aos amigos da graduação e do mestrado, alguns deles que por consequências da vida ME SEGUIRAM até o Chile (Priscila, Camila, Douglas), e outros que ainda sigo os passos por redes sociais e eventuais mensagens (Eduardo, Yanna, Stephane, Aline, Natasha, Rodrigo, e muitos outros). 

Agradeço ao BR chilean gang, grupo esse que reuniu brasileiros únicos porém duvidosos (Priscilla, Camila, Douglas, Carol, Tuila, Danielle, Flor, Johnny, Thallis e Larissa). Vocês fizeram com que minha estadia aqui não fosse tão dolorosa como poderia ter sido. É normal que um estrangeiro sinta um grande impacto ao estar longe de casa, cultura e família. Porém vocês reduziram demais esse possível dano com sua maravilhosa companhia, risadas, diversões, fofocas, memes e muito mais! Amo vocês, lindos!

Agradezco mucho las amistades que hice en Santiago, especialmente a los grupos de tenis de mesa en los que participé. Recordé mi vieja pasión por el deporte y realmente me dediqué mucho a mejorar. Esa experiencia fue increíble gracias a la compañía de gente maravillosa y muy amable. Gracias a los grandes maestros Juan Papic, Francisco Mendoza, Roberto, Luis Peña y Luis ``Loco'' Fredes por ayudarme a mejorar y desarrollar esa pasión que tenía, y por la paciencia que tuvieron con mi español limitado. Agradezco a mis amigos Milio, Leo, Lucas y Gabriel. Agradezco muchísimo a Milio, que fue un gran amigo, me invitó a su casa para jugar muchas veces, y es una persona increíble, de corazón gigante. Agradezco de corazón a Gabriel, con quien además del tenis de mesa, compartió sus logros, se interesó por los míos, me llevó a cafés y me enseñó acerca de la cultura chilena y de cosas muy raras como películas underground, weás filosóficas/políticas y la participación de Platón en todo eso. Descubrí cuánto me gustaba la comunidad chilena gracias a él. Lamentablemente, dejaré su compañía por ahora, pero dejo aquí constancia de que fue el chileno que más me hizo apreciar su país y su cultura. Un día volveré y te sacaré la chucha en un partido.



I profoundly thank the people from my office. From the veterans who have moved on to other offices across lands and oceans (Aldo, Kacper, Rafi, Emily), to the core members of the Why??? group (Keerthana, Pablo, Trisha, Tatevik, Emilio, Florence, Kevin, and Simon). I don't know if you guys realize how important you were/are to me and how crucial you were for me to keep my sanity during my Ph.D. The never-ending conversations, laughs, and gossips (while of course keeping the working pace) made my stay and my working hours incredibly enjoyable. Every day I was excited to go to the university because of your presence. I love you guys! I will really miss you, and your close family members who I also  love (Ranajit and the beautiful Rohini, Karen and the long-awaited Vaneh).


Agora agradeço à minha família mais próxima nesses anos: minha amada esposa, Carol, e meu primogênito felino, Morpheu. Eu amo tanto vocês! São tão parte de mim que não consigo imaginar uma vida sem suas presenças. Estiveram comigo nos momentos difíceis, tristes, felizes, durante derrotas e conquistas! No filme ``Dança Comigo'', há um trecho que acho super pertinente a esse parágrafo. A personagem de Susan Sarandon questiona, ``Why do people get married?'' e logo em seguida diz: ``There are billions of people on the planet. What does any one life mean? In a marriage, you are promising to care about everything: the good things, the bad things, the terrible things, the mundane things. All of it. All the time, every day. You are saying, -- Your life will not go unnoticed, because I will notice it. Your life will not go unwitnessed, because I will be your witness.'' Portanto, deixo aqui palavras mais pertinente à Carol: Te amo, meu amor! Obrigado por ser minha âncora quando precisei frear, meu motor quando precisei acelerar, mas principalmente minha principal testemunha e cúmplice! Agora, deixo palavras mais pertinentes ao Morpheuzinho: miau, miau, miau, miau, miau, miau, miau, miau, miau. Você sempre foi e sempre será nosso filhinho amado, nosso reizinho, o líder dos gatinhos.

Por fim, agradeço a mim mesmo. Deixo essa nota pessoal para relembrar-me que no final das contas, o carro-chefe de mudanças e progresso é você mesmo. É você quem colhe os frutos que planta, e você é quem deveria ser seu maior motivador e quem mais acredita no seu potencial. Eu consegui! Sou doutor! 
\thispagestyle{empty}
\vspace*{\fill}

\coffeestainB{0.7}{1}{-30}{18 pt}{-135 pt}

\epigraph{\itshape ``When you feel yourself hitting up against your limit, remember for what cause you clench your fists... remember why you started down this path, and let that memory carry you beyond your limit.''}{--- All Might, \textit{in My Hero Academia} -- Kōhei Horikoshi}


  \selectlanguage{english}
  
  \begin{foreignabstract}

We are currently in a golden era of observing and characterizing Young Stellar Objects (YSOs), protoplanetary disks, and substellar objects. This period provides essential data that helps unveil the most complex aspects of the formation and evolution of these celestial bodies. In this thesis, we explore two binary systems at different stages of formation. 

Firstly, we investigate an eruptive YSO system with ALMA band 6 (1.3 mm) observations, revealing its binarity and dust/gas properties. HBC 494, a FUor object in the Orion Molecular Cloud, displays disk asymmetry at 0.2\arcsec resolution, resolved into two components, HBC 494 N (primary) and HBC 494 S (secondary), at $\sim$0.03\arcsec. The disks, quasi-aligned and with similar inclinations give hints about its formation scenario. The system exhibits a projected separation of $\sim$0.18\arcsec (75 au). HBC 494 N is $\sim$5 times brighter and $\sim$2 times larger than HBC 494 S. We notice that the northern component has a similar mass to the FUors, while the southern has to EXors. The HBC 494 disks show individual sizes that are smaller than single eruptive YSOs. The $^{12}$CO, $^{13}$CO, and C$^{18}$O molecular line observations reveal bipolar outflows and rotating, infalling envelopes. Cavity features within the continuum disks' area suggest continuum over-subtraction or slow-moving jets and chemical destruction along the line-of-sight. 

Secondly, we explored a young binary system with the advent of high-contrast imaging observations performed with the instrument VLT/SPHERE in the H band. We performed astrometric, photometric, and orbital analysis. $\eta$ Tel, a 18 Myr system, comprises a 2.09 M$_{\odot}$ A-type star and a M7/M8 brown dwarf companion, $\eta$ Tel B, separated by $\sim$4.2\arcsec (208 au). New SPHERE/IRDIS coronagraphic observations with a contrast of 1.0$\times 10^{-5}$ at the location of the companion are presented in this work. Alongside with previous literature astrometric measurements, a $\eta$ Tel B astrometric follow-up of over a 19-year baseline is provided. The astrometric positions of the companion are calculated with an uncertainty of 4 milliarcseconds (mas) in separation and 0.2 degrees in position angle. The companion's contrast is 6.8 magnitudes in the H band, with a separation of 4.218\arcsec and a position angle of 167.3 degrees. Orbital analysis reveals a low eccentric orbit (e $\sim$ 0.34) with an inclination of 81.9 degrees (nearly edge-on) and a semi-major axis of 218 au. The mass of $\eta$ Tel B is determined to be 48 M$_{Jup}$, consistent with previous literature. No significant residual indicative of a satellite or disk surrounding the companion is detected, with limits ruling out massive objects around $\eta$ Tel B at separations down to 33 au with masses as low as 1.6M$_{Jup}$.

The two studies presented in this thesis establish a couple of different analyses and methodologies, ranging from sub-mm to near-IR wavelength observations. I hope it can help elucidate how diverse, complex, and intriguing the field of (sub)stellar formation/evolution can be.

\vspace{\baselineskip}
 \noindent

\end{foreignabstract}

  \tableofcontents   
  \listoffigures     
  \listoftables      


  \mainmatter
  \chapter{Introduction}
\label{ch:intro}

Look up, and then look further. Astronomers are adventurers seeking to discover and unveil the secrets that extend beyond our clouds. Embedded by mighty minds and advanced tools, we are equipped with knowledge meant to be shared.

In this thesis, I am pleased to elaborate on the work conducted during my doctoral stage. My goal is to offer valuable perspectives on stellar and planetary formation by studying a binary eruptive system. I also aim to provide an exhaustive analysis of a young stellar system composed of a giant star and a brown dwarf, on its first Myrs of life. In this chapter, I introduce
the scientific background required to provide a comprehensive basis for the subsequent analysis of our research findings.

\section{Protoplanetary disks - theory and observations}
\label{sec:intro/disks}
A protoplanetary disk is a rotating circumstellar disk composed of gas and dust. This structure is a direct outcome of angular momentum conservation from a rotating chunk of a molecular cloud, which undergoes flattening as part of its evolutionary process. The protoplanetary disk is formed by the leftover materials that were not accreted into the forming protostar at the center or were evaporated/expelled through other physical mechanisms.

The investigation of protoplanetary disks plays a fundamental role in our comprehension of planetary formation. In an era where observations offer comprehensive insights into the structure and physical mechanisms of disks at various developmental stages, the theoretical aspects of planetary formation are continually refining, yielding physically more accurate models over time. Nevertheless, many questions remain unanswered such as: How do physical and chemical mechanisms vary across different environments and stellar masses? To what extent is chemical complexity necessary for the formation of mature planetary systems? At what frequency rate do protostars accrete and disks expel gas and dust? How and when do planets form, and how do they interact with disks? What processes contribute to the formation of a system like the Solar System, and how common is such an occurrence? These and other questions underscore the complexity of our understanding of this field.


\subsection{YSOs classification and evolutionary stages}

\label{sec:YSO_classification}
Young stellar objects (YSOs) can be denoted as objects composed of a protostar and its surroundings, mainly the protoplanetary disk and its substructures. A YSO delimits a phase between the formation of a rotating quasi-spherical molecular core around a forming protostar and a main-sequence star and its system.

\begin{figure}[!htb]
    \centering
    \includegraphics[width=0.9\textwidth]{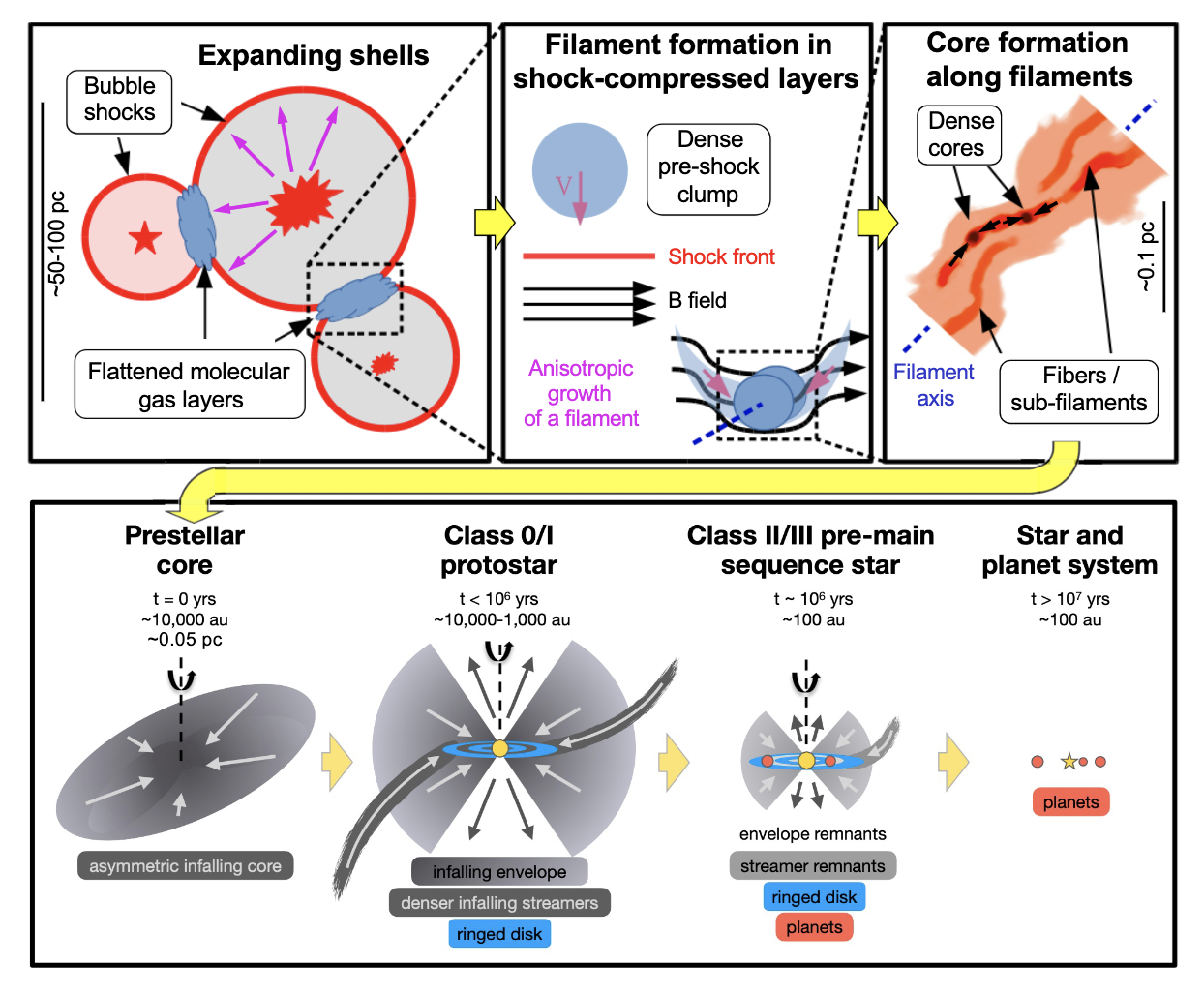}
    \caption[Schematic representation of the stages of star/planetary formation]{Schematic representation of the stages of star/planetary formation, since the formation of filaments through the shock of expanding bubbles mature star/planetary systems. Credits: \citet{pineda2023}.}
    \label{allstages}
\end{figure}

The connections between stellar/planetary theory and observations allow us to trace how a YSO evolves. According to the most recent compilation/description of protoplanetary disk formation and evolution \citep{pineda2023}, it can be summarized as follows:


According to observations, bubbles of both neutral and ionized gas
are ubiquitously observed in the interstellar medium across the Galaxy. Also, according to models (e.g., \citealp{padoan2017,2018ApJ...853..173K,2021MNRAS.504.1039R}),  bubbles can be formed by supernova explosions, radiation, and winds from massive stars. Therefore, first, on a scale of 1-100 pc, neutral and ionized hydrogen gas in the shape of quasi-spherical bubbles will expand. The shock between two or more bubbles will generate layers of magnetized molecular gas ($\sim$100 cm$^{-3}$), which can be described as filaments (scales up to tenths of pc). The filaments will continuously accrete gas and dust and compress it media transverse velocity gradients arising from large-scale inflow, rotation, shearing motions, or a combination
of these types of motions. These filaments may further break down into filamentary substructures, called fibers, which can also co-locate with dense cores. These cores will collapse if their local mass per unit length ($M_{line}$) overcomes the critical $M_{line/crit}$, equals to $2 c^{2}_{s}
/G$, where $c_{s}$ it is the sound speed at the environment, and $G$ the gravitational constant. If the core starts to collapse, it will trigger gas and dust infalling towards its center. This process will form a hydrostatic core. Consequently, this core, surrounded by a gas and dust thick quasi-spherical shell, will get more dense and hot, thus forming a protostar. Then, due to the conservation of angular momentum, the rotating shell will turn into a disk, becoming flattened and flared. This stage is characterized by episodic accretion and magnetically driven bipolar outflows that are launched from the inner disk. After, the envelope will start to be cleared where the central star, now called pre-main sequence star (PMS). Finally, the primordial disk (protoplanetary disk) will dissipate due to accretion, winds, outbursts, and photoevaporation. However, a debris disk, where second-generation dust will continuously be replenished by the collision of planetesimals,  will remain. The evolutionary path here presented usually involves the formation of multiple planets and minor bodies (see figure \ref{allstages}). But when and how these planets form still is not well understood.

The presence of circumstellar material was first inferred through infrared excess observations due to the thermal emission of warm dust and through Spectral Energy Distribution analysis (SEDs; \citealp{1966ApJ...143.1010M,1979ApJS...41..743C,1987ApJ...323..714K,1989AJ.....97.1451S}). This happened before radio observations assumed a fundamental role. However, even in the current landscape, where interferometry radio data is the primary means for observing disks, allowing for the resolution of substructures, determination of geometries, and identification of multi-components, as well as exploring the involved physical and chemical mechanisms, there remains a common practice of classifying evolutionary stages. This approach is reminiscent of the earlier SED analysis era, measuring the slope between about 2 and 25 $\mathrm{\mu m}$ and first formalized by \citet{lada1987} and \citet{greene1994}:
\begin{equation}
 \alpha_{\mathrm{IR}}=\frac{d \log v F_v}{d \log v}=\frac{d \log \lambda F_\lambda}{d \log \lambda}   
\end{equation}

\begin{figure}[htb!]
    \centering
    \includegraphics[width=1\textwidth]{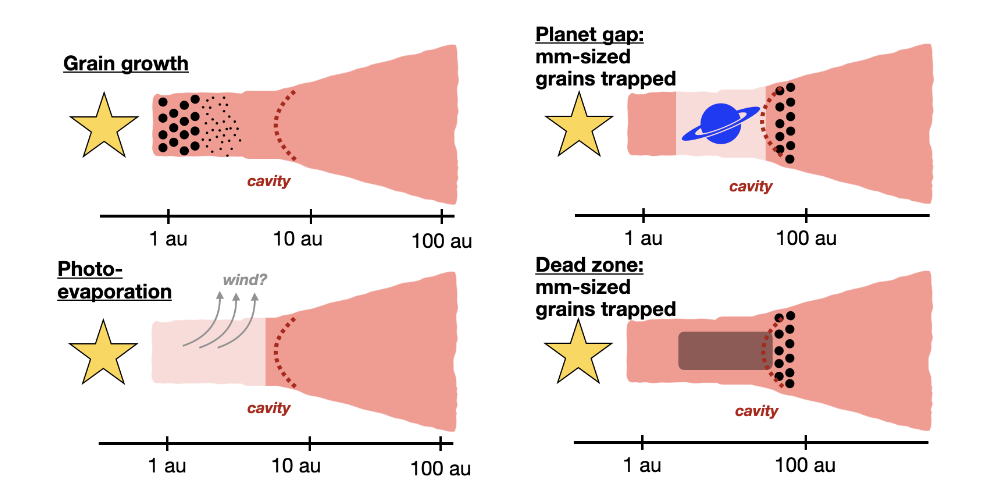} 
    \caption[Four main mechanisms proposed to explain transition disks: grain growth, photoevaporation, clearing by a companion and dead zones]{Four main mechanisms proposed to explain transition disks: grain growth, photoevaporation, clearing by a companion and dead zones. Credits: \citet{2023EPJP..138..225V}}
    \label{cavitiesmechanisms}
\end{figure}

These stages include\footnote{It is important to state that this classification can lead to evolutionary misinterpretation depending on the disk's inclination/dust extinction. For example, a class II YSO at high inclination has a similar SED as a class I YSO  
\citep{2006ApJS..167..256R,william2011}}: 

\begin{itemize}
    \item Class 0: These sources are fully embedded by the envelope throughout this short stage (t $<$ 10$^{4}$ yrs), and they show almost no emission in optical or near-IR wavelengths. Their SEDs are similar to a cold black body and their emission can be detected at longer wavelengths (far-IR to (sub)-mm). 
    \item Class I: In this stage, it starts the clearing of the envelope. The SEDs will be constituted by the black body of the protostar, radiation scattered into the line of sight from dust grains in the disk, emission from the collapsing outer envelope, and thermal emission from a circumstellar disk. The infrared excess slope is positive ($\alpha_{\mathrm{IR}}$ $>$ 0.3) in this stage and the SED peaks in the mid to far infrared. The presence of the cold envelope absorbing emission from the dust, gas, and protostar, will produce absorption lines. In this stage, jets and bipolar outflows, playing a fundamental role in clearing the remaining envelope, can be observed. The age of these objects is between 10$^{4}$ and 10$^{5}$ years. 
    \item Flat-spectrum: This stage is an intermediate between Class I and II and has slopes between -0.3 and 0.3. This stage shows less envelope obscuration and has a greater contribution to the accretion into the SED.
    \item Class II: In this phase, the disks exhibit a slope of -1.6 $<$ $\alpha_{\mathrm{IR}}$ $<$ 0.3. The dissipation of the envelope is a defining characteristic, leading to an SED dominated by the stellar photosphere and the circumstellar accretion disk. The initiation of clearing in the accreting circumstellar disk surrounding the now-called pre-main sequence star (PMS) shifts the SED towards the near-IR to the optical range.  The SED, wider than a blackbody, reflects the disk's contribution. Concurrently, the YSO's activity begins to diminish, marked by fewer outbursts, jets, and accretion events. These systems typically have an age range of approximately 10$^{6}$ to 10$^{7}$ years and are commonly referred to as classical T Tauri stars (YSOs leading to a lower mass stellar evolution, up to 2 M$_{\odot}$), or Herbig Ae/Be stars (YSOS with spectral type earlier than F0). Moreover, classical T Tauri stars are also known to have strong H$\alpha$ and UV emission, while Herbig Ae/Be stars do not show a so prominent UV excess.
    \item Transition disks: The disks in this stage are known to have large inner gaps due to pressure bumps and/or due to planet formation. They also represent the last stages of inner disk clearing due to accretion onto the PMS. Transition disks are known as having inner cavities between the PMS+inner disk and an outer disk. The process of halting the dust and causing the cavities can be many (see \citet{2023EPJP..138..225V} and Fig. \ref{cavitiesmechanisms} for further references).
    
    \item Class III: In this phase, the disk exhibits near-complete passivity with either no or very weak accretion, and the sources demonstrate negative slopes ($\alpha_{\mathrm{IR}}$ $<$ -1.6). The spectral energy distribution (SED) closely resembles the blackbody spectrum of the pre-main sequence (PMS) or main-sequence central star. During this stage, the gas content diminishes, leading to the gradual disappearance of the circumstellar accretion disk. These disks host newborn planets, comets, asteroids, and residual pebbles, earning them the designation of debris disks. When the central star remains in the pre-main sequence phase, it is termed a Weak-lined T Tauri, indicating a minimal or absent accretion presence.
\end{itemize}

\begin{figure}[htb!]
    \centering
    \includegraphics[width=1\textwidth]{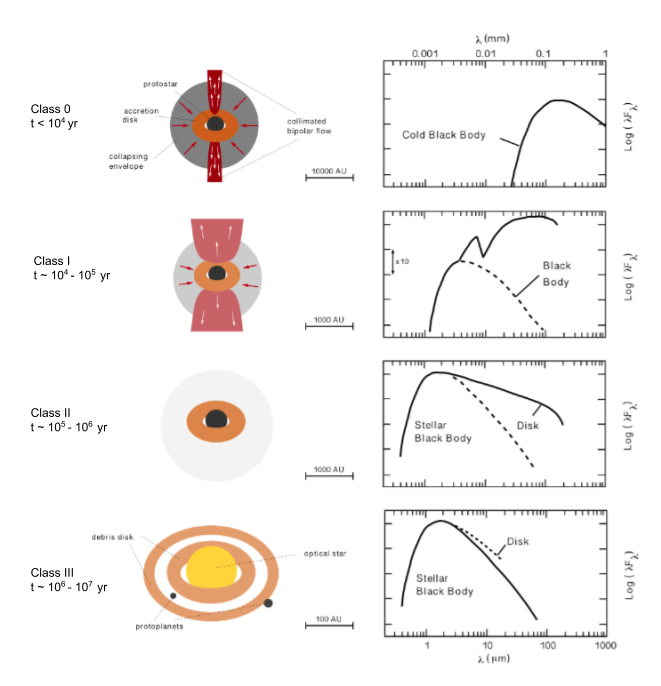} 
    \caption[Classification of YSOs according to SED classification]{Classification of YSOs according to SED classification. The stages represent the standard low-mass star formation. Credits: Adapted from \citet{thesismaciasquevedo}.}
    \label{sedstages}
\end{figure}

A representation of Class 0, I, II, and III YSOs is presented in Fig. \ref{sedstages}. 

\subsection{Eruptive stars - a ``special'' case}
\label{sec:eruptive}
Young Stellar Objects (YSOs) have photometric and spectroscopic variability across multiple evolutionary stages. Although understanding the precise mechanisms driving these variations can be challenging in certain cases, it is established that some fluctuations arise from changes in accretion rates.

\begin{figure}[htb!]
    \centering
    \includegraphics[width=1\textwidth]{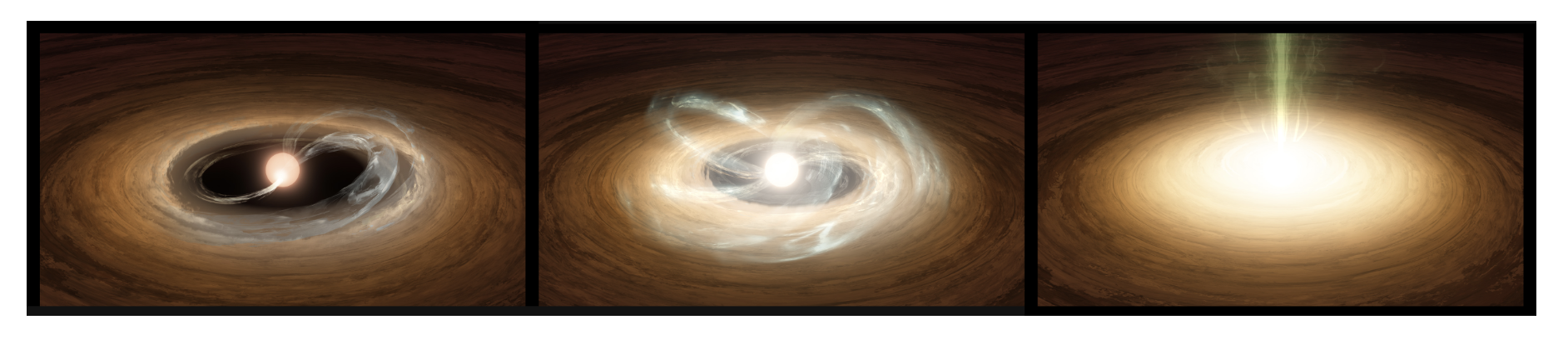} 
    \caption[Artistic representation of a YSO having an outbursting accretion and rise in luminosity]{Artistic representation of a YSO having an outbursting accretion and rise in luminosity, becoming an eruptive star. Credit: T. Pyle (Caltech/IPAC).}
    \label{accretion}
\end{figure}

When a protostar or pre-main sequence (PMS) star experiences a large enhanced phase of accretion, that may last for months to decades, it can be called an eruptive star. During this stage, the accretion disk will transit from a magnetospheric accretion, typical from Class 0-2 YSOs, towards an outbursting accretion. Roughly, the luminosity of the YSO can be expressed as:
\begin{equation}
    L_{\text {proto }}=L_{\text {photospheric }}+L_{\text {accretion }}=L_{\text {phot }}+G M_* \dot{M}_{\mathrm{acc}} / R_*, 
\end{equation}
where $M_*$ and $R_*t$ are the mass and radius of the star, thus stating that an enhancement of accretion will raise the total luminosity (see Fig. \ref{accretion}). This process will also trigger more violent outbursts and jets. 

Changes in the accretion rate can be measured through different physical components and mechanisms of a YSO. If low-extinct and with a favorable geometry, a direct diagnosis from the heated gas can be done through the UV continuum and several emission lines in UV and optical wavelengths. In the case of more embedded sources, the analysis also can be done up to near-IR. In other cases, a reprocessed emission that occurs at infrared or millimeter wavelengths needs to be used.

For classical T Tauri stars, the accreting gas is conducted by stellar magnetic fields from the disk to the star. The accretion shock heats the photosphere, mostly with X-rays. The heated photosphere and the accretion flow will produce hydrogen recombination, $H^{-}$ continuum, and line emission, processes observed in optical and UV wavelengths. 

For passively heated dust disks (low-accretion, the heating of the dust is essentially from the starlight), the innermost dust emits in the near and mid-infrared. The outer disk, however, dominates long-wavelength emission.

For rapidly accreting disks, viscous processes heat efficiently the disk to higher temperatures, which can surpass the stellar one. Therefore, the luminosity of the YSO is dominated by the accretion, which can be measured at optical and infrared wavelengths. If a typical low-state accretion disk temporarily turns to be a high-state viscously heated disk, as expected for eruptive disks, a large increase in brightness will be detectable at optical and infrared wavelengths.

Historically, since the review from \citet{audard2014}, eruptive stars are mainly classified into two groups, the EX-Lupi-like stars or EXors and the FU-Ori-like stars or FUors. Nowadays, however, due to several eruptive stars having different spectroscopic signatures, or having peak luminosities and different light curves than the regular FUOrs and EXOrs, this classification is being challenged (for an updated review, see \citealt{fischer2023}). For the sake of simplicity, I will proceed to explain these two categories. 

EXors are YSOs that present 2.5-5 magnitude optical outbursts that last for months to years and may recur after a few years. These objects were inspired by Ex Lupi, a T-Tauri star in the constellation of Lupus. During the periods of quiescence, Ex-Lupi, and other EXors can't be distinguished from a regular T-Tauri in optical and near-IR wavelengths, and spectroscopically in absorption lines of Na, Ca, K, Fe, Ti, Si, and CO. However, in outbursts, most absorption lines turn into emission and many emission lines become stronger. Also, the emission line fluxes concerning accretion, as Paschen $\beta$ and Brackett $\gamma$, show accretion rates on the order of 10$^{-7}$ to 10$^{-6}$ M$_{\odot}$ yr $^{-1}$.

FUors, from the precursor FU Orionis, are a type of Young Stellar Objects (YSOs) distinguished by more intense and prolonged outbursts compared to EXor variables. During an outburst, the bolometric luminosity of a FUor undergoes a substantial increase, reaching tens to hundreds of solar luminosities. Their accretion rates can vary from 10$^{-6}$ to 10$^{-4}$ M$_{\odot}$ yr$^{-1}$. Spectroscopically, FUors in outburst exhibit features typical of a viscously heated accretion disk overshadowing the central protostar across all wavelengths. However, their behavior differs at various wavelengths—longer wavelengths reveal colder and slower-rotating regions of the disk. In the optical, they display absorption spectra resembling F or G supergiants, featuring broad lines indicating strong winds, a P-Cygni profile, and notable Li absorption \citep{1996ARA&A..34..207H}. In the infrared, their characteristics are more similar to M or K giants and supergiants, showing strong CO absorption, water, TiO, or VO absorption bands, and prominent He I absorption \citep{connelley2018}.

\begin{figure}[!htb]
    \centering
    \includegraphics[width=1\textwidth]{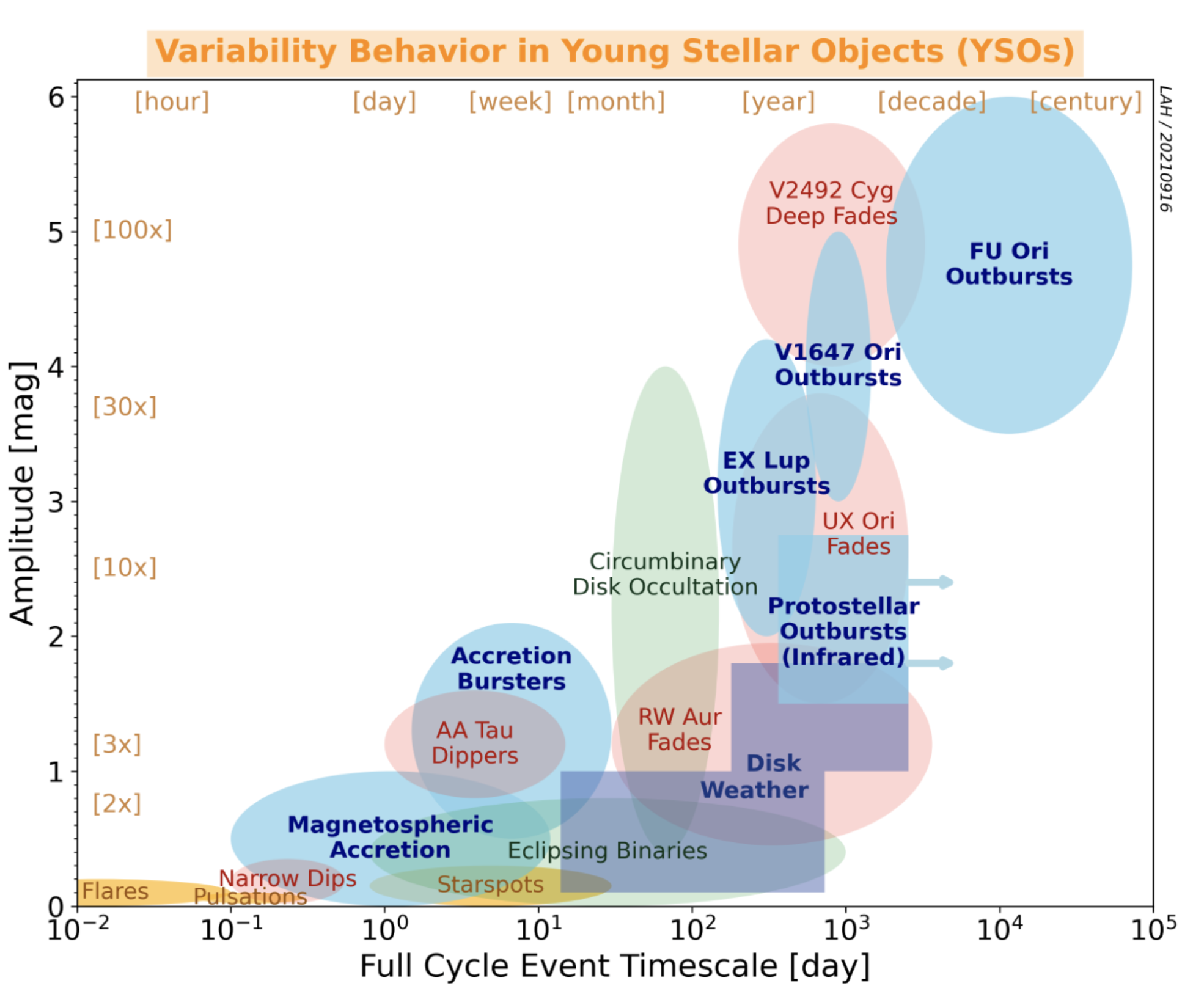} 
    \caption[YSOs variabilities in magnitudes due to bursts and outbursts and their timescales]{YSOs variabilities in magnitudes due to bursts and outbursts and their timescales. I call attention to the already mentioned EX Lup and FU Ori outbursts, being V1647 Ori one intermediate case between both. Credits: \citet{fischer2023}.}
    \label{vari}
\end{figure}

In an early investigation of protostars within the Taurus-Auriga region, \citet{kenyon1990} introduced the "luminosity problem." This concept pointed out a statistical anomaly, indicating that protostars displayed lower luminosity than anticipated, considering a steady accretion scenario and the expected duration for star formation. At the time, two explanations were proposed: protostellar lifetimes last longer than $10^{5}$ yrs proposed by the authors, or episodic bursts of accretion, akin to EXors and/or FUors, accelerating the process. Later, it was inferred from mid-infrared surveys that protostellar lifetimes are $\sim$5 times longer than expected \citep{evans2009,dunham2014, dunham2015}, reducing the luminosity necessary to be observed in YSOs in case of steady accretion. Although some authors like \citet{offner2011}, consider the ``problem'' solved, discrepancies in luminosity persist, ranging from 2 to 10 times lower than expected from a steady accretion scenario. Following, \citet{fischer2017} have shown that protostellar luminosities have different accretion rates in different evolutionary stages, lowering the rate between Class 0 and I YSOs. This reinforced that a variable scenario, not necessarily episodic or with regular frequency in nature, can match observed protostellar luminosities. Therefore, even though episodic accretion was detected in several sources, it is hard to establish they occur in all YSOs and during all of their evolution. A contemporary understanding of the ``luminosity problem'' diverges from its initial meaning, now more fittingly used to emphasize the wide range of protostellar luminosities observed in different sources, environments, and stages of evolution. To better represent the variabilities in magnitudes and timescales where a YSO can undergo, see Fig. \ref{vari}.

Up to now, a few tens of eruptive stars have been detected and spectroscopically characterized. Considering that variability in accretion is fundamental to connecting models to observations and inferring important physical parameters such as lifetime and efficiency of mass bulking towards the star, a lot of questions remain unanswered. One of the most crucial ones refers to triggering mechanisms. What drives FUors/EXors accretion variability in a YSO? And how do they stop? In literature, the triggering processes are usually summarized in 3: 1) disk fragmentation \citep{vorobyov2005, vorobyov2015, zhu2012}; 2) magneto-rotational instability (MRI) \citep{armitage2001, zhu2009}; and 3) tidal interactions between a disk and a companion \citep{bonnell1992,lodato2004, borchert2022}.

The mechanism of disk fragmentation is based on gravitational instabilities usually producing clumps, which will be rapidly inward driven towards the inner disk, and possibly tidally disrupted replenishing material in the viscously heated part. Disk fragmentation can also induce magneto-rotational instabilities (MRI) in the inner disk, another important triggering process.

\begin{figure}[!htb]
    \centering
    \includegraphics[width=1\textwidth]{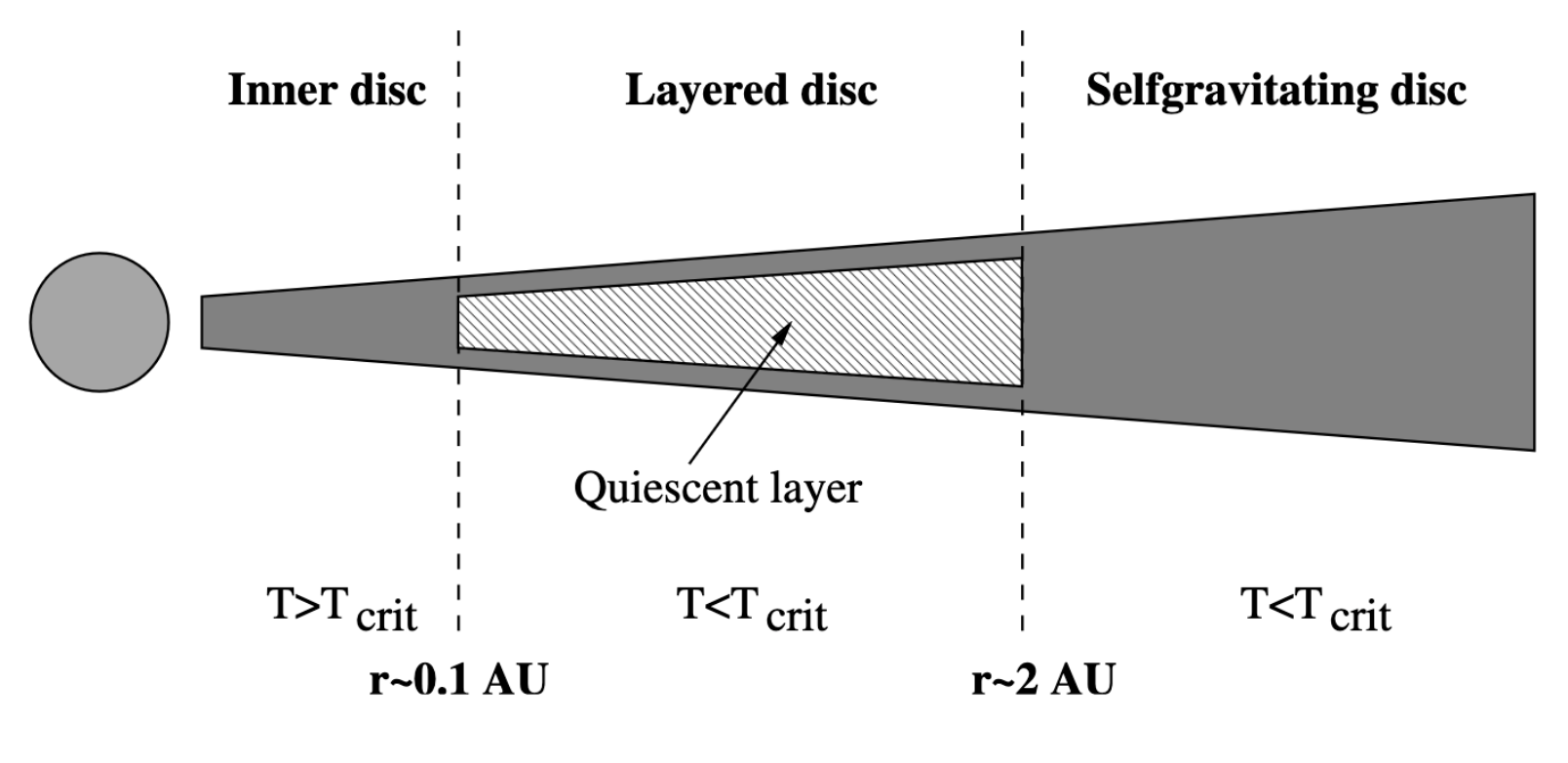} 
    \caption[Schematic representation of regions of an accreting disk]{Schematic representation of regions of an accreting disk, showing the inner disk coupled to MRI; a quiescent layer, with a temperature lower than the required to couple material to MRI, thus stalling inward transport; and the outer disk which has its material transported by self-gravity. Credits: \citet{armitage2001}.}
    \label{mri}
\end{figure}

Magneto Rotational Instability (MRI) is an efficient triggering mechanism. As YSOs undergo radial rotation at different angular velocities, the embedded magnetic fields become twisted and distorted, thereby initiating the MRI. This instability induces turbulence in the disk, enabling the efficient inward transport of gas and dust. Consequently, the MRI redistributes angular momentum, preventing its excessive accumulation in specific regions. However, a certain level of ionization is necessary to couple the material to the magnetic field, making it susceptible to MRI. The dynamically cold regions in a disk, thus, stall the inward transport, accumulating self-gravitating material in the outer disk (see Fig. \ref{mri}). All of the potential solutions to break this flow discontinuity point towards episodic mass accretion events, generated by heat-inducing of the outer disk to the dynamically cold region, re-engaging the transport of material to the inner disk.

The tidal disruption due to a companion is a mechanism, not as efficient as MRI since depends on close encounters with external companions, which may be or not periodic and may be less common to happen. The gravitational disturbance, thus, can form spiral arms, density waves, or other features, redistributing the material in the disk, or even causing disk fragmentation, leading to another mechanism already mentioned and its consequent episodic accretion. 

All these mechanisms are not constant and, thus, could explain the return to quiescence stages.

\subsection{Protoplanetary disks - properties, structure, and principal physical mechanisms}

\begin{figure}[htb!]
    \centering
    \includegraphics[width=0.9\textwidth]{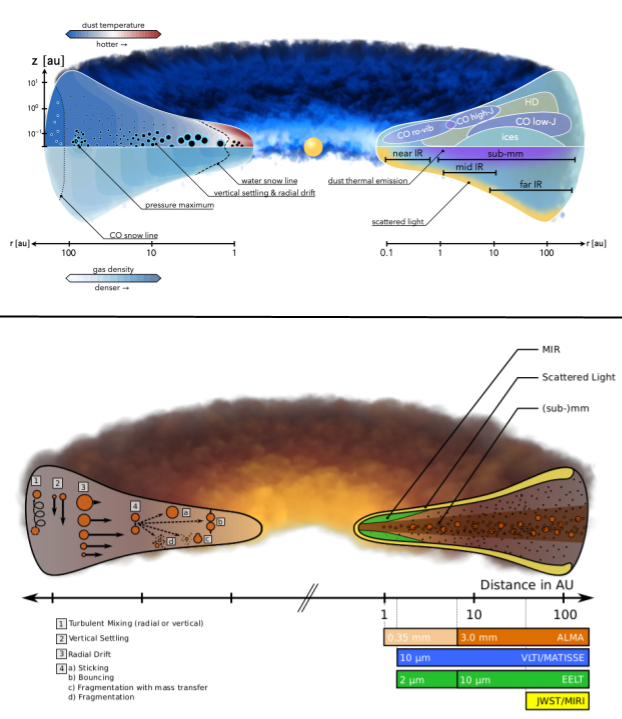}
    \caption[Schematic representation of the structure of protoplanetary disks and how they can be observed depending on the wavelength]{Schematic representation of the structure of protoplanetary disks and how they can be observed depending on the wavelength. \textit{Top panel}: dust temperature profile and gas density structure. On the left, the grain sizes are represented according to the size of black-filled circles, when blue contours representing icy water, and white contours CO-coat. On the right, different gas emission regions of the main simple molecules are shown, and dust thermal and scattered emissions are highlighted in purple and yellow, respectively. Credits: \citet{miotello2023}. \textit{Bottom panel}: dust evolution main transport and collision mechanisms are represented on the left. On the right side, areas of the disk that can be observed by different wavelengths and current/future instruments to detect them are shown. Credits: \citet{testi2014}.}
    \label{disk_comp}
\end{figure}

Here, I characterize the protoplanetary disk (PPD) as the circumstellar disk, excluding consideration of the envelope. A PPD primarily consists of gas and dust inherited from the natal envelope, with molecular hydrogen (H$_{2}$) and helium (He) dominating the gas composition, alongside (sub)micron-sized dust particles. The gas constitutes approximately 99\% of the PPD's total mass \citep{2009psdf.book.....O}. Typically, the temperature distribution exhibits higher temperatures in the innermost regions (closer to the star), gradually decreasing radially towards the outermost regions. However, the disk's vertical structure also significantly influences temperature distribution and the distribution of grain sizes. Additionally, different regions within the disk exhibit variations in the movement velocity of both dust and gas, guided by distinct physical mechanisms. Subsequently, I provide a more detailed exploration of the principal parameters and physical mechanisms inherent in a PPD. A schematic representation of the different disk components probed by different wavelength ranges and instruments can be seen in Fig. \ref{disk_comp}, both panels. Specifically, the temperature distribution is depicted at the top panel, while the bottom panel shows different grain evolution processes.

\subsubsection{Disk mass}

Knowing the total content of dust in a PPD is crucial to determine accretion rates, lifetime, and which substructures and objects can be formed within. Additionally, the grain size distribution is important to determine which grain size can be observed in different wavelengths and also helps to disentangle problems in planetary formation models regarding where, when, and how rocky planets/cores of giant planets are formed. Therefore, it is crucial to determine masses and surface densities, which describe how the mass is distributed per unit area on the observed surface (this distribution can vary with depth depending on the wavelength used for observation).

Disk masses are typically derived from observations of thermally heated dust grains. This involves measuring the dust mass and, based on an assumed dust-to-mass ratio, inferring the total mass of the disk. A frequently adopted assumption is a gas-to-dust ratio of 100, assuming a fixed Interstellar Medium (ISM) \citep{andrews2005,andrews2007}. The continuum emission is optically thin, except for the innermost regions of the disks (approximately less than 10 astronomical units, at 1 mm observations), where the column density becomes too high. The determination of the disk mass from the optically thin reprocessed emission involves employing the following equation:

\begin{equation}
    \label{eq_mass}
    M(\text {gas}+ \text {dust})=\frac{F_\nu d^2}{\kappa_\nu B_\nu(T)},
\end{equation}

\noindent where $F_\nu$ is the measured flux at frequency $\nu$, d is the distance to the source,
$\kappa_\nu$ is the dust opacity, and $B_\nu(T)$ is the Planck function. At (sub)mm wavelengths, the Planck function approximates to the Rayleigh-Jeans limit, and the following approximation can be done:

\begin{equation}
    B_\nu(T)=\frac{2 h \nu^3}{c^2} \frac{1}{e^{\frac{h \nu}{k_{\mathrm{B}} T}}-1} \approx \frac{2 h \nu^3}{c^2} \cdot \frac{k_{\mathrm{B}} T}{h \nu}=\frac{2 \nu^2 k_{\mathrm{B}} T}{c^2},
\end{equation}

\noindent where h in Planck's constant, c is the speed of light and $k_{\mathrm{B}}$ is the Boltzmann constant. Therefore, observing at these wavelengths provides a dust temperature profile that linearly varies with radius, contrasting with an exponential trend. 

Although equation \ref{eq_mass} is countless times used in literature, it is highly dependent on the determination of average temperature and dust opacity. A good and classical approximation for both is a temperature of 20 K, consistent with IR-millimeter modeled SEDs, CO observations, and theoretical expectations (\citealp{andrews2005,qi2004, chiang1997} and references therein) and a dust opacity for mm wavelengths from \citet{beckwith1990}, expressed as:

\begin{equation}
    \kappa_\nu=0.1\left(\frac{\nu}{10^{12} \mathrm{~Hz}}\right)^\beta \mathrm{cm}^2 \mathrm{~g}^{-1},
\end{equation}

\noindent where the power-law index, $\beta$ is related to the size distribution and composition of the dust grains \citep{1994A&A...291..943O,1994ApJ...421..615P}. It is commonly assumed a $\beta$ is equal to 1 for 1.3 mm observations. In addition, the assumptions do not provide good estimations for mass accountability of larger grains, which remains a source of great uncertainty, offering the idea that mass estimations through equation \ref{eq_mass} may be underestimated. A more complicated scenario is driven by the more recent studies, stating that not only the spectral index $\beta$ can vary a lot depending on the properties of the grains, such as how porous it is but also estimates big mass content in disks arising from optical thick regions. Additional analyses and methodologies have been explored and suggested for estimating disk masses based on gas content, primarily utilizing CO isotopologues and HD emission. This approach is taken because H$_{2}$ lacks an electric dipole moment as a symmetric molecule, resulting in rare transitions capable of emitting radiation (for a review about mass measurements from gas, see e.g. \citealp{2017ASSL..445....1B,miotello2023}).



\subsubsection{Disks structure and temperature distribution}
\label{sec:substructures}
As aforementioned, how disk material and temperature are distributed in a PPD, is of major importance. During the lifetime of a disk, those distributions can highly vary, as part of the disk will be accreted onto the protostar/PMS, dissipated, and/or radially and vertically. In particular, which process dominates will impact the surface density distribution, $\Sigma (R)$, and the disk outer radius, $R_{out}$\footnote{There is no total consensus and a robust physical definition for $R_{out}$, besides as being the radius that encloses a chosen percentage of the emission. However many authors recently have been using $R_{out}$ = $R_{68\%}$.}. Therefore, these two parameters are priorities whenever PPDs are been analyzed.

\begin{figure}[htb!]
    \centering
    \includegraphics[width=0.9\textwidth]{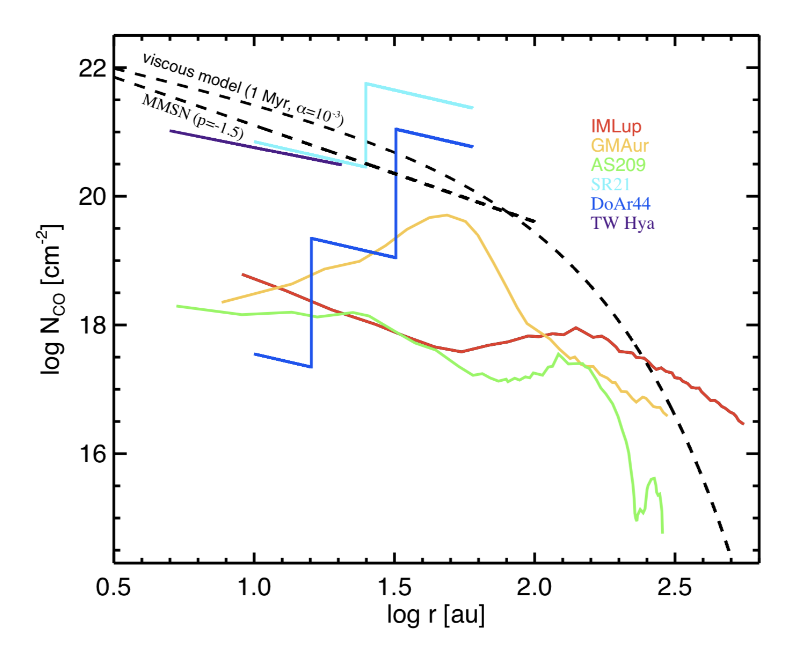}
    \caption[Gas surface density profiles for edge-on and large cavity transition disks, with higher CO-based gas-to-dust ratio]{Gas surface density profiles for edge-on and large cavity transition disks, with higher CO-based gas-to-dust ratio. The dashed black lines provide the Minimum Mass Solar Nebula (extrapolated to 100 au) and a viscous disk profile for comparison. Credits: \citet{miotello2023}.}
    \label{gas_surf_dens}
\end{figure}

The surface density of a disk can be modelled and analyzed in total or split into gas and dust counterparts, depending on the estimations you are accounting for. A common prescription to describe the radial distribution of the dust and the gas in a viscous disk, and which fits interferometric observations is given by: 

\begin{equation}
    \Sigma(R)=(2-\gamma) \frac{M_{\text {disk }}}{2 \pi R_{\mathrm{c}}^2}\left(\frac{R}{R_{\mathrm{c}}}\right)^{-\gamma} \exp \left[\left(-\frac{R}{R_{\mathrm{c}}}\right)^{2-\gamma}\right],
\end{equation}

\noindent where $M_{disk }$ is the disk mass, $R_{\mathrm{c}}$ is the characteristic radius, and $\gamma$ is the power-law exponent that describes the radial dependence of the disk viscosity \citep{1974MNRAS.168..603L,1998ApJ...495..385H}. In this equation, the characteristic radius refers to the region where the surface density starts to deeply steepen from the power law, where it indicates at which part of the disk is viscously dominated. More recently, mainly due to ALMA observations, a recurring observation of continuum substructures, such as gaps and rings, has emerged in most disks observed at sufficient spatial resolution. Additionally, faint optically thin molecular lines could be detected in several cases, tracing disk masses and gas distributions. These findings indicate considerable variability in surface density profiles, deviating from the assumptions of a simple model (see an example in Fig. \ref{gas_surf_dens}). To improve surface density models, multi-wavelength observations of resolved disks become crucial. Such observations may not only disentangle the degeneracy among temperature, dust surface density, and dust opacity but also provide valuable insights for accurately modeling radial variations in dust particle sizes and scattering opacities \citep{miotello2023}. As for gas direct detections to model the surface densities, it is necessary to find optically thin emissions, which are usually faint and rarely show emission from the outer disk in small disks ($R_{c} <$ 100 au; \citealp{2016ApJ...830...32W,2018A&A...619A.113M}). Detecting them implies observations with high sensitivity and well-known ratios of the tracer to other more abundant tracers. Usually CO gas mass is traced using $^{13}$CO or C$^{18}$O, and adopting a ratio CO-to-H$_{2}$.

Outer gas disk radii, as observed in many $^{12}$CO observations, are generally larger than the radii of (sub)mm-sized dust particles. The reason behind this difference is currently under discussion. It could be due to optical depth effects, where CO remains optically thick and bright in the outer disk, while the dust becomes optically thin, and thus faint and challenging to detect. Another possibility is the efficient radial drifting of dust, leading to a reduction in dust radii. The ongoing debate revolves around determining whether optical depth effects, radial drifting, or a combination of both factors contribute to this observed difference. Moreover, according to recent models (e.g., \citealp{2021MNRAS.507..818T}), if radial drifting is the main process, it is extremely efficient, halting the disk size way more quickly than expected from observations. Thus, it is needed in such cases to invoke substructures that locally slow down or stop radial drifting, as gaps. 

\begin{figure}[htb!]
    \centering
    \includegraphics[width=1\textwidth]{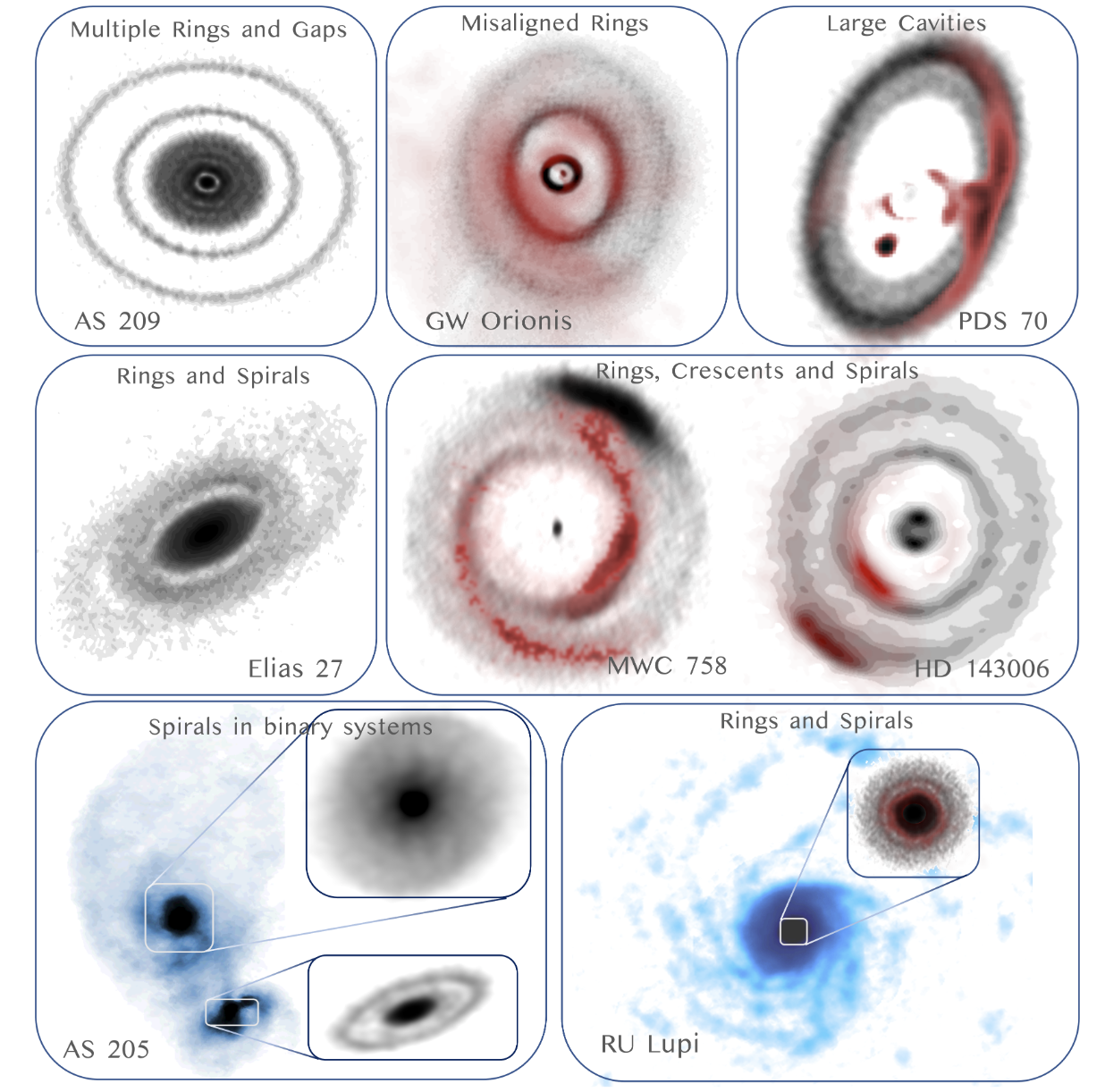}
    \caption[Substructures observed in PPDs]{Example of substructures observed in PPDs. Black and white contours show millimeter continuum observations, blue contours show millimeter line observations and red contours show near-IR observations. Credits: \citet{bae2023}.}
    \label{substructures}
\end{figure}

Analyzing a sample of 470 YSOs observed at NIR and/or mm wavelengths, \citet{bae2023} offered a comprehensive statistical and morphological overview of substructures. From this examination and prior analyses, it is accurate to assert that Class II-to-Class III disks can exhibit three primary classes of substructures: rings and gaps, spirals, and crescents (see Fig. \ref{substructures}).

\begin{figure}[htb!]
    \centering
    \includegraphics[width=1\textwidth]{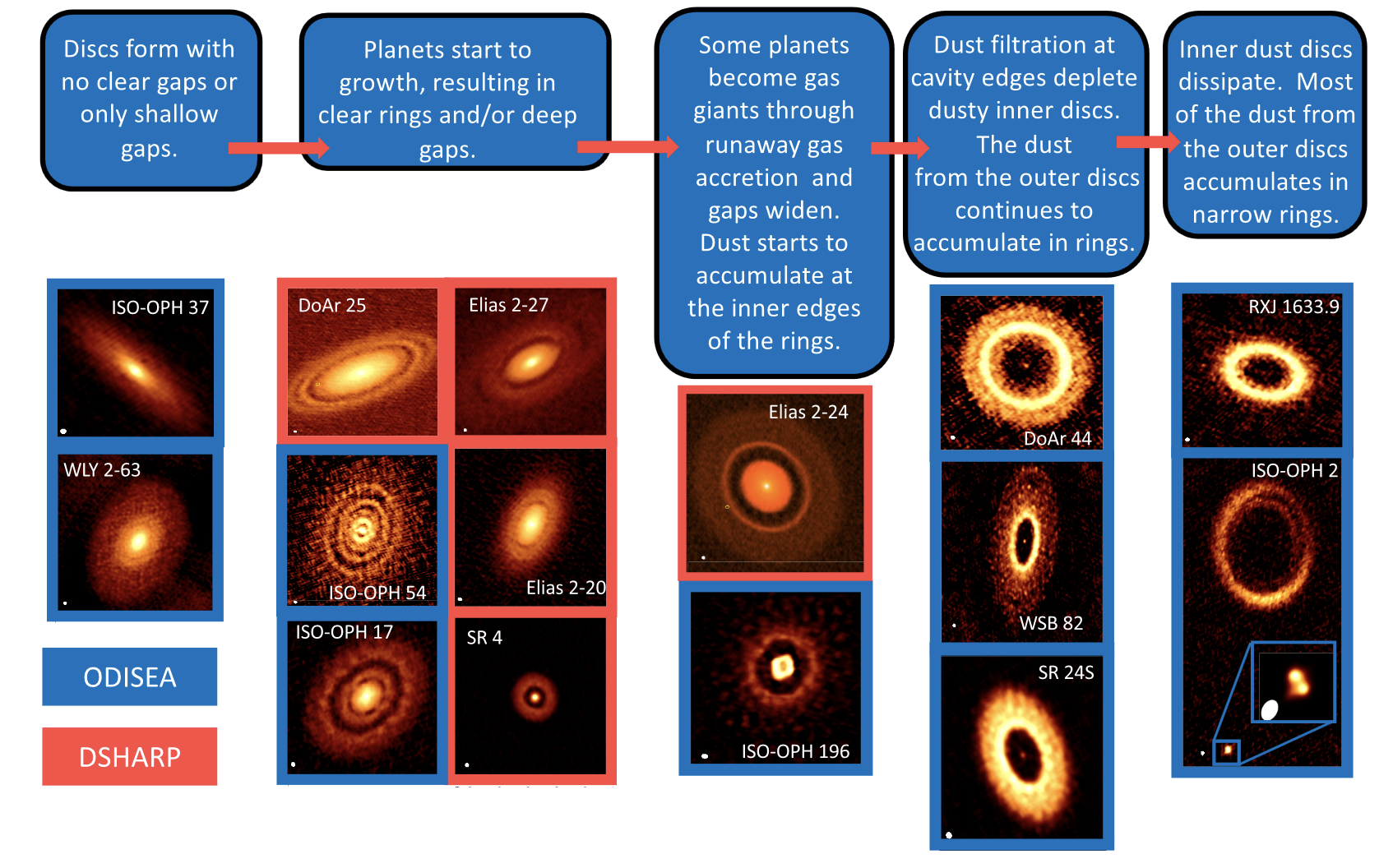}
    \caption[Proposed scenario of substructures evolution driven by giant planets formation]{Proposed scenario of substructures evolution driven by giant planets formation using massive disks from ODISEA sample. Credits: \citet{cieza2021}.}
    \label{substructuresevolution}
\end{figure}

Rings are generally defined as circular and azimuthally symmetric. They are the most common type of substructure detected until the moment. Gaps are the regions of reduced surface density between denser resolved rings. Rings and gaps are most commonly observed in mm observations rather than near-IR. The apparent discrepancy between both wavelengths can be explained by considering the possibility that the observed rings correspond to pressure maxima within the protoplanetary disk. In such regions, larger particles may be efficiently trapped, leading to their enhanced detection in mm observations. Rings and gaps play a crucial role in the study of planetary formation, as they can be carved by the presence of giant planets forming and opening gaps in the protoplanetary disk. In recent studies, there has been a focus on understanding the evolution of these substructures over time. Some of the proposed scenarios associate these substructures with the formation of giant planets and their consequent migration (see e.g. \citealp{cieza2021} and Fig. \ref{substructuresevolution}).

Spirals are more commonly detected at NIR than in mm observations, mainly because they are linked to smaller grains that are coupled to the gas. While several mechanisms have been suggested to explain the formation of these substructures, the only one that has been conclusive confirmed involves tidal interactions with companions, whether gravitationally bound or not (see discussion and references in \citealp{bae2023}). Detecting stronger spirals is expected to be associated with higher companion masses, making it challenging to observe these substructures in the presence of giant planets using current observational capabilities. Therefore, spiral observations are more commonly associated with brown dwarfs or larger objects. Another popular scenario that arises to explain spirals is gravitational instability (GI), where pitch angles are
inversely proportional to disks' mass. When the disk is sufficiently massive ($M_{disk}/M_{*} \geq 0.25$, based on numerical simulations to explain Elias 2-27 spirals morphology), such that the Toomre parameter $Q = c_{s}\Omega / (\pi G \Sigma)$ is $\leq$ 1, perturbations at intermediate scales are subject to GI \citep{toomre1964,kratter2016}. This may lead to the formation of non-axisymmetric spirals which can efficiently transport angular momentum through gravitational torques. Finally, infalling streamers and sound wave propagation can also be invoked as possible causes for spirals.

Crescents, the less commonly observed substructure, are rings that have an azimuthal variation in intensity. Only two leading mechanisms can explain crescents: vortices, created by hydrodynamic instabilities trapping particles at the region, and lumps created by angular differential velocities in eccentric disks or by trapping material in Lagrangian points. Different from the other two substructures, there is no clear evidence that companions or planet formation are connected to crescents.

\begin{figure}[htb!]
    \centering
    \includegraphics[width=0.6\textwidth]{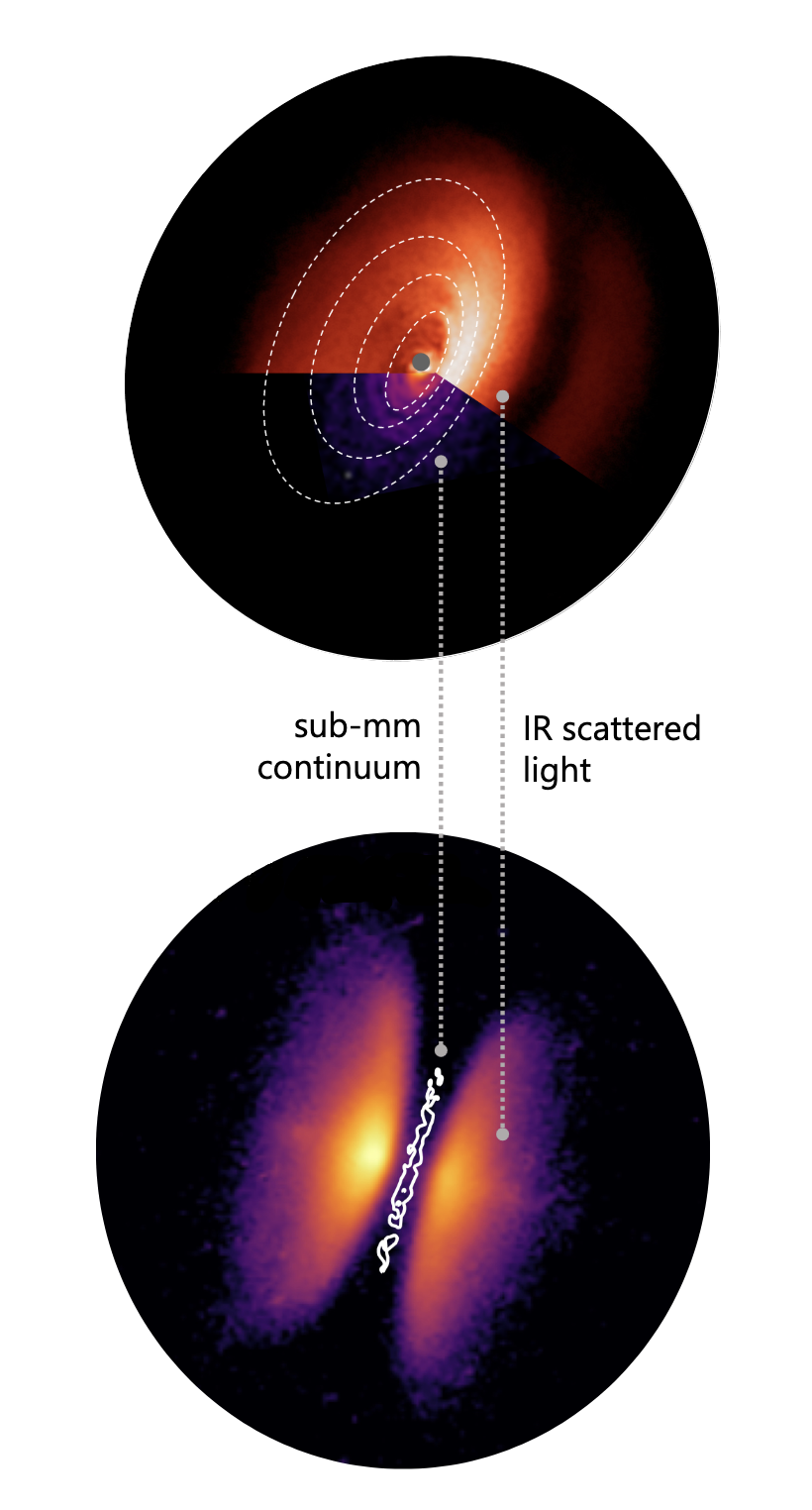}
    \caption[Representation of different geometries and structures observed in scattered light and sub-mm continuum]{Representation of different geometries and structures observed in scattered light and sub-mm continuum. \textit{Top}: Collage of observations of IM Lup, where the white dashed lines represent the disk surface layer. Credits: \citet{avenhaus2018,2018ApJ...869L..41A}. \textit{Bottom}: Combined observations of the edge-on disk Tau 042021. Credits: \citet{2020A&A...642A.164V}.}
    \label{ir_vs_mm}
\end{figure}

Protoplanetary disks have finite thickness and are flared mainly due to pressure support from the gas disk balanced by the gravitational force of the star. In a disk, at a radius $r$ from the star, and at a height of $z$ above the mid-plane, the pressure $p$ of the gas is given as:

\begin{equation}
\label{eq_press}
    \frac{d p}{p}=\frac{\Omega_{\mathrm{K}}^2}{v_{\mathrm{s}}^2} z d z
\end{equation}

\noindent where $v_{s}$ is the speed of sound, and $\Omega_{K}$ is the keplerian angular velocity defined by $\Omega_{\mathrm{K}}=\left(\frac{G M_{\star}}{r^3}\right)^{1 / 2}$. From equation \ref{eq_press}, I derive the scale height $h_{\mathrm{g}}$ of the disk \citep{1987ApJ...323..714K}:

\begin{equation}
    \frac{h_{\mathrm{g}}}{r}=\left(\frac{v_{\mathrm{s}}^2 r}{G M_{\star}}\right)^{1 / 2} ,
\end{equation}

\noindent or simply, $h_{\mathrm{g}}=v_{\mathrm{s}} / \Omega_{K}$. If the internal temperature of the disk, which is proportional to $v_{\mathrm{s}}^2$ decreases more slowly than $r^{-1}$, the height scale will increase outwards. And the warmer the disk, the higher will be its maximum height - it will be thicker. How the dust vertical structure is deployed, however, depends on the dust coupling to the gas, which also depends on the thermal speed of the gas, grain sizes, and relative velocity between gas and dust and their interaction timescales. The mechanism responsible for this process is aerodynamic drag, which proves to be more effective with smaller dust grains, in less dense regions. Consequently, to capture the emissions and flared surface characteristics of a disk through dust observations, a more effective strategy involves observing at shorter wavelengths, specifically in the infrared range. This approach targets the scattering emissions of smaller grains. On the other hand, sub-mm continuum emission access larger grains decoupled from the gas, which are settled in the mid-plane. Thus, mm-sized grains settle to a thin layer as smaller grains provide a greater vertical distribution (see Fig. \ref{ir_vs_mm}). 

Understanding the distribution of temperature is crucial for modeling dust dynamics, gas flow, and chemical mixing within protoplanetary disks. In disks maintaining a steady state (radiative equilibrium), the balance between cooling through radiation emission and heating through absorption of starlight (in passive disks) or a combination of starlight absorption and accretional heating due to viscous dissipation (in active disks) is essential. Theoretical models suggest that 2D temperature profiles for both types of disks should exhibit vertical increases from edges to mid-plane and radial decreases from inner to outer regions (see Fig. \ref{disk_comp}). In the innermost part of the disk, stellar radiation significantly affects the gas and dust at the surface. Smaller grains at the surface scatter, absorb, and re-emit radiation to larger grains in the mid-plane.  For dust particles, thermal and scattered continuum emission at a wavelength of $\lambda$ are dominated by grains with sizes similar to $\lambda /$2$\pi$. Thus, at a certain distance from the star, temperature profiles are dominated by reprocessed emission throughout the vertical distribution. Radially, from the region where dust is not sublimated towards the outer regions, the inner disk experiences a mixture of stellar irradiation and viscous accretional heating until it encounters ``dead zones'' where low ionization levels inhibit MRI, locally halting viscous transport and heating toward the star. Further from the star, self-scattering and re-emission dominate, implying that a higher scale height at a specific distance results in a colder mid-plane. Vertically, uppermost disk layers with sufficient stellar irradiation exhibit higher temperatures, while intermediate layers, shielded by the upper layers, remain warm ($>$30-40 K). The heavily shielded disk mid-plane is cold ($<$30 K) and harbors the highest gas densities. Despite more complex models considering additional mechanisms and deviating from standard vertically isothermal assumptions, these common simplifications persist due to their practical simplicity.

\begin{figure}[!htb]
    \centering
    \includegraphics[width= 1\textwidth]{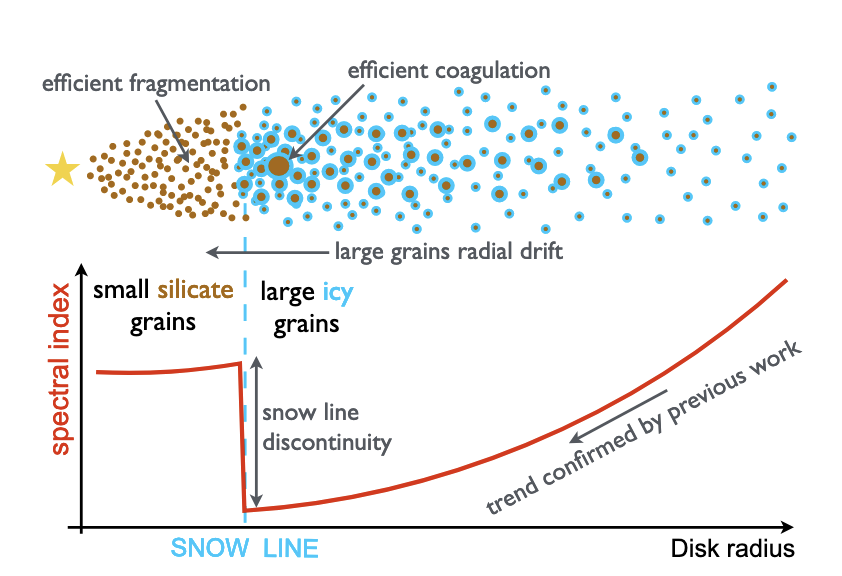}
    \caption[Illustration of the spectral index as a function of radius in a protoplanetary disk]{Illustration of the spectral index as a function of radius in a protoplanetary disk. Credits: \citet{banzatti2015}.}
    \label{transition}
\end{figure}

An essential feature in protoplanetary disks (PPDs), driven by the radial temperature distribution in the mid-plane, is the presence of ice lines -- a key element in the context of planetary formation. Icelines mark the radial positions where volatiles undergo sublimation and deposition. These locations are composition-dependent, with different volatile species exhibiting ice lines at distinct radial distances due to their varying sublimation temperatures. For example, N$_{2}$, CO,
CO$_{2}$, and H$_{2}$O have sublimation temperatures of $\sim$10-20, 20-30, 60-70, and 130-160 K at typical disk densities, respectively \citep{2015ApJ...806L...7Z}. Ice formation within and beyond these regions introduces a new dynamic to dust interactions and the distribution of volatiles. The presence of icy caps can enhance the stickiness of grains, promoting the coagulation and lumpiness essential for the formation of giant gas planet cores in the outer regions of disks (see, e.g., \citealp{2014Icar..233...83C, banzatti2015}). Furthermore, the inward transport of ice or icy caps, upon sublimation, releases vapor that can be transported outward by turbulence. Additionally, larger grains, upon losing their icy mantles, may fragment, causing a sudden increase in optical depth in inner regions—a scenario proposed as one hypothesis to explain gaps. As small fragments drift more slowly than larger ones, they may accumulate near the ice lines (see, e.g., \citealp{2011ApJ...728...20S, 2016A&A...596L...3I, 2019A&A...629A..90H} and Fig. \ref{transition}). However, current observations do not show a clear correlation between rings and the expected locations of ice lines, challenging the notion of ice lines co-locating with these substructures \citep{bae2023}.


To determine disk temperature profiles from observations, optically thick tracers as $^{12}$CO were used. More recently,  with the help of high-resolution observations, better results can be obtained. For example, to study the \textit{molecular layer} of the disk (the region between which molecules are shielded from dissociating radiation but are warm enough to remain in the gas phase), three main techniques can be used:

\begin{itemize}
    \item Chemistry: Through freeze-out regions for each molecule. For example, observing sharp drops in a sub-mm emission may indicate a freeze-out of specific molecules.
    \item Line excitation: Through multiple transitions of optically thin lines, the excitation temperatures can be inferred.  
    \item Line optical depth: If the spectral line emission is optically thick, the brightness temperature at a certain region can be interpreted as gas temperature.
\end{itemize}

For the \textit{mid-plane}, the same techniques can be used. However, as this region contains larger dust grains, additional techniques may come into play:
\begin{itemize}
    \item Dust optical depth: Millimeter dust grains may have optically thick emission in the inner disk. In such a case, the brightness temperature reveals the dust temperature, which is equal to the gas temperature where collisions counterbalance these two components' temperatures.
    \item Polarization fractions: Temperature gradients can manifest in the observed polarization fraction due to distinct polarization angles probing varying line-of-sight optical depths. This methodology has been developed by \citet{2020MNRAS.493.4868L}, where conversions from polarized flux patterns to line-of-sight temperature gradients.
    \item Optically thick gas lines in highly inclined disks: This method utilizes projection effects to derive temperature differentials between the near and far sides of the disk \citep{2020A&A...633A.137D}. It is effective for warmer regions of low-mass star disks or disks around more massive stars with extended CO ice lines.

\end{itemize}






\subsubsection{Grain growth}
\label{sec:graingrowth}

\begin{figure}[!htb]
    \centering
    \includegraphics[width= 0.8\textwidth]{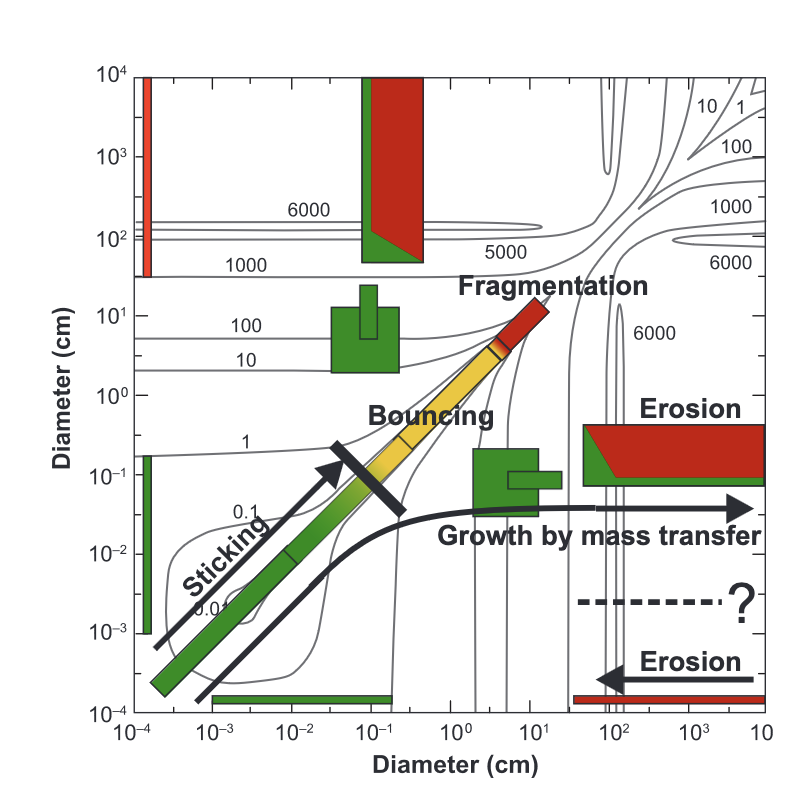}
    \caption[Schematic representation of the outcomes of dust collisions in PPDs]{Schematic representation of the outcomes of dust collisions in PPDs. It displays collision velocities (denoted by contours, in units of cm/s) between two non-fractal dust agglomerates in a minimum-mass solar nebula model at 1 AU. The green, yellow, and red boxes represent the explored parameter space and results of laboratory experiments, where green signifies sticking or mass transfer, yellow bouncing, and red fragmentation or erosion. The arrow labeled ``Sticking'' indicates the direct growth of mm-sized dust aggregates, as observed in simulations by Zsom et al. (2010). Further growth is impeded by bouncing. The arrow labeled ``Growth by mass transfer'' suggests a potential direct path to the formation of planetesimals. Credits: \citet{testi2014}.}
    \label{graingrowth}
\end{figure}

The grain growth process within a PPD is predominantly influenced by collisions among smaller dust particles. Despite persistent efforts in experiments and modeling, challenges arise in explaining the emergence of dust particles exceeding (sub)$\mu$m sizes to mm/cm sizes or larger (see e.g. \citealp{2022MNRAS.515.4780T}). The collisional processes leading to particles of this scale can be categorized into four main mechanisms, dependent on factors such as the relative size/velocities and composition of the grains involved (see, for example, \citealp{testi2014}, Fig. \ref{disk_comp}, bottom panel, and Fig \ref{graingrowth}):

\begin{itemize}
    \item Sticking: It is the dominant mechanism for smaller scales between grain sizes and relative velocities, thus being the dominant outcome from collisions at early stages. Assuming minimal influence from electrostatic charges or magnetic materials during collisions between dust grains, their interaction can lead to sticking through van der Waals forces \citep{1999PhRvL..83.3328H,2011Icar..214..717G}. At lower velocities, grains can grow up to mm sizes.
    \item Bouncing: Even at the same relative velocities as the previous sticking regime, if grains are of similar sizes and are too large (mm-cm sizes), their collisional outcome will be bouncing instead of sticking. This stage/size scale is also known as the "bouncing barrier" \citep{2010A&A...513A..57Z}.
    \item Fragmentation/erosion: If the grains surpass the bouncing barrier, getting larger, or if the grains have higher relative velocities, they may undergo fragmentation. This process is important to replenish the disk with smaller grains and provide slower grain growth and slower inward migration. 
    \item Mass transfer: If grains with different sizes collide at a relative velocity approaching the fragmentation velocity, the collision can lead to substantial mass transfer from the smaller to the larger particle. This mechanism can surpass the bouncing barrier, enabling grains to grow to the size of planetesimals.
\end{itemize}

Therefore, small dust grains can efficiently grow up to mm sizes through sticking. Subsequently, mass transfer becomes a dominant factor. Throughout this process, fragmentation/erosion continually replenishes the population of small grains in the disk \citep{2009A&A...503L...5B}. However, this straightforward model may exhibit discrepant size scales and relative velocities depending on the composition, porosity, and the presence of ice or icy shells.

\subsubsection{Coupling between gas and dust}
\label{sec:gascoupling}


Different processes are associated with dust transport and migration, mainly driven by dust-to-gas interaction through aerodynamic drag forces and turbulence (see Fig. \ref{disk_comp}, bottom panel).

Aerodynamic drag forces are caused by the relative motion between gas and dust particles, which produces a friction force. When the grains radii $a$ are smaller than the mean free path between gas particles, they will be subject to Epstein drag \citep{1924PhRv...23..710E}. In the Epstein regime, the drag force $F_{D}$ is given by:

\begin{equation}
    F_D=-\frac{4}{3} \pi a^2 \rho_g \Delta v v_{\mathrm{th}},
\end{equation}
\noindent where $\rho_g$ is the gas density, $\Delta v$ is the relative velocity between the particle and the gas, $v_{\mathrm{th}} \equiv \sqrt{8 / \pi} c_s$ is the thermal speed of the gas molecules and $c_s$ is the sound speed.

The drag force acts opposite to $\Delta v$, so the solid particle loses angular momentum. The timescale over which this happens is called stopping time $t_{\text {stop }} \equiv m \Delta v /\left|F_D\right|$, where $m$ is the mass of the solid particle of interest. The stopping time can be converted to a dimensionless number, called Stokes number ($St$):
\begin{equation}
    \mathrm{St}=\Omega_K t_{\text {stop }},
\end{equation}
\noindent where $\Omega_K$ is the local Keplerian frequency. In the context of dust dynamics in protoplanetary disks, the $St$ plays a crucial role in determining the behavior of dust particles relative to the gas. Mathematically, when $t_{\text{stop}}$ is less than the dynamical time $t_{\text{dyn}}$ or when $St \ll 1$, the term $\Delta v$ vanishes, indicating that the dust particles follow the gas motion (more aerodynamic drag force). When $St \gg 1$, aerodynamic drag becomes negligible, and the particle decouples from the gas. Physically, the Stokes number reflects how well particles are coupled to the gas and how they respond to thermal pressure in less dense environments. In the case of small particles (sub-micron sizes), they are significantly influenced by gas thermal pressure and remain coupled to the gas, with little acceleration produced by drag forces. On the other hand, large particles (mm - cm sizes) experience higher drag force acceleration until they grow to approximately km sizes, beyond which the drag force becomes negligible \citep{weidenschilling1977}. 

Aerodynamic drag forces are fundamental to explain two dynamical processes, horizontal drift and vertical settling. Due to a balance between centrifugal force, stellar gravity, and a radial gas pressure gradient that increases towards inner regions, the gas in protoplanetary disks (PPDs) rotates at sub-Keplerian velocities. Conversely, large enough particles, not entirely dragged by the gas, would rotate at Keplerian velocities. Consequently, these particles experience a headwind from the gas, leading to the removal of their angular momentum and causing inward drift, commonly referred to as radial drift. In terms of the vertical structure, grains located above the midplane experience the vertical component of stellar gravity pulling them toward the midplane. This gravitational force is counteracted by the pressure gradient force pointing outward from the midplane. Unlike lighter and smaller particles that are primarily influenced by gas pressure and are less affected by gravity, larger particles undergo more efficient vertical settling due to the gravitational forces acting upon them. Thus, there is a larger concentration of larger dust particles in the mid-plane, while smaller particles are more vertically scattered. Moreover, PPDs are marked by significant turbulence. Turbulence is characterized by chaotic changes in pressure and flow velocity of the gas, facilitating the mixing and diffusion of particles in all directions. This turbulent setting proves especially efficient for grains with stronger coupling, characterized by lower Stokes numbers.

The interplay between turbulent mixing, affecting principally smaller dust particles, and drag forces, primarily influencing larger particles, governs the overall size distribution of dust grains within the disk.

\section[Substellar objects: formation and detection via HCI]{Substellar objects: formation and detection via High-Contrast Imaging}

A star is defined as a self-luminous astronomical object, maintained by the balance between gravity and gas pressure (known as hydrostatic equilibrium), and possessing sufficient mass to initiate thermonuclear fusion of hydrogen isotopes, particularly $^{1}$H.

In contrast, a substellar object is a celestial body with less mass, insufficient for triggering $^{1}$H fusion. Its luminosity may vary, and this term is commonly associated with brown dwarfs, possessing enough mass for deuterium ($^{2}$H) fusion, and giant planets. Importantly, giant planets do not exhibit self-luminosity; they depend on reflected sunlight for visibility, or, when young, on the formation heat.
 
In this Section, I intend to detail how substellar objects can be formed and consequently detected and characterized by high-contrast imaging observations and post-processing techniques. 

\subsection{The main scenarios for substellar objects formation}
\label{sec:introformationscenario}

Among several possible scenarios for planetary formation, the two widely discussed and accepted models are the gravitational instability (henceforth GI) (\citealp{boss1997} and references therein) and the core-accretion (henceforth CA) (e.g., \citealp{pollack1996} and references therein). Both theories require different initial conditions to explain the formation of planets and minor bodies observed to date. Also, each one presents its pros and cons while explaining observed systems; none can be discarded. Below, I highlight their more important characteristics and their type of planetary outputs. 

\subsubsection{Core-accretion: the bottom-up process}

The core accretion (CA) model serves as a paradigm for the formation of giant gas planets, particularly for objects orbiting within the range of 10-50 astronomical units (au) \citep{perryman2014}. Moreover, it extends its applicability to the formation processes of rocky planets and minor celestial bodies. It works under the assumption that smaller grain interactions will eventually form larger bodies and, possibly, planetary cores. The model steps can be summarized as the following. 

First, dust and ice particles (sub-micron to micron) will start to agglutinate when they hit each other due to van der Wall forces. With these initial sizes, the dust grains are still well coupled to the gas. Thus, as the solids grow, they eventually will settle to the disk mid-plane \citep{kusaka1970}.

%
The higher surface densities in mid-plane promote collisional growth, leading to the formation of larger (m-km sized) rocky/metal structures. However, the growth of meter-sized objects is yet to be fully explained, since they are expected to quickly drift towards the star and also fragment when colliding at high speed (meter-size barrier; \citealp{adachi1976,weidenschilling1977,blum2008} and references therein). 

Next, km-sized objects, also known as planetesimals, will induce gravitational interactions, quickly attracting more dust and other planetesimals. This particular stage, heavily dependent on the masses of the objects, is called runaway growth \citep{stewart1988,wetherill1989} and it precedes the oligarchic growth, the second to last stage of planetary formation in the CA model. 

After reaching $\sim$100-1000 km, the bodies, now called planetary embryos, will move on their planetary feeding zone, thus accreting matter while perturbing their vicinities. They will achieve an isolation mass, thus, causing a lower growth rate at the time. 

The previous stages are thought to cease in less than 1-10 Myr. The last stage, called post-oligarchic or chaotic growth, is a slow stage that lasts as much as 100 Myr or more. This stage leads to the formation of planets larger than Mars. Additionally, during this step, the remaining planetary embryos may cross other orbits, leading to an increase in eccentricities, inclinations, and numbers of catastrophic collisions.

\begin{figure}
    \centering
    \includegraphics[width=0.9\textwidth]{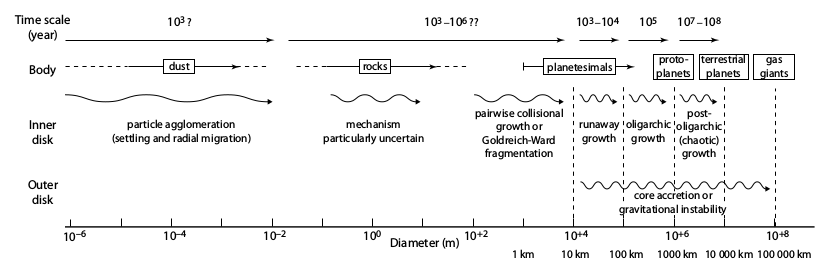}
    \caption[Schematic representation of the grain growth towards planetary formation]{Schematic representation of the grain growth towards planetary formation, starting with sub-micron dust, and extending up to the terrestrial planets in the inner disk, and the gas giants in the outer disk. Some indicative time scales are given, although some intervals, especially around
the meter-size barrier, remain highly uncertain. Credits: \citet{perryman2011}}
    \label{ca}
\end{figure}

From the continuous interactions and growth of some planetary embryos, a critical core mass \citep{bodenheimer1986,pollack1996}, thought to be between $\sim$3.5 M$_{earth}$ and $\sim$8.5 M$_{earth}$ \citep{piso2014}, will trigger a gas accretion rate that exceeds the ones that planetary embryos induce. Consequently, this process may lead to gas giant planets' formation (for further details, see figure \ref{ca} and \citealt{perryman2014}).

Finally, it is possible to cite some characteristics of CA: it can efficiently produce metal-enriched objects, rocky planets, and minor bodies. However, the classical CA process, fed principally by km-sized objects (planetesimals), faces two major problems. Firstly, the model requires extremely high column densities of planetesimals to construct the observed core masses of gas giants within their respective feeding zones. This raises concerns about the fate of the majority of planetesimals that are not utilized in core formation. Additionally, the accretion rates of planetesimals decrease significantly with distance from the star, since they migrate fast, making the formation of cold giant cores and subsequent atmosphere formation particularly challenging within the protoplanetary disk's lifetime. Secondly, the assumption that cores grow in isolation and can accrete planetesimals with negligible velocity dispersion has been challenged by dynamic simulations. Consequently, these challenges have been the inspiration to alternative perspectives.

To address the challenges encountered in the classical Core Accretion (CA) through planetesimal accretion, a more efficient model has been developed -- pebble accretion coupled with streaming instabilities (see reviews by \citealp{2010AREPS..38..493C, 2014prpl.conf..547J, 2017AREPS..45..359J}). Pebbles, typically ranging from millimeter to meter sizes, are scattered radially throughout the entire disk, vertically confined on its majority in the mid-plane, and exhibit partial decoupling from the gas (see subsection \ref{sec:substructures}). The aerodynamic drag can reduce the speed difference between pebbles and larger bodies as they reach closer. This slowdown prevents some pebbles from escaping the gravitational pull of the larger bodies or from having collisional speeds leading to fragmentation (see Fig. \ref{graingrowth}). As a result, these pebbles spiral or settle toward the surface of the larger body and become accreted. This entire process enhances the effective cross-sectional area for the accumulation of material by larger bodies, accelerating their growth. Conversely, streaming instability is a process where pebbles within the disk undergo collective gravitational collapse due to aerodynamic drag. This leads to the formation of dense filaments, providing favorable sites for rapid particle growth and subsequent planetesimal formation, up to objects of $\sim$ 100 km. According to pebble accretion models, they thus become the dominant growth mechanism over planetesimals accretion, at a starting point from the mass of Ceres/mass of the Moon. Consequently, the process is halted at the pebble isolation masses of approximately 10 M$_{\oplus}$, where contraction of the gas envelope to form giant planets starts. Therefore, streaming instability, working in conjunction with pebble accretion, contributes to the quicker and more efficient formation of planetary building blocks, even at larger distances from the star.


\subsubsection{Gravitational instability: the top-down process}
\label{gi}
The planetary formation by GI, derived from the Nebula Hypothesis \citep{kant,laplace}, works under the assumption of large clumps of gas and dust collapsing and forming giant planets. This process, which starts similarly to the formation of chunks by instabilities in molecular clouds, can be summarized as follows.

During the protoplanetary disk evolution, instabilities can cause over-densities and clumps. However, interactions with another over-density or tidal shredding due to slow cooling/rapid migration can easily disrupt these clumps and stop the process. Therefore, if a massive clump subject to self-gravity survives to dispersal, a giant body (planet, brown dwarf, or low-mass star) will begin to form.


In the context of planetary formation, GI modeling may be challenging to explain observations. 
GI planets' cores don't need to be dense nor reach a minimum value to explain the extended atmospheres that gas/ice giants show. Nevertheless, disks and clumps susceptible to GI show that the initial disk must be very massive ($M_{disk} / M_{star} \geq 0.1$; \citealp{kratter2016}), which is not so observed\footnote{For a list of references measuring the percentage of observations that have disk masses prone to GI, check section 2.2.1, \citet{kratter2016}.}. Also, it is known that the amount of gas on a clump that lasts attached until the formation of a denser structure usually leads to brown dwarfs or low-mass star objects (e.g., \citealp{kratter2006,stamatellos2009,kratter2010,zhu2012,forgan2013}). 

\begin{figure}[htb!]
    \centering
    \includegraphics[width=0.95\textwidth]{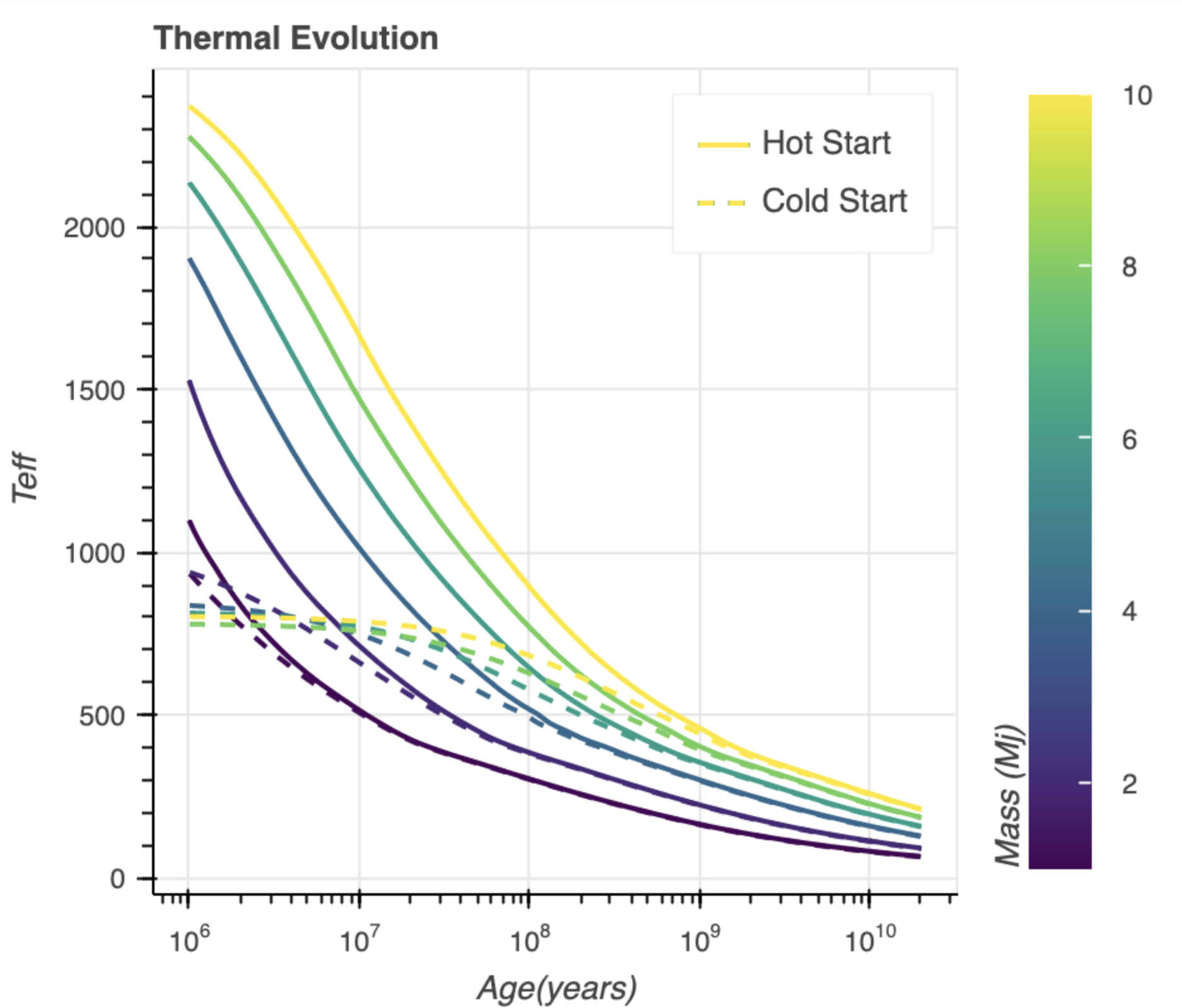}
    \caption[Luminosity-time evolutionary tracks computed using the \citet{marley2007} methodology]{Luminosity-time evolutionary tracks computed using the \citet{marley2007} methodology. The hot start, credited to gravitational instability planets (solid), and the cold start, representing the core-accretion planets (dashed) are shown, according to the initial planetary masses (1-10 $M_{jup}$). Credits: Natasha Batalha; \url{https://natashabatalha.github.io/picaso/notebooks/workshops/SaganSchool2021/1_Spectroscopy.html}}
    \label{starts}
\end{figure}

Finally, when comparing GI and CA processes, it is possible to affirm that the first is rather faster ($\sim$10$^{3}$-10$^{4}$ years) than the latter ($\sim$10$^{7}$-10$^{8}$ years), and young GI planets are hotter and irradiate more than recently formed CA planets with similar conditions. It is common to say that GI planets have a ``hot start" and CA planets have a ``cold start" (see figure \ref{starts}).

\subsection{(Sub)stellar multiplicity formation}
\label{sec:intromultipleformation}

Multiple (sub)stellar systems can be defined as systems with two or more gravitationally bound stars/brown dwarfs with separations $\lesssim$ 0.1 pc. Multiple formation requires two or more events of gravitational collapse that will thus form a bound system. The process by which self-gravitating objects develop substructures that independently collapse is called fragmentation. If formed at a certain distance, within time, they can gradually approach each other and become gravitationally bound. Following, the main models for multiple star formation are better detailed.

The multiple formation models can be categorized in three: 1) fragmentation of a core or filament; 2) fragmentation of a massive accretion disk; or 3) through dynamic interactions (see Fig. \ref{multiple}). Following, these processes are better detailed: 

\begin{figure}[htb!]
    \centering
    \includegraphics[width=0.95\textwidth]{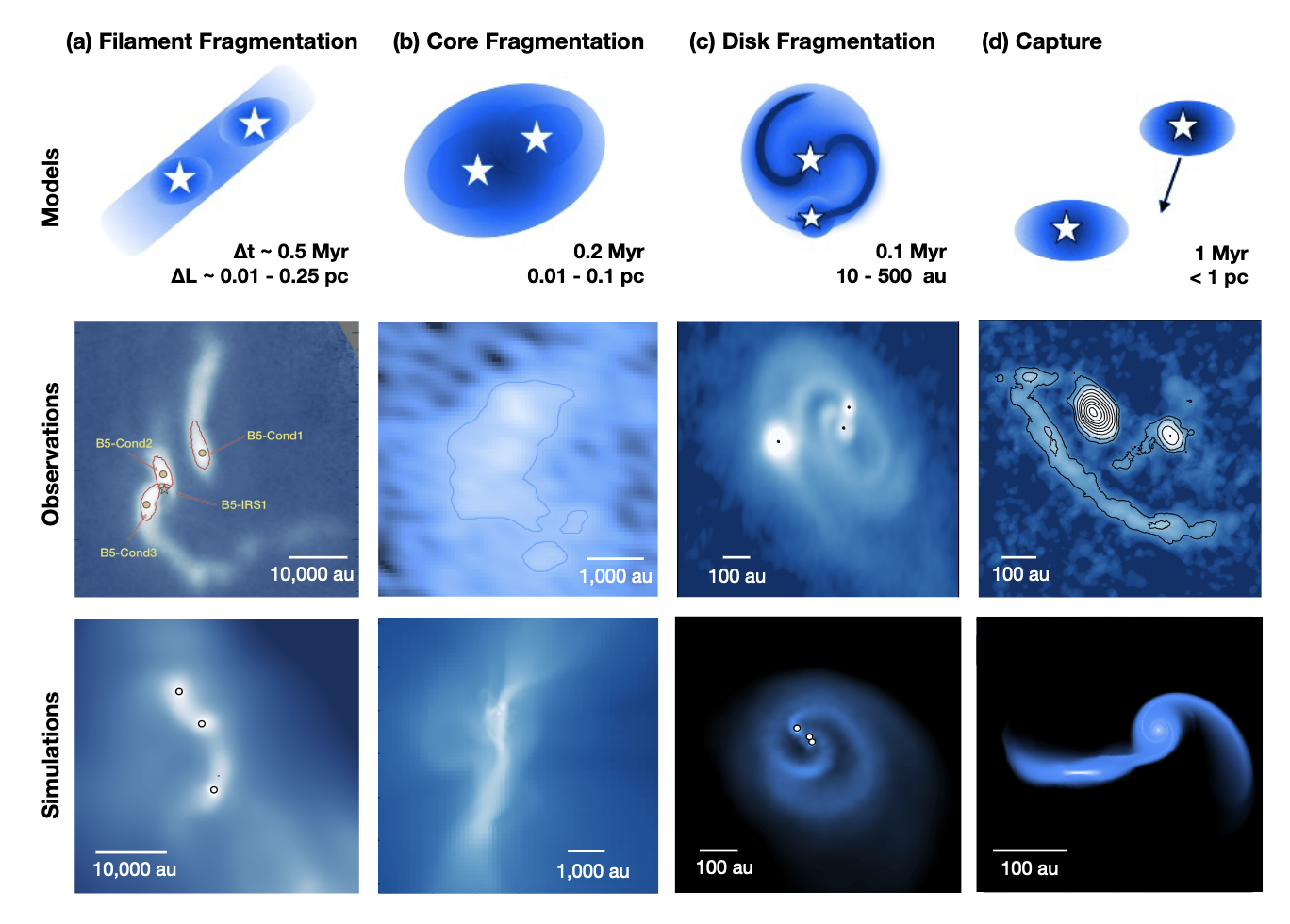}
    \caption[Mechanisms for multiple (sub)stellar formation]{Mechanisms for multiple (sub)stellar formation. The model (top row) presents an approximate range of time and length scales for each process. The middle row shows proposed observational examples for each case. The bottom row shows examples of each formation model from numerical simulations. Credits: \citet{offner2022}. The observations/simulations references are listed in Figure 5 caption of that work.}
    \label{multiple}
\end{figure}
\begin{itemize}
    \item \textbf{Core and filament fragmentation}: The processes of core and filament fragmentation occur in dense embedded structures, with each object developing and collapsing within the same parent structure. Filaments are the second larger structures in the stellar formation sequence, formed by the encounter of expanding shells (see subsection \ref{sec:YSO_classification}). However, hereafter I define cores and filaments as morphological structures, rather than physical ones, which explains the similar scales between the two modes of fragmentation in Fig. \ref{multiple}. Therefore, the two presented fragmentation modes are defined by an over-density in a structure formed, before the formation of a disk. Two main mechanisms trigger core/filamentary fragmentation, rotation, and turbulence:
    \begin{itemize}
        \item \textbf{Rotation-driven fragmentation}: 
        
        When collapsing cores rotate rapidly, they become susceptible to fragmentation if specific criteria, ensuring a balance among thermal, gravitational, and rotational energies, are satisfied. Excluding fragmentation as the initial trigger, the simplified expression by \citet{1992ApJ...388..392I} establishes the relationship between the initial ratio of thermal to gravitational energy, $\alpha_{\mathrm{vir}}$, and the ratio of rotational to gravitational energy, $\beta_{\text {rot }}$:
        \begin{equation}
          \alpha_{\text {vir }} \beta_{\mathrm{rot}}=0.02\left(\frac{T}{10 \mathrm{~K}}\right)\left(\frac{M}{1 M_{\odot}}\right)^{-2}\left(\frac{R}{0.1 \mathrm{pc}}\right)^4\left(\frac{\Omega}{10^{-14} \mathrm{~s}^{-1}}\right)^2  .
        \end{equation}
        

        Here, $T$, $M$, $R$, and $\Omega$ represent the local temperature, mass, radius, and Keplerian frequency of the collapsing structure, respectively. The conditions for fragmentation in this expression are $\alpha_{\text {vir }} \beta_{\text {rot}} > 0.12$ and $\alpha_{\text {vir }} < 0.5$. Rotational fragmentation often results in the formation of multiple systems within the same plane, tending towards spin alignment and coplanarity \citep{2016ApJ...827L..11O, 2018MNRAS.475.5618B}. These characteristics serve as signatures of the formation scenario, which can be used as an educated guess in the case of evolved multiple systems (see chapters \ref{chap:2} and \ref{chap:3}).

    \item \textbf{Turbulent fragmentation}: Turbulence is an effective trigger for structure formation, encompassing both stars and brown dwarfs within the same system. Regions within a molecular cloud can fragment when the turbulence seeds collapse and provide pressure support -- when the local gravitational energy exceeds the contribution of both turbulent and thermal energy. To initiate fragmentation via turbulence, a critical density $\rho_{\text{crit}}$ must be achieved or exceeded. For an isothermal cloud, a relatively straightforward expression is provided by:
    \begin{equation}
        \rho_{\text {crit }}=\frac{\rho_0}{1+\mathcal{M}^2}\left(\frac{R}{R_0}\right)^{-2}\left[1+\mathcal{M}^2\left(\frac{R}{R_0}\right)^{p-1}\right],
    \end{equation}

where $\rho_0, R_0$ and $\mathcal{M}$ are the initial density, radius, and Mach number of the parent cloud ( ratio of local flow velocity to the sound speed of the medium), respectively \citep{2015MNRAS.450.4137G,2017MNRAS.468.4093G}. In this expression, p denotes a term equal to 2 for supersonic turbulence and 5/3 for subsonic scales. Additional contributions can also increase the complexity. For example, an isothermal
filament collapses if its mass per unit length, $M_{\mathrm{l}}$, exceeds a critical value. A filament threaded laterally by a magnetic field, $B_{\|}$, will become unstable when $M_{\mathrm{l}}$ exceeds \citep{1997ApJ...480..681I,2014ApJ...785...24T}:
\begin{equation}
\begin{aligned}
M_{l, \mathrm{crit}} & \simeq \frac{2 c_s^2}{G}\left(1+\beta_{\mathrm{mag}}^{-1}\right) \\
& \simeq 16.4\left(\frac{T}{10 \mathrm{~K}}\right)+7.8\left(\frac{n_{\mathrm{H}}}{10^4 \mathrm{~cm}^{-3}}\right)^{-1}\left(\frac{B}{10 \mu \mathrm{G}}\right)^2 M_{\odot} \mathrm{pc}^{-1} ,
\end{aligned}
\end{equation}

where $\beta_{\text {mag }}=8 \pi c_s^2 \rho / B_{\|}^2$ is the ratio of thermal to magnetic pressure. Furthermore, within filaments, if turbulent support after triggering the collapse is negligible, the critical mass can be stated as: 
\begin{equation}
\begin{aligned}
    M_{crit} & \simeq 1.3 \frac{c_s^4}{G^2 \Sigma_{\mathrm{fil}}} \\
& \simeq 1.1\left(\frac{T}{10 \mathrm{~K}}\right)^2\left(\frac{M_{1, \text { crit }}}{40 M_{\odot} \mathrm{pc}}\right)^{-1}\left(\frac{W_{\mathrm{fwhm}}}{10^4 \mathrm{au}}\right) M_{\odot} ,
\end{aligned}
\end{equation}

where the filament surface density, $\Sigma_{\mathrm{fil}}$, is the ratio of the filament mass-per-length to the width of the filament $\Sigma_{\mathrm{fil}}=M_{\text {line }} / W_{\text {fwhm }}$ \citep{2019A&A...629L...4A}. Multiple systems originating from the same parent structure through turbulent fragmentation are expected to share similar ages and typically exhibit a modest number of members, typically two to three within a core. They are usually formed at wider separations (set to a scale of 102 au to 0.1 pc; \citealp{offner2022}) compared to objects formed through rotational-driven fragmentation. As a result, they accrete gas at varying rates and have distinct spin alignments and rotational speeds. Consequently, (sub)stars resulting from turbulence-driven fragmentation tend to have misaligned stellar spins, accretion disks, and protostellar outflows \citep{2000MNRAS.314...33B, 2016ApJ...827L..11O, 2018MNRAS.475.5618B, 2019ApJ...887..232L}.


    \end{itemize} 
    \item \textbf{Disk fragmentation}: Circumstellar disks can undergo fragmentation at later stages compared to the previously mentioned fragmentation modes. This particular type of fragmentation is expected to occur in massive forming disks through gravitational instability (GI), as discussed in subsection \ref{gi}. Referenced in \ref{sec:substructures} as a mechanism for triggering spirals in circumstellar disks, GI can also cause instabilities that may lead to fragmentations when the Toomre parameter $Q$ is $\leq$ 1 \citep{toomre1964,kratter2016} in razor-thin disk. The Toomre parameter criteria can also be described in terms of the mass of the protostar/PMS and the disk, aspect ratio and radial distance: 

    \begin{equation}
        Q=\frac{c_s \Omega}{\pi G \Sigma}=f \frac{M_*}{M_d} \frac{H}{r} \leq 1
    \end{equation}
    
where $c_s$ is the disk sound speed, $\Omega$ the disk orbital frequency,  $\Sigma$ the surface density, $M_* / M_d$ the star-disk mass ratio, $H$ the disk scale height, $r$ the radial distance, and $f$ the order unity scale factor which depends on the assumed surface density power law \citep{offner2022}. However, gravitational instability (GI) does not inevitably lead to fragmentations; it serves as the mechanism causing local instabilities. For fragmentation to occur after GI is triggered, effective cooling must counteract the inflation caused by higher temperatures. In the case of thick disks, the critical $Q$ significantly drops below unity due to the extended vertical distribution of matter \citep{1978ApJ...226..508L}. This implies that gravitational instability (GI) in such disks can be triggered in outer regions, specifically when the disk is not excessively flared and is located at a distance with sufficient gas density -- in these outer regions, lower temperatures are associated with higher radial distances. However, the probability of gravitational instability fragmentation increases during the early stages of disk formation, particularly when the infall rate dominates over viscous accretion (typically in the Class 0 phase) since during this phase, an increase in surface density does not necessarily correspond to higher temperatures that could inhibit gravitational instability \citep{2010ApJ...708.1585K, 2011MNRAS.413..423H, zhu2012}. This scenario is likely linked to the rapid formation of binaries sharing approximately the same mass \citep{1997MNRAS.285...33B, 2005ApJ...623..922O, 2015MNRAS.452.3085Y}. 
An alternative pathway to GI fragmentation arises when the disk temperature decreases more rapidly than accretion can modify the surface density. A plausible explanation for this scenario is variations in stellar luminosity resulting from accretion changes (see section \ref{sec:eruptive}). Fragmentation may occur suddenly if there is a significant decrease in central luminosity, provided that the temperature drop does not coincide with an increase in the cooling timescale \citep{2014MNRAS.444..887D, 2018MNRAS.475.2642K}. Regardless of the process that triggers gravitational instability (GI) fragmentation, the binary companion must overcome interactions with the disk and subsequent tidal forces induced by inward migration.
    \item \textbf{Dynamic interactions}: A more effective mechanism to elucidate the formation of close binaries is through capture and subsequent orbital evolution. 
    
    
    The effectiveness of the capture mechanism depends on the environment, whether it is gas-dense or has less gas. In gas-dense environments, objects within a stellar cluster may transition between unbound and bound states due to energy and angular momentum exchange with the surrounding gas cloud or individual circumstellar disks. This process, known as gas-mediated capture, is observed in various scales in numerical simulations and can lead to the formation of bound systems. However, the net frequency of capture through this mechanism is likely low, making these systems observationally indistinguishable from those formed via fragmentation, except in cases where two disks seem to be on a collision course. On the contrary, in regions of high stellar density without much gas, interactions between truly unbound stars can lead to the formation of new binaries and partner exchange. Recent work emphasizes the contributions of higher order multiples to overall interaction rates in moderately dense clusters, producing a more diverse set of outcomes, including the formation of very compact binaries \citep{2015ApJ...808L..25G,2017MNRAS.465.2198D,2020MNRAS.494..850H}. These processes are more relevant for dense open clusters and globular clusters but are sub-dominant in gas-rich environments. In addition, widely separated bound binaries can reduce their separations through the dynamical friction force, caused by the torque provoked by accreting systems trailing their material during close encounters \citep{1997MNRAS.285...33B,2010MNRAS.402.1758S,2011MNRAS.416.3177L}, by numerical viscosity and mergers \citep{vorobyov2005, zhu2012} or by  Kozai-Lidov cycles coupled with tidal friction \citep{1962AJ.....67..591K,1962P&SS....9..719L,2007ApJ...669.1298F}. The latter scenario occurs when binaries exchange angular momentum with an inclined tertiary object, which thus leads to a periodic exchange between eccentricity and inclination of the binary, therefore leading to circularization and a hardening of their orbits.



\end{itemize}



\subsection{High-Contrast Imaging: instrumentation, observing strategies and data-processing}
\label{sec:instrumentation}
In the Galaxy, more than 5500 exoplanets have been identified to date\footnote{For more information, refer to sources such as \url{https://exoplanetarchive.ipac.caltech.edu/} and \url{http://www.exoplanet.eu/catalog/}}, alongside a few thousand brown dwarfs\footnote{For details, see resources like \url{http://www.johnstonsarchive.net/astro/browndwarflist.html} and \url{https://simple-bd-archive.org}} (hereafter called BDs). While various techniques can unveil the existence and characteristics of these objects by observing their interactions with the brighter stars they orbit, high-contrast imaging (HCI), also known as direct imaging (DI), stands out as the sole method capable of directly revealing their presence. Thus, the HCI offers the advantages of extracting the flux and astrometry of the detected source. With a few observations of the latter, orbital characterization can also be performed. Lastly, HCI provides a recovery of the photometric spectra, which can provide key parameters such as effective temperature, surface gravity, mass of the object, and presence of clouds and certain molecules.

\begin{figure}[htb!]
    \centering
    \includegraphics[width=0.95\textwidth]{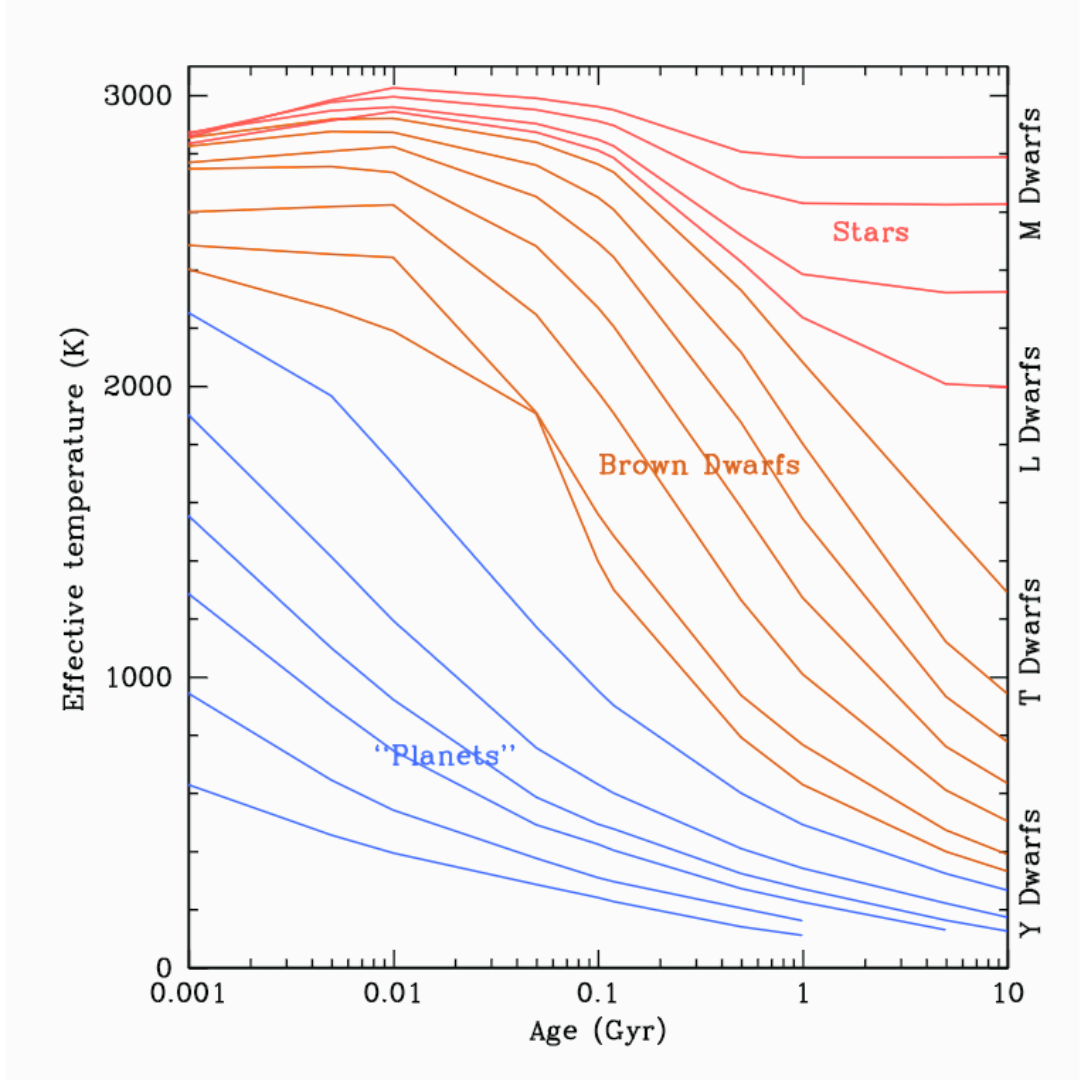}
    \caption[Evolution of effective temperature for substellar and low mass stars]{Evolution of effective temperature for substellar and low mass stars. The red lines trace stars (masses above the hydrogen-burning limit), the orange track brown dwarfs, and the blue trace objects below the deuterium-burning limit. Credits: Adaptation of \citet{bailey2014}.}
    \label{subs_profile}
\end{figure}

Planck’s law defines that a black body, an object that is a perfectly efficient energy absorber and emitter, will have its radiation characterized by its temperature. Assuming that stars and substellar objects can emit approximately black bodies, an observational instrument will be more susceptible to detect them using broadband filters covering wavelengths close to the peak emission. The detectability of such objects will also depend on the instrument’s sensibility, resolution power, and possible interference from background sources. Black-body objects with a radiation peak in the optical/IR can emit in two different ways. They can produce their radiation through nuclear reactions (as in the case of low-mass stars and deuterium-burning objects) or they can emit thermal radiation due to atomic and molecular agitation. The latter applies to newly formed planets and moons, which have enough primordial heat derived from their origin. Brown dwarfs are a special case, where their deuterium-burning phase does not last a long time and therefore, similarly to giant planets, they will cool down gradually with time (see Fig. \ref{subs_profile}).

\begin{figure}[htb!]
    \centering
    \includegraphics[width=0.95\textwidth]{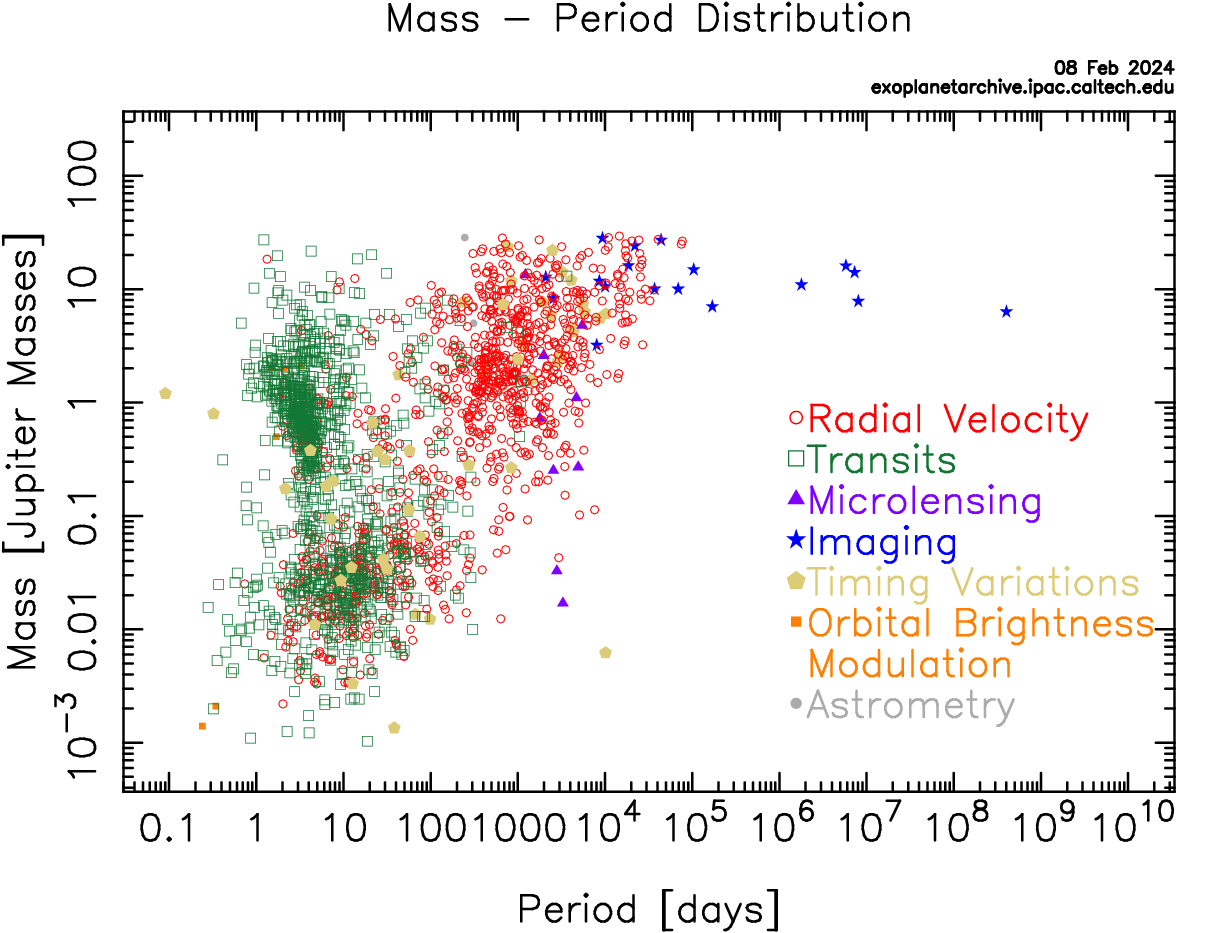}
    \caption[Exoplanets population and their respective discovery techniques]{The diagram depicts the population of exoplanets and the corresponding discovery techniques. It illustrates masses in Jupiter masses versus orbital periods in days. The high-contrast imaging/direct imaging planets are specifically highlighted by blue-filled stars. Credits: \url{https://exoplanetarchive.ipac.caltech.edu/exoplanetplots/exo_massperiod.png}.}
    \label{massperiod}
\end{figure}

The HCI technique relies on enhancing the contrast of substellar objects, by suppressing the starlight from the host star. This technique focuses on detecting objects that still retain heat from their formation, which are faint in comparison with the host star. In summary, the technique is optimized for detecting massive substellar objects on their first millions of years (Myrs) and at wider separations from their host star using Optical/near-IR wavelengths (see Fig. \ref{massperiod}, where discovered by imaging planets (HCI/DI planets) are compared to planets detected by other techniques).

\begin{figure}[htb!]
    \centering
    \includegraphics[width=0.95\textwidth]{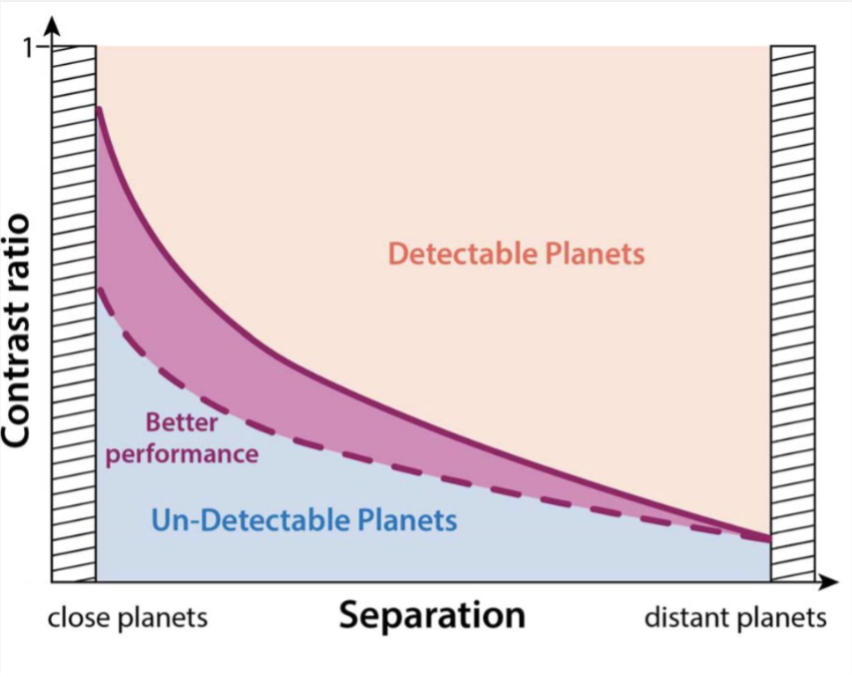}
    \caption[Schematic representation of how a contrast curve is interpreted]{The chart illustrates how to interpret a contrast curve, where a companion becomes detectable when positioned above the solid curve at a particular contrast and separation. Notably, the achieved contrast strongly depends on the separation from the central star, making only the brightest planets detectable at close separations. The dashed line represents a deeper contrast curve compared with the solid line. A deeper contrast curve improves the effectiveness of high-contrast imaging, making it more probable to detect faint objects at closer separations. Credits: Adapted from \citet{2023PASP..135i3001F}.}
    \label{contrastscheme}
\end{figure}

Despite its potential, HCI effectivity has significant challenges as it requires a certain level of contrast between star-companion, and also a certain angular resolution between both. The contrast is quantified by the flux ratio $F_{object} / F_{host\_star}$, a value influenced by both the angular separation of the sources and their intrinsic brightness corresponding to the observed wavelength. In the most favorable cases, with the best instruments and post-processing algorithms at the moment, contrasts can be reached in the order of 10$^{-6}$ to 10$^{-5}$. Greater contrast values result in brighter companions compared to the host star, hence aiming for lower values (fainter objects -- less massive or older companions) is the objective for the new generation of telescopes and post-processing methodologies dedicated to HCI (see Fig. \ref{contrastscheme}). The angular resolution is determined by a combination of the orbital separation between the objects, linked to the distance observer system. 

\subsubsection{Instrumentation}

In order to perform HCI/DI, it is required a proper instrumentation, characterized by adaptive optics (AO) in the case of terrestrial telescopes, and coronagraphs for all cases. 

\begin{figure}[htb!]
    \centering
    \includegraphics[width=0.95\textwidth]{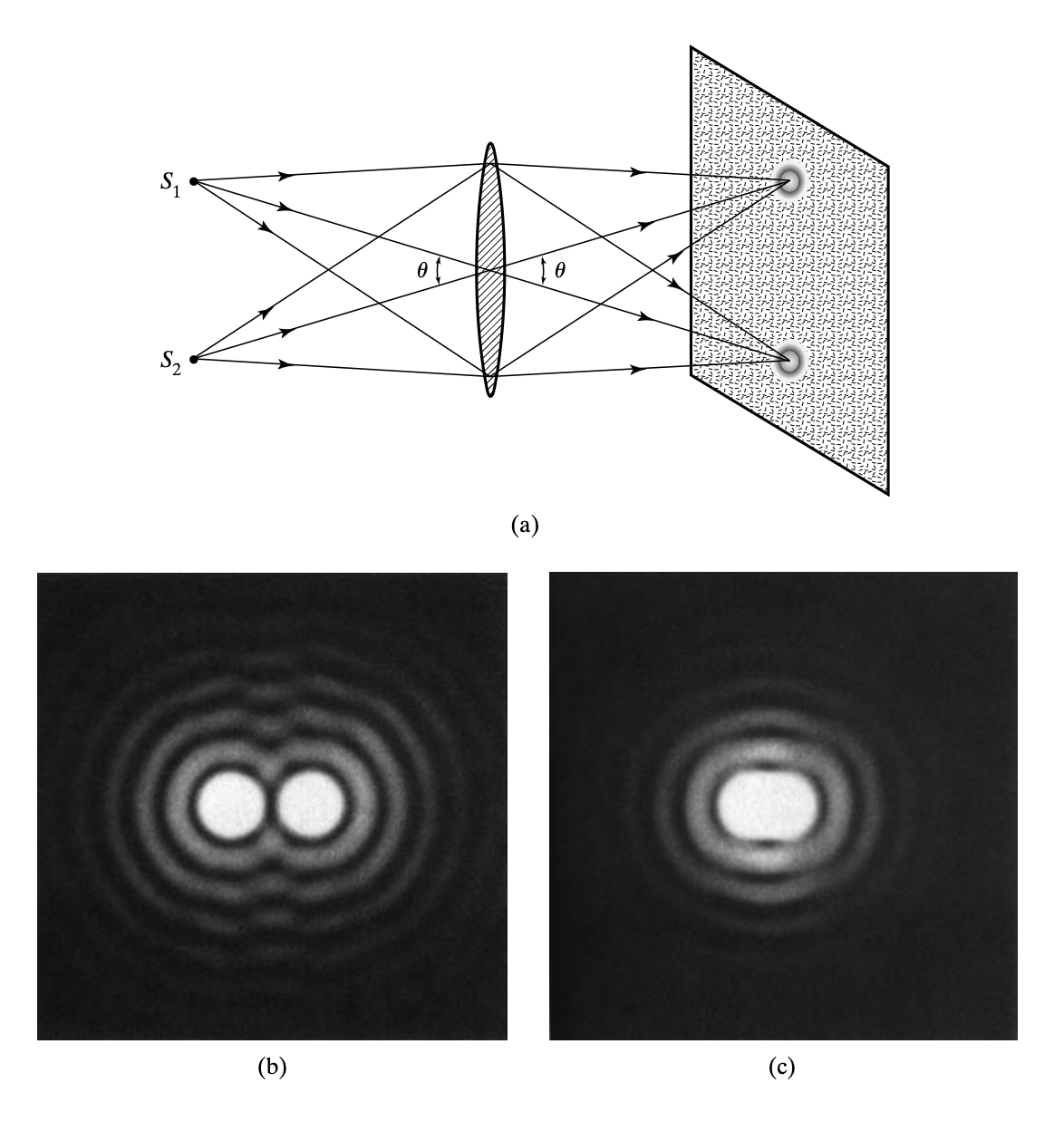}
    \caption[Diffraction-limited images of two point objects formed by a lens.]{Diffraction-limited images of two point objects produced by a lens are illustrated. In the schematic depiction shown in Figure (a), two airy disks are formed upon crossing a lens, originating from two distinct points $S_1$ and $S_2$. Figures (b) and (c) respectively depict scenarios where the sources are resolved and unresolved. Credits: \citet{2017inop.book.....P}.}
    \label{airy}
\end{figure}

\begin{figure}[htb!]
    \centering
    \includegraphics[width=0.65\textwidth]{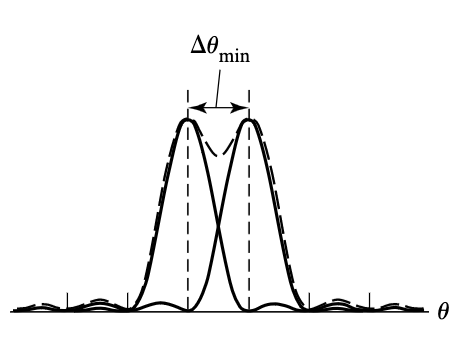}
    \caption[Rayleigh’s criterion for just-resolvable diffraction patterns.]{Rayleigh’s criterion for just-resolvable diffraction patterns. The dashed curve is the observed sum of independent diffraction peaks. Credits: \citet{2017inop.book.....P}.}
    \label{rayleigh}
\end{figure}

The light coming from celestial objects, which travels approximately a quasi-infinite distance, should be a perfect plane wave when entering the pupil of the telescope. In such case, without interference from the atmosphere, these spherical waves should ideally form an ``Airy Disk'' diffraction pattern at the telescope focal plane (see Fig. \ref{airy}). This diffraction pattern will thus form the ideal Point Spread Function (PSF), with an angular resolution $\theta_{diff}$ (angular radius of the central airy disk maximum) given by:

\begin{equation}
\theta_{\text {diff }}=1.22 \frac{\lambda}{D}    
\end{equation}

\noindent where $D$ is the diameter of the telescope's aperture and $\lambda$ is the wavelength of the observation (see e.g., \citealp{2017inop.book.....P} and also Fig. \ref{rayleigh}). $\theta_{\text {diff }}$ is of the order of 30 mas for $\lambda=1 \mu \mathrm{m}$ and $D \sim 8 \mathrm{~m}$.

\begin{figure}[htb!]
    \centering
    \includegraphics[width=0.75\textwidth]{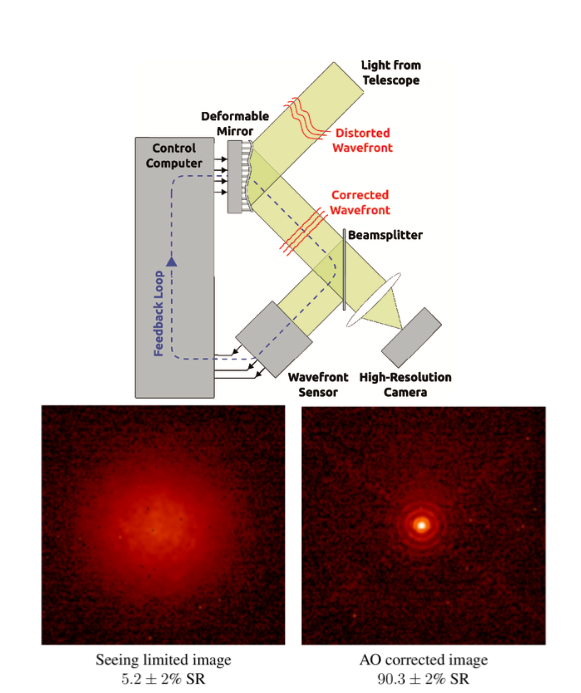}
    \caption[Simplified AO configuration and its impact in observations]{Simplified AO configuration and its impact in observations. Upper panel: a diagrammatic representation of an Adaptive Optics (AO) system according to \citet{2015PASP..127.1197R}. Real-time correction of the distorted wavefront involves a wavefront sensor, a feedback loop, and a deformable mirror. Lower panel: illustration of an image with a distorted wavefront before AO correction (left) and the same image after AO correction (right), revealing the Airy pattern \citep{2016JATIS...2b5003S}. Credits: \citet{2020PhDT.........1B}.}
    \label{aosystem}
\end{figure}

AO is the technology used in the instrument to compensate for the effects of atmospheric turbulence. The wavefronts coming from the sources are susceptible to unpredictable Earth's atmospheric variations. The atmosphere is formed by layers of turbulent ``bubbles''.  Before reaching the telescope's pupil, the incoming light is often divided into sub-apertures determined by the radius of these bubbles. The interference of these sub-apertures creates fringes-patterns of constructive and destructive interference. In regions where the sub-apertures interfere constructively, bright regions (speckles) are formed. On the contrary, in regions of destructive interference, dark regions appear. The resulting pattern is the speckle pattern. The speckles exhibit dimensions of approximately $\lambda / d$, determined by the separation $d$ between the two sources of interference. When $d \sim D$, these speckles will share identical dimensions with the PSF of the star, making it challenging to differentiate them from an astrophysical source with equivalent brightness. To mitigate these wavefront variations, a control system must continuously measure the impact of the fluctuations at a high frequency during observations, and subsequently provide corrections as/when needed. To simulate atmospheric behavior, a reference point source, usually a star, is employed. This source should be sufficiently bright to ensure a high signal-to-noise ratio (SNR) in the wavefront sensor. Certain instruments offer the option of employing an artificial reference, specifically the laser guide star (LGS) or natural point sources when a bright star is unavailable, or even the host star. The instrument responsible for quantifying these fluctuations is known as a wavefront sensor, providing valuable information about phase variations. A real-time calculator then analyzes this data and determines the precise adjustments required to compensate for wavefront errors, transmitting instructions to deformable mirrors in a closed-loop system (see \citealp{2012SPIE.8447E..05M} and Fig. \ref{aosystem}). More recently, Extreme Adaptive Optics (ExAO or XAO) systems were developed in high-performance high-contrast imagers as Spectro-Polarimetric High-contrast Exoplanet Research (VLT/SPHERE, \citealp{beuzit2019}) and Gemini Planet Imager (GPI, \citealp{2008SPIE.7015E..18M}). They are able to provide Strehl ratios (ratio of maximum diffraction intensities of an aberrated wavefront to a perfect wavefront) up to 95\%.

Coronagraphs are optical devices designed to suppress the brightness of stars, thereby improving the contrast for observing faint substellar objects and structures in the context of HCI. According to \citet{2006ApJS..167...81G}, the classical coronagraphic designs can be classified into four, 1) the interferometric coronagraphs, 2) the pupil apodized coronagraphs, 3) the Lyot coronagraphs, and 4) hybrid designs combining elements from the aforementioned categories.  Modern coronagraphic designs often integrate elements from these categories, with variations centered around the apodizer concept, mask properties, and the complexity of the coupled Adaptive Optics (AO) system and mirrors' disposition. 

\begin{figure}[!htb]
    \centering
    \includegraphics[width=0.75\textwidth]{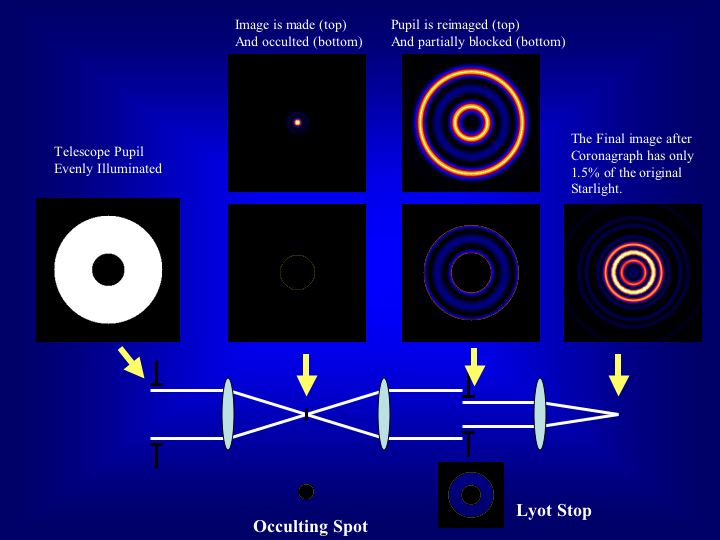}
    \caption[Schematic representation of how a classical Lyot coronagraphic works]{Schematic representation of how a classical Lyot coronagraphic works. The figure also shows how a Lyot stop can enhance the signal from regions closer to the borders of a physical mask.  Credits: Ben R. Oppenheimer -- \url{https://lyot.org/background/coronagraphy.html}.}
    \label{lyot}
\end{figure}

\begin{figure}[htb!]
    \centering
    \includegraphics[width=0.95\textwidth]{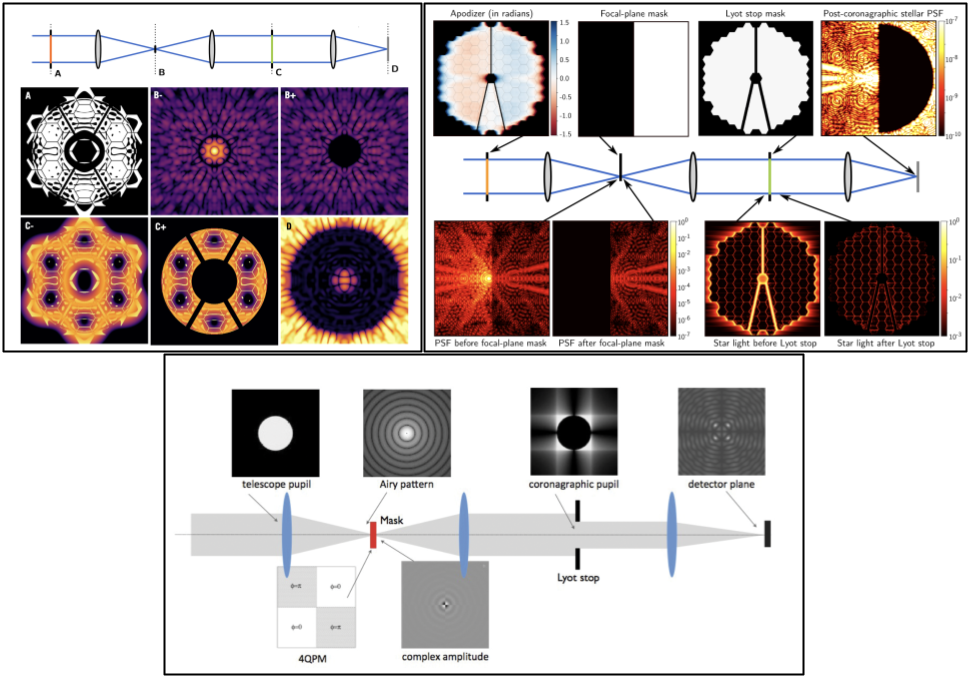}
    \caption[Representative examples of possible coronagraphic designs with Lyot stops]{Representative examples of possible coronagraphic designs with Lyot stops. Top-left panel: Schematic layout of an apodized pupil Lyot coronagraph (APLC). The incoming light follows the sequence from A to D. The labels ``-'' and ``+'' represent before and after the stage/instrument utilization, represented by the letters. Credits: \citet{2020SPIE11443E..3PP}. Top-right panel:  Schematic layout of a phase-apodized pupil Lyot coronagraph (PAPLC). In that case, the apodizer has a phase shift in different regions, and the mask only covers the brighter half-side of incoming light. Credits: \citet{2020ApJ...888..127P}. Bottom panel: a schematic representation of a four Quadrant Phase-Mask (4QPM). In this case, no apodizer is presented, and a phase mask is utilized. Credits: \citet{2015PASP..127..633B}.}
    \label{coroconfig}
\end{figure}

The most simple description of such a system, following the incoming light towards the detector focal plane unfolds as follows: to attenuate incoming light, an apodizer may be introduced into the telescope pupil, modifying the amplitude or intensity profile -- typically gradually attenuating the intensity towards the apodizer's edges. Following the pupil, and potentially the apodizer, the light advances through a lens, converging at the initial focal plane. At this stage, a mask is implemented, serving either as a physical occulting mask or a phase mask. The latter works through destructive interference, thus also blocking the starlight as the former. Following another passage through a lens and the alignment of the wavefront, a component known as the Lyot stop is strategically positioned. This device partially obstructs the light and serves to attenuate the diffraction caused by the mask. The Lyot stop will thus permit the incoming of a predominant portion of light from surrounding sources close to the limits of the mask, therefore contributing to the final image as it converges on the detector plane. A schematic representation of how a Lyot stop works is presented in Fig. \ref{lyot}. Three representative examples of possible coronagraphic designs with Lyot stops are presented in Fig. \ref{coroconfig}. The instrument used in the work presented in chapter \ref{chap:3}, VLT/SPHERE, provides a possibility of using three coronagraphic designs, a classical Lyot coronagraph (CLC, without an apodizer), an apodized pupil Lyot coronagraph (ALC or APLC), and a four-quadrant phase-mask (4QPM, with a phase mask instead of an amplitude mask as the previous mentioned configurations).

\subsubsection{Observing strategies}
\label{observingstrategy}
In addition to the speckle patterns produced by the atmospheric variations, a second type of speckle pattern that changes relatively slowly over time and arises from residual aberrations and imperfections in the optical system is called quasi-static speckle. They are mostly permanent during an exposure of tens of minutes (see e.g. \citealp{2007ApJ...669..642S}). These speckles can be partially suppressed using post-processing techniques after the observation sequence is completed.
To do so, an observing strategy, determining how data is acquired at the instrument level, provides the raw data, which is subsequently processed through a post-processing technique. The main observing strategies to detect substellar companions can be categorized into three: 1) angular differential imaging (ADI; \citealp{marois2006}), 2) spectral differential imaging (SDI; \citealp{1999PASP..111..587R,2002ApJ...578..543S}), and 3) reference differential imaging (RDI; e.g. \citealp{2009ApJ...694L.148L,2011ApJ...741...55S,2019AJ....157..118R}). They are briefly described as follows:

\begin{itemize}
    \item \textbf{ADI}:
    Angular differential imaging (ADI) is one of the most commonly used observing strategies to mitigate quasi-static speckles in HCI. The ADI process requires the science target to pass near the meridian passage. The procedure follows as the pupil of the telescope is fixed in the host star, and its derotator is switched off. Consequently, the coronagraphic frames will maintain the center of the mask fixed with the host star position, while the Field-of-View (FoV) will be noticed rotating. This will result in the acquisition of frames with different parallactic angles throughout the observation sequence. Moreover, the stabilization of the pupil ensures that the aberrations it induces remain relatively constant across all parallactic angles. This is crucial, as a significant portion of quasi-static speckles originates from the pupil and its supporting structures (spiders). By having the frame sequence, a model of the speckle halo will be created and then subtracted from all frames, thus removing substantially the quasi-static speckles. The frames will thus be derotate, aligning whichever possible companion/structure at the same position, thus enhancing its signal (see a general schematic representation of ADI sequence, independently of the post-processing model used in, Fig. \ref{adigeneral}). The post-processing technique used will define how the speckle field models will be created and how the signal of the companion will be reconstructed (see subsection \ref{sec:postprocessing} and Fig. \ref{adipostprocessing}).

    \begin{figure}[htb!]
    \centering
    \includegraphics[width=0.95\textwidth]{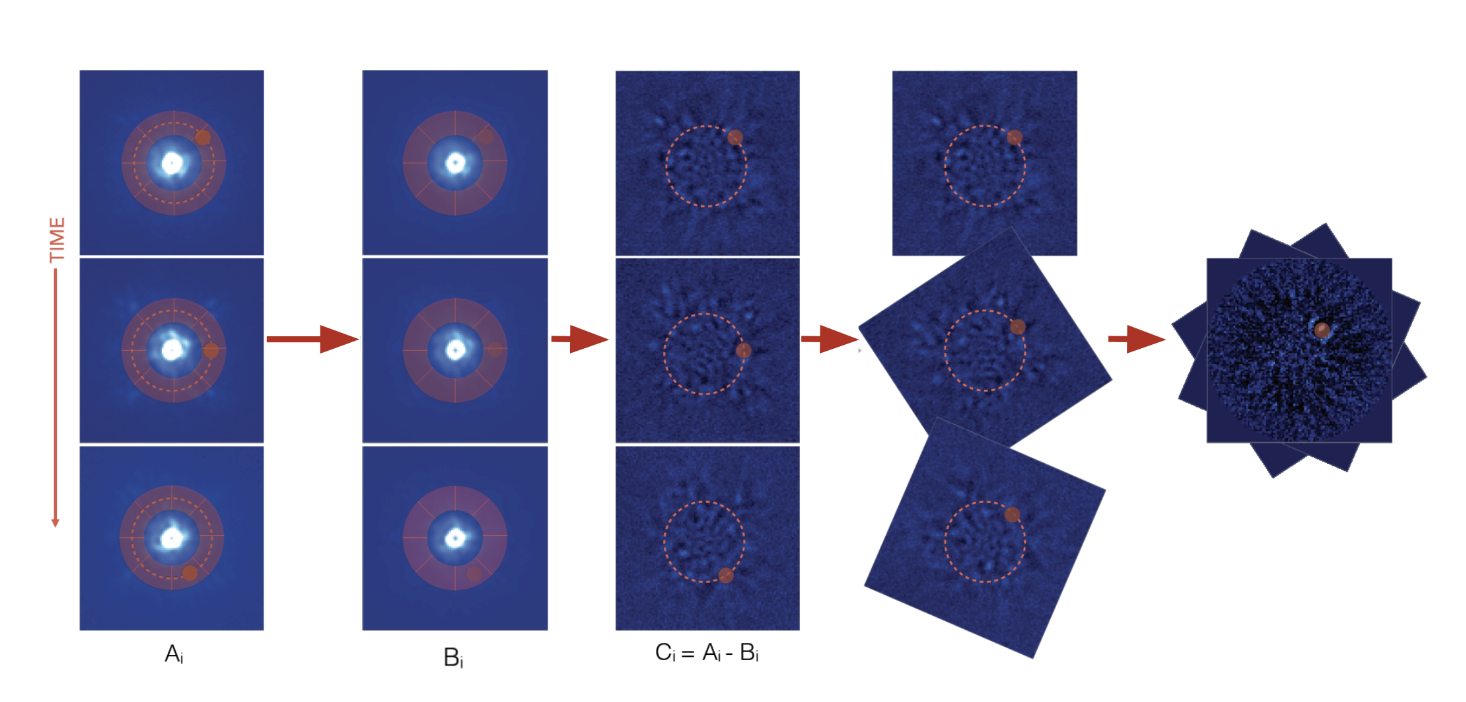}
    \caption[Illustration outlining the key stages of ADI observing strategy]{Illustration outlining the key stages of ADI observing strategy. From left to right, the sequence includes the calibrated science images (A), the frame-wise reference PSF (B), the residual images (C), the derotated residual images, and the final residual map. Credits: \citet{dahlqvistthesis}.}
    \label{adigeneral}
    \end{figure}

    \item \textbf{SDI}: The spectral differential imaging (SDI) consists of decoupling the companion/structure from the speckles using different wavelengths (rather than azimuthal rotation as ADI). This observing strategy was first proposed by \citet{1999PASP..111..587R}. The researchers relied on two images taken in adjacent bands, aiming to exploit the presence/absence of an exoplanet's signal in two or more adjacent bands. Their idea was to carefully select two wavelengths to ensure the exoplanet's presence in one image while limiting its brightness in the other, often utilizing the methane absorption line. This approach effectively reduced speckle noise without causing self-cancellation of the planetary signal during subtraction. One example, where the methane absorption is more pronounced in one band than in the other is presented in Fig. \ref{dualbandsdi}.
    \begin{figure}[htb!]
    \centering
    \includegraphics[width=1\textwidth]{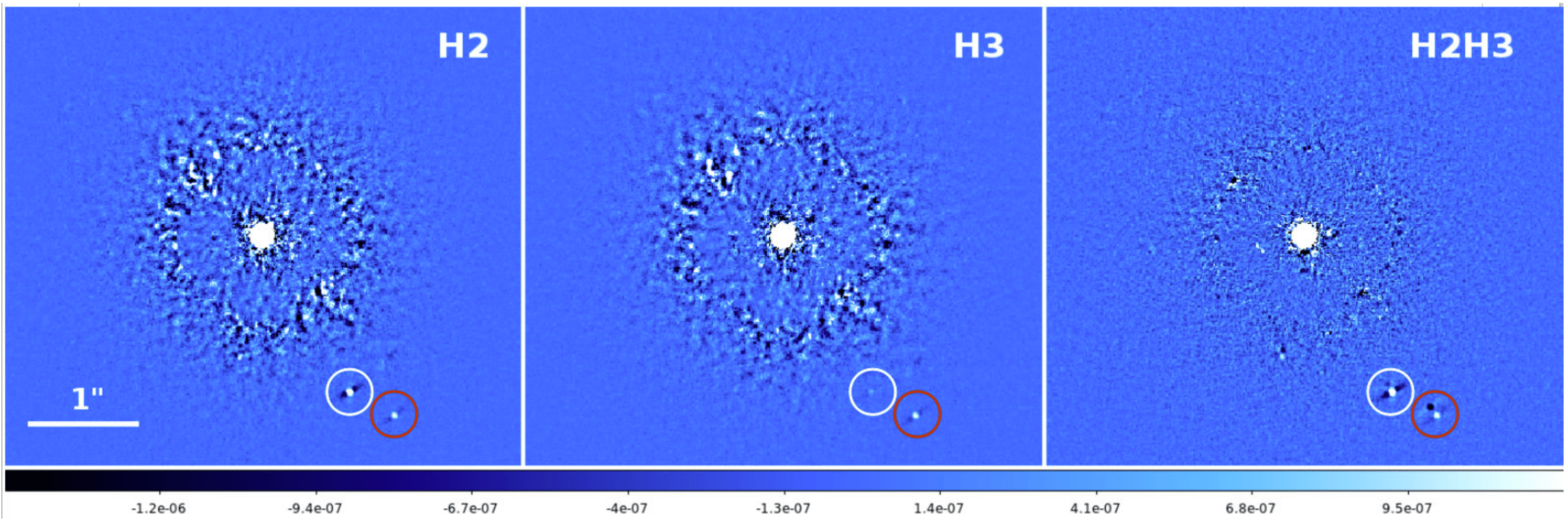}
    \caption[Images of the system around GJ 758]{Images of the GJ 758 system, highlighting the late T-dwarf GJ 758 B within the white circle. A background star is indicated in the red circle. The methane absorption in the H3 filter is evident, showcasing lower flux in the object. In the SDI combination of the two filters, the T-dwarf displays two distinct peaks, while the neighboring star exhibits equal peaks. Credits: Adapted from \citet{2016A&A...587A..55V}.}
    \label{dualbandsdi}
    \end{figure}
    A more advanced SDI approach, also known as the spectral deconvolution technique (SD), was introduced by \citet{2002ApJ...578..543S}. Their concept exploits the fact that quasi-static speckles exhibit radial scaling proportional to $\lambda$ (appearing to move radially from the image center), while real objects remain stationary. This difference becomes noticeable when frames observed with different bands are appropriately rescaled (see Fig. \ref{sdi}). To ensure a more robust analysis, it is essential to observe multiple bands simultaneously, invoking the utilization of Integral Field Spectrograph (IFS) datacubes, as suggested later by \citet{2007MNRAS.378.1229T}.

    \begin{figure}[htb!]
    \centering
    \includegraphics[width=1\textwidth]{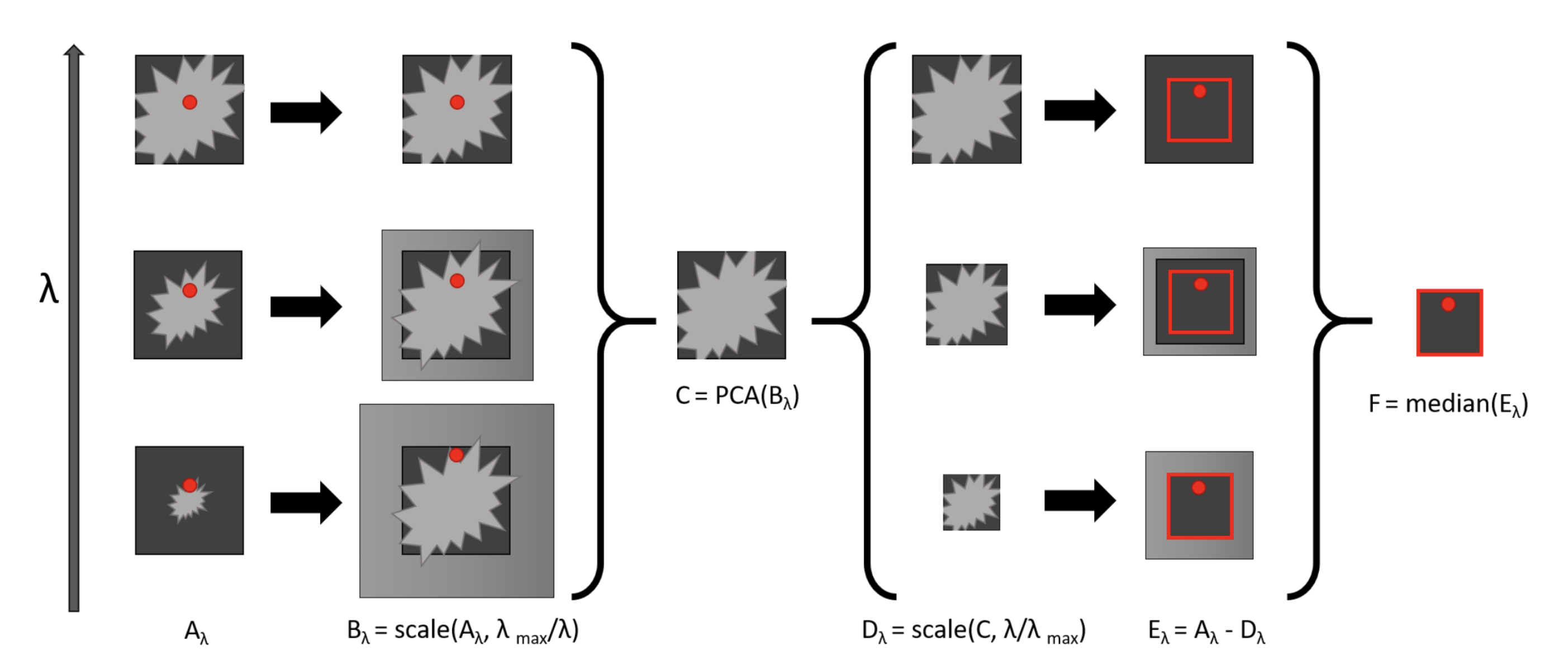}
    \caption[Schematic representation of SDI observing strategy coupled with PCA post-processing]{Schematic illustration of the Spectral Differential Imaging (SDI) observing strategy in combination with Principal Component Analysis (PCA) post-processing. The process involves scaling the images (A$\lambda$) inversely proportional to their wavelength (B$\lambda$), followed by the creation of a stellar PSF model (C) using PCA analysis. The stellar PSF model (C) is then rescaled for each wavelength channel proportional to its wavelength (D$\lambda$) to match the original images' scaling (A$\lambda$). This results in a reduced Field of View (FoV) marked by a red border, with the outer region of small wavelength images not covered by the scaled stellar PSF models (D$\lambda$). The companion is denoted by a red dot. Credits: \citet{2021A&A...652A..33K}.}
    \label{sdi}
\end{figure}

    \item \textbf{RDI}: Reference-star differential imaging (RDI) is an observing strategy that consists of observing different stars to build a model of the speckle field. This procedure uses the same setup as the one used for the target of interest. After scaling the flux/brightness, the PSF model can be used to subtract the quasi-static speckles of the science data frames, which are mainly created by the instrumentation (see Fig. \ref{rdi}). RDI has the detectability advantage under a small rotation of FoV in comparison to ADI or SDI. However, this technique requires high stability of the instrument, short duty cycles, and relatively similar magnitudes between the target and the reference star. Some contemporary ground-based HCI systems have an observing mode called ``star-hopping'' (see e.g. \citealp{2021A&A...648A..26W}). This functionality enables the AO loop to pause temporarily on a target, like the science target, and resume when the telescope points to another nearby target, such as the PSF reference star. This ensures a close match in their PSFs. Additionally, RDI can be carried out to improve performance, especially using space telescopes like Hubble space telescope (see e.g., \citealp{2006AJ....131.3109G,2009AJ....137...53S}).

    \begin{figure}[htb!]
    \centering
    \includegraphics[width=1\textwidth]{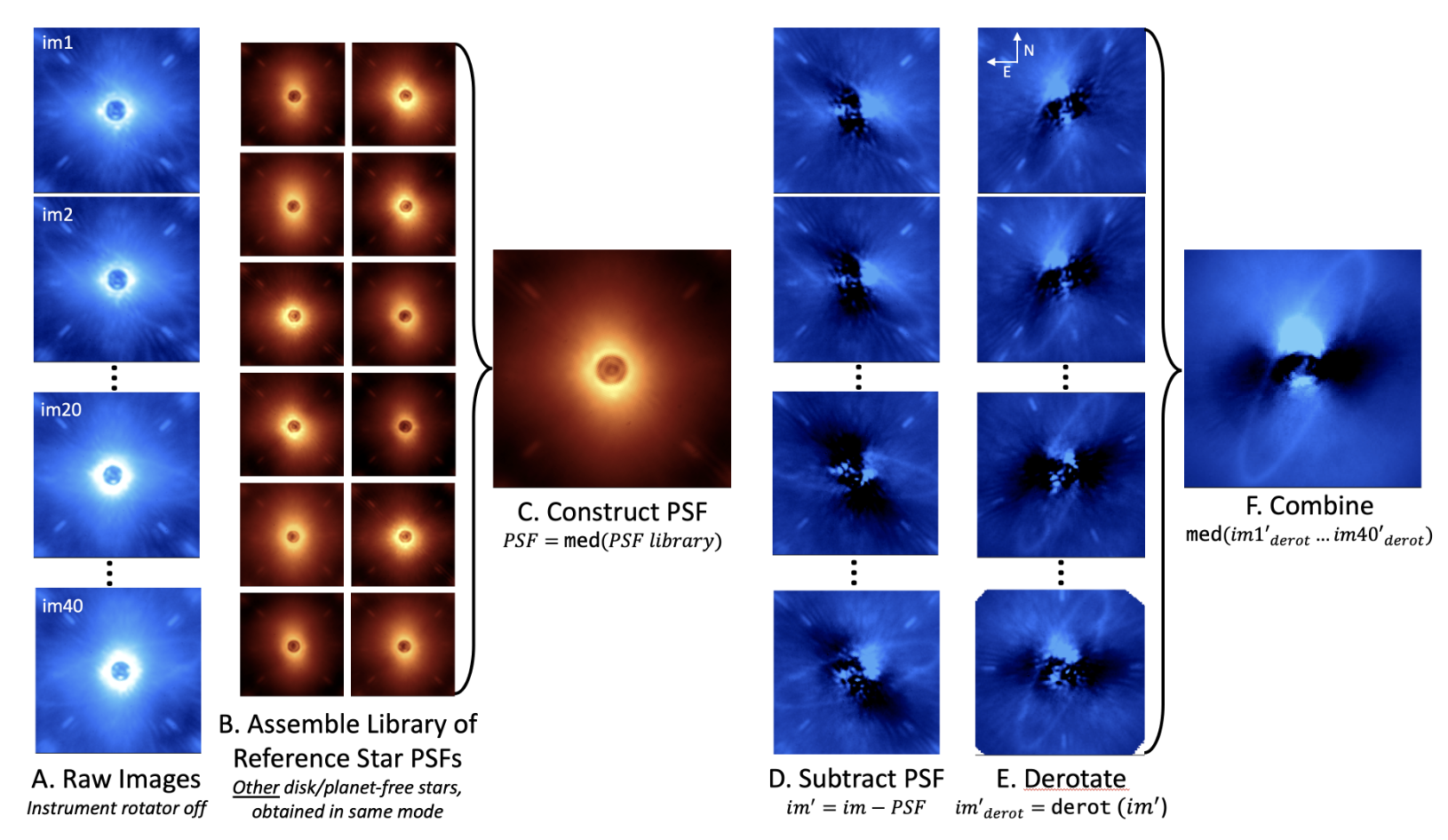}
    \caption[Schematic representation of RDI sequence]{The figure depicts the Reference Differential Imaging (RDI) process using Gemini Planet Imager H-band images of HR 4796A. RDI involves a library of images of stars without known disk or planet signals, serving as references. These reference images can be straightforwardly combined or employed to create a customized PSF for each target image in the sequence. In cases where the instrument derotator is turned off, subtracted images are rotated to a common on-sky orientation and combined, as in the ADI procedure. Credits: \citet{2023PASP..135i3001F}}
    \label{rdi}
    \end{figure}
\end{itemize}

\subsubsection{Post-processing techniques}
\label{sec:postprocessing}

Following the acquisition of data through an observing strategy executed in the instrument/coronagraphic observation, a post-processing procedure designed to enhance the contrast of faint companions/substructures will be employed. Post-processing techniques can be categorized into three primary groups: (1) speckle subtraction techniques, (2) maximum likelihood techniques, and (3) supervised machine learning techniques \citep{dahlqvistthesis}. 

\begin{figure}[htb!]
    \centering
    \includegraphics[width=0.95\textwidth]{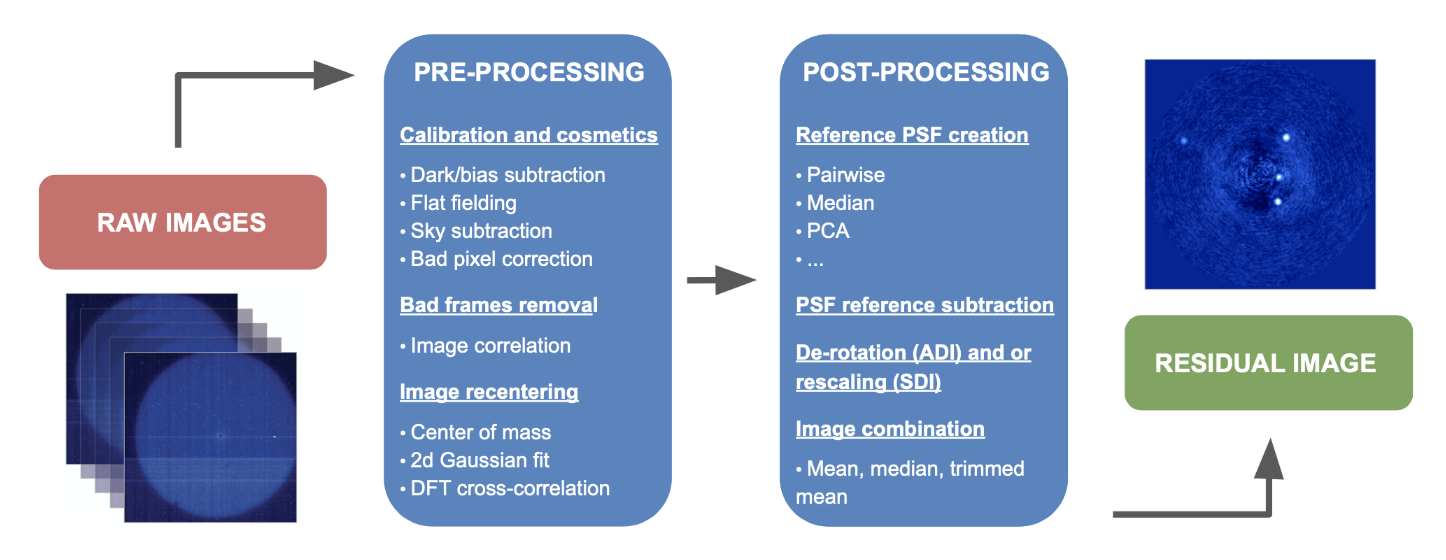}
    \caption[Typical differential imaging pipeline for speckle/PSF subtraction observing strategies coupled with post-processing techniques]{Typical differential imaging pipeline for speckle/PSF subtraction observing strategies coupled with post-processing techniques.Credits: \citet{dahlqvistthesis}.}
    \label{pipeline}
\end{figure}

Speckle subtraction techniques, often referred to as PSF-subtraction techniques, are widely utilized within the High-Contrast Imaging (HCI) community. The process typically involves the following steps outlined in Fig. \ref{pipeline}. Initially, the process involves estimating a model of the speckle field (also referred to as the reference PSF), subtracting this model from each science image, and performing de-rotation (ADI) and/or rescaling (SDI) of the PSF-subtracted images. Subsequently, these processed images are combined to produce a residual image. 

The maximum likelihood post-processing techniques involve addressing the potential distortion of the planetary signal due to the subtraction of the speckle field model. Instead of subtracting the reference PSF or decomposing it into multiple components, these techniques replace or complement reference PSF subtraction with forward modeling of the planetary companion. This approach includes a maximum likelihood-based estimation of the planetary flux. By modeling and tracking the planetary signal based on knowledge of speckle noise statistics, the expected planetary movement (depending on the observing strategy), and the impact of speckle field modeling, these techniques employ a maximum likelihood approach to provide estimated contrast values for each position within the field of view. Additionally, they may provide a detection map based on SNR.

The post-processing techniques employing supervised machine learning can be categorized into two main methods: decision trees and deep neural networks. These approaches, exemplified by the SODINN (Supervised exOplanet detection via Direct Imaging with deep Neural Networks) algorithm proposed by \citet{2018A&A...613A..71G}, redefine the task of exoplanet detection as a binary classification problem (is it a astrophysical source or a speckle?). The primary challenge lies in obtaining a substantial number of labeled samples for effective training since few planets were detected with HCI. The negative classes (speckle noise) can be generated by standard machine learning
data-augmentation techniques \citep{2018A&A...613A..71G} or through the use of Generative Adversarial Networks (GANs; \citep{2019arXiv190406155H}). On the contrary, the positive classes can be generated by injecting synthetic substellar companions at various SNR levels, using an off-axis PSF of the host star as model.

Although an exhaustive list of post-processing techniques and their variations, associated with various observing strategies could be provided and described, I will narrow the focus to speckle subtraction techniques which can be coupled with ADI. Specifically, I will briefly mention the median subtraction, Principal Component Analysis (PCA), and Locally Optimized Combination of Images (LOCI) techniques. These three were employed throughout the course of the work outlined in Chapter \ref{chap:3}:

\begin{itemize}
    \item \textbf{Median subtraction}: The most simple approach to create a speckle model is to perform a median frame with the pre-processed frames from an observing sequence. This technique, originally developed alongside ADI \citep{marois2006}, relies on the fact that a moving planet is not visible in the median image of the sequence, which can be used as an initial estimation of the speckle model. Subsequently, after subtracting the median, the residual noise averages incoherently when the images are aligned to a common north. This initial approximation, however, is less effective in the immediate vicinity of the host star. In this region, residual speckle noise is more concentrated and intense, and the companion's differential angular displacement is lower, leading to more pronounced self-subtraction. Median subtraction, thus, performs better for brighter and more angular separated sources.
    \item \textbf{Locally Optimised Combination of Images (LOCI)}: The LOCI technique, similar to median subtraction, generates speckle model frames through a combination of pre-processed frames. However, LOCI's models are more robust as they are constructed using a linear combination of selected images. In LOCI, reference frames are segmented into subsections, both radially and azimuthally. Coefficients, derived from a reference frame and through residual minimization, are calculated to weight each section across all frames. Consequently, the model frames utilize these weighted linear combinations of images, resulting in an enhanced speckle model. Since its original development by \citep{2007ApJ...660..770L}, various adaptations of LOCI have been introduced (see e.g., \citealp{2012ApJS..199....6P, 2014SPIE.9148E..0UM, 2012ApJ...755L..34C}).
        
    \item \textbf{Principal Component Analysis (PCA)}: PCA-based techniques \citep{2012ApJ...755L..28S, 2012MNRAS.427..948A} are widely utilized as a statistical tool in post-processing for HCI images. In HCI, PCA serves as the builder of speckle models for each pre-processed frame. Mathematically, PCA techniques involve the projection of images onto a lower-dimensional orthogonal basis. By vectorizing pre-processed images along their uncorrelated dimensions, such as position, rotation, wavelength, brightness, and other n-parameters, a covariance matrix can be built. If there is more variation in specific aspects of information across different images, more vectors for a complete representation are needed. These vectors, within the orthogonal basis, are termed principal components -- the number of principal components needed to express the required information is selected by the PCA user. The initial components of a PCA encapsulate more information than higher-order components. Therefore, the principal components can be defined as a truncation of the orthogonal basis, where the higher the number of principal components, the closer the information expressed to the original data. In the context of PCA for HCI applications, the conceptual idea involves identifying patterns across multiple pixels present in both the target image and a certain number of reference images. The initial principal components typically capture large-scale PSF structures, while the higher-order principal components resemble various manifestations of the speckle pattern. For the aforementioned observing strategies, it can be used as a tool to select just a part of the more correlated information, taking advantage of the fact that the companion rotates in the FoV but not the quasi-static speckles. The addition of components to the model enhances its ``aggressiveness'', increasing the probability of a well-matched PSF model. However, this also raises the likelihood of over-subtracting or self-subtracting the companion's signal.
    
\end{itemize}

\begin{figure}[htb!]
    \centering
    \includegraphics[width=0.95\textwidth]{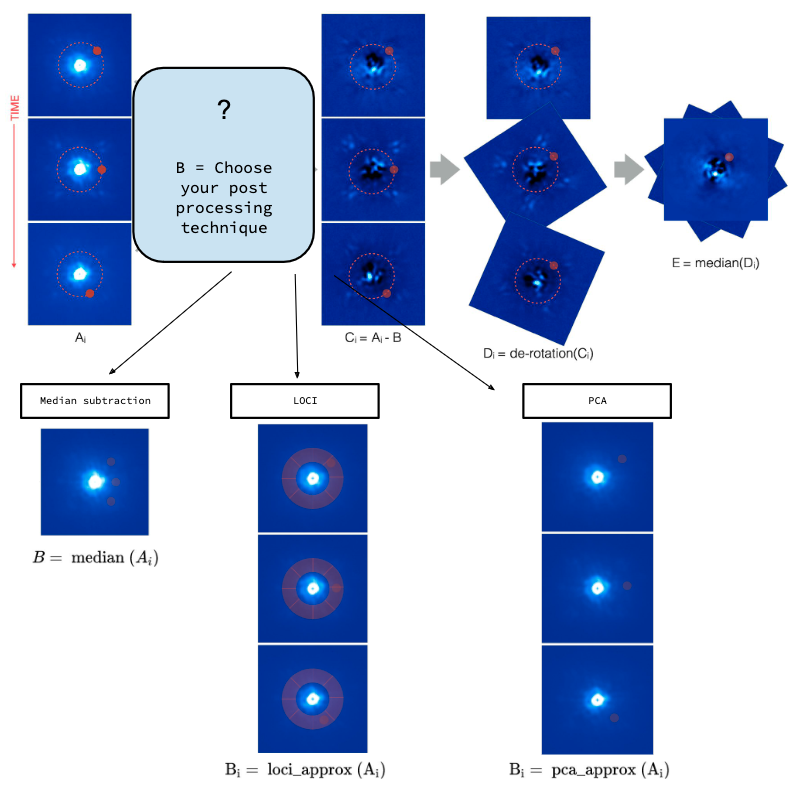}
    \caption[Diagram illustrating the ADI observing strategy with three distinct post-processing techniques for building the speckle model]{Diagram illustrating the ADI observing strategy with three distinct post-processing techniques for building the speckle model: 1) median subtraction (bottom-left), 2) LOCI (bottom-center), and 3) PCA (bottom-right). The subsequent steps follow the procedure outlined in subsection \ref{observingstrategy} and Fig. \ref{adigeneral}. Credits: Adapted from \citet{gomezgonzalezthesis}.}
    \label{adipostprocessing}
\end{figure}

A diagram illustrating the ADI observing strategy and the functionality of the three described post-processing techniques is presented in Fig. \ref{adipostprocessing}.

\subsection{High-Contrast Imaging: observing/characterizing substellar companions}
\label{chap:hci-obs}

The HCI technique, as previously mentioned, is an adequate tool to detect and study substellar companions. Usually, companions detected by HCI play a fundamental role in connecting the formation of these objects in disks (see section \ref{sec:intro/disks}) and their later evolution into mature stellar/planetary systems. This section provides a brief description of substellar companions observations, encapsulating the most recent and remarking exoplanetary observations, and highlighting how they were/can be characterized.


\subsubsection{Taxonomy of substellar objects}

Earlier in this chapter, substellar objects were classified as those lacking sufficient mass to trigger $^{1}$H mass fusion. However, within this category, further subdivisions can be made. Based on their nature and properties, substellar objects can be categorized into planetary objects, and brown dwarfs, which can exist either as free-floating objects or companions.

According to the ``classical'' definition, a BD can be defined as a deuterium-burning object, whereas a threshold between planets and BDs should occur at $\sim$ 13 $M_{Jup}$, and between a BD and star $\sim$ 80 $M_{Jup}$, with metallicities involved on the uncertainties (see e.g., \citealp{1997A&A...327.1039C, 2011ApJ...727...57S} and further references). When defining BDs and planets, is still commonly used the definition proposed by the IAU Working Group Definition \citep{2007IAUTA..26..183B}, which states:

\textit{1) Objects with true masses below the limiting mass for thermonuclear fusion of deuterium (currently calculated to be 13 Jupiter masses for objects of solar metallicity) that orbit stars or stellar remnants are ``planets'' (no matter how they formed). The minimum mass/size required for an extrasolar object to be considered a planet should be the same as that used in our Solar System.}

\textit{2) Substellar objects with true masses above the limiting mass for thermonuclear fusion of deuterium are ``brown dwarfs'', no matter how they formed nor where they are located.}

\textit{3) Free-floating objects in young star clusters with masses below the limiting mass for thermonuclear fusion of deuterium are not ``planets'', but are ``sub-brown dwarfs'' (or whatever name is most appropriate).}

However, the classical mass classification has been challenged regarding the deuterium burning criteria and formation scenario of these objects. For example, some below the deuterium burning mass limit show indicative of cloud fragmentation formation, including members of quadruple systems with inferred masses as low as 5 Jupiter masses and free-floating objects with sub-deuterium-burning masses \citep{2010ApJ...714L..84T,2013ApJ...777L..20L}. Also, some observations revealed systems with companions above the deuterium limit but orbiting at low separations and exhibiting characteristics indicating formation is a disk (see e.g. \citealp{2019A&A...624A..18Q}). Therefore, more recent criteria have been proposed.

It is possible to categorize substellar objects based on the presence and distance from a host star, and the mass ratio $q = M_{*} / M_{sub\_obj}$. By definition, protoplanets and planetary objects form within a protoplanetary disk. In contrast, a BD can either form within a circumstellar disk (without being the object formed at the center) or exist as an isolated object, forming separately from the clump that eventually becomes the closest star (refer to subsections \ref{sec:introformationscenario} and \ref{sec:intromultipleformation}). 

Given that the distributions of PPDs decline to low frequencies at approximately 300 au \citep{andrews2007}, a tentative characterization limit for the separation between substellar objects detected after the dissipation of the disk can be established. Specifically, if an object presents a semi-major axis $a_{sub\_obj} >$ 300 au from a star, is probably a brown dwarf, while lower separations can denote any of the mentioned objects, including BDs. 

Furthermore, a substellar object can be ejected from the formation system, or in the case of BDs, they can form in isolation. In a detection, if no stellar host is present or if the substellar object is not gravitationally bound to any discernible star, it can be referred to as a free-floating or field object. Conversely, if it is gravitationally bound to a star, it will be classified as a companion.  In the context of HCI, providing evidence of companionship, especially for wide-separation objects, remains challenging as it requires multi-year observations to disentangle it from a background source.

Demographic analyses of substellar objects also offer criteria for distinguishing planets from brown dwarfs. The substellar mass function, as observed in both previous and recent radial velocity surveys (see e.g. \citealp{2011IAUS..276..117S, 2019A&A...631A.125K}), reveals a local minimum around msin(i) $\sim$ 16-30 $M_{Jup}$. This minimum in the companion mass function may be linked to the primary mass, suggesting that the companion mass ratio $q$ could be a crucial discriminator --  binary companions to more massive stars with $q \geq $ 0.025 are exceptionally rare \citep{2008ApJ...679..762K,2016A&A...586A.147R}.

Following all these constrainings, \citet{2023ASPC..534..799C} proposed a modern classification of BDs and planets: planetary companions should have mass $<$ 25 $M_{Jup}$, $q <$ 0.025, and $a \lesssim$ 300 au. However, I highlight that a unified and accepted classification is lacking at the moment.

\subsubsection{Characterization of HCI companions}

Ultimately, an HCI object offers three main observable features that can be used to infer other complex properties: astrometric position, photometry, and spectra.

\begin{figure}[htb!]
    \centering
    \includegraphics[width=0.65\textwidth]{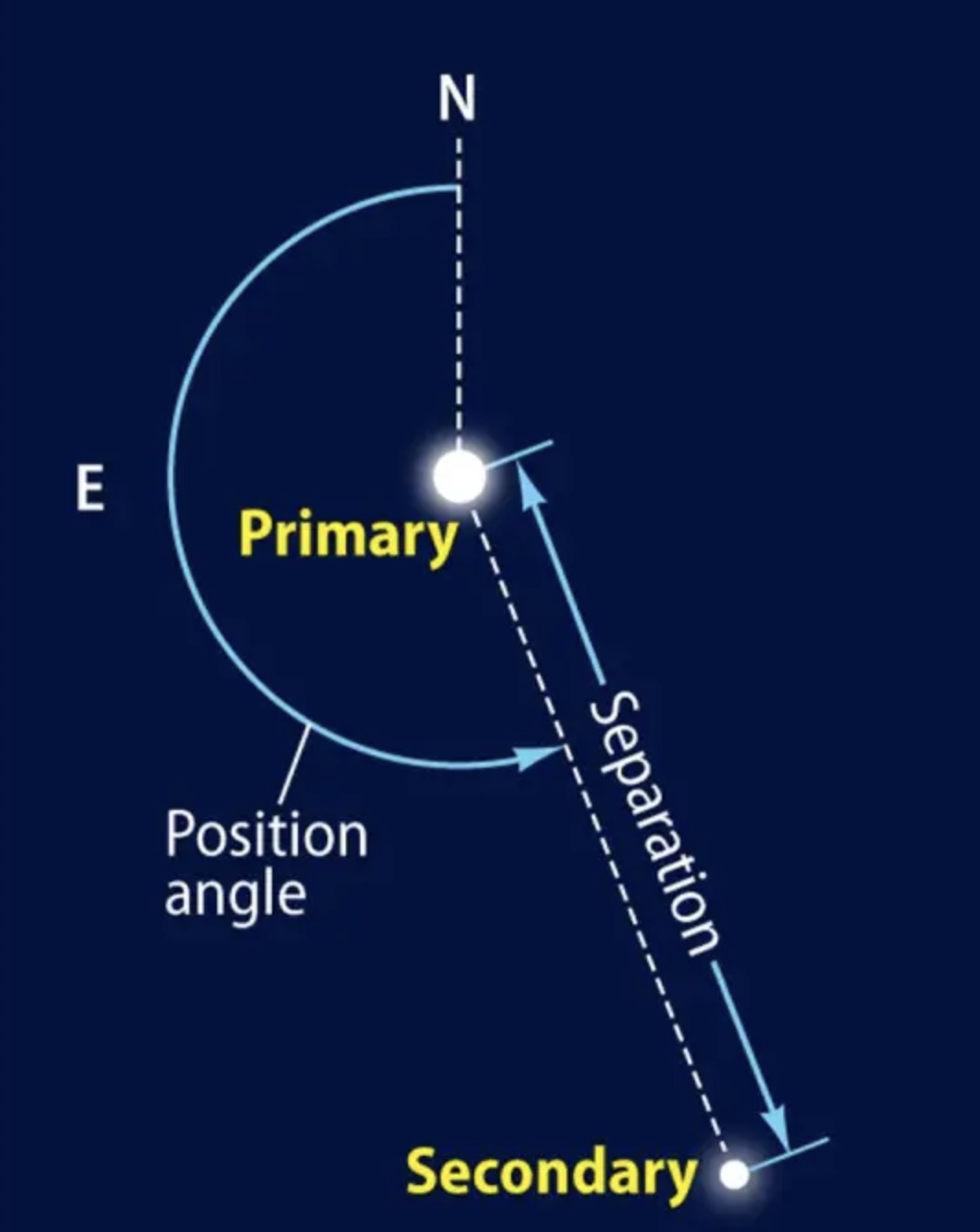}
    \caption[Representation of the conventional establishment of the position angle (PA), along with the parameter separation ($\rho$) for companion astrometry]{Representation of the conventional establishment of the position angle (PA), along with the parameter separation ($\rho$) for companion astrometry. Credits: \url{https://www.astronomy.com/science/double-star-fever/}.}
    \label{sep_pa}
\end{figure}

Firstly, astrometry is established by the position of the HCI companion, normally associated with the host star. Knowing the astrometric position of the stellar host and thus measuring the position in right ascension (RA) and declination (DEC) of the companion at the sub-pixel level, the relative position between the objects can be established. Alternatively, two variables can be measured, the separation $\rho$ and the position angle (PA). The separation can be directly measured in arcseconds ($''$) during observations, and it is then converted to physical separation, typically expressed in astronomical units (au). The position angle, also known as the parallactic angle, is the conventional angle measured from the north celestial pole through the east, firstly increasing into the direction of the right ascension. The angle ranges between 0$\degree$ and 180$\degree$ when measured east of true north, and 180$\degree$ to 360$\degree$ when measured west (see Fig. \ref{sep_pa}). In coronagraphic images, a notable challenge lies in precisely determining the image center, corresponding to the position of the central host star behind the mask. Advanced high-contrast imagers such as VLT/SPHERE and GPI provide an effective solution for accurately establishing the center using ``waffle spots''. These spots are replicas of the central PSF strategically positioned in symmetrical locations around the stellar center, facilitating the determination of the center during the pre-processing stages.

As mentioned earlier, achieving comprehensive orbital coverage for HCI companions requires observation periods on the order of years, given their substantial separations from the host star. By having astrometric points, an orbital analysis can be performed, increasing its accuracy as more percentage of the orbital is covered. To precisely characterize the orbit of a companion in relation to its stellar host, it is essential to determine the companion's velocity vectors and positions over time. The conventional mathematical representation of the orbit, typically an ellipse in relation to a reference plane, employs six Keplerian elements (refer to Fig. \ref{keporbit}):

\begin{figure}[htb!]
    \centering
    \includegraphics[width=0.85\textwidth]{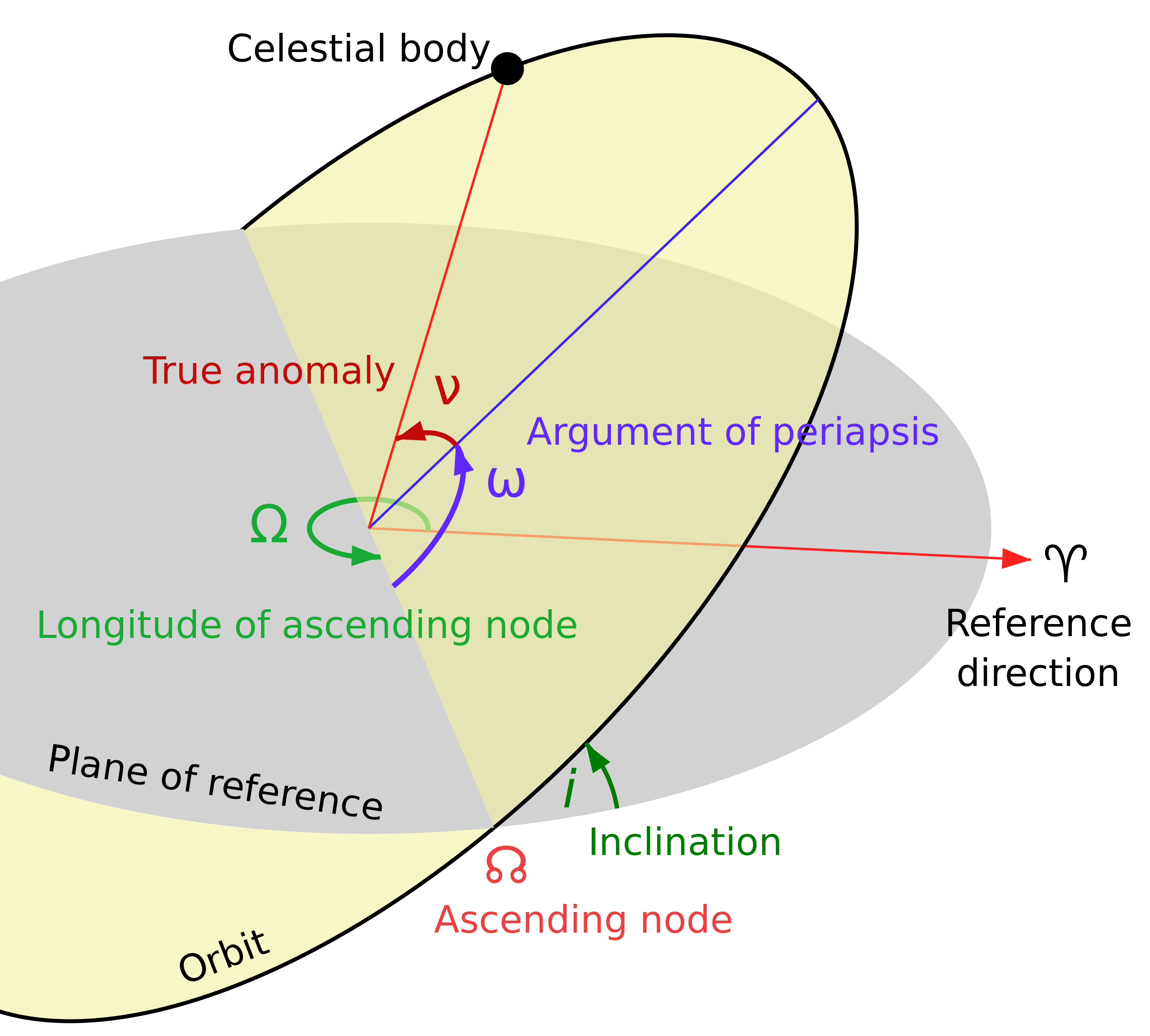}
    \caption[Representation of the six Keplerian orbital elements to uniquely describe the orbit of a companion]{Representation of the six Keplerian orbital elements to uniquely describe the orbit of a companion. Credits: By Lasunncty at the English Wikipedia, CC BY-SA 3.0, \url{https://commons.wikimedia.org/w/index.php?curid=8971052}}
    \label{keporbit}
\end{figure}

\begin{itemize}
    \item Eccentricity ($e$): Describes the shape of the ellipse, indicating how much it is elongated compared to a circle.
    \item Semi-major axis ($a$): Represents half the sum of the periapsis and apoapsis distances. For classic two-body orbits, it is the distance between the centers of the bodies, not the distance of the bodies from the center of mass.
    \item Inclination ($i$): Denotes the vertical tilt of the ellipse concerning the reference plane, measured at the ascending node. The tilt angle is measured perpendicular to the line of intersection between the orbital plane and the reference plane.
    \item Longitude of the ascending node ($\Omega$): Horizontally orients the ascending node of the ellipse (where the orbit passes from south to north through the reference plane) concerning the reference frame's vernal point. 
    \item Argument of the periapsis ($\omega$): Defines the orientation of the ellipse in the orbital plane, representing the angle measured from the ascending node to the periapsis (the closest point the satellite object comes to the primary object around which it orbits).
    \item True anomaly ($\nu$) at a specific time $t_{0}$: Specifies the position of the orbiting body along the ellipse at a specific time.
\end{itemize}

Using the orbital elements, it is also possible to determine key aspects of the orbit, such as the period, periapsis, and apoapsis. Typically, for companion objects, orbital characterization (orbit fitting) can be accomplished through various algorithms. Some of these algorithms utilize astrometric data of the companion, astrometric tracking of the star through surveys like Hipparcos and Gaia, and additional information such as radial velocity data points.

Other key information extracted from HCI observations is the photometry. Photometry is determined by measuring the contrast between the PSF of the host star and that of the companion. To achieve this, off-axis observations of the star (without the mask) are valuable throughout the observing sequence. Utilizing the flux ratio, the magnitude difference $\Delta \text{m}_{x}$ in a specific band $x$ can be computed as $\Delta \text{m}_{x} = -2.5 \text{log}_{10}(F_{x_{comp}}/F_{x_{*}})$. Thus, knowing the absolute magnitude of the host star, which represents its intrinsic brightness at a standard distance of 10 parsecs, allows for the direct determination of the absolute magnitude of the companion. 

The fluxes can also be converted to luminosities using the relationship $F_{comp}=L / 4 \pi r_*^2$, where $L$ is the luminosity, and $r$ is the distance of the object from the observer. Assuming the companion behaves like a black body, the Stefan-Boltzmann law, $F_*=\sigma T^4$, where $T$ is the temperature in Kelvin and $\sigma$ is the Stefan-Boltzmann constant ($\sigma=5.67 \times 10^{-8} \mathrm{~W} \mathrm{~m}^{-2} \mathrm{~K}^{-4}$), can be applied to determine the flux at the surface of the object. Consequently, a direct relation between luminosity and temperature can be established. As illustrated in Fig. \ref{starts}, models can utilize the temperature, indirectly inferred through photometric measurements, along with the age of the companion (assumed to be the same as the system/stellar host's age) to estimate an approximate mass. However, precise measurements of photometry and the stellar age are essential for accurate estimations. Furthermore, the accuracy of the estimation heavily relies on the model used and the assumed formation origin, whether it is a \textit{hot start} (associated with objects formed from gravitational instability) or a \textit{cold start} (objects formed from core accretion). 

Similar to astrometry, observing variations in photometric measurements of substellar objects over several years provides significant insights. Variability in these objects is often associated with factors such as rotation and the presence of clouds. Objects with rapid rotation and uneven atmospheric structures, featuring patchy thin, and thick clouds, tend to exhibit noticeable changes in brightness during different rotational phases. Large-scale surveys of brown dwarfs have consistently identified common variability patterns, with amplitudes reaching up to approximately 20\% in the near-infrared \citep{2014ApJ...793...75R,2014ApJ...797..120R,2014A&A...566A.111W,2015ApJ...799..154M}. Young, directly imaged giant planets are expected to display even more pronounced variability. The extent and nature of variability of substellar objects, however, is linked to their spectral types. For a list of references concerning variability in giant planets and BDs, encompassing both field objects and companions with various spectral types, please refer to \citet{2023ASPC..534..799C}.

\begin{figure}[htb!]
    \centering
    \includegraphics[width=0.95\textwidth]{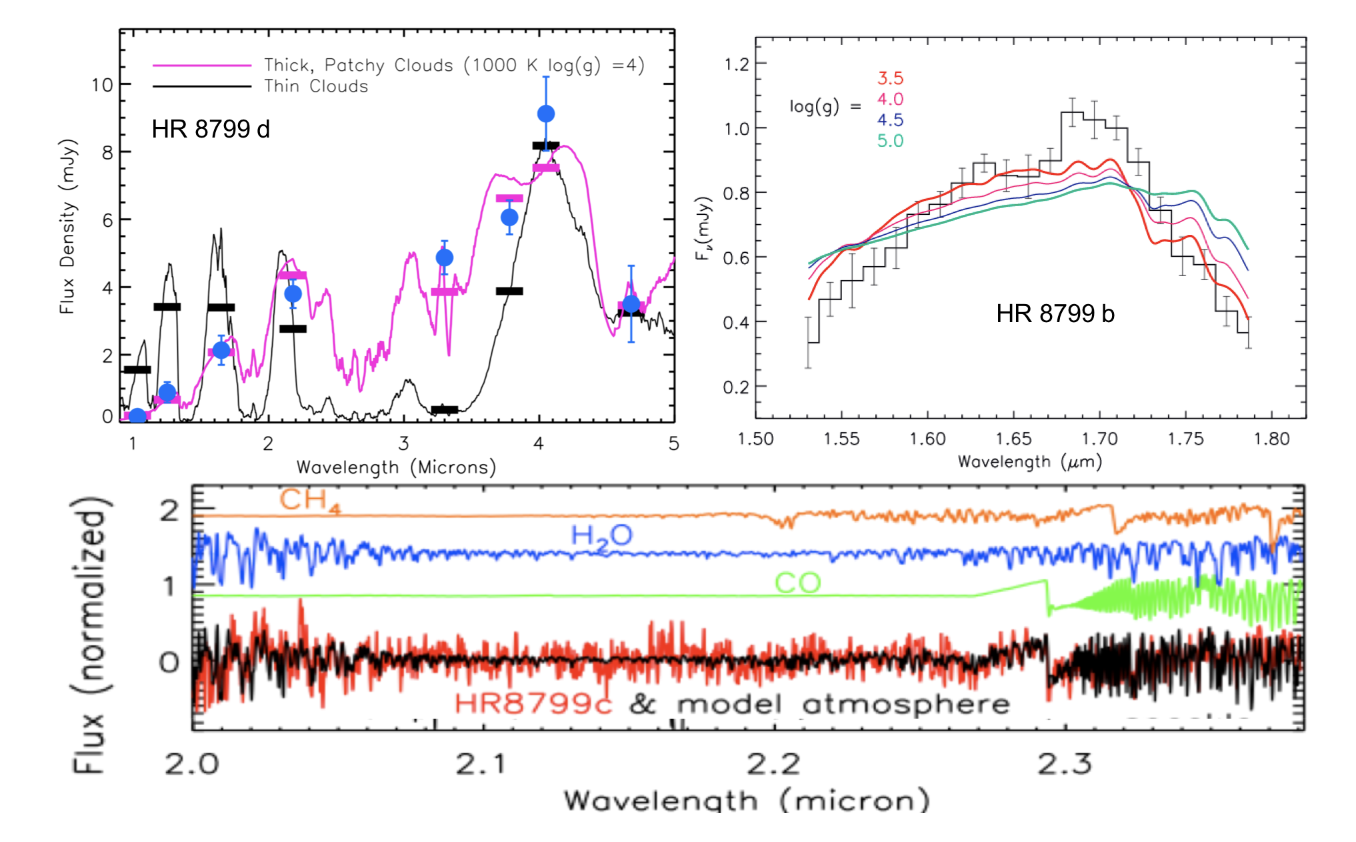}
    \caption[Demonstration of key properties inferred from direct imaging observations]{Demonstration of key properties inferred from direct imaging observations. Top-left: Fit to HR8799d 1–5 $\mu$m photometry showing that models with thick clouds better reproduce the planet data (adapted from \citealp{2011ApJ...729..128C}). Top-right: Keck/OSIRIS low-resolution H-band spectrum for HR8799b: a sharply-peaked H-band spectrum is a signpost of low surface gravity \citep{2011ApJ...733...65B}. Bottom: Keck/OSIRIS medium-resolution HR8799 c spectrum, revealing lines of CO and H$_{2}$O \citep{2013Sci...339.1398K}. Credits: \citet{2023ASPC..534..799C}.}
    \label{HCIspectra}
\end{figure}

Multi-band photometric bands HCI observations can provide hints on the spectral type of substellar objects. While most spectra to date for HCI planets were obtained from low-resolution NIR observations (R $<$ 100), higher resolution powers (R = 2000-6000) are offered in some modern IFS instruments in 8-meter class telescopes (e.g, Gemini/NIFS, VLT/MUSE, VLT/SINFONI, and Keck/OSIRIS). The resolution of the obtained spectra gives how well-resolved, less-spaced, are the points in the spectra. Thus, more resolvable power offers better feature analysis in the HCI object, such as molecular bands/chemical abundances, rotational velocity, presence of clouds, and surface gravity (see Fig. \ref{HCIspectra}). 

The category of substellar objects encompasses spectral types from late (redder) M to Y \citep{1999ApJ...519..802K,2011ApJ...743...50C}. The Fig. \ref{bdspectra} illustrates some of the main features of substellar objects along the spectral sequence from M to T. Key properties of each spectral class are as follows:

\begin{figure}[htb!]
    \centering
    \includegraphics[width=0.85\textwidth]{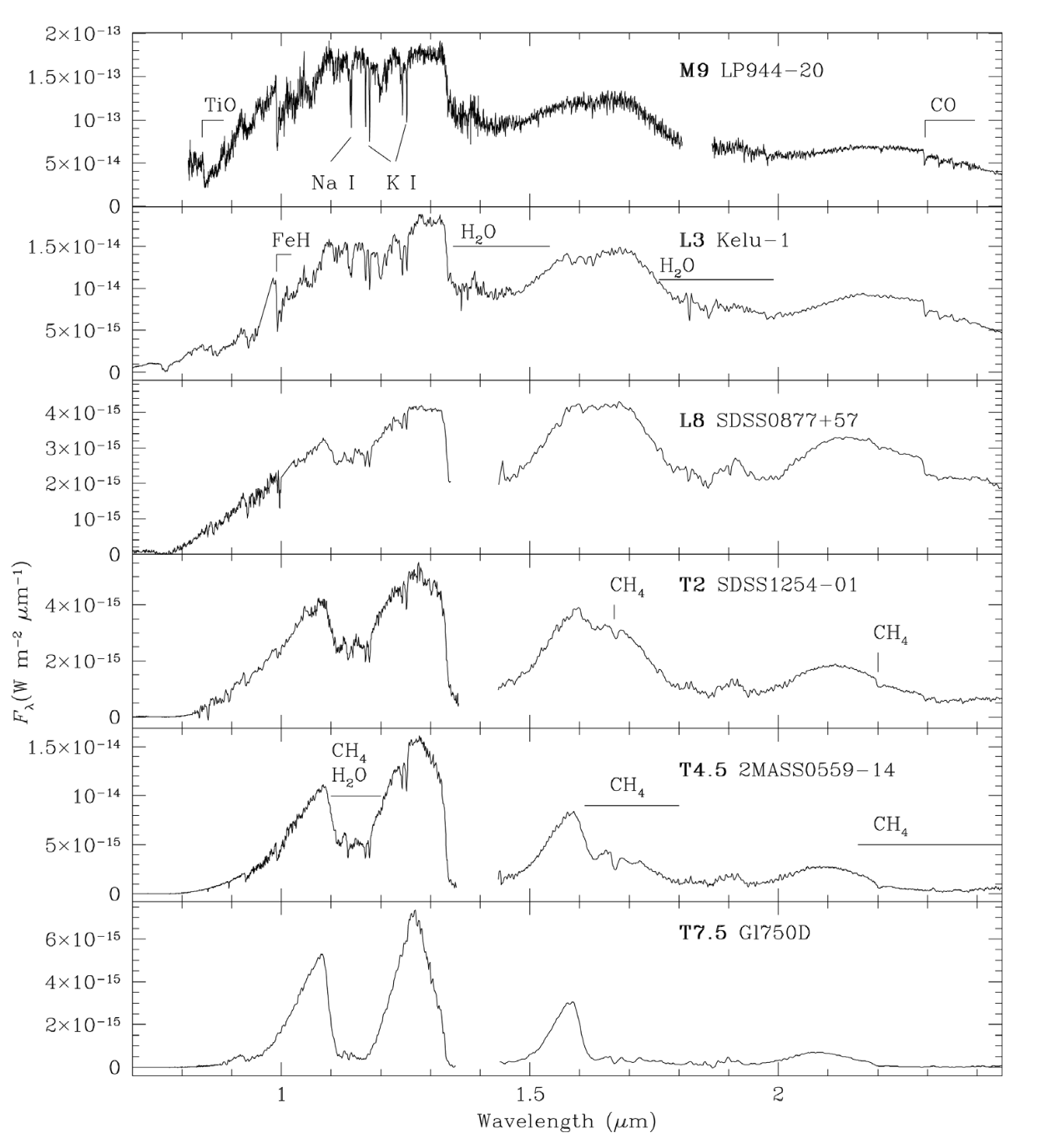}
    \caption[Spectra of ultracool dwarfs from M9 to T7.5]{Spectra of ultracool dwarfs from M9 to T7.5. Credits: \citet{bailey2014}.}
    \label{bdspectra}
\end{figure}

\begin{itemize}
    \item \textbf{M-dwarfs}: Serving as a transitional link between stellar objects and brown dwarfs, the latest types in the M sequence include young objects classified as brown dwarfs. The spectrum is characterized by the presence of TiO and VO. TiO bands increase in strength up to spectral type M6, and VO becomes prominent in the latest types \citep{bailey2014}. Broad absorptions due to H$_{2}$O are found around 1.4 and 1.9 $\mu$m, particularly in later spectral types. Other molecules with strong absorption include FeH, CrH, and MgH.
    
    \item \textbf{L-dwarfs}: TiO and vanadium monoxide diminish in L-dwarfs, replaced by bands of CrH, FeH, and alkaline metals in the optical. In the near-infrared, bands of CO, CrH, and FeH are strong until the mid-L range, after which they become weaker. The L-dwarfs present dusty clouds of species such as enstatite, forsterite, spinel, and solid iron condensing in the upper layers of the atmosphere (see e.g. \citealp{2001ApJ...556..357A}).
    
    \item \textbf{T-dwarfs}: Characterized by methane (CH$_{4}$) absorption features in the near-IR region (1-2.5 $\mu$m) and water absorption bands, T-dwarfs exhibit blue shifts in the mag/color diagram due to methane absorptions at 1.6 and 2.2 $\mu$m (H and K bands). They bear a closer resemblance to the giant planets in the solar system. 
    
    \item \textbf{Y-dwarfs}: Y dwarfs are the coldest class of brown dwarfs (T $\lesssim$ 450K), which can present a rich collection of molecular bands in their atmospheres such as CH$_{4}$, CO$_{2}$, H$_{2}$O, and NH$_{4}$. In the specific case of water, a cutoff of molecular absorption is expected to occur at temperatures lower than $\sim$ 500K, where no water clouds will be formed (see e.g. \citealp{2001ApJ...556..872A,2003ApJ...596..587B}).

\end{itemize}

Therefore, robust characterization of HCI substellar objects can be achieved through astrometric, photometric, and spectroscopic observations. Continuous monitoring and the utilization of multi-wavelength observations, facilitated by IFS instruments, contribute to a more comprehensive analysis and yield detailed results.



\subsubsection{HCI exoplanetary detections}
Since the first detection of a planetary-mass companion via HCI \citep{2005A&A...438L..25C}, fewer than 40 additional detections have been confirmed (see Fig. \ref{diplanets}, top-panel). These detections highlight the challenges associated with the technique, as discussed in section \ref{sec:instrumentation}. Furthermore, confirming their association with the stellar host and ruling out the possibility of background sources necessitates extensive observation periods, often spanning several years due to their considerable separation. However, the new generations of high contrast imagers, improving contrast power, may provide a notable increase in the number of objects in the upcoming years. Following, a non-exhaustive list of emblematic systems analyzed through HCI is presented:

\begin{figure}[htb!]
    \centering
    \includegraphics[width=0.75\textwidth]{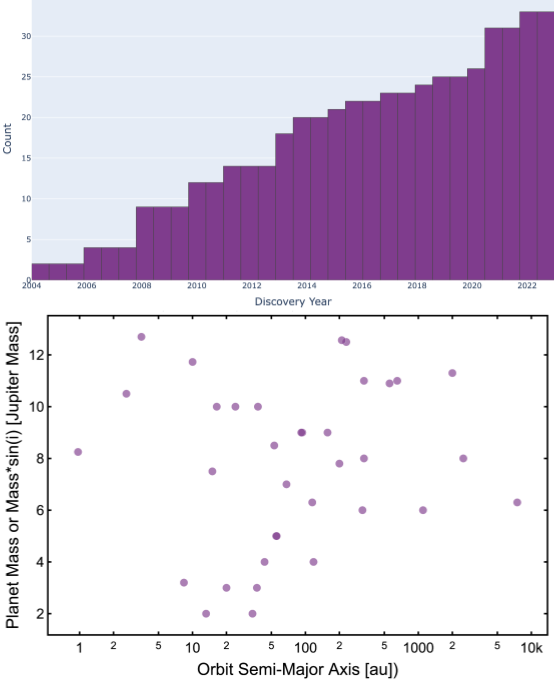}
    \caption[Exoplanets discovered by HCI up to 25-02-2024]{Exoplanets discovered by HCI up to 25-02-2024. They were selected through the ``classical'' mass threshold -- objects with less than 13 $M_{Jup}$. Top panel: cumulative distribution of discoveries throughout years. Bottom-panel: Mass of the companions in Jupiter masses vs log of semi-major axis in au. Credits: NASA Exoplanet Archive: \url{exoplanetarchive.ipac.caltech.edu}.}
    \label{diplanets}
\end{figure}


\begin{itemize}
    \item \textbf{2M1207}: 2M1207 b was the first exoplanet detected via HCI \citep{2005A&A...438L..25C}. It was detected using the instrument NACO/VLT. The host is an M8 type star part of the TW Hydra association (age of $\sim$ 10 Myr; \citealt{2006A&A...459..511B,2015MNRAS.454..593B}), at a distance of 64.5 $\pm$ 0.4 pc \citep{gaiadr3}. The companion has a separation of 0.70'' (55 au). According to HCI observations and evolutionary models, the estimated mass of the companion is approximately 5-6 $M_{Jup}$, and it exhibits an effective temperature of 1200-1300 K \citep{2023ApJ...949L..36L}.  The characteristics of this object are unique, as the formation of the planet within the protoplanetary disk is highly improbable due to its considerable distance from the host star. It is often regarded more as a binary system of low-mass objects rather than a conventional planetary system \citep{2007ApJ...657.1064M}.
    
    \item \textbf{HR8799}: HR8799 is the first multiplanetary system discovered by HCI \citep{2008Sci...322.1348M,2010Natur.468.1080M}. HR8799 has 4 planets, which were discovered through Keck and Gemini telescopes. HR8799 is a young system ($\sim$42 Myr; \citealt{2010ApJ...716..417H,2011ApJ...732...61Z,2015MNRAS.454..593B}) at a distance of 40.88 $\pm$ 0.08 pc \citep{2020yCat.1350....0G}. The host star has a mass of ${1.47}_{-0.17}^{+0.12}$ M$_{\odot}$ \citep{2022AJ....163...52S}. This particular system stands out as one of the most frequently studied subject using HCI, serving as a standard reference for analysis of planetary atmospheres, the dynamics inherent in multi-planetary systems, and the mechanisms guiding their formation. The system also offers an outstanding scenario to study the formation and dynamics of these planets since they orbit at wide separations (the closest planet has a separation of $\sim$390 mas $=$ 15.94 au). The masses of the companions, inferred through evolutionary models and dynamical studies, range between 5-10 $M_{Jup}$ \citep{2010Natur.468.1080M,2011ApJ...729..128C,2012ApJ...755...38S,2018AJ....156..192W,2021ApJ...915L..16B}. 
    %
    
    \item \textbf{Fomalhaut}: Fomalhaut is a young system (approximately 500 Myr) at a distance of 7.7 pc, composed of a bright A3V star surrounded by a quasi-face-on debris disk \citep{2006AJ....132..161G,2012ApJ...754L..20M,2019AJ....158...13N}. \citet{2008Sci...322.1345K}, using Hubble Space Telescope (HST) observations, revealed a co-moving source in visible light, quasi-nested within the dust belt and at a separation of approximately 119 au from the star. The source was then classified as a planetary object. Fomalhaut b is the lowest mass planet ever imaged (3 $M_{Jup}$ or lower; \citealp{2009ApJ...693..734C,2009ApJ...700.1647M}). However, subsequent studies have cast doubt on its classification as a planet. Several factors contribute to this skepticism, including the absence of infrared emission proportional to its visible brightness, thus casting the lack of thermal emission and just presenting reflected light \citep{2013ApJ...777L...6C}, its high eccentric orbit crossing the ring inconsistent with the stability expected of a planetary object \citep{2014A&A...561A..43B}, and the failure to detect Fomalhaut b in multiple epochs of Spitzer imaging \citep{2009ApJ...700.1647M,2015A&A...574A.120J}. Moreover, later observations with HST revealed the object expanding in size while decreasing in brightness \citep{2020PNAS..117.9712G}. The prevailing explanation for the nature of Fomalhaut b leans toward it being a slowly dissipating, expanding remnant resulting from the collision of two planetesimals \citep{2015ApJ...802L..20L,2020PNAS..117.9712G}. Despite the controversy about the planetary nature of Fomalhaut b, the system is still investigated concerning planets using HCI, where indicative of other objects has been spotted \citep{2024AJ....167...26Y}.
    
    \item \textbf{$\beta$ Pictoris}: $\beta$ Pic is an intriguing system composed of a young A6V-type star, an edge-on debris disk, and two planets. The system has an age of $\sim$18.5 $\pm$ 2 Myr \citep{2020A&A...642A.179M} and is situated at a distance of 19.45 $\pm$ 0.05 pc \citep{2007A&A...474..653V}. $\beta$ Pic b was first imaged by \cite{2009A&A...493L..21L,2010Sci...329...57L}, using VLT/NaCo observations. The planet has a mass of $\sim$9-10 $M_{Jup}$ and a semi-major axis orbit of approximately 10 au \citep{2020A&A...642L...2N,2021AJ....161..179B}. More recently, a closer planet was discovered with European Southern Observatory/High Accuracy Radial Velocity Planet Searcher (HARPS) observations, using the radial velocity technique \citep{2019NatAs...3.1135L}. It was later confirmed with VLTI/GRAVITY observations \citep{2020A&A...642L...2N}. $\beta$ Pic c has a dynamical mass of $\sim$ 8 $M_{Jup}$ with a semi-major axis of $\sim$ 3 au \citep{2020A&A...642L...2N, 2021AJ....161..179B}. Both planets are on eccentric orbits ($\sim$0.3 and $\sim$0.121 for $\beta$ Pic b and $\beta$ Pic c, respectively; \citealp{2021AJ....161..179B}). Overall, the $\beta$ Pic system has proven to be an interesting subject for investigating its dynamics and planetary formation scenarios.
    
    
    \item \textbf{AF Leporis}: AF Lep constitutes a system featuring an F8V star \citep{2006AJ....132..161G} located at a distance of 26.8 pc \citep{2023A&A...674A...1G}, accompanied by a debris disk located at $\sim$54 $\pm$ 6 au from the star \citep{2021MNRAS.502.5390P}, and a Jupiter-like planet. The system has an age of 24 $\pm$ 3 Myr \citep{2015MNRAS.454..593B}. AF Lep b drew attention through its remarkable discovery in three independent and simultaneous studies. Anomalies in the astrometric accelerations observed in the HIPPARCOS-Gaia data (HGCA; \citealp{2021ApJS..254...42B}) hinted at the potential presence of an undetected orbiting substellar object. Detected through HCI with VLT/SPHERE IRDIS/IFS in Y, J, H, and K bands \citep{mesa, 2023A&A...672A..94D}, and Keck/NIRC2 in the L' band \citep{2023ApJ...950L..19F}, AF Lep b showcased a mass range of 3-7 $M_{Jup}$, deduced from photometric measurements and astrometric fitting. The planet is situated at a separation of $\sim$340 mas = 9 au, possesses an eccentricity ranging from 0.2 to 0.5, and its orbit aligns with the stellar inclination. Its distance from the innermost region of the debris belt suggests the potential existence of other undetected planets within the system. AF Lep b is one of the least massive planets imaged thus far, exhibiting intriguing and promising features for further imaging and spectral characterization.
    
    \item \textbf{51 Eri}: 51 Eri is a young system (10-20 Myr; \citep{2015Sci...350...64M,2022MNRAS.511.6179L}) positioned at a distance of 29.90$\pm$0.06 pc \citep{2021A&A...649A...1G}. This system revolves around a F0IV star, a member of the $\beta$ Pictoris moving group \citep{2001ApJ...562L..87Z,2015MNRAS.454..593B}, and is known for hosting the planet 51 Eridani b. The star is part of a hierarchical triple system, accompanied by the M-dwarf binary GJ 3305AB, situated approximately 2000 au away from 51 Eri \citep{2006AJ....131.1730F,2007A&A...472..321K}. Additionally, the system possesses two debris belts, located at 5.5 au and 82 au \citep{2014ApJS..212...10P,2014A&A...565A..68R}. The discovery of 51 Eri was initially made by \citet{2015Sci...350...64M} through GPI and Keck/NIRC2 J, H, and L${p}$ observations, and its companionship was later confirmed by \citet{2015ApJ...814L...3D}. The planet orbits at a separation of 449 mas = 13 au. Various studies have attempted to constrain its mass, indicating a range between 2-12 $M{Jup}$, with the actual value highly dependent on the assumed formation scenario \citep{2017A&A...603A..57S,2020AJ....159....1D,2021MNRAS.507.2094M,2022MNRAS.509.4411D}. Interestingly, the planet appears to not be coplanar with the distant binary \citep{2015ApJ...813L..11M}. Spectral studies reveal an effective temperature of up to $\sim$800 K (see e.g. \citealp{2023A&A...673A..98B}). The planet also shows strong methane spectral signatures, an unusual feature in most directly imaged exoplanets. 


    
    \item \textbf{PDS 70}: PDS 70 is a young system ($\sim$5.4 Myr), with a K7-type pre-main sequence member of the Upper Centaurus-Lupus group \citep{2006A&A...458..317R,2016MNRAS.461..794P} at a distance of 113.43 $\pm$ 0.52 pc \citep{2016A&A...595A...1G,2018A&A...616A...1G}. The system has two protoplanets located within the cavity, PDS 70 b and c. PDS 70 b was first discovered with SPHERE, NaCo, and Gemini/NICI, H to L' bands' observations \citep{2018A&A...617A..44K}. Later, it was imaged with SPHERE K band and Y-H spectroscopic data, imposing an improved characterization \citep{2018A&A...617L...2M}. Its H$\alpha$ accreting nature was later revealed \citep{2018ApJ...863L...8W}. PDS 70 c was later discovered with VLT/MUSE observations in the H$\alpha$ line \citep{2019NatAs...3..749H}, also showing to be an actively accreting protoplanet. PDS 70 b and c have separations of $\sim$180 mas = 20 au and $\sim$240 mas = 34 au, respectively \citep{2020AJ....159..263W}. Subsequently, ALMA sub-mm observations have revealed the first circumplanetary disk detected co-locating with a protoplanet -- in this case, PDS 70 c \citep{2019ApJ...879L..25I,2021ApJ...916L...2B}. The masses of the two planets are still uncertain, although both planets are likely lighter than 10 $M_{Jup}$ to ensure dynamical stability \citep{2021AJ....161..148W} and a non-eccentric outer disk \citep{2019ApJ...884L..41B}. Masses inferred from SED modeling remain inconclusive but suggest planet masses between 1 and a few $M_{Jup}$ \citep{2020A&A...644A..13S}. Consequently, PDS 70 remains one of the most intriguing and extensively studied systems, holding the promise of unveiling crucial properties of its protoplanets and disk.

\end{itemize}

\begin{figure}[htb!]
    \centering
    \includegraphics[width=0.95\textwidth]{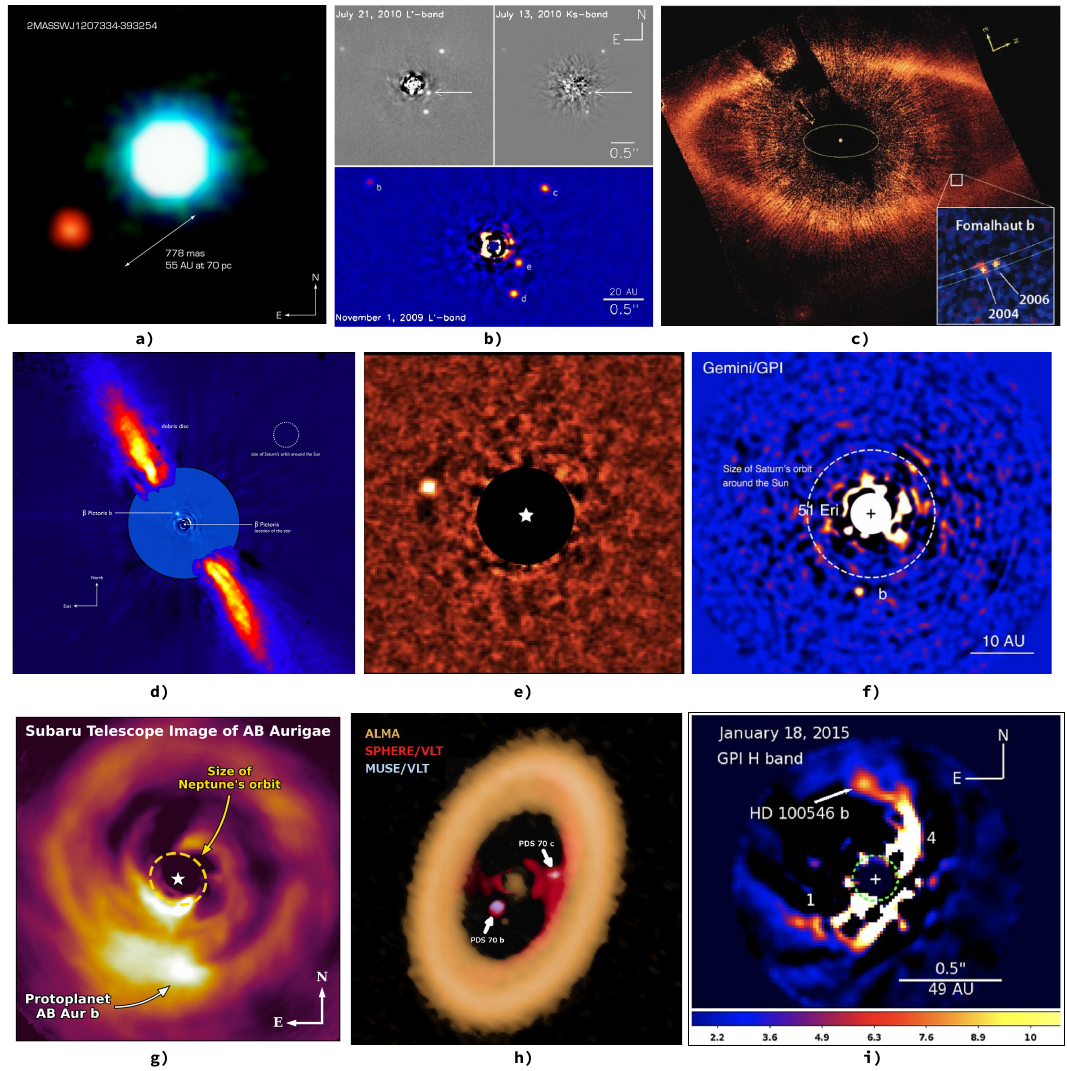}
    \caption[Mosaic of remarkable companions and protoplanet candidates detected via HCI]{Mosaic of remarkable companions and protoplanet candidates detected via HCI. \textit{Panel a}: 2M1207. Credits: \citet{chauvin2004}. \textit{Panel b}: HR8799. Credits: \citet{2010Natur.468.1080M}. \textit{Panel c}: Fomalhaut. Credits: \citet{2008Sci...322.1345K}. \textit{Panel d}: $\beta$ Pic. Credits: ESO/A.-M. Lagrange et al. \textit{Panel e}: AF Lep. Credits: ESO/Mesa, De Rosa et al. \textit{Panel f}: 51 Eri. Credits: J. Rameau, University of Montreal / C. Marois, Herzberg Institute of Astrophysics. \textit{Panel g}: AB Aur. Credits: \citet{2022NatAs...6..751C}. \textit{Panel h}: PDS 70. Credits: ALMA (ESO/NAOJ/NRAO)/A. Isella/ESO. \textit{Panel i}: HD 100546. Credits: \citet{2015ApJ...814L..27C}.}
    \label{nicepics}
\end{figure}

The systems described above are depicted in Fig. \ref{nicepics}.

\clearpage
\section{Outline of this thesis}
\coffeestainC{1}{1}{180}{0}{-5 mm}

This dissertation describes the results obtained during my PhD research program. In this thesis, I aim to contribute with new perspectives on how (multi)stellar systems form in eruptive environments, along with the detection and characterization of multiple (sub)stellar objects in later stages. To achieve this, I conduct a detailed analysis using multiple wavelengths and exploring various stages of (sub)stellar formation.

The scientific targets under investigation include the eruptive binary HBC 494, analyzed through ALMA band 6 (1.6 mm) observations, and the $\eta$ Tel system, consisting of a giant star and a brown dwarf companion, examined through near-infrared high-contrast imaging observations, more specifically with VLT/SPHERE IRDIS in H band.

The following chapters are organized as follows: In Chapter ~\ref{chap:2}, I explore the properties of HBC 494, regarding its dust, gas, and dynamics. This work was published and can be found in \citet{nogueira2023}. In Chapter ~\ref{chap:3}, I provide an intricate study regarding the astrometry, photometry, and orbital properties of $\eta$ Tel B system, also providing a detailed analysis of its surroundings for looking for satellites or circumplanetary disks. This work is under corrections following the reviewer's comments/suggestions. Finally, in Chapter~\ref{chap:conclusions}, I summarize the conclusions and final remarks of the projects. I discuss the immediate future steps to keep investigating the (sub)stellar formation and evolution of these sources.

  \chapter[ALMA resolves the binary HBC 494]{Resolving the Binary Components of the Outbursting Protostar HBC 494 with ALMA}
\label{chap:2}
\textit{The content of this chapter is published in the journal Monthly Notices of the Royal Astronomical Society (MNRA), see \citet{nogueira2023} - \url{https://ui.adsabs.harvard.edu/abs/2023MNRAS.523.4970N/abstract}.}

\section{Introduction}

\label{sec:intro}
During their evolution, young stellar objects (YSOs) dissipate their envelopes while feeding their developing protostars through accretion via a disk. However, YSOs are underluminous compared to the luminosity and accretion rates expected from steady disk accretion. This discrepancy has been established as the ``luminosity problem" \citep{kenyon1990, evans2009}. One potential solution to the luminosity problem is that young stars undergo episodes of high accretion interspersed by quiescent phases. During the episodes of enhanced accretion, large amounts of material are accreted very quickly (e.g., \citealp{dunham2012}). Such phenomena have been observed in low-mass pre-main sequence stars like FU Orionis and EX Lupi, the prototypes of the two subclasses that display the episodic accretion phenomenon. 

FU Orionis type stars, a.k.a. FUors, are pre-main sequence low-mass stars that show variability in both luminosity and spectral type due to variation in the accretion mass rate, on a short timescale \citep{herbig1966}. Their optical brightness can dramatically increase due to enhanced mass accretion by more than five magnitudes over a few months. After this ``outburst'', FUors can remain in this active state for decades. This occurrence is considered episodic and suspected to be common at the early stages of star formation. Throughout an outburst, the star can accrete $\sim$ 0.01 M$_{\odot}$ of material, roughly the mass of a typical T-Tauri disk \citep{andrews2005}. The bolometric luminosity of the FUors during the outburst is 100--300 $L_{\odot}$ and the accretion rate is between $10^{-6}$ and $10^{-4}$ M$_{\odot}$ yr$^{-1}$ \citep{audard2014}. EX Lupi type stars, a.k.a. EXors, are a scaled-down version of the FUors, with shorter and less intense outbursts \citep{herbig2007}. The EXors enhanced accretion stage can last months to years, with accretion rates ranging from 10$^{-7}$ to 10$^{-6}$ M$_{\odot}$ yr$^{-1}$, which are the order up to 5 magnitudes brighter than quiescent periods. The episodic recurrence is also on the order of years (e.g., \citealp{2018A&A...614A...9J, 2022ApJ...929..129G}).

The mechanisms producing the eruptions in systems like FUors and EXors have yet to be understood \citep{cieza2018}. Several different triggers have been proposed to explain this phenomenon, such as disk fragmentation \citep{vorobyov2005, vorobyov2015, zhu2012}, coupling of gravitational and magneto-rotational instability (MRI) \citep{armitage2001, zhu2009}, and tidal interaction between the disk and a companion \citep{bonnell1992,lodato2004, borchert2022}. In terms of evolution, a scenario where FUors are understood as an earlier phase followed by an EXor phase could explain some of the observed properties of these systems \citep{cieza2018}. EXors have less prominent (or lack of) outflows, and smaller masses/luminosities than FUors since the mass loss, accretion, and outbursts will significantly remove the gas and dust material during their evolutionary stages.  

If most stars undergo FUor/EXor-like episodic accretions during their evolution, imaging the circumstellar disks of FUors at sub-mm/mm wavelengths can inform or constrain the underlying outburst mechanisms. For class 0/I objects (e.g., those still accreting from their circumstellar disks and envelopes), the massive disk could be expected to be prone to be gravitationally unstable, which can, consequently, trigger the MRI and/or disk fragmentation. Additionally, close encounters of stars can shape both gas and dust disk morphology \citep{2019MNRAS.483.4114C}. Inferring which scenario plays in each disk requests an analysis that strongly depends on how well-resolved and which features one can extract from the observations. Consequently, when eruptive disks give a hint of irregular morphology or kinematics, follow-up observations at higher resolution are needed to resolve the substructures and infer the nature of the system.

HBC 494 is a Class I protostar, located in the Orion molecular cloud at a distance of 414$\pm$7 pc \citep{2007A&A...474..515M}, and has been classified as a FUor\footnote{The FU Ori classification is still controversial and has been contested by \citet{connelley2018}.}. This young eruptive object (also called Reipurth 50) was discovered during an optical survey and described as a bright, conical, and large nebula reflecting light from a 1.5 arcmin away from a 250 L${\odot}$ infrared source. Both were located at the southern part of the L1641 cloud in Orion \citep{reipurth1985,reipurth1986}. This nebula was claimed to appear after 1955, and its first detailed study showed high variability between 1982 and 1985, a consequence of a primordial IR source variability. Posterior studies have also confirmed its variability. For example, \citet{chiang2015} reported a dramatic brightening (thus clearing of part of the nebula) that occurred between 2006 and 2014. These events can be explained by an episode of outflow coming from HBC 494. More recently, \citet{2019A&A...631A..30P}, using archival photometry data along with Herschel and Spitzer spectra presented the detection of several molecular lines and the spectral energy distribution (SED) of HBC 494. The SED presented strong continuum emission in the mid and far-infrared, which is indicative of envelope emission. Such violent outflows were observed and described in \citet{ruizrodriguez2017}.

ALMA Cycle-2 observations in the millimeter continuum ($\sim$0.25\arcsec angular resolution) show that the disk is elongated with an apparent asymmetry \citep{cieza2018}, indicating the presence of an unresolved structure or a secondary disk. In this work, we present ALMA Cycle 4 observations at $\sim$0.03\arcsec angular resolution that reveal the binary nature of the HBC 494 system. Only Cycle 4 data were used in this work. The paper is structured as follows. The ALMA observations and data reduction are described in section \ref{sec:obs}. The results from the continuum and line analysis are described in section \ref{sec:results}. The discussion is presented in section \ref{sec:dis}, while we conclude with a summary of our results in section \ref{sec:sum}. 

\section{Observations and data reduction}
\label{sec:obs}

HBC 494 has been observed during the ALMA Cycle-4 in band 6 (program 2016.1.00630.S: PI Zurlo) for two nights, one on the 9\textsuperscript{th} October 2016 and the other on the 26\textsuperscript{th} September 2017 (see table \ref{table}). On the first night, the short baseline configuration was acquired, with a precipitable water vapor of 0.42 mm. The ALMA configuration of antennas was composed of 40 12-m antennas with a baseline from 19 to 3144 m. The flux calibrator and bandpass calibrator were both J0522-3627, and the phase calibrator was J0607-0834. The total integration on the target was 8.39 min. The second night, the long baseline dataset, had precipitable water vapor of 1.08 mm. The ALMA configuration consisted of 40 12-m antennas with a baseline from 42 m to 14851 m. The flux calibrator was J0423-0120, the bandpass calibrator was J0510+1800, and the phase calibrator was J0541-0541. The total integration on the target was 25.55 min. 
\begin{table}
\centering
\caption{List of the Cycle 4 ALMA observations of the target HBC 494, used in this work}
\begin{tabular}{cccccc}
\hline
PI & Proj. ID                                     & Ang. Res.                                 & Date of obs. & Integration \\
& & (beam minor axis) & & & \\
\hline
\hline
 Zurlo & 2016.1.00630.S
 & 0.142\arcsec               
 & 09/10/2016            & 8.39 min                         \\
Zurlo & 2016.1.00630.S
 & 0.027\arcsec               
 & 26/09/2017            & 25.55 min                         \\
\end{tabular}
\label{table}
\end{table}

The central frequencies of each spectral window were: 230.543 (to cover the transition $^{12}$CO J=2-1), 233.010 (continuum), 220.403 ($^{13}$CO J=2-1), 219.565 (C$^{18}$O J=2-1), and 218.010 (continuum) GHz. The minimum spectral resolution achieved (second night) for $^{12}$CO was 15.259 kHz ($\sim$ 0.02 km/s), while for $^{13}$CO and C$^{18}$CO, it was 30.518 kHz ($\sim$ 0.04 km/s). Both continuum bands presented 2 GHz bandwidths. The data were calibrated with the Common Astronomy Software Applications package \citep[\texttt{CASA} v.5.5,][]{2007ASPC..376..127M}, and the python modules from CASA API, \textit{casatasks} and \textit{casatools} \citep{2022PASP..134k4501C}. The standard calibrations include water vapor calibration, temperature correction, and phase, amplitude, and bandpass calibrations.

\subsection{Continuum imaging}
\label{cont_imag}
We started our analysis by fixing the visibilities' phase center on J2000 05h40m27.448s -07d27m29.65s. The image synthesis of the 1.3 mm continuum emission was performed with the \textit{tclean} task of CASA. For the high-resolution data, we used the Briggs weighting scheme with a robust parameter of 0.5, resulting in a synthesized beam size of 41.2 $\times$ 29.8 mas, and a position angle of 40.6. The pixel scale of the image was set to 1.5 mas. One iteration step of self-calibration was applied to both observations. For the first night (low-resolution image: 142 mas), it resulted in an improvement of $\sim$6 SNR. For the second night (high-resolution image: 27 mas), it resulted in an SNR improvement of $\sim$ 1.6. 

Our final continuum image was produced from a concatenated measurement set combining both, the short and the long baseline visibilities, using the \textit{concat} task of CASA. We used the same image synthesis parameters used for the second night of observation.

\subsection{Line imaging}

\begin{figure}[!htb]
\begin{center}
    \includegraphics[width=0.8\textwidth]{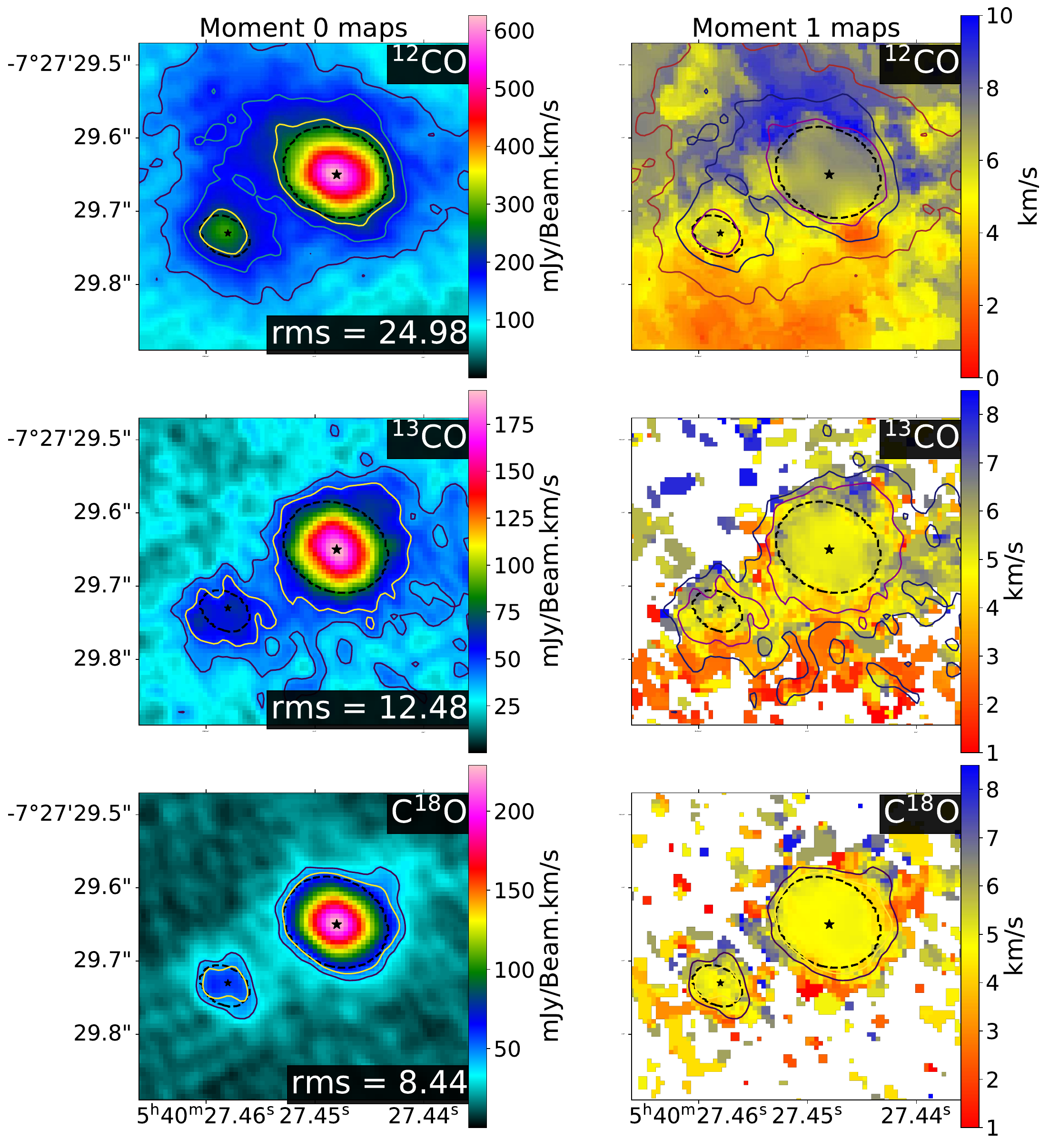}
    \caption[Small-scale moment maps for the analyzed CO isotopologues, $^{12}$CO (top), $^{13}$CO (middle) and C$^{18}$O (bottom), without subtraction of the continuum]{Small-scale moment maps for the analyzed CO isotopologues, $^{12}$CO (top), $^{13}$CO (middle) and C$^{18}$O (bottom), without subtraction of the continuum. The black stars mark the peak flux positions of the continuum disks. The dashed black contours correspond to 100$\times$rms from the continuum emission (100$\times$34$\mu$Jy/beam). Left: Moment 0 maps for each mentioned molecular line. The rms values at the bottom of each panel are in units (mJy/beam)(km/s). The $^{12}$CO moment 0 map (top-left) has contour levels of 5, 7, and 9$\times$rms showed at the bottom of the panel. $^{13}$CO and C$^{18}$O moment 0 maps (middle and bottom - left) have contour levels of 3 and 4$\times$rms showed at the bottom of each panel. Right: Moment 1 maps for each mentioned molecular lines. The contours are the same as each left panel respectively. The maps include only pixels above 3 times the rms measured from the set of channels used for each molecular line (-6 km/s to 18 km/s for $^{12}$CO and 0 km/s to 11 km/s for $^{13}$CO and C$^{18}$O).}
    \label{moments_small_with_cont}
\end{center}
\end{figure}

The visibilities were fixed at the same phase center as was done for the continuum coordinates (see subsection \ref{cont_imag}). After producing a dirty image from the visibilities, we noticed that the emission mimicked the presence of gaseous envelopes around the disks (moment 0 maps, see Figure \ref{moments_small_with_cont}, left column) but no gas dynamic signatures were revealed (moment 1 map, see Figure \ref{moments_small_with_cont}, right column). Therefore, we proceeded to remove the visibility's continuum contributions by using the CASA task \textit{uvcontsub}. 

Next, we performed the \textit{tclean} process using Briggs weighting with a robust parameter of 0.5, Hogbom deconvolver, and pixel scale of 6 mas. The total velocity ranged between -6 km/s to 18 km/s, although a smaller range (0 km/s to 11 km/s) was used to create the $^{13}$CO and C$^{18}$O moment maps. We used spectral resolution widths of 1 km/s for $^{12}$CO and of 0.3 km/s for $^{13}$CO and C$^{18}$O. The rms values, obtained for the channel without a clear signal on the channel maps (4 km/s), were 1.9, 3.6, and 2.6 mJy/beam for $^{12}$CO, $^{13}$CO and C$^{18}$O, respectively.

\section{Results}
\label{sec:results}
\subsection{Continuum analysis}
\label{sec:con}

\begin{figure*}[!htb]
\includegraphics[width=1\textwidth]{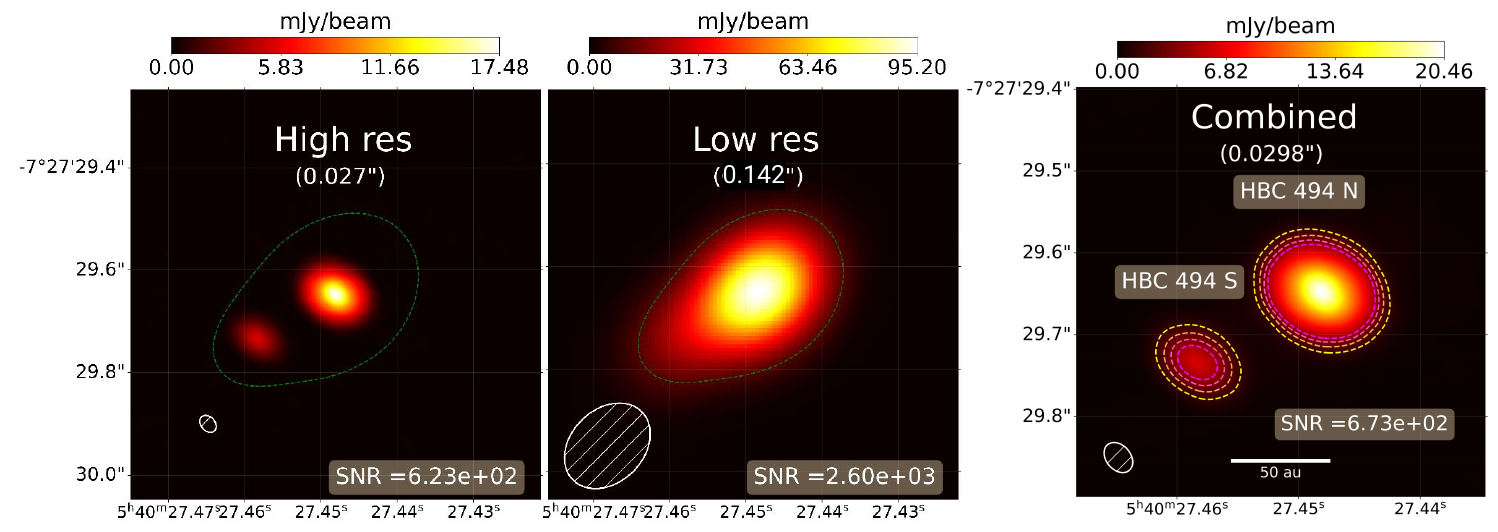}
\caption[ALMA observations at 1.3 mm of the HBC 494 system]{ALMA observations at 1.3 mm of the HBC 494 system. The images reveal the binary nature of HBC 494 with two components surrounded by continuum emission associated with circumstellar dust. The respective signal-to-noise ratios of each image are shown in each panel. We show the resulting images from the long baseline observation with a resolution of 0.027\arcsec (\textit{left}), the short baseline data at 0.142\arcsec resolution (\textit{middle}), and the combination of the two observations with 0.0298\arcsec (\textit{right}). Dashed green lines in the left and middle images represent a 3-sigma contour associated with the lower-resolution continuum image. The combined image has a beam size of $\sim$ 41$\times$30 mas, shown in the lower-left corner of the image. Its contours correspond, from the yellow (more external) to the pink dashed lines, to 10, 40, 70, and 100 times the rms value, respectively. The rms is $\sim$0.034 mJy/beam.}
\label{stages_cont}
\end{figure*}

With the high-resolution data shown in Figure~\ref{stages_cont}, we could reveal the binary nature of the HBC 494 system. Each of the components in HBC 494 is surrounded by continuum emission, most likely associated with circumstellar disks.

\begin{figure*}[!ht]
\begin{center}
\includegraphics[width=1\textwidth]{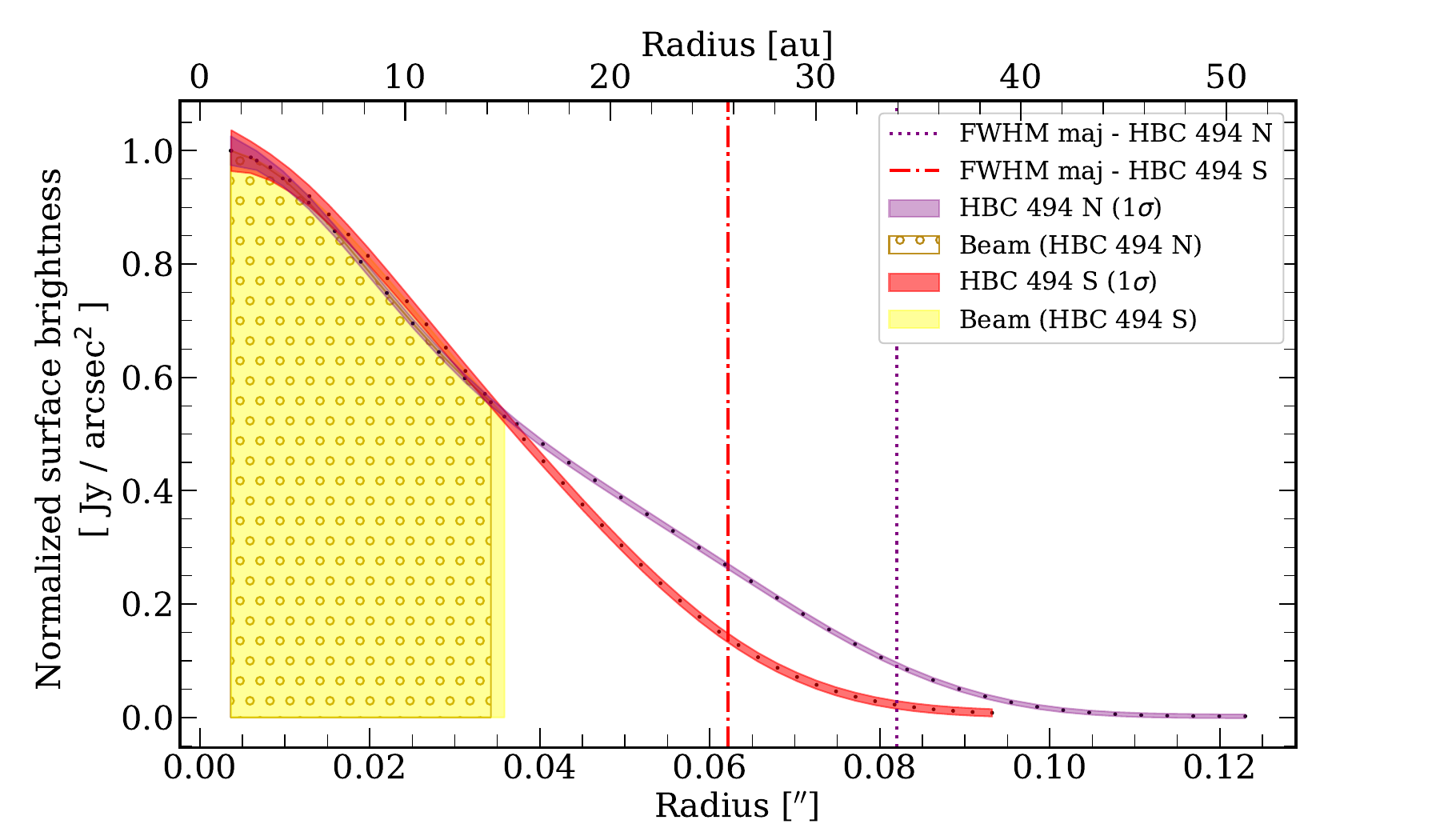}

\caption[Radial brightness profile of HBC 494 N and HBC 494 S]{Radial brightness profile of HBC 494 N and HBC 494 S. The surface brightness (dots), measured in apertures ranging from the beam's major axis to 3 times the semi-major axis FWHM of each continuum disk, were normalized to the peak fluxes. The shaded regions indicate 1$\sigma$ error measured in each aperture. The dashed lines correspond to the ALMA imfit FWHM of each disk.}

\label{rad_prof}
\end{center}
\end{figure*}

We performed a 2D Gaussian fitting with the \textit{imfit} tool of CASA in order to characterize the newly resolved components. The projected separation between the two sources is 0.18\arcsec (75 au). We name each component with the usual convention, i.e. ``N'' and ``S'' components, referring to the northern and the southern source, respectively. We assume the northern component to be the primary as it is five times brighter than the southern (secondary) disk. Both individual disks are resolved according to our Gaussian fit. For the primary component, we find a major axis FWHM of 84.00 $\pm$ 1.82 mas and a minor axis of 66.94 $\pm$ 1.50 mas (34.8 $\pm$ 0.7 $\times$ 27.8 $\pm$ 0.6 au), with a position angle of 70.01 $\pm$ 4.46 deg (values deconvolved from the beam). For the secondary, the major axis FWHM is 64.60 $\pm$ 2.49 mas and the minor axis FWHM is 45.96 $\pm$ 1.89 mas (26.7 $\pm$ 1.0 $\times$ 19.9 $\pm$ 0.8 au), and the position angle is 65.38 $\pm$ 5.32 deg. The integrated fluxes for ``N'' and ``S'' are 105.17 $\pm$ 1.89 mJy and 21.06 $\pm$ 0.63 mJy, respectively. The peak fluxes are 22.96 $\pm$ 0.35 mJy/beam and 8.71 $\pm$ 0.21 mJy/beam for the primary and secondary sources, respectively. The rms is $\sim$0.034 mJy/beam, as observed in a circular region with a radius of 225 mas without emission. The inclinations are calculated using the aspect ratios of the disks. These are assumed as projected circular disks when face-on, thus elliptical when inclined:
\begin{center}
    $i = \arccos{\left( \frac{b}{a} \right)}$,
\end{center}
\noindent where ``b" and ``a" are the semi-minor and semi-major axes, respectively. It resulted in 37.16 $\pm$ 2.36 degrees inclination for HBC 494 N and 44.65 $\pm$ 3.27 for HBC 494 S. At 30 mas resolution, no substructures were detected (see the radial profiles in Figure~\ref{rad_prof}).  

A large fraction of the mm-sized grains in protoplanetary disks resides in the midplane. They are usually optically thin to their own radiation. Thus, assuming standard disks, we can use the observed fluxes to estimate the dust masses of HBC 494 N and S. For that, we use the following formula \citep[as in][]{beckwith1990, andrews2005, cieza2018}: 

\begin{equation}
    M_{\rm dust} =\frac{F_{\nu}d^{2}}{k_{\nu}B_{\nu} (T_{\rm dust})},
\end{equation}
\noindent where $F_{\nu}$ is the observed flux, \textit{d} is the distance to the source, $B_{\nu}$ is the Planck function and $\kappa_{\nu}$ is the dust opacity. Adopting the distance of 414 pc, isothermal dust temperature ($T_{\rm dust}$ = 20 K) and dust opacity assuming a $M_{\rm gas}/M_{\rm dust}$ fraction of 100, ($\kappa_{\nu}$ =  10 ($\nu /$ 10$^{12}$Hz) cm$^{2}$ g$^{-1}$; \citealt{beckwith1990}) we get dust masses of 1.43 $M_{\rm Jup}$ (HBC 494 N) and 0.29 $M_{\rm Jup}$ (HBC 494 S). Following the standard gas-to-dust mass ratio of 100, we report disk gas masses estimations of 143.46 $M_{\rm Jup}$ for HBC 494 N, and 28.68 $M_{\rm Jup}$ for HBC 494 S. The physical parameters based on the CASA analysis are listed in Table \ref{compiled_disks}.

\begin{table}
\centering
\caption{ALMA imfit analysis results and physical parameters inferred from the HBC 494 continuum disks.}
\label{compiled_disks}

\small\addtolength{\tabcolsep}{-2pt}
\begin{tabular}{lcc}
                                        & HBC 494 N             & HBC 494 S            \\
\hline
\hline
Flux density (mJy)                      & 105.2 $\pm$ 1.9  & 21.1 $\pm$ 0.6 \\
Peak flux (mJy / beam)  & 23.0 $\pm$ 0.4 & 8.7 $\pm$ 0.2                        \\
Major axis (mas)                        & 84.0 $\pm$ 1.8  & 64.6 $\pm$ 2.5 \\
Minor axis (mas)                        & 66.9 $\pm$ 1.5  & 46.0 $\pm$ 1.9 \\
Major axis (au)                        & 34.8 $\pm$ 0.7  & 26.7 $\pm$ 1.0 \\
Minor axis (au)                        & 27.8 $\pm$ 0.6  & 19.9 $\pm$ 0.8 \\
Position angle (deg)                    & 70.0 $\pm$  4.5 & 65.4 $\pm$ 5.3 \\

Beam major axis (mas)    & 41.2 & 41.2 \\
Beam minor axis (mas)    & 29.8 & 29.8  \\
Beam Position angle (deg)               & 40.6                         & 40.6                        \\
Beam area (sr)                          & 3.3e-14                      & 3.3e-14   \\
rms ($\mu$Jy / beam)                          & 34.0                      & 34.0 \\
\hline
Dust mass $\, (M_{\rm Jup})$ & 1.4 & 0.3 \\
Gas mass $\, (M_{\rm Jup})$ & 143.5 & 28.7 \\
Inclination (deg) & 37.2$\pm$2.4 & 44.7$\pm$3.3 \\
FWHM Radius (au) & 34.8$\pm$0.7 & 26.7$\pm$1.0 \\
\end{tabular}%
\end{table}

\subsection{Line analysis}
\label{sec:gas}

The next two subsections describe the large and small-scale structure analyses. The large scale, using the extent of 8000 au, shows the results of outflows and envelopes. At such a scale, there is not enough resolution to display both disks' dynamics and their possible interactions. The small scale (150 au) fulfills this role, and thus, both gas scenarios must be taken into account.

\subsubsection{Large-scale structures (8000 au)}

\begin{figure*}[!htb]
\begin{center}
\includegraphics[width=1\textwidth]{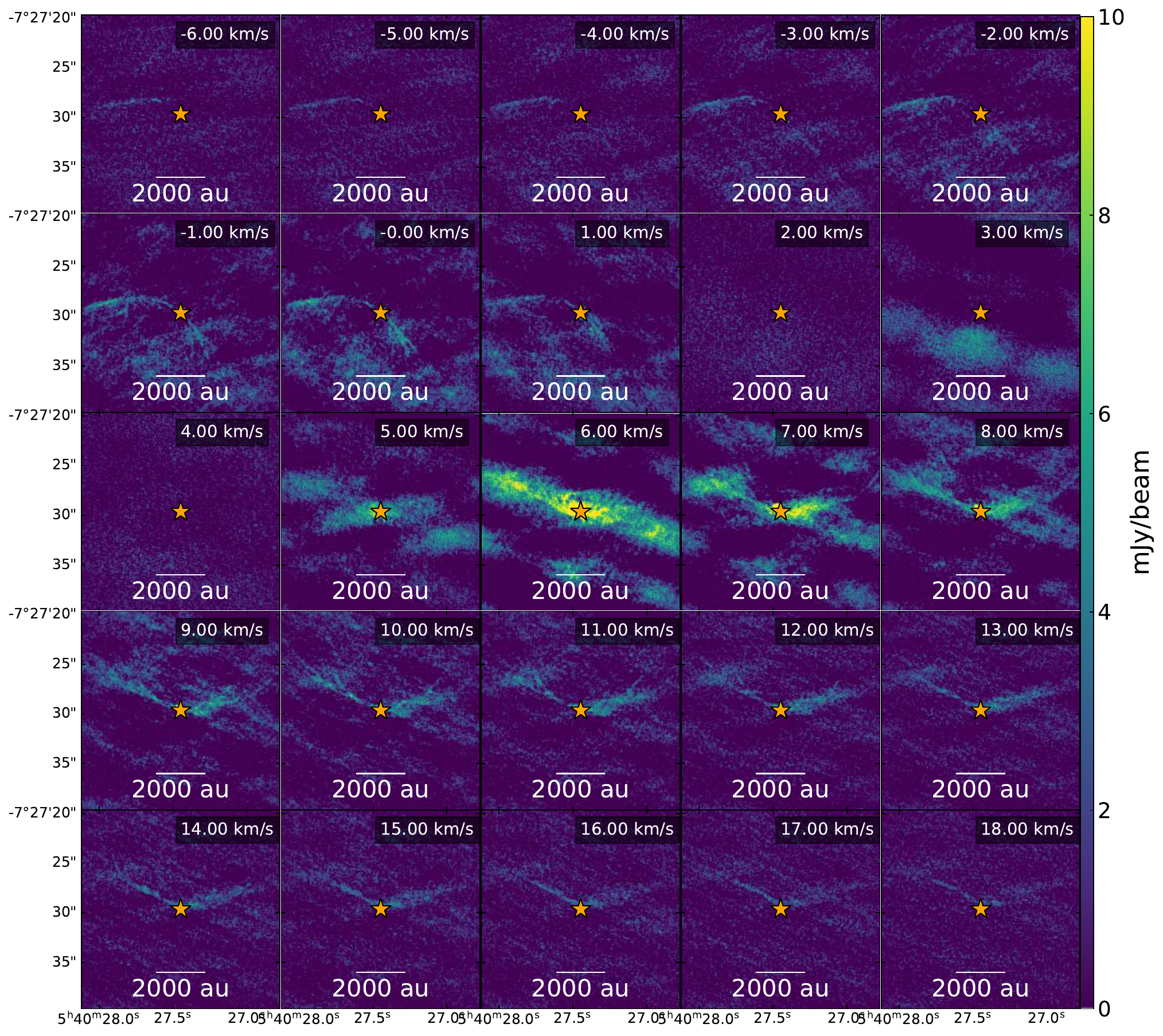}
\caption[$^{12}$CO channel maps of the HBC 494 system]{$^{12}$CO channel maps of the HBC 494 system. The star in the center marks the position of the continuum disks. The southern arc is more evident between the velocities -6 km/s to 1 km/s, while the northern is more easily seen between 7 km/s and 18 km/s.}
\label{12cochannel}
\end{center}
\end{figure*}

\begin{figure*}[!htb]
\begin{center}
\includegraphics[width=1\textwidth]{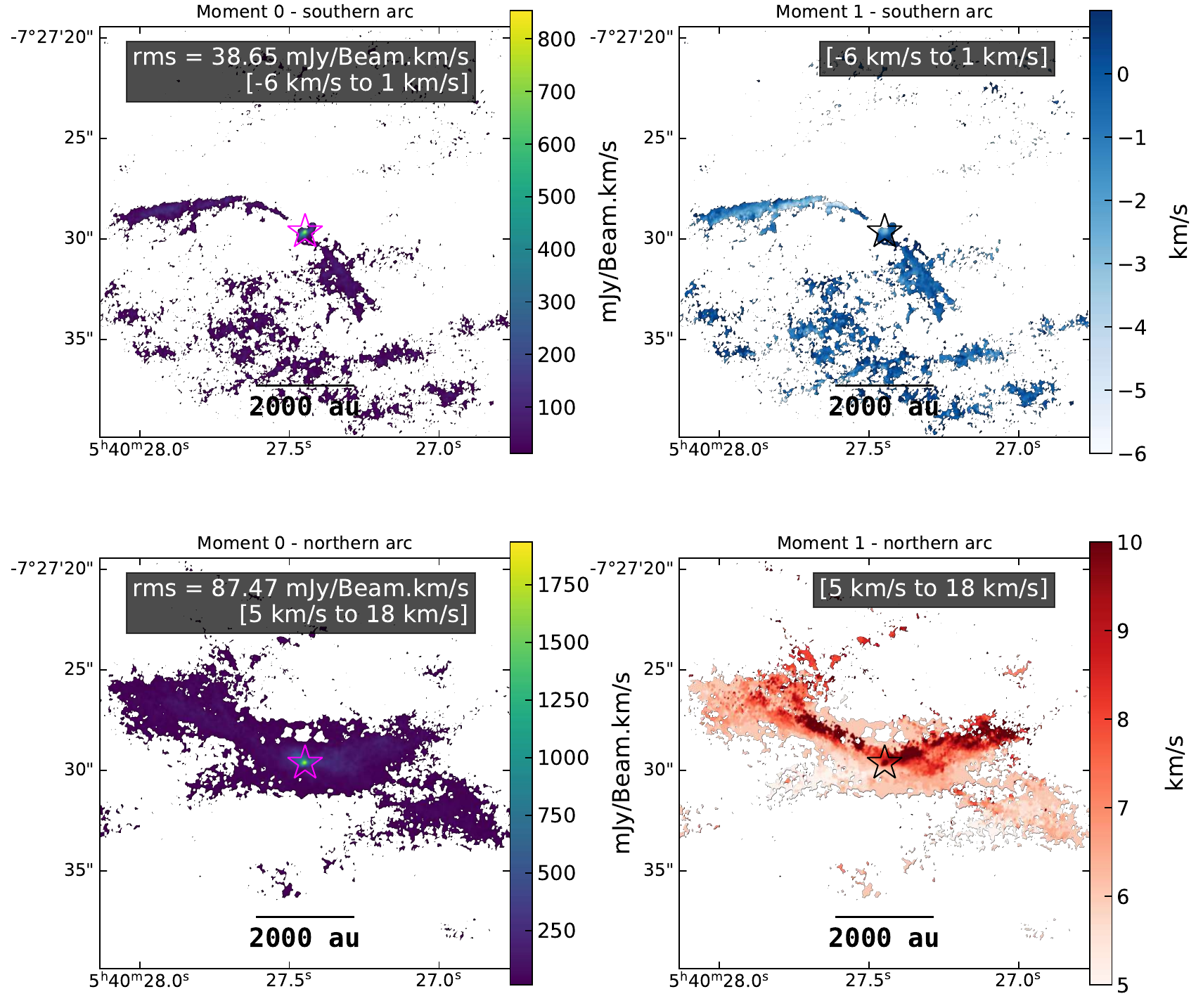}
\caption[Large-scale $^{12}$CO moment maps split in each one of the bipolar outflow structures]{Large-scale $^{12}$CO moment maps split in each one of the bipolar outflow structures. The left column corresponds to moment 0 maps, and the right column to moment 1 maps. The plots in the same row refer to the same structure, which is the southern arc (\textit{upper row}), and the northern arc (\textit{bottom row}).  The position of the HBC 494 N is marked with a star. The maps include only pixels above 2 times the rms measured from the set of channels shown at the top of each panel.}
\label{arcs}
\end{center}
\end{figure*}

\begin{figure*}[!htb]
\includegraphics[width=0.95\textwidth]{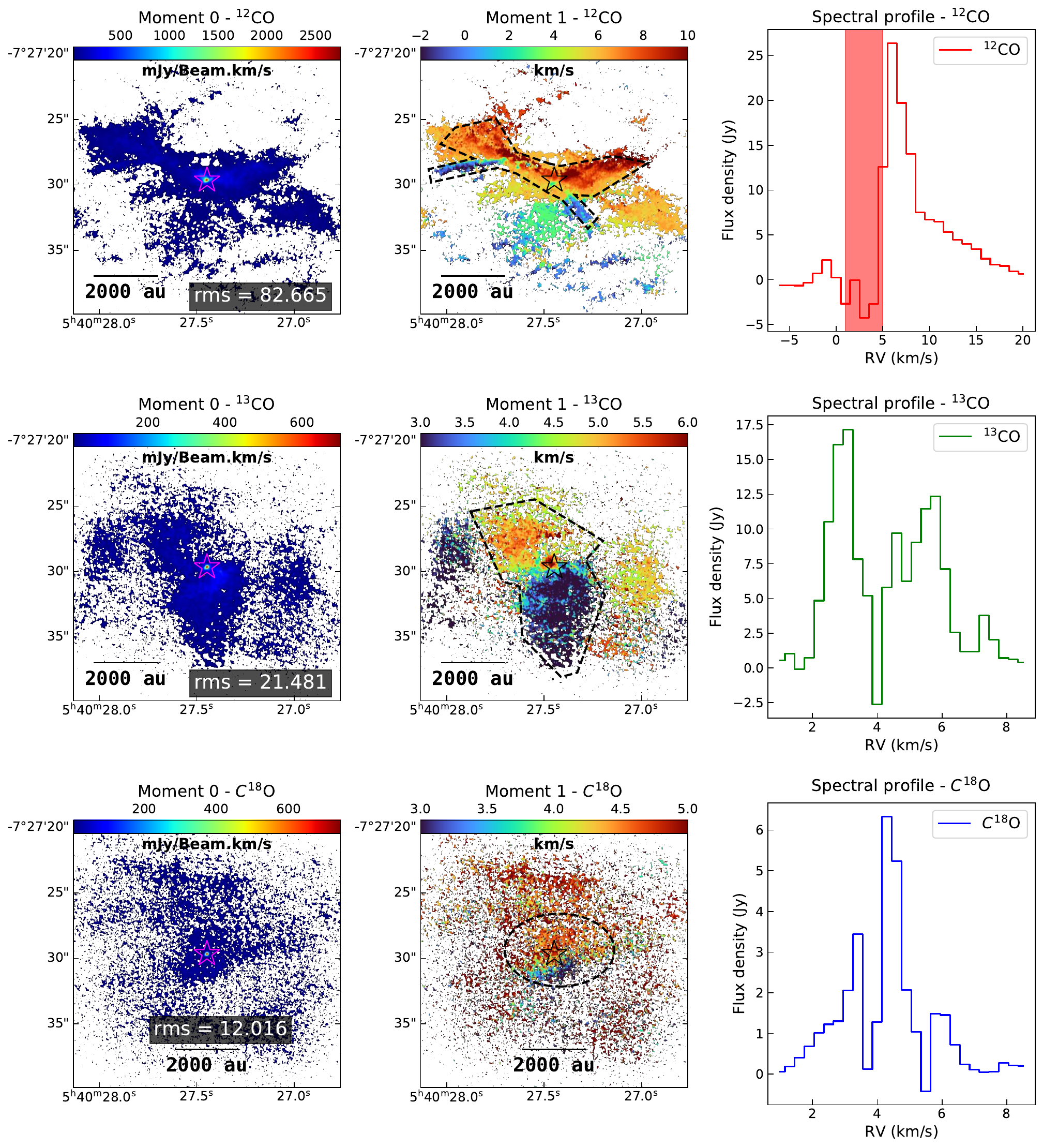}
\caption[Large-scale moment maps for the analyzed CO isotopologues and spectral fluxes]{Large-scale moment maps for the analyzed CO isotopologues and spectral fluxes. The left column corresponds to moment 0 maps, the middle one to moment 1 maps, and the third column to spectral profiles. The rms values at the bottom of each panel in the left column are in units (mJy/beam)(km/s). The plots in the same row refer to the same molecular lines, which are $^{12}$CO (\textit{upper row}), $^{13}$CO (\textit{middle row}), and C$^{18}$O (\textit{bottom row}). The dashed black lines in the moment 1 column correspond to the regions used to produce the spectral profiles. The red shaded area in $^{12}$CO profile represents the channels we excluded before creating the moment maps lying in the same row (to avoid cloud contamination). The center of stars, in the middle of each figure, marks the position of both disks. The maps include only pixels above 2 times the rms measured from the set of channels used for each molecular line (-6 km/s to 18 km/s for 12CO and 0 km/s to 11 km/s for 13CO and C18O).}
\label{largescalemoments}
\end{figure*}

The channel maps (Figures \ref{12cochannel}, \ref{13cochannel} and \ref{c18ochannel}) show southern and northern $^{12}$CO wide-angle arcs ($\sim150^{\circ}$) and no clear signal of large structures from the other molecular lines. This bipolar outflow was previously described in \citet{ruizrodriguez2017}. The wide angular structure shows an outflow morphology expected for Class I disks \citep{arce2006}. In our observations, the southern and northern arcs are defined by the velocity ranges of [-6 km/s to 1 km/s] and [5 km/s to 18 km/s], respectively. The moment 0 and 1 $^{12}$CO arcs are shown in Figure~\ref{arcs}. However, it is noticeable that the northern arc (redshifted emission) dominates the displayed velocities. The lack of signal in the southern arc can be explained by lower molecular density, or due to intracloud absorption. Figure~\ref{largescalemoments} shows the three molecular lines' concentration and dynamics (moment 0 and 1 maps), as well as the spectral profile measured within the dashed regions. 

The $^{13}$CO emission traces the rotating, infalling, and expanding envelope surrounding the system, showing a small deviation from the $^{12}$CO outflowing arcs' rotation axis. The blue-shifted emission concentrates around 3 km/s, while the red-shifted emission is around 5 km/s. It is noticeable that the $^{13}$CO emission has a rotation axis almost perpendicular to the $^{12}$CO outflows axis, ensuring that the physical processes behind the two are different. The spectral $^{13}$CO also shows a dip around 4 km/s, but this may not trace an absorption but a lack of signal due to the analyzed uv-coverage or cloud contamination.
 
The C$^{18}$O emission, however, is faint and hardly distinguishable from the surrounding gas in moment 0 maps. It has also the lowest abundance of the three analyzed isotopologues, being a good tracer of the higher gas densities in the innermost regions of the cloud. Its velocity maps show that a faint blueshifted motion was detected around 3 km/s, very weak when compared to the redshifted emission detected principally around 5 km/s. Due to the higher concentration of gas in the northern part, it is reasonable that we observed a weak gas counterpart in the southern region. At large-scale, our results are comparable with the ones presented in \citep{ruizrodriguez2017}. Furthermore, a lower limit of the envelope material of $\sim$600 Jy.km/s towards HBC 494 could be traced using Total Power (TP), ALMA compact array (ACA), and the main array ALMA observations with C$^{18}$O data in the range of 4.3$\pm$0.5 km s$^{-1}$ (Ru\'iz-Rodr\'iguez private communication).

\subsubsection{Small scale structures (150 au resolution)}

We started by removing the continuum contribution from all molecular lines observed ($^{12}$CO, $^{13}$CO, C$^{18}$O). We already expected that $^{12}$CO would trace the gas in larger scales due to its optical depth and, thus, could blend with smaller gas signatures. To avoid cloud contamination during the creation of $^{12}$CO small-scale moment maps, we removed the channels that were more affected. We considered velocities in the range of -6 km/s to 1 km/s, and from 5 km/s to 18 km/s, as similarly done for the different large-scale arcs described in the previous subsection. 

\begin{figure*}[!htb]

\begin{center}
    \centering
    \includegraphics[width=1\textwidth]{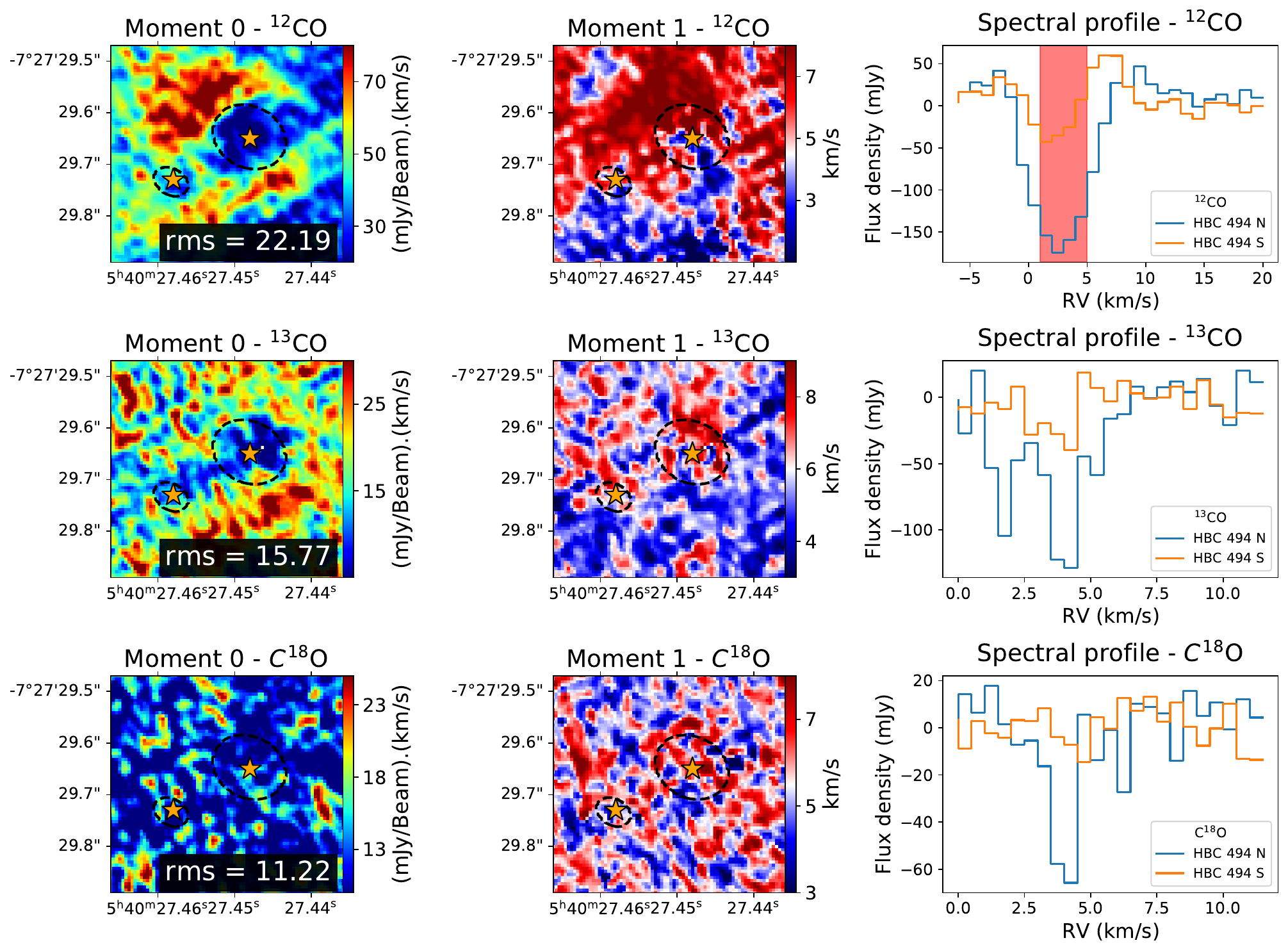}
    \caption[Small-scale moment maps for the analyzed CO isotopologues, after subtraction of the continuum, and their spectral profiles]{Small-scale moment maps for the analyzed CO isotopologues, after subtraction of the continuum, and their spectral profiles. The first column corresponds to moment 0 maps, and the middle one to moment 1 maps. The rms values at the bottom of each panel in the left column are in units (mJy/beam)(km/s). The molecular lines correspond to $^{12}$CO (\textit{upper row}), $^{13}$CO (\textit{middle row}) and C$^{18}$O (\textit{bottom row}). The dashed black lines spatially correspond to the contour level equal to 100 times the continuum rms (100$\times$34$\mu$Jy/beam), in the continuum combined image (Figure \ref{stages_cont}, \textit{right}). The orange stars mark the peak flux positions of the continuum disks. The right column represents the spectral profiles measured from velocity maps taken within each disk. The red shaded area in $^{12}$CO profile represents the channels we excluded before creating the moment maps lying in the same row (to avoid cloud contamination).}
    \label{moments_small}
\end{center}
\end{figure*}

Following, CO channel maps and moment 0 and 1 maps were produced to explore potential hints of interaction between the stars and the two circumstellar disks. In particular, we looked for disk substructures and perturbed rotation patterns. The moment maps can indeed provide valuable information about the binary orbit and the disks' geometry. However, by looking at the molecular lines in this scale (see the moment maps in Figure~\ref{moments_small}, left and middle columns), we could not clearly detect the presence of the disks or their rotational signatures. We can only notice that the $^{12}$CO and $^{13}$CO moment 0 maps show an interesting pattern, where less gas emission is found where the continuum disks are. Additionally, the area within the continuum disks highlights regions similar to cavities or ``holes" (Figure~\ref{moments_small}, left column, upper and middle row). Interestingly, \citet{ruizrodriguez2022} observed similar features from the same molecular lines, in addition to HCO$^{+}$, for another FUor system, V883 Ori. The work also showed that the origin of cavities and ring emission around disks can be interpreted in two ways, one based on gas removal and the other on optical depth effects. 

If the signal traces the lack of gas surrounding the disks, the leading mechanism may be slow-moving outflows. The central regions of young continuum disks are expected to be constantly carved by the influence of magneto-hydrodynamical jets \citep[][]{frank2014}. In addition, it is also expected that the lower emission observed comes from the chemical destruction by high-energy radiation (more active disks will create bigger cavities). By looking at the size of the gaps, we can observe that HBC 494 S exerts a smaller influence on the gas than HBC 494 N and, thus, HBC 494 N is the more active disk in the system if this hypothesis is correct.

Another interpretation of the negatives in our gas maps is that this is a spurious result of continuum subtraction. Usually, the continuum emission is inferred from channels devoid of gas line emission. The usual procedure does not consider the line emission where these channels overlap. Suppose the emission lines are optically thick above the continuum. In that case, the foreground gas can significantly absorb the photons coming from the underlying continuum, which may lead to a continuum overestimation. Consequently, in such cases, the continuum may be over-subtracted, causing the ``hole" feature that is stronger where the line peaks (see e.g., \citealp{boehler2017,weaver2018}).

\section{Discussion}
\label{sec:dis}
\subsection{Multiplicity and triggering mechanisms}

Besides HBC 494, there are $\sim$ 25 known FUor objects within 1 kpc \citep{audard2014}. As previously commented, many triggering mechanisms can cause episodic accretion events. From this sample, only HBC 494 and a few other FUor systems are known binaries (e.g., FU Orionis, L1551 IRS, RNO 1B/C, AR 6A/B; \citealp{pueyo2012} and references therein). 

Given the high occurrence rate of stellar binaries harboring disks, the lack of detection of multiple YSO systems remains a case of investigation. Additionally, young disks evolve in crowded star-forming regions, enhancing the hypothesis that the outbursts of Class 0/I  disks may affect other disks, driving episodic accretion events. Also, surveys and studies using ALMA and NIR data for Class I-III disks, show that there is a lack of detection when only visual multiple systems (separations of 20-4800 au) are considered (e.g., \citealp{zurlo2020b,zurlo2021}). When all the possible separations are taken into account, the multiplicity frequencies considerably increase. For Taurus, for example, it goes up to 70\% when spectroscopic binaries are included \citep{kraus2011}. In the case of Orion, 30\% of the systems are multiple with separations between 20 and, 10000 au \citep[][]{tobin2022}. The latter study also noticed that the separations decrease with time (when comparing Class 0, I, and flat-spectrum disks) and that the multiplicity frequency in Class 0 is higher than in the later evolutionary stages. Therefore, multiple eruptive young systems must be common, and more detectable within evolutionary time. They probably are not frequently detected due to the short time duration of enhanced accretion events in comparison with the lifetime of disks in class 0/I stages (see \citealt{audard2014} for a FUors review). Still, it is not clear if disks belonging to close-separation systems are more susceptible to eruptive events than isolated ones. 

For each HBC 494 disk, no clue of trigger mechanisms was found, since the gas and continuum data did not reveal clear spiral or clumpy features indicating a case of infall or GI (see e.g., \citealp{zhu2012,kratter2016}). Looking at the dust continuum (Figure \ref{stages_cont}), CO isotopologues moment maps (Figure \ref{moments_small}), and the scattered $^{12}$CO emission (Figure 12 of \citealt{ruizrodriguez2017}), the hypothesis of stellar flybys is also not encouraged since no trace of perturbation was detected (\citealt{cuello2020, Cuello+2023}). Moreover, dynamical data (small-scale moment 1 maps) also can provide clues for the triggering mechanism \citep{vorobyov2021}. However, it requires clear detection of quasi-keplerian rotation, which was not observed. 

In the case of binaries, the first component to ignite the FUor outbursts can quickly trigger the secondary one by inducing perturbations and mutual gravitational interactions \citep{bonnell1992, Reipurth+2004, vorobyov2021, Borchert+2022a, Borchert+2022b}. Therefore, we can assume that all the disks in FUor multiple systems like HBC 494 might have undergone successive enhanced eruptive stages, despite the differences in mass and radius between each component. However, it is not clear how the outbursts in HBC 494 N and HBC 494 S affect each other since the moment maps do not show the connection between disks. Although, on larger scales, we detected the difference in the amount of gas mass detected in the northern regions compared to the southern. This scenario may be described if the more massive disk (i.e. HBC 494 N) has a higher contribution to the outbursts and winds.

With astrometric data, the eccentricity of orbits can be determined. If both disks are still accreting gas from the environment, it is expected that quasi-circular orbits trigger enhanced accretion every few binary periods. Eccentric orbits, on the other hand, can induce eruptive events in every orbit, preferentially when the binaries reach the pericentre (see e.g. \citealp{2015MNRAS.448.3545D,2022arXiv221100028L}). Observational evidence of circumbinary disks in different YSO stages was found (see e.g. \citealp{1994A&A...286..149D,1997AJ....113.1841M,2016Natur.538..483T,2020ApJ...897...59M}) and are seen as a common counterpart of young binaries in formation. However, no HBC 494 circumbinary gas disk was observed due to cloud contamination and optical depth effects.
\subsection{Disk sizes and masses}
FU Orionis, the precursor of the classification FUor, is also a binary system. HBC 494 system, however, has smaller disks separation compared to the 210 au between FU Orionis north and south components \citep[][]{perez2020}. FU Orionis components have similar sizes and masses between their disks, contrary to the HBC 494 components. We can argue that, due to HBC 494 being a close-packed system with substantially different masses between their components, a scenario of radius truncation can be tested. However, the FU Orionis disks are exceptional compared to other FUors, which are generally more massive and larger. Nevertheless, it is important to check if HBC 494 follows the trend which shows that FUor objects are more massive than class 0/I disks and that disks from multiple systems are smaller compared to those from non-multiple systems \citep{cieza2018,hales2020,tobin2020,zurlo2020}. First, we will describe our results regarding masses and radius. Then, we will compare HBC 494 disks to FUors and Orion YSOs in the literature.

\begin{figure*}[!htb]
\begin{center}
\includegraphics[width=1\textwidth]{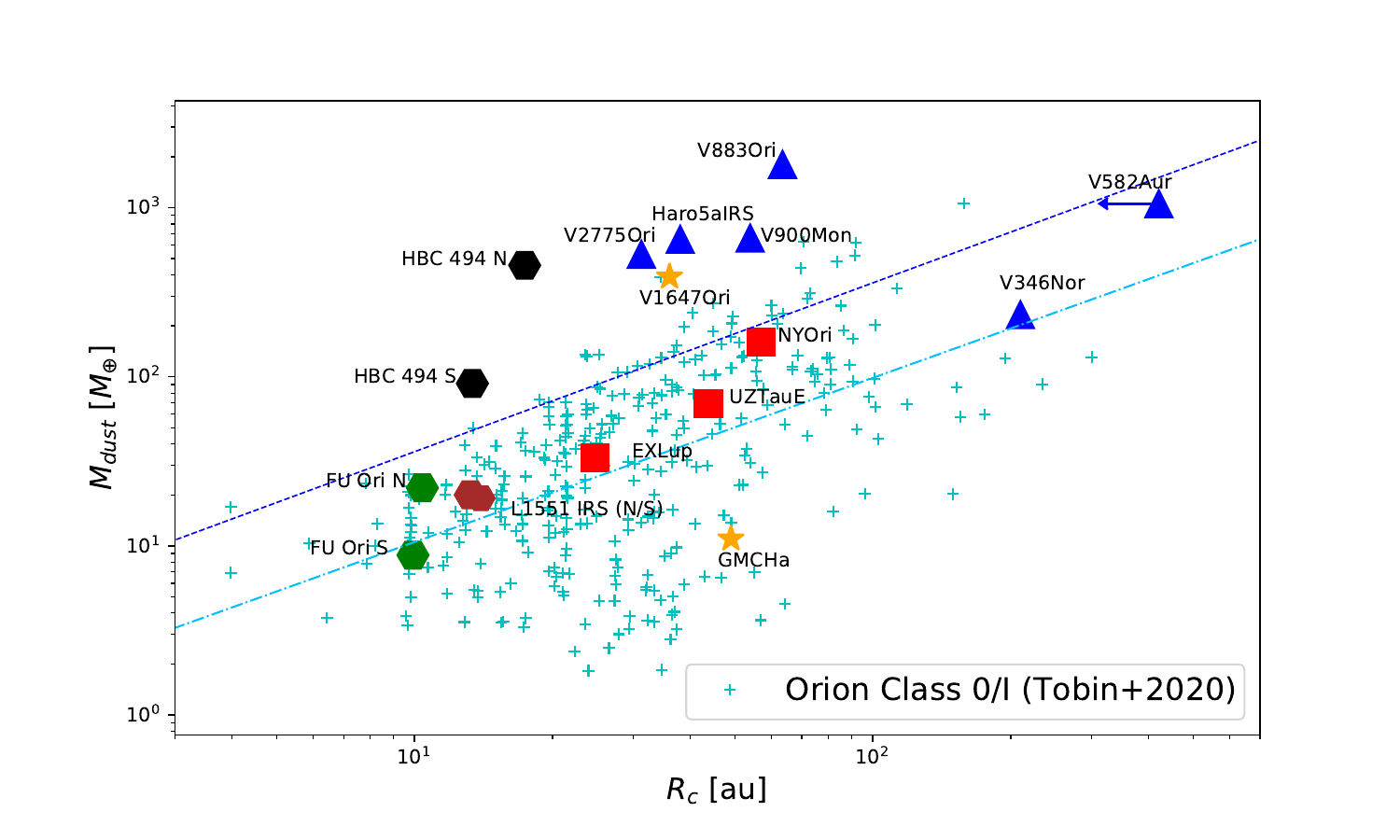}
\caption[Dust disk masses as a function of characteristic radii (FWHM/2)]{Dust disk masses as a function of characteristic radii (FWHM/2) for FUors (blue triangles), EXors (red squares), for double FUor and EXor classification (yellow stars), and Class 0/I objects (cyan crosses), as similarly presented in \citet{hales2020}. The horizontal arrow on V852 Aur denotes the disk radius in an upper limit. The hexagons represent eruptive/multiple systems, with matching colors for disks lying in the same system. The data for HBC 494 disks was generated in this work. The blue dashed line corresponds to the power-fit law to the FU/EXor data (Spearman correlation coefficient: 0.63), and the cyan dash-dotted line, to the Class 0/I data (Spearman correlation coefficient: 0.55). To generate this plot, we obtained data from \citet{kospal2017,hales2018,cieza2018,takami2019,cruz2019,perez2020,hales2020}.}.
\label{radiusxmass}
\end{center}
\end{figure*}

We assumed dust masses (assuming the gas-to-dust ratio of 100) inferred by optically thin approximation (described in section \ref{sec:con}). However, it is worth mentioning that the optically thin analysis can not take into account the innermost regions of the disk (optically thick). This may lead to an underestimation of the dust masses. Also, an underlying miscalculated dust grain temperature can alter the results, overestimating the masses if the dust is warmer than expected. The chosen assumptions of temperatures (fixed 20 K for dust grains) and opacities for both disks lead to 1.43 $M_{\rm Jup}$ and 0.29 $M_{\rm Jup}$ for HBC 494 N and HBC 494 S, respectively. Based on these results, we can say that the HBC 494 N is comparable to other FUor sources, but HBC 494 S has its dust mass comparable to EXor disks \citep{cieza2018}. The disk sizes were calculated using the deconvolved Gaussian FWHM/2 radius obtained from 2D Gaussian fits. This methodology was also used for other ALMA datasets (e.g., \citealp{hales2020, tobin2020}). Therefore, it allows a consistent comparison with other disks. The dust mass vs radius comparison between HBC 494 N and HBC 494 S, a sample of 14 resolved FUors and EXors (extracted from \citealp{hales2020}) and Class 0/I systems \citep[][]{tobin2020}, can be seen in figure \ref{radiusxmass}. Here, as previously stated, we notice that HBC 494 disks (black, hexagon symbols), as other eruptive systems (hexagon, square, and triangle symbols), present higher masses than young systems, which are not in the stage of episodic accretion. However, there is no obvious distinction between the disk sizes of individual systems, eruptive or not. A clear difference is seen when we compare the disks from multiple systems (hexagon symbols) and single systems. The latter present bigger sizes as they are not targets for tidal truncation, enhanced radial drift, and more aggressive photoevaporation --- processes known to rule close-systems dynamics and evolution (see e.g., \citealp{kraus2012,harris2012,rosotti2018,zurlo2020b, zurlo2021}).

\subsection{Binary formation and alignment}

Different scenarios have been proposed to explain 
the formation of binary systems. The main one for close-separated systems is fragmentation, followed by dynamic interactions (for a review, see \citealp{offner2022}). 

The fragmentation process consists of partial gravitational collapse from self-gravitating objects. To be successful, many initial physical conditions are relevant as thermal pressure, density, turbulence, and magnetic fields. In addition, the fragmentation can be divided into two main classes: direct/turbulent (e.g., \citealp{1979ApJ...234..289B,1997MNRAS.288.1060B}) and rotational (e.g., \citealp{larson1972,1994MNRAS.269..837B,1994MNRAS.269L..45B,1994MNRAS.271..999B,1997MNRAS.289..497B}). 

The direct/turbulent fragmentation from a collapsing core is highly dependent on the initial density distribution. It can form wide-separated multiple YSO systems with uncorrelated angular momentum. Thus, the direct/turbulent fragmentation may lead to preferentially misaligned multiple systems \citep{2000MNRAS.314...33B, 2016ApJ...827L..11O, 2018MNRAS.475.5618B, 2019ApJ...887..232L}. The rotational fragmentation, instead, is caused by instabilities in rotating disks and leads to preferentially spin-aligned and coplanar systems \citep{2016ApJ...827L..11O, 2018MNRAS.475.5618B}. Moreover, later dynamical processes such as stellar flybys \citep{ClarkePringle1993, Nealon+2020, Cuello+2023} and misaligned accretion from the environment \citep{Dullemond+2019, Kuffmeier+2020, Kuffmeier+2021} can also induce alignment or misalignment.

The HBC 494 disks have similar inclinations ($\Delta i$ = 7.5 $\pm$ 4.1 degrees) and similar PAs ($\Delta PA$ = 4.6 $\pm$ 7.0 degrees). The relative orientations of the disks suggest that the system is coplanar. Assuming they are quasi-coplanar, with the same PA, the unprojected separation would be $\sim$0.24\arcsec (99 au). Therefore, the HBC 494 disks may show a spin-alignment situation, more typical to observed close-separated systems (see, e.g., \citealp{2016ApJ...818...73T,tobin2020}). Thus, we tentatively suggest that HBC 494 was formed by rotational disk fragmentation rather than direct collapsing. However, more precise observation of the kinematics of the surrounding gas on a small scale \citep{vorobyov2021}, complemented with orbital information coming from astrometrical measurements, is required to evaluate this hypothesis. 

In this context, IRAS 04158+2805 is of interest as \cite{Ragusa+2021} reported the detection of two circumstellar disks, similar to HBC 494 in terms of small to moderate misalignment, and an external circumbinary disk. However, in contrast to IRAS 04158+2805, no circumbinary disk was detected in HBC 494. This difference could be explained either by the absorption of the circumbinary disk emission due to cloud contamination, or because HBC 494 had enough time to sufficiently empty the circumbinary mass reservoir. Since most hydrodynamical models of star formation naturally produce young binaries with individual disks and a surrounding circumbinary disk \citep{2018MNRAS.475.5618B, Bate2019, Kuruwita+2020}, it is likely that a circumbinary disk formed at some earlier evolutionary stage of HBC 494. If the circumbinary disk is actually there, but remains undetected, the supply of material could extend the disk lifetime of the circumstellar disks in HBC 494. Deeper multi-wavelength observations would provide more information, enabling us to solve this binary riddle.

\section{Conclusions} 

In this work, we presented high-resolution 1.3 mm observations of HBC 494 with ALMA. The unprecedented angular resolution in this source (0.027\arcsec), reveals that HBC 494 is a binary that can be resolved into two disks, HBC 494 N and HBC 494 S. The disks have a projected separation of $\sim$75 au. We derived sizes, orientations, inclinations, and dust masses for both components. Both objects appear to be in a quasi-coplanar configuration and with their sizes halted by dynamical evolution, besides preserving high masses, common to FUors and EXors than typical (not eruptive) Class 0/I disks. Comparing the two sources, we noticed that HBC 494 N is $\sim$5 times brighter/more massive and  $\sim$2 times bigger than HBC 494 S. 

The gas kinematics was analyzed at two different spatial scales: 8000 au and 150 au. At large-scale, we have obtained similar results as those presented in \citet[][]{ruizrodriguez2017}, revealing the wide outflow arcs (traced by the $^{12}$CO) and the infalling envelopes ($^{13}$CO and C$^{18}$O). At the small-scale, we detected depleted gas emission (cavities) for both disks  ($^{12}$CO and $^{13}$CO, moment 0). They were probably formed by slow outflows and jets coming from the disks along the line of sight or by optical depth effects and over-subtraction of the continuum emission. The dynamics of both disks in such scales were masked principally by cloud contamination. 

Further observations with similar resolution but of optically thinner molecular lines may lead to the characterization of the dynamical interaction of the two components. For example, the identification of rotational signatures from the disks can be used to identify the dynamical masses of their central stars. With this information, we could also constrain truncation radii. Additionally, observations of small-scale gas structures, allied to astrometric measurements, can be used to constrain the possible formation scenarios.

We conclude that the HBC 494 system constitutes a test bed for eruptive binary-disk interactions and the connection between stellar multiplicity and accretion/luminosity processes (such as outbursts). Also, it provides a statistics enhancement about the rarely observed systems FUors and Exors.
\label{sec:sum}

\section{Appendix}
\subsection{Effects of continuum over-subtraction}

\begin{figure*}[!htb]
\begin{center}
\includegraphics[width=1\textwidth]{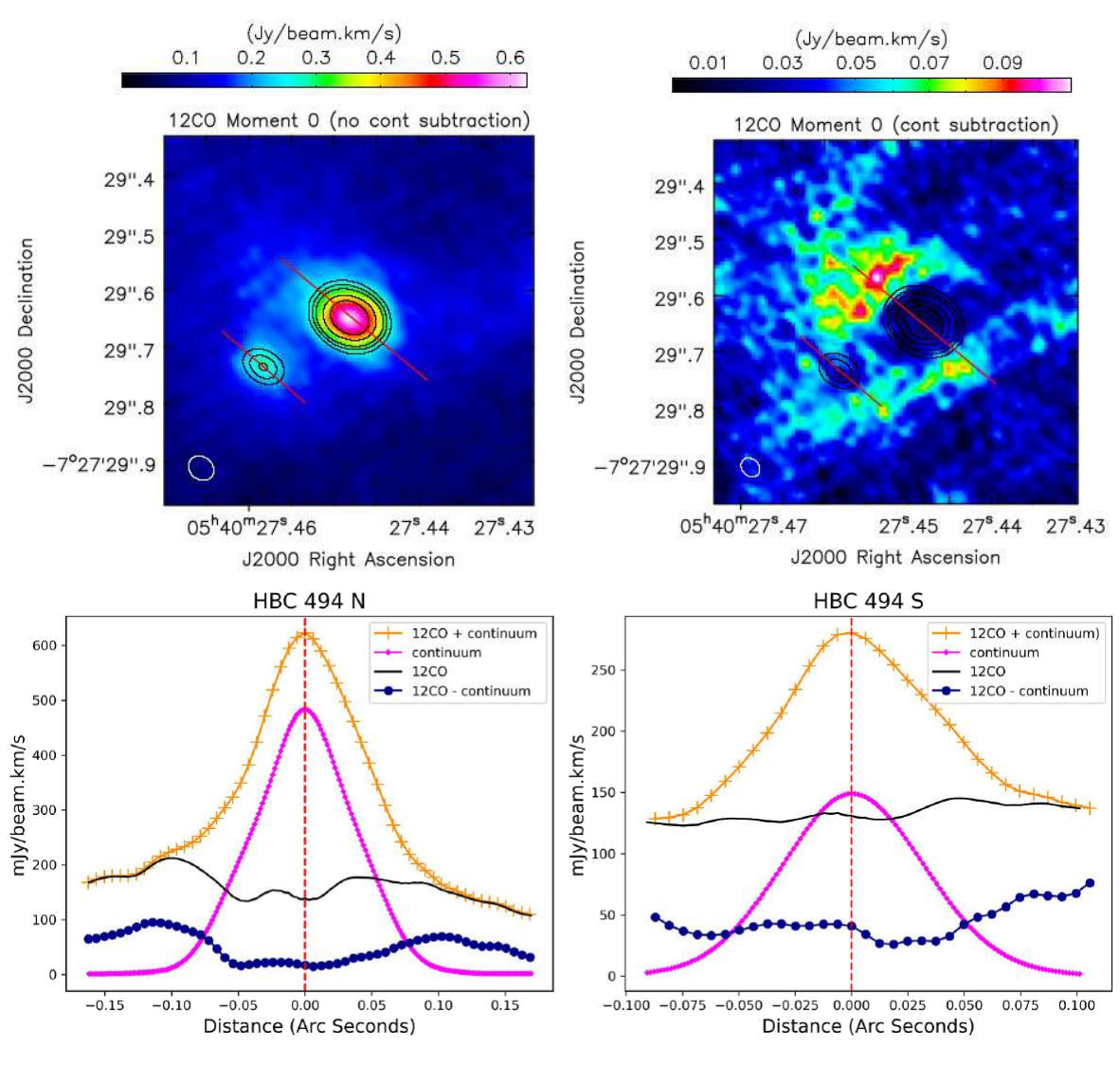}
\caption[$^{12}$CO moment 0 maps before and after continuum removal, along with their spatial profiles]{The top panels show the $^{12}$CO moment 0 maps before (left) and after (right) continuum removal, along with their spatial profiles. The moment 0 contours represent values of 50, 75, 100, 150, and 200 times the rms ($\sim$0.06 mJy/beam.km/s). The red lines trace the axis used to create spatial profiles, which can be seen in the bottom panels. The bottom left panel shows the spatial profile of the HBC 494 N disk, while the bottom right panel shows that of the HBC 494 S disk. The vertical black dashed lines indicate the center of the disks. The orange crosses/lines represent spatial profiles before the continuum subtraction, created using the red line axis in the top panel, left. The magenta diamonds/lines represent the continuum spatial profile. The black lines represent the subtraction between the $^{12}$CO+continuum and the continuum spatial profiles for each disk. The dark blue filled circles/lines represent spatial profiles after the continuum subtraction, created using the red line axis in the top panel, right.}
\label{spatialprofiles}
\end{center}
\end{figure*}

To evaluate the effects of continuum over-subtraction, we generated spatial profiles for each disk using $^{12}$CO moment 0 maps at small scales before and after continuum subtraction. We created cuts that crossed the semi-major axis of each disk (see red lines in Figure \ref{spatialprofiles}, top panels) for this purpose. Figure \ref{spatialprofiles}, top-left panel displays moment 0 maps before continuum removal, while the top-right panel shows the same maps after this process. The comparison between these effects is better visualized in the bottom panels, where continuum emission (magenta diamonds/lines) has a similar shape as $^{12}$CO before continuum subtraction (orange crosses/lines), but shifted in flux, for both disks. The spatial profile subtractions are displayed by black lines. However, they display higher fluxes when compared to the spatial profiles produced after continuum subtraction (blue-filled circles/lines). Additionally, the continuum was overestimated by foreground optical depth molecular lines that absorbed continuum emission and then propagated to the \textit{uvcontsub} CASA task. The "holes" or cavities seen in the top-right panel could potentially be diminished by manually adding back some of the lost line signal, but this would only create a fake signal at the level of noise. Thus, optically thinner lines must be used to obtain the real signal of disks in moment 0 and 1 maps.

\subsection{$^{12}$CO, $^{13}$CO and C$^{18}$O channel maps with contours and the effect of cloud contamination}

\begin{figure*}[!htb]
\begin{center}
\includegraphics[width=1\textwidth]{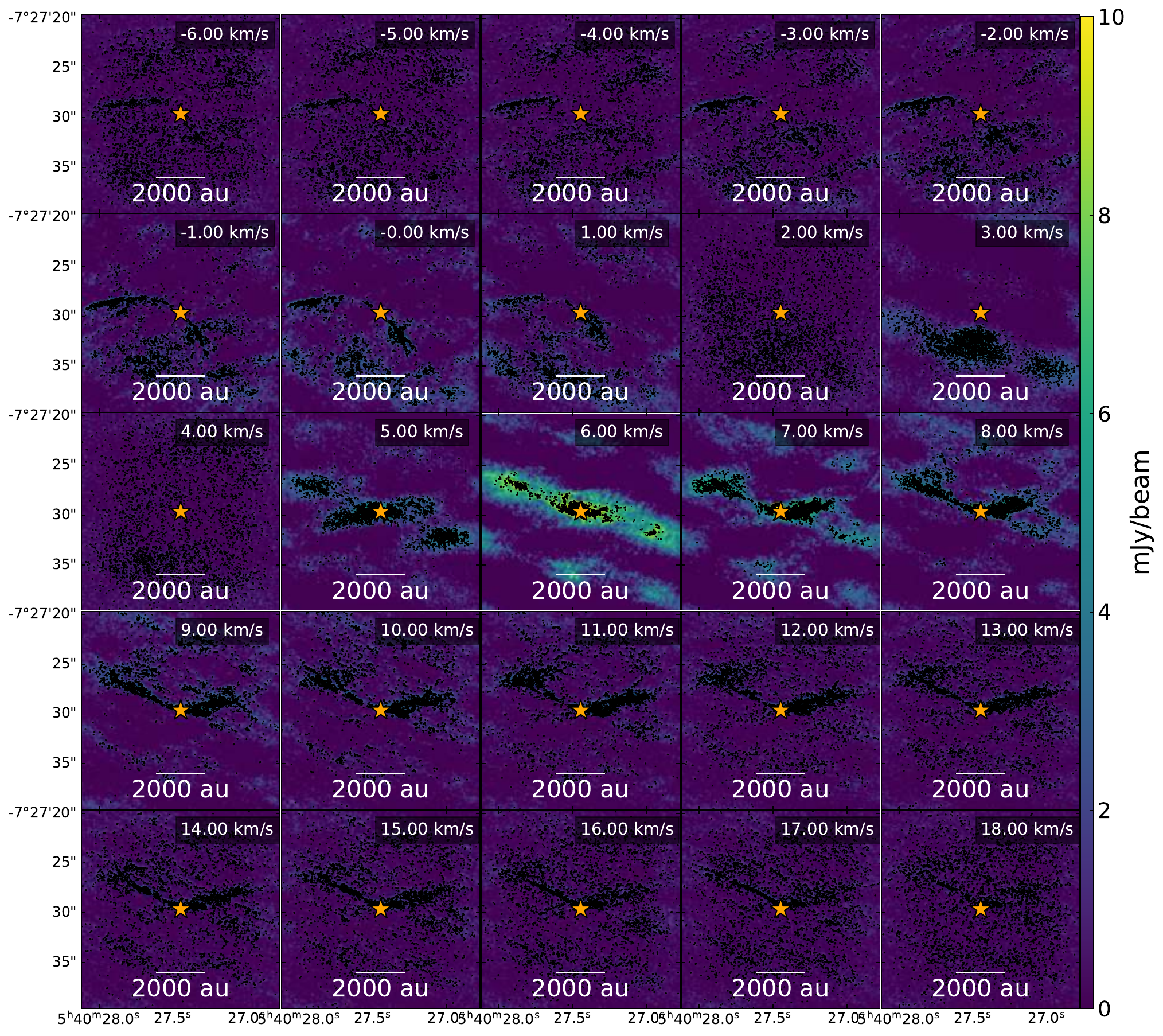}
\caption[$^{12}$CO channel maps of HBC 494 system with contours]{$^{12}$CO channel maps of HBC 494 system with contours. The contours were displayed in black and they are equivalent to 3 times the rms of every channel. The rms values were calculated using all data in each channel, through the CASA task \textit{imstat}. The peak flux values have reached up to 19 times the rms values. The star in the center marks the position of the continuum disk around HBC 494 N.}
\label{12cochannelcontours}
\end{center}
\end{figure*}

\begin{figure*}[!htb]
\begin{center}
\includegraphics[width=1\textwidth]{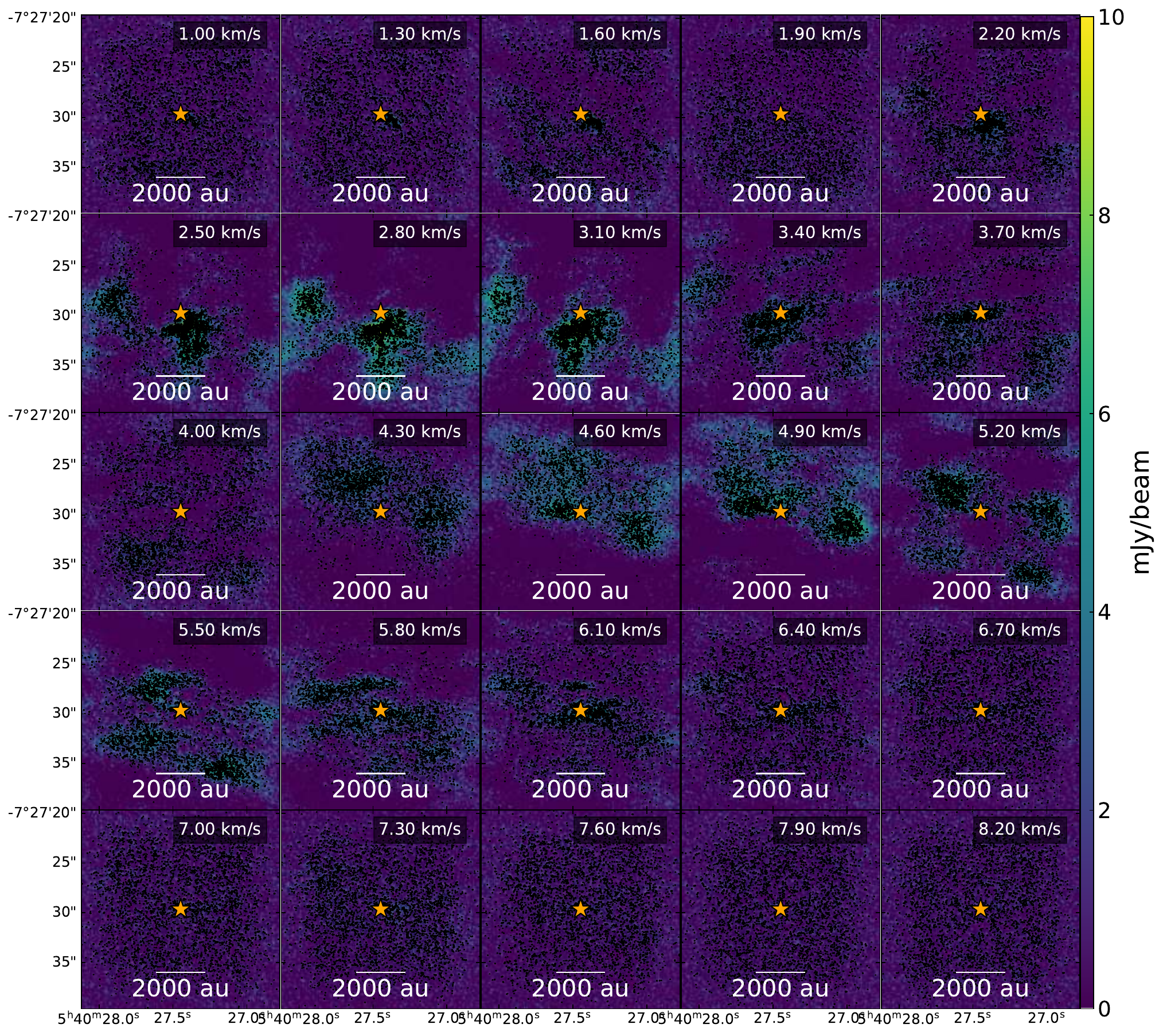}
\caption[$^{13}$CO channel maps of HBC 494 system with contours]{$^{13}$CO channel maps of HBC 494 system with contours. The contours were displayed in black and they are equivalent to 3 times the rms of every channel. The rms values were calculated using all data in each channel, through the CASA task \textit{imstat}. The peak flux values have reached up to 11 times the rms values. The star in the center marks the position of the continuum disk around HBC 494 N.}
\label{13cochannelcontours}
\end{center}
\end{figure*}

This appendix presents the $^{12}$CO, $^{13}$CO, and C$^{18}$O channel maps with contour plots (see Figures \ref{12cochannelcontours} and \ref{13cochannelcontours}). To visualize the signals and corresponding flux values using a color bar, we also provide versions of the same channel maps without contours (see Figures \ref{12cochannel} and \ref{13cochannel}).

Figure \ref{12cochannelcontours} shows the 3-rms significant southern arcs (velocity channels from -6 km/s to 1 km/s) and northern arcs (velocity channels from 5 km/s to 18 km/s). However, cloud contamination affects channels between 2 km/s and 6 km/s. These channels were not used in the creation of large-scale moment maps. Channel 6 km/s shows the most evidence of cloud contamination, with contour plots showing innermost flux values near the disks (marked by the central orange star). It also shows some significant contours on both horizontal extremes, which can be tracers of the filtered signals using only the main array data. The data partially trace features that, due to their large extension, require a combination of the ALMA main and single dish arrays to recover extended flux emission.

In Figure \ref{13cochannelcontours}, the $^{13}$CO channel maps are heavily marked by 3-rms black contours, which show the significant gas signal on each channel. Specifically, channels from 2.20 km/s to 6.10 km/s indicate gas infalling clumps near where the disks are located, but also show random/scattered signals far from the disks due to cloud contamination.

\begin{figure*}[!htb]
\begin{center}
\includegraphics[width=1\textwidth]{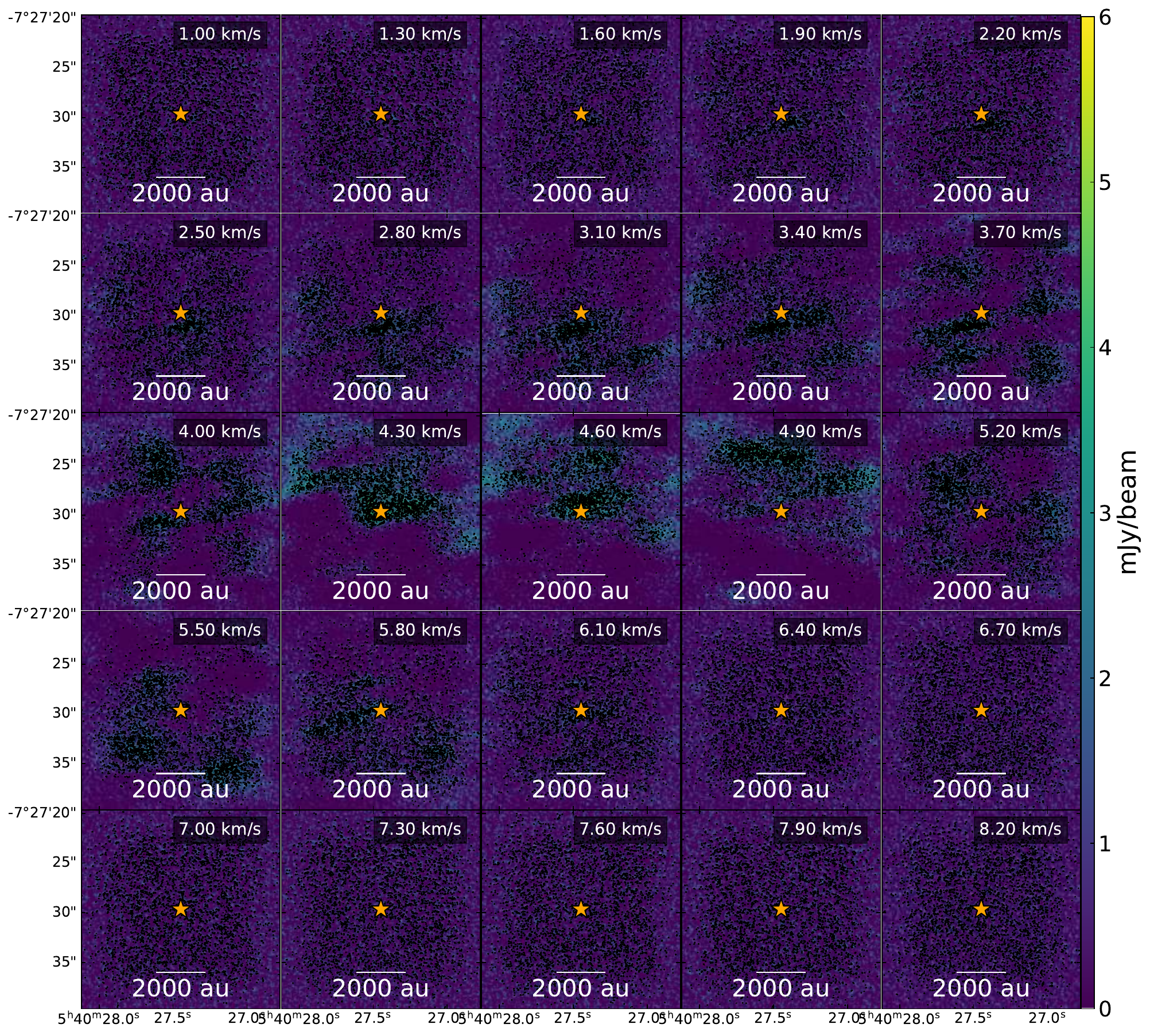}
\caption[C$^{18}$O channel maps of HBC 494 system with contours]{C$^{18}$O channel maps of HBC 494 system with contours. The contours were displayed in black and they are equivalent to 3 times the rms of every channel. The rms values were calculated using all data in each channel, through the CASA task \textit{imstat}. The peak flux values have reached up to 13 times the rms values. The star in the center marks the position of the continuum disk around HBC 494 N.}
\label{c18ochannelcontours}
\end{center}
\end{figure*}

The C$^{18}$O channel maps represented in Figure \ref{c18ochannelcontours} are also heavily marked by 3-rms black contours. Channels from 2.80 km/s to 4.60 km/s indicate gas infalling clumps near where the disks are located, but also show random/scattered signals far from the disks due to cloud contamination. Other channels are affected by cloud contamination or lack a significant signal.

\subsection{Extra channel maps}

\begin{figure*}[!htb]
\begin{center}
\includegraphics[width=1\textwidth]{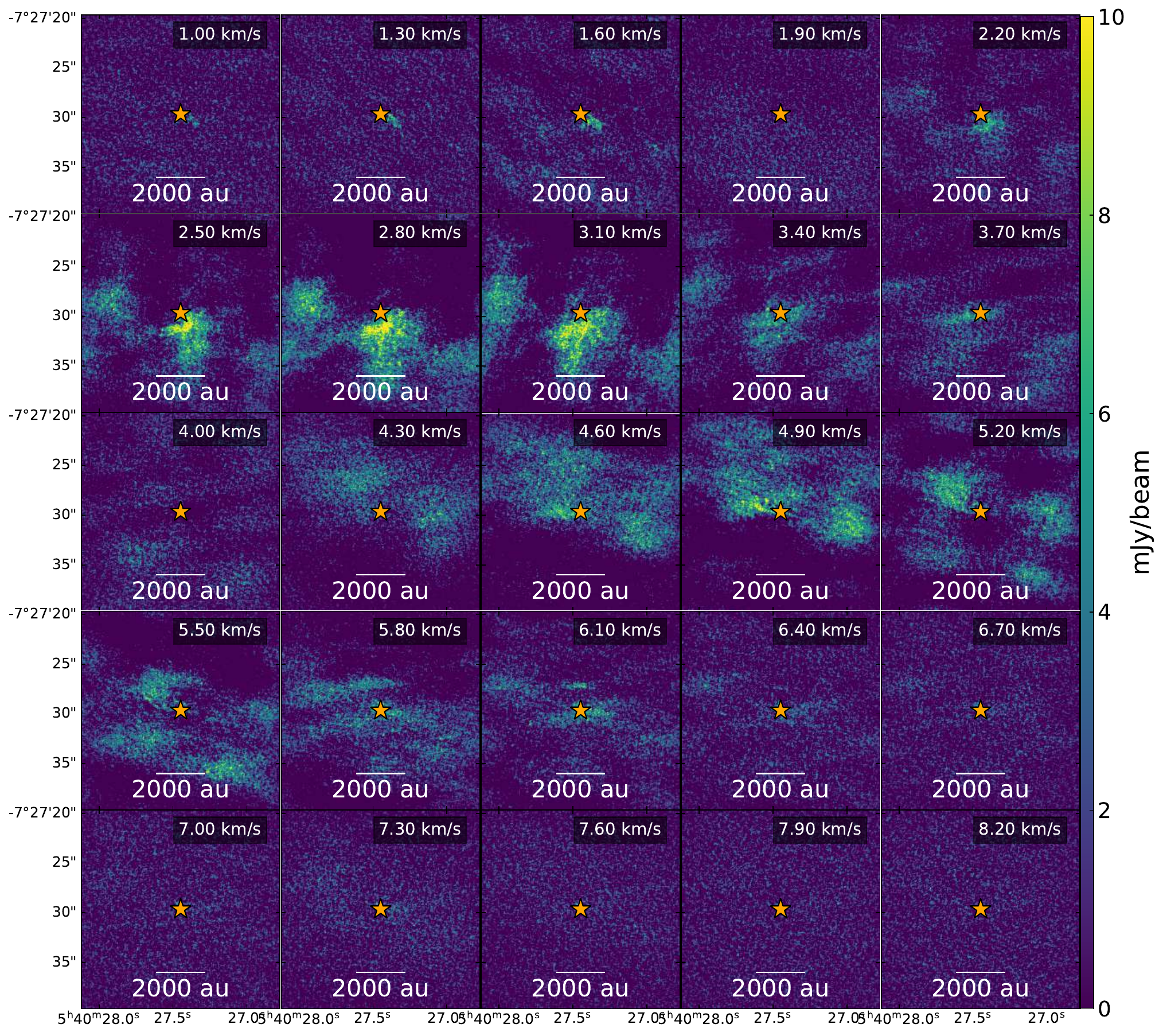}
\caption[$^{13}$CO channel maps of HBC 494 system]{$^{13}$CO channel maps of HBC 494 system. The star in the center marks the position of the continuum disk around HBC 494 N.}
\label{13cochannel}
\end{center}
\end{figure*}

\begin{figure*}[!htb]
\begin{center}
\includegraphics[width=1\textwidth]{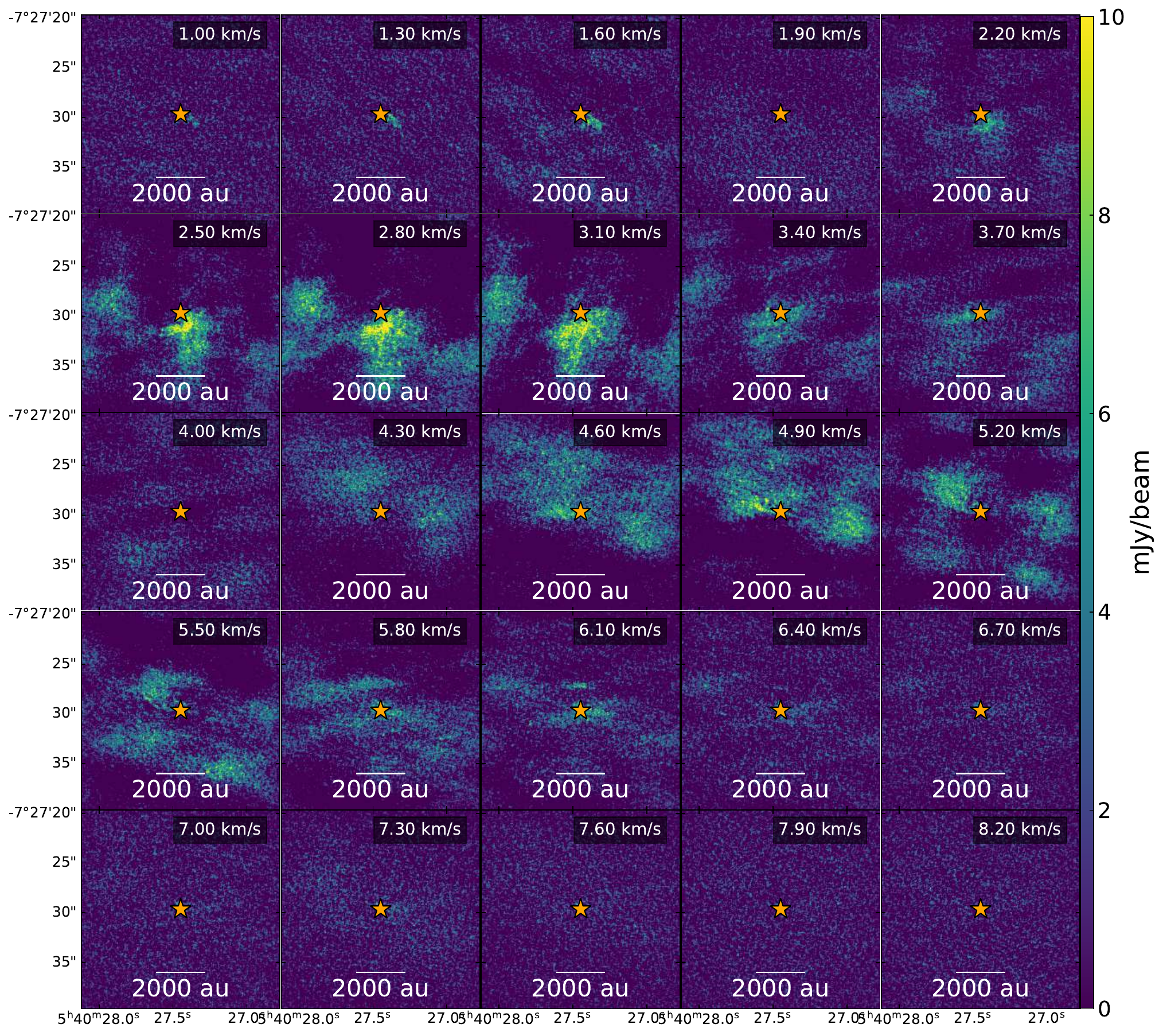}
\caption[C$^{18}$O channel maps of HBC 494 system]{C$^{18}$O channel maps of HBC 494 system.}
\label{c18ochannel}
\end{center}
\end{figure*}

\begin{figure*}[!htb]
\begin{center}
\includegraphics[width=1\textwidth]{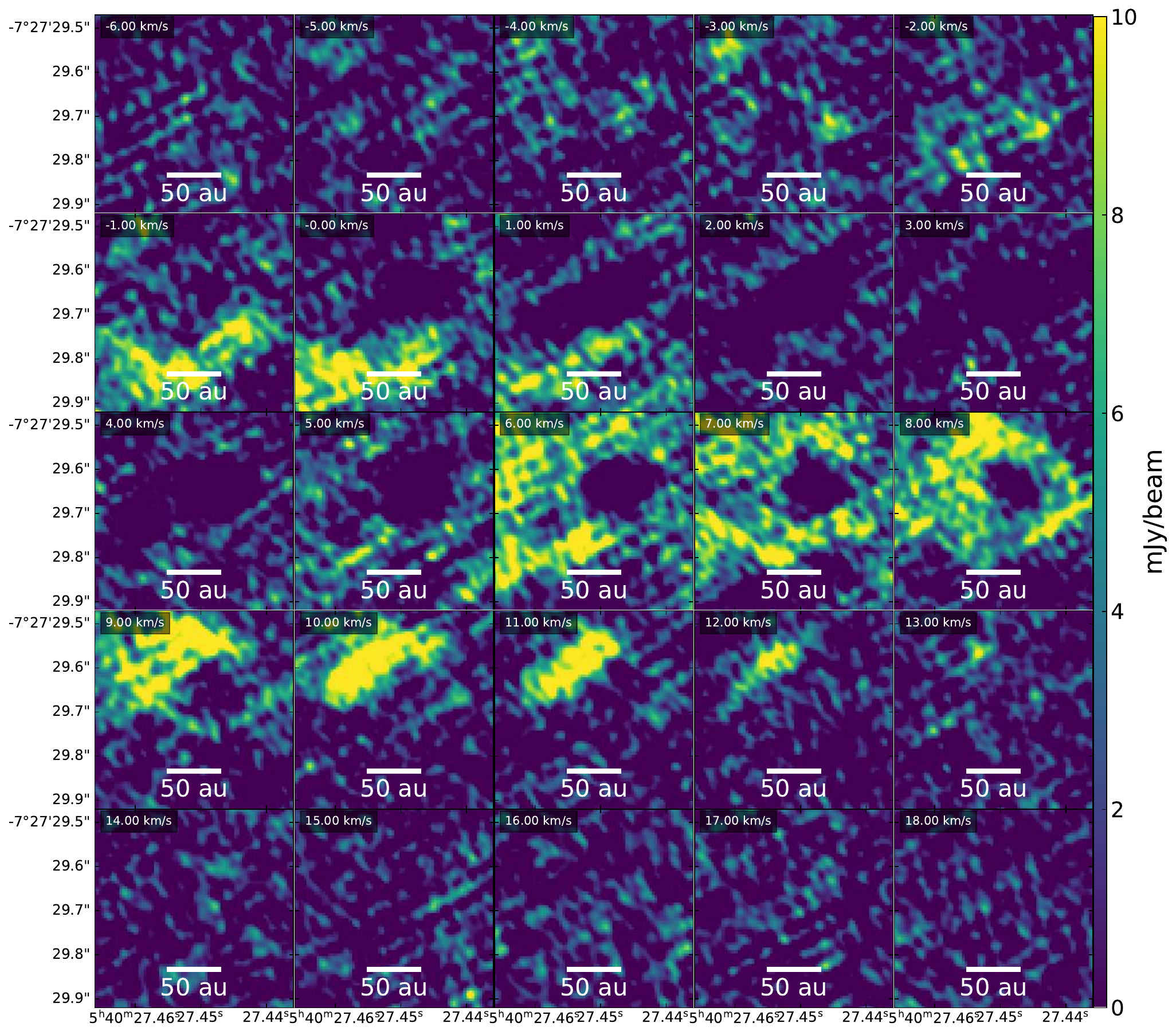}
\caption[$^{12}$CO channel maps of HBC 494 system, small-scale, after removing the continuum contribution]{$^{12}$CO channel maps of HBC 494 system, small-scale, after removing the continuum contribution.}
\label{12cochannel_nocont}
\end{center}
\end{figure*}

\begin{figure*}[!htb]
\begin{center}
\includegraphics[width=1\textwidth]{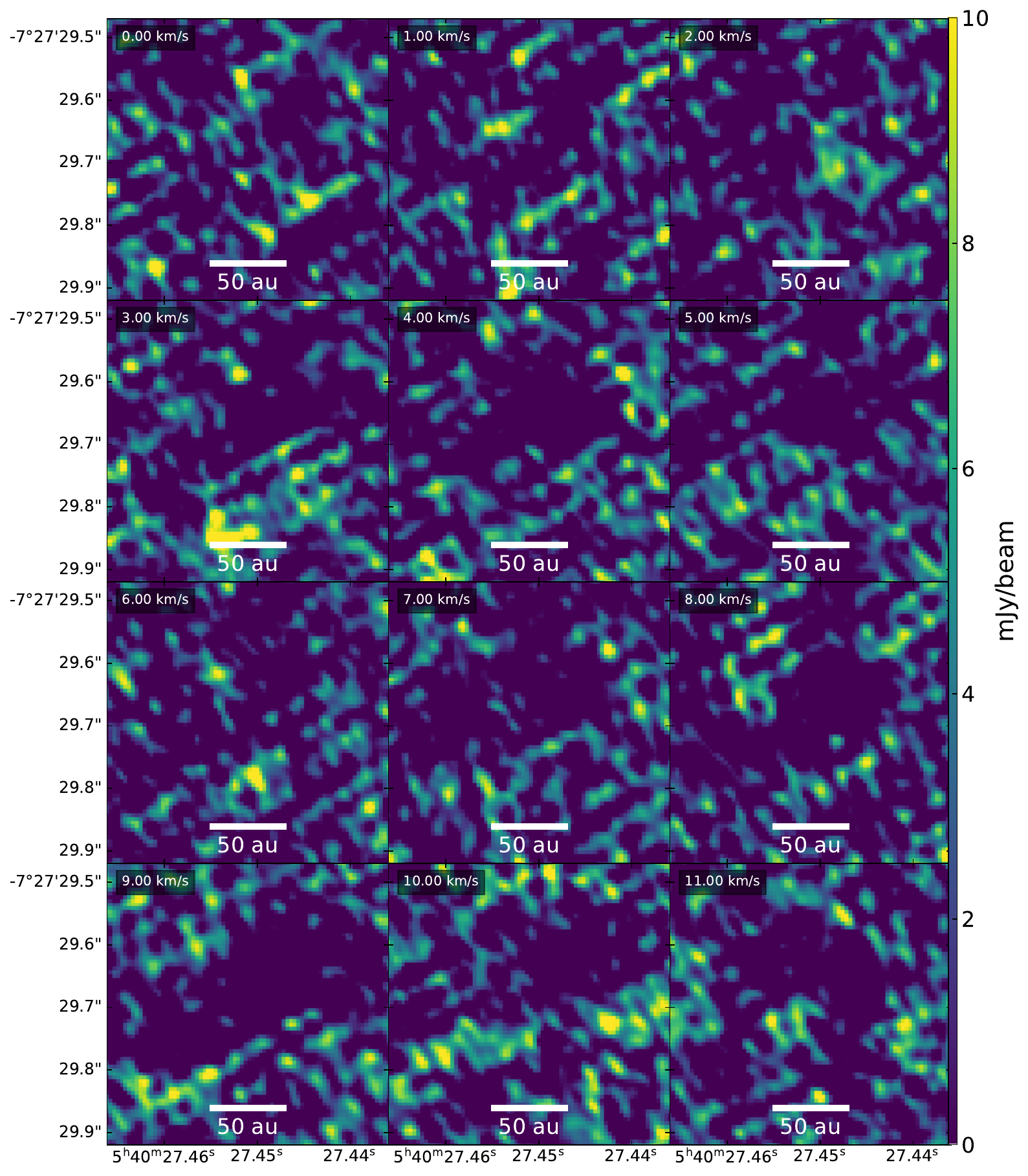}
\caption[$^{13}$CO channel maps of HBC 494 system, small-scale, after removing the continuum contribution]{$^{13}$CO channel maps of HBC 494 system, small-scale, after removing the continuum contribution.}
\label{13cochannel_nocont}
\end{center}
\end{figure*}

\begin{figure*}[!htb]
\begin{center}
\includegraphics[width=1\textwidth]{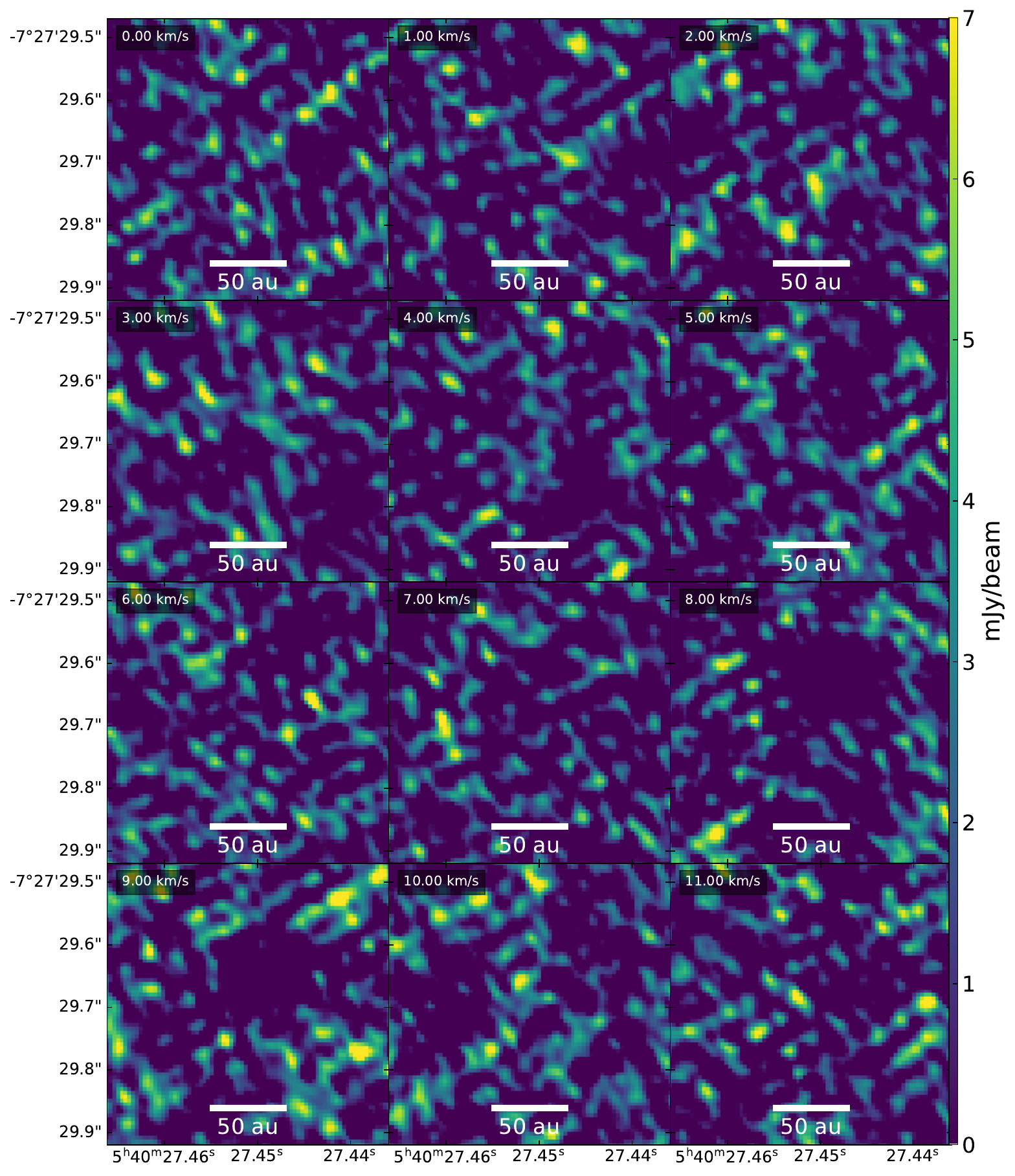}
\caption[C$^{18}$O channel maps of HBC 494 system, small-scale, after removing the continuum contribution]{C$^{18}$O channel maps of HBC 494 system, small-scale, after removing the continuum contribution.}
\label{c18ochannel_nocont}
\end{center}
\end{figure*}

The $^{13}$CO (large scale), C$^{18}$O (large scale), $^{12}$CO (small scale and with continuum subtracted), $^{13}$CO (small scale and with continuum subtracted), and C$^{18}$O (small scale and with continuum subtracted) channel maps are displayed in Figs. \ref{13cochannel}, \ref{c18ochannel}, \ref{12cochannel_nocont}, \ref{13cochannel_nocont}, and \ref{c18ochannel_nocont}, respectively.

  \chapter[Characterization of the $\eta$\,Tel system]{Astrometric and photometric characterization of $\eta$ Tel B combining two decades of observations}
\label{chap:3}
\textit{The content of this chapter was accepted for publication in the journal Astronomy \& Astrophysics (A\&A), see \citet{nogueira2024} - \url{https://ui.adsabs.harvard.edu/abs/2024arXiv240504723N/abstract}.}

\section{Introduction}

Since the first discovery of a planet around a main-sequence star \citep{mayorequeloz1995}, techniques and instruments for discovering and characterizing substellar objects have advanced exponentially. Today, more than 5500 exoplanets have been confirmed\footnote{Information extracted from the NASA Exoplanet archive database \url{https://exoplanetarchive.ipac.caltech.edu/} \citep{akeson2013}.}, and more than 19,000 candidates of ultracool dwarfs (spectral type later than M7) are known \citep{dalPonte2023}. However, little is known about where, when, and how substellar objects form. Thus, investigating newborn substellar objects in a stellar system is essential to address this knowledge gap, since they are expected to retain signatures of their formation pathway through the system architecture (e.g., \citealp{2004ApJ...609.1045M,2014MNRAS.443.2541P,2014MNRAS.438L..31G,2019MNRAS.484.1926D,2020AJ....159...63B,2023AJ....166...48D}) and from their atmosphere (e.g., \citealp{2007prpl.conf..733M,2011ApJ...743L..16O,2014ApJ...790..133D,2015A&A...577A..42B,2018ApJ...853..192C,2023AJ....166..198Z,2024ApJ...966L..11P}).

In particular, the high-contrast imaging technique (hereafter called HCI) is a capable tool to discover and characterize young giant planets and brown dwarfs. HCI is sensitive to recently formed objects that conserve some formation heat. Indeed, substellar objects have been discovered using HCI techniques in the past decade (e.g., 2MASSWJ 1207334-393254: \citealp{chauvin2004}; $\beta$ Pic b: \citealp{lagrange2010}; HD 95086 b: \citealp{rameau2013discovery}; PDS 70 b: \citealp{keppler2018}; YSES 2b: \citealp{bohn2021}). More recently, with the release of the Gaia catalogs and an update on astrometric precision, HCI has also been used to follow up on stars that showed anomalous accelerations (difference between their long-term HIPPARCOS-Gaia and short-term Gaia proper motion vectors), pointing towards the presence of companions in the system. One remarkable example is the discovery of AF Lep b through proper motion anomalies and HCI \citep{mesa,2023A&A...672A..94D,2023ApJ...950L..19F}.

Near-infrared substellar companions observed by High-Contrast Imaging (HCI) may offer valuable multiwavelength and multi-analysis data for the detection of circumplanetary disks (CPDs) or satellites in their surroundings. For instance, \citet{perez2019} analyzed ALMA band 6 observations featuring directly imaged companions, establishing upper limits on CPD detectabilities. Similarly, using SPHERE near-infrared (NIR) images, \citet{lazzoni2020} set upper limits for satellite detections around substellar companions. Notably, \citet{lazzoni2020} also identified a potential satellite candidate around DH Tau B, although confirmation of its nature is still pending. Subsequently, \citet{lazzoni2022} provided a more robust analysis on the detectability of satellites through all standard exoplanet discovery techniques. Additionally, \citet{ruffio2023} proposed to direct efforts towards detecting potential candidates through radial velocity monitoring, a method that can be implemented by monitoring companions with high-resolution spectroscopy. Another technique for searching for satellites is spectroastrometry, which consists of the fine measurement of any deviation from the position of the center of light \citep{2015ApJ...812....5A}. If an integral field spectrograph targets an unresolved planet-satellite system, it is expected that the center of the light shifts position depending on the wavelength. The movement of the centroid of the PSF would reveal if a satellite is present.

One system that comes to attention based on its age, being targetable from HCI and satellite analysis, and its astrometric follow-up is $\eta$ Telescopii (hereafter called $\eta$ Tel). $\eta$ Tel is part of the $\beta$ Pic moving group, with an estimated age of 18 Myr \citet{2020A&A...642A.179M} and a distance of 49.5 pc \citep{gaiadr3}. It is composed of an A0V, 2.09 M$_{\odot}$ primary star ($\eta$ Tel A; \citealp{houk1975, desidera2021,desidera2021catalogue}) and an M7-8, brown dwarf companion ($\eta$ Tel B) at $\sim$4.2\arcsec separation and position angle of $\sim$169$^{\circ}$ (\citealp{lowrance2000, guenther2001}, and references therein). $\eta$ Tel A presents a likely warm debris belt at 4 au (unresolved, only inferred from the SED) and an edge-on cold debris belt at 24 au discovered via infrared excess \citep{backman1993,mannings1998}. The outer debris disk was later resolved with T-ReCS (Thermal-Region Camera Spectrograph) on Gemini South \citep{2009A&A...493..299S}. Moreover, despite gas absorption towards $\eta$ Tel debris disk was previously attributed to radiatively driven debris consistent with C/O solar ratio \citep{youngblood2021}, more recent findings revealed that the origin of absorption lines is more likely of an interstellar cloud traversing the line of sight of $\eta$ Tel A \citep{Iglesias2023}.
$\eta$ Tel B is a bright substellar companion with a contrast of 6.7 magnitudes in the VLT/NACO H band \citep{neuhauser2011}, or 11.85 in apparent magnitude. Its astrometrical points and orbital constraints were first compiled and analyzed in \citet{neuhauser2011}, which used 11 years of imaging data (1998-2009). An additional NACO observation was presented in \cite{rameau2013}, which broadened the time baseline to 2011.

In this paper, we present a characterization of the companion, including astrometrical and photometrical follow-up and orbital constraint analysis. We present three new epochs from the Spectro-Polarimetric High-contrast Exoplanet REsearch (SPHERE) instrument \citep{beuzit2019}. With the addition of the new SPHERE data and a baseline of observations from 1998 to 2017, we present the most recent and complete orbital characterization of the system. Furthermore, we performed an analysis of the surroundings of $\eta$ Tel B to constrain the possible presence of features such as satellites or circumplanetary disks. 

The manuscript is structured in the following order. Observations and data reduction are described in Section \ref{sec:obs}. The photometric and astrometric measurements of the system of the new SPHERE observations are reported in Section \ref{sec:astrometry}. The orbital fitting analysis, taking into account the new data and the literature can be found in Section \ref{sec:orbfit}. The description of the study of the close vicinity of the substellar companion and the contrast curves around $\eta$ Tel B are presented in Section \ref{sec:sathunt}. Final remarks and conclusions are presented in Section \ref{sec:sum}. 

\section{Observations and data reduction} 
\label{sec:obs}
\subsection{Observations}
We present new SPHERE/IRDIS coronagraphic data of the system around $\eta$ Tel A. SPHERE is a VLT (Very Large Telescope) planet-finder instrument located at Paranal, Chile - UT3. It is an instrument dedicated to HCI. SPHERE is composed of four main scientific parts: the Common Path and Infrastructure (CPI), which includes an extreme adaptive optics system (SAXO, \citealp{fusco2006,petit2014}) and coronagraph systems; the Infrared Dual-band Imager and Spectrograph (IRDIS, \citealp{dohlen2008}) with a pixel scale of 12.25 mas and field of view (FoV) of 11$\times$12.5 arcsec; the integral field spectrograph (IFS, \citealp{claudi2008}) with FoV of 1.73\arcsec$\times$1.73\arcsec; and the Zurich Imaging Polarimeter (ZIMPOL, \citealp{schmid2018}), the visible light imager and polarimeter of SPHERE. Both IRDIS and IFS belong to the NIR branch and can be used concomitantly if required during observations. Given the separation of $\eta$ Tel B and its brightness in the near-infrared, IRDIS observations are the most suitable for the purpose of our paper. 

The observations were performed with IRDIS on three different nights. Since the stellar companion is outside the field of view (FoV) of the integral field spectrograph (IFS) instrument, we do not report its data in this manuscript. The first observation (program 095.C-0298(A); PI Beuzit) was taken on 2015-05-05, generally under average conditions, with variable seeing ranging between 1 and 2 arcseconds (this epoch had photometric sky transparency). The second observation (program 097.C-0394(A); PI Milli) was taken under good conditions on 2016-06-15, and the seeing was mostly stable during the period of observation, varying between 1-1.4 arcseconds (this epoch had the sky transparency declared as thin). This program was performed as part of the SPHERE High-Angular Resolution Debris Disks Survey (SHARDDS; \citealp{dahlqvist2022}), designed to image circumstellar disks around bright nearby stars (within 100 pc from the Earth). The third observation (program 198.C-0209(H); PI Beuzit) was taken on 2017-06-15, with a slightly higher value of the seeing and poorer conditions (``thin'' sky transparency). An apodized Lyot coronagraph (N\_ALC\_YJH\_S; inner working angle $\sim$0.15\arcsec) was used on all three nights. A dual-band filter H2H3 was set ($\lambda$= 1.593 and 1.667 $\mu$m for H2 and H3, respectively; $\Delta$$\lambda$= $\sim$0.053 $\mu$m for both filters) on the first and third nights, while on the second night, a broadband filter H ($\lambda$= 1.625 $\mu$m; $\Delta$$\lambda$= 0.29 $\mu$m) was used instead. A fourth IRDIS sequence was taken (2018-05-08; ID: 1100.C-0481(G); PI: Beuzit), but it was discarded from our analysis due to bad observing conditions. The observations and their specifications are summarized in Table~\ref{table_obs}.

\begin{sidewaystable}
\caption{List of the SPHERE/IRDIS $\eta$ Telescopii observations used in this work.}
\begin{center}
\begin{tabular}{ccccccccc}
\hline
\hline
Date (UT)  & ESO ID - PI            & IRDIS filter & DIT $\times$ NDIT & $\Delta$PA & Seeing ["]& Avg. coherence time [s]\\
\hline
2015-05-05 & 095.C-0298(A)- Beuzit & DB\_H2H3     & 32 $\times$ 8               & 46.83  & Mostly 0.9-1.5$^a$  &   0.0011    \\
2016-06-15 & 097.C-0394(A)- Milli  & BB\_H        & 8 $\times$ 8                & 18.17  &  1-1.4 &   0.0024       \\
2017-06-15 & 198.C-0209(H)- Beuzit & DB\_H2H3     & 32 $\times$ 12              & 2.04   & 1.5-2  &   0.0023    \\
2018-05-08$^b$ & 1100.C-0481(G)- Beuzit &  --    & -- & --     & --  &   --   \\
\hline
\end{tabular}
\end{center}
\Centering
 \raggedright {\footnotesize $^a$A subtle increase in seeing above 1.5 was registered between 08h40m-9h00m UTC at that night.}
 \\
\raggedright {\footnotesize $^b$Discarded from the analysis for bad observing conditions.  }
\label{table_obs}
\end{sidewaystable}
\subsection{Data reduction}
The reduction of the three epochs was performed by the High Contrast Data Center pipeline\footnote{Formerly known as SPHERE Data Center or SPHERE DC.} (hereafter called as HC DC; \citealp{delorme2017}), which utilizes the Data Reduction and Handling software \citep[v0.15.0;][]{pavlov2008} and routines presented in \cite{2018A&A...615A..92G}. The process includes standard pre-reduction steps such as background subtraction, flat-fielding, and bad pixel correction. The frames are recentered using the SPHERE waffle pattern, followed by corrections for the anamorphism of the instrument and astrometric calibration (pixel scale and True north correction), as described in \citet{maire2016}. The final products comprise a master cube that contains all frames, the position angle (PA) values of each frame, and an off-axis PSF reference cube. The off-axis PSF is an unsaturated image of the central star taken before and after the coronagraphic sequence for flux calibration.  

\begin{figure}[!h]
\begin{center}
    \includegraphics[width=0.78\textwidth]{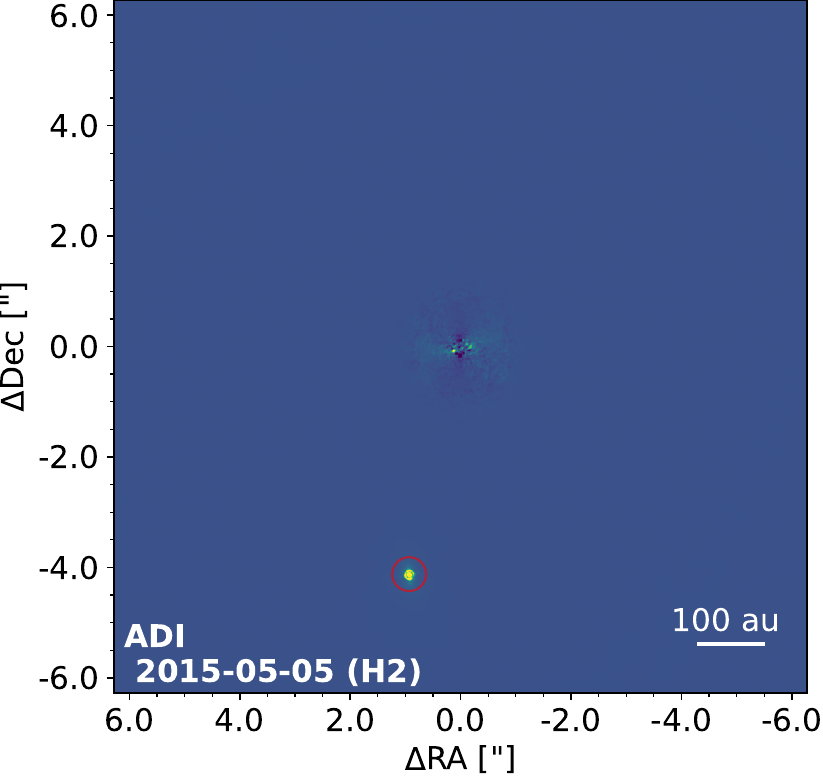}
    \caption{ADI-processed from IRDIS coronagraphic sequence (epoch: 2015-05-05; filter: H2) for the $\eta$ Tel system. The red circle marks the position of $\eta$ Tel B. The star is masked under the coronagraph at the center of the image. North is up, East is left.}
    \label{adis}
\end{center}
\end{figure}

We made use of the VIP code (Vortex Image Processing, \citealp{vip2017}; v.1.5.1) to reject the bad frames in each master cube, considering only frames with a Spearman correlation above 0.85 compared to the first frame taken under good conditions. After this step, we created post-processing images, employing the Angular Differential Imaging \citep[ADI;][]{marois2006} and principal component analysis (PCA; \citealp{soummer2012, amaraequanz2012}) with varied numbers of principal components.
Since the PCA introduces deep over-subtraction due to the FoV rotation even for a few components, we favored the classical ADI images for our analysis. The result of this post-processing technique is presented in Fig. \ref{adis}.

\section{Methodology and results}

\subsection{Astrometry and photometry}
\label{sec:astrometry}

\subsubsection{The NEGFC technique}
\label{sec:astrometry_negfc}

To precisely determine the position and flux of $\eta$ Tel B, we employed the NEGative Fake Companion Technique (NEGFC), as described in studies such as \citet{lagrange2010} and \citet{zurlo2014}. This technique involves modeling the target source by introducing a negative model of the instrumental PSF into the pre-processed data. The procedure aims to minimize residuals in the final image, adjusting the flux and position of the model to align with the source's properties (astrometry and photometry). We implemented the NEGFC using the VIP package.

For this purpose, we adapted the {\tt single\_framebyframe} routine proposed by \citet{lazzoni2020}. This routine provides estimates of separation, position angle, and photometry for the companion in each frame from the coronagraphic sequence. $\eta$ Tel B's brightness (signal-to-noise ratio $\geq$100) and relative distance from the speckle-dominated region are sufficient for this approach to be applicable.

As a model PSF, we utilized the off-axis image of the central star captured before and after the coronagraphic sequence. For each set of coordinates, a negative flux was introduced, and the set of positions and fluxes yielding the lowest residual (standard deviation) in each frame was selected. Consequently, for each night and filter, the routine's results were represented by the median values of the parameters measured across all frames.

\subsubsection{Measurement of the photometry}
\label{sec:phot}
\begin{figure}[!htb]
\begin{center}
    \includegraphics[width=0.8\textwidth]{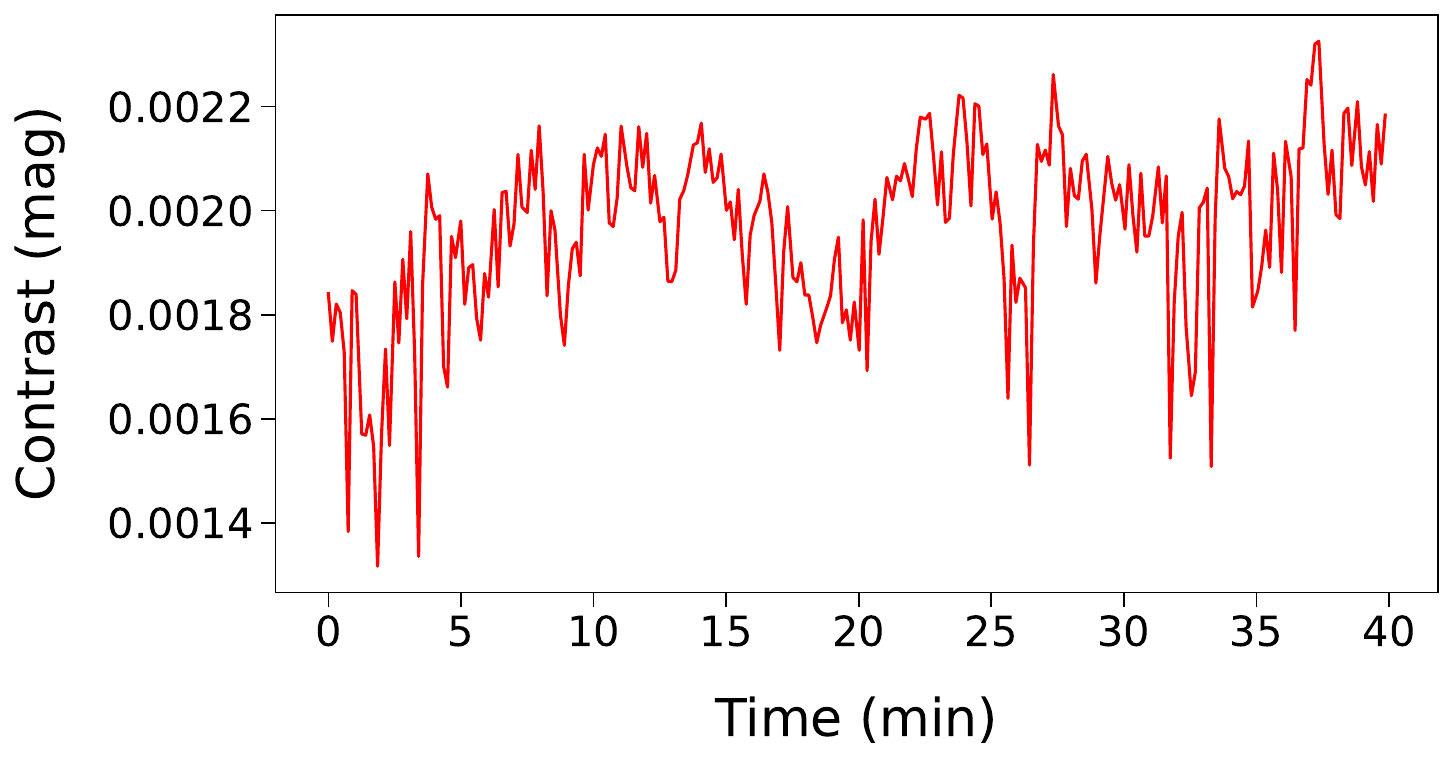}
    \caption{Contrast flux variations for the companion with respect to the star (filter: BB\_H).}
    \label{flux_variations}
\end{center}
\end{figure}

The contrast of the brown dwarf was calculated as the median of the NEGFC technique values on each frame of the coronographic sequence. Flux contrast uncertainties were derived from the standard deviation of the fluxes measured by the NEGFC technique. Fig. \ref{flux_variations} sets flux measurements throughout the sequence using BB\_H filter, showing how the values can vary. The contrast in flux and magnitude with respect to the central star for each filter is shown in Table \ref{table_phot}. The results are consistent with the measurements presented in the literature for the companion.

\begin{table}[!h]
\centering
\caption{IRDIS flux contrast in flux and magnitude with respect to the central star for $\eta$ Tel B.}
\begin{tabular}{cccc}
\hline
\hline
Epoch & Filter        & Contrast [e-3]                                & $\Delta$ magnitude           
\\
\hline

2015-05-05 & H2 & 1.52 $\pm$ 0.60   & 7.05 $\pm$ 0.45   \\
2015-05-05 & H3 & 1.92 $\pm$ 0.12   & 6.79 $\pm$ 0.07   \\

2016-06-15 & BB\_H & 2.00 $\pm$ 0.17 & 6.75 $\pm$ 0.09 \\

2017-06-15 & H2    & 1.53 $\pm$ 0.78 & 7.04 $\pm$ 0.61  \\
2017-06-15 & H3    & 1.92 $\pm$ 0.84 & 6.79 $\pm$ 0.51

\end{tabular}
\label{table_phot}
\end{table}

\begin{figure}[!h]
\begin{center}
    \includegraphics[width=0.8\textwidth]{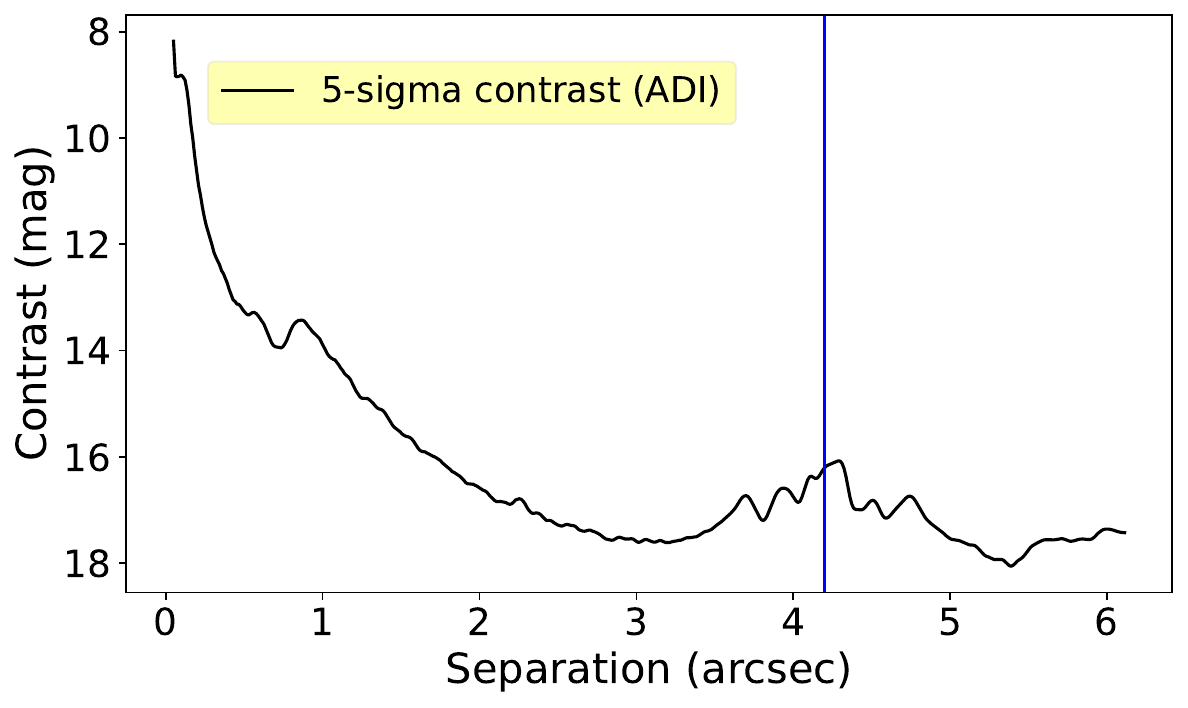}
    \caption{Contrast curve with respect to $\eta$ Tel A, using the datacube corresponding to the 2016-06-15 (BB\_H) observation. The vertical blue line corresponds to the position of the companion.}
    \label{contrastcurve_aroundstar}
\end{center}
\end{figure}

After using the NEGFC technique, the signal of the companion was removed in each frame of the datacubes, creating therefore empty datacubes. Following, a 5-sigma contrast curve with respect to and around $\eta$ Tel A was produced using the ADI processed data. The achieved contrast for epoch 2016-06-15 is shown in Fig. \ref{contrastcurve_aroundstar}.

\subsubsection{Measurement of the astrometric positions}
\label{astro_determination}

To robustly determine the separation and position angle at each epoch, we applied multiple methods while concurrently estimating the disparities between these algorithms. Specifically, we employed: 

\begin{itemize}
    \item the NEGFC method detailed in Section \ref{sec:astrometry_negfc};
    \item a 2D-Gaussian fitting encompassing the companion. We optimized Gaussian statistics using two stochastic algorithms: Adam \citep{kingma2014adam} and the Levenberg–Marquardt algorithm \citep{more2006levenberg};
    \item the peak intensity pixel location within a FWHM of the companion.
\end{itemize} 


All the algorithms were applied to each frame of the coronagraphic (frame-by-frame) sequence as well as to their median-collapsed reduction. When using the frame-by-frame method, it is important to note that a sequence of positions is obtained and it is necessary to reduce them by using the median. Finally, the astrometric points for each night and filter were determined by calculating the median among the outputs of the algorithms. Similarly, the uncertainties of the astrometrical fitting were calculated as the standard deviation across all results obtained from the various methods. We justify applying different methods as we observed that the final results of each method could differ by a maximum of 10 mas in separation and approximately 0.15 degrees in position angle. Therefore, we opted for implementing a median instead of an average to filter out values that may otherwise skew the results.

In addition to the uncertainties on the astrometric fitting of the companion, we have to consider other factors in the error budget, as previously stated by \citet{wertz2017}:
\begin{itemize}
    \item Instrumental calibration, where the most relevant errors come from the orientation of the True North, pupil offset, plate-scale, and anamorphism;
    \item Determination of the position of the central star behind the coronagraph;
    \item Systematic error due to residual speckles;
    \item Statistical error due to planet position determination.
\end{itemize}

Therefore, considering R as the final expression for radial separation in arcseconds and $\Theta$ as the final expression for position angle in degrees, we applied the following approximated equations:

\begin{equation}\label{eq:R}
    R = PS ( R_{*} \pm R_{spec} \pm r R_{AF} \pm r)\\ \text{and}
\end{equation}

\begin{equation}\label{eq:theta}
 \Theta =  \Theta_{*} \pm \Theta_{spec} \pm \Theta_{AF} \theta \pm \Theta_{PO} \pm \Theta_{TN} \pm \theta \\ \text{, }
\end{equation}

\noindent  
 where r is the radial distance and $\theta$ the position angle, $R_{*}$ and $\Theta_{*}$ the radial and azimuthal values related to stellar centering; $R_{spec}$ and $\Theta_{spec}$ the radial and azimuthal values related to speckle noise;  $R_{AF}$ and $\Theta_{AF}$ values related to the anamorphic factor expressed in percentage; TN related to the true north, PS to plate scale (\as/pixel) and PO to the pupil offset. All distances are measured in pixels and all angles are measured in degrees.

Consequently, the Equations \ref{eq:R} and \ref{eq:theta} can be used to propagate the errors:

\begin{equation}
    \sigma _{R}^{2}= PS^{2}[ \sigma _{R_{spec}}^{2} + \sigma _{R_{*}}^{2}  + r^{2}\sigma _{R_{AF}}^{2} + ( R_{AF} + 1 )^{2} \sigma _{r}^{2}] + \frac{R^{2}}{PS^{2}}\sigma_{PS}^{2} 
\end{equation}

\noindent and,

\begin{equation}
    \sigma _{\Theta}^{2}=\sigma _{\Theta_{spec}}^{2} + \sigma _{\Theta_{*}}^{2} +  \theta^{2} \sigma _{\Theta_{AF}}^{2} + \sigma _{PO}^{2} + \sigma _{TN}^{2} + (\Theta_{AF} + 1)^ {2}\sigma _{\theta}^{2} 
    \\ \text{. }
\end{equation}

The instrumental calibration uncertainties were determined through astrometric calibrations outlined in \citet{maire2016}, \citet{2021JATIS...7c5004M}, and the last version of the SPHERE manual, 18th release. Before July 2016, an issue with the synchronization between SPHERE and VLT internal clocks led to abnormal fluctuations in True North measurements. Consequently, for the initial two $\eta$ Tel observations, the True North uncertainties were extracted from Table 3 of \citet{maire2016}. For the third observation, we adopted a fixed value of 0.04$^{\circ}$, representing the stabilized uncertainty following calibrations and corrections.

The plate (pixel) scale uncertainties, were extracted from close-in-time coronagraphic observations, obtained with the SPHERE-SHINE GTO data, using the globular stellar cluster
47 Tuc as field reference (Table 7 of \citealt{2021JATIS...7c5004M}). 
The pupil offset uncertainty, derived from commissioning and guaranteed time observations, is 0.11 degrees. Distortion is predominantly influenced by a 0.60\% $\pm$ 0.02\% anamorphism between the horizontal and vertical axes of the detector. Since each frame has undergone correction by the HC DC, rescaling each image by 1.006 along the axis, the uncertainty of 0.02\% (for both $R_{AF}$ and $\Theta_{AF}$) was incorporated into the error budget analysis.

The radial stellar centering uncertainty per dithering is 1.2 mas, derived from observations of bright stars during commissioning runs \citep{zurlo2014, zurlo2016}. This value was then adjusted by dividing it by the square root of the number of frames per observation. Subsequently, the latter was translated into an uncertainty on position angle by division by the separation $r$ of the companion. 

\begin{figure*}[!htb]
\begin{center}
\includegraphics[width=0.98\textwidth]{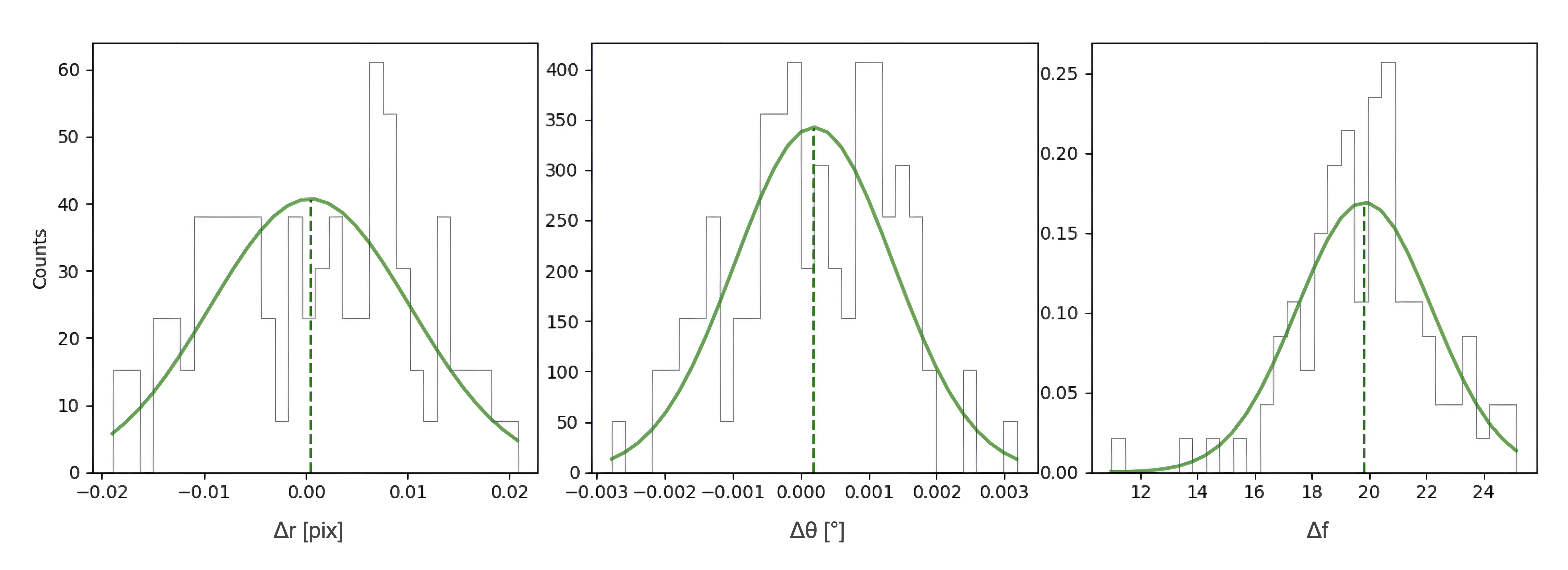}
    \caption{Speckle noise estimation for $\eta$ Tel B in the data set of 2015-05-05. The histograms illustrate the offsets between the true position and flux of a fake companion and its position and flux obtained from the NEGFC technique. The dashed lines correspond to the 1D Gaussian fit from which we determined the speckle noise.}
    \label{speckleuncert}
\end{center}
\end{figure*}

Uncertainties arising from speckles may persist even after ADI post-processing and have an impact on photometric and astrometric measurements \citep{guyon2012,wertz2017}. To address this, we employed the {\tt speckle\_noise\_uncertainty} function from VIP, injecting multiple simulated companions into companion-free cubes at the same radial distance and flux as the actual companion. The positions of these simulated companions were determined using Nelder-Mead optimization, and then the values of separation and position angle were measured. By comparing the offsets between the input values and the estimations by the code, a distribution of parameters was generated. A Gaussian function was then fitted to the distribution, and the uncertainties in $R$ and $\Theta$ were estimated as the standard deviations of the fitting. In this instance, a total of 100 simulated companions, equally spaced azimuthally, were utilized. A similar methodology was employed by \citet{maire2015, wertz2017}. In Fig.~\ref{speckleuncert}, three histograms illustrate the distribution of separation, position angle, and flux of the companion observed on 2015-05-05. The separation and position angle values for the epoch are shown in Table~\ref{tab_framebyframe}, and the detailed uncertainties used to calculate the error budget are compiled in Table~\ref{tabuncert}. 

\begin{table}[!h]
\caption{List of astrometric positions of $\eta$ Tel B obtained from the IRDIS observations. The error budget is listed in Table~\ref{tabuncert}.}
\begin{center}
\begin{tabular}{cccccc}
\hline
Epoch & Filter$^a$ & Sep. (\arcsec) & PA ($^{\circ}$) \\

\hline
\hline
2015-05-05  & H3 & 4.215$\pm$0.004                                                     &         167.326$\pm$0.197                                                               &                                                        \\
2016-06-15  &  BB\_H & 4.218$\pm$0.004                                                  &                167.260$\pm$0.130                                                        &                                                        \\
2017-06-15 &  H2 & 4.218$\pm$0.004                                                 &             167.346$\pm$0.142                                                           &                                                       
\end{tabular}
\end{center}
\label{tab_framebyframe}
\footnotesize{\raggedright{$^{a}$The table only presents values related to the filters where the lowest uncertainties, retrieved from NEGFC, on each epoch, were achieved.}}
\end{table}




\begin{sidewaystable}
    \caption{Uncertainties used to calculate the error budget for the astrometry of the companion. The values correspond to errors associated with individual frames.}
\centering
\begin{tabular}{c|ccccccccccc}
\hline
Epoch      & \begin{tabular}[c]{@{}c@{}}$\sigma_{r}$\\ (pixels)\end{tabular} & \begin{tabular}[c]{@{}c@{}}$\sigma_{R_{spec}}$\\ (pixels)\end{tabular} & \begin{tabular}[c]{@{}c@{}}$\sigma_{R_{*}}$\\ (pixels)\end{tabular} & \begin{tabular}[c]{@{}c@{}}$\sigma_{R_{AF}}$\\ (\%)\end{tabular} & \begin{tabular}[c]{@{}c@{}}$\sigma_{PS}$\\ (mas/pixel)\end{tabular} & \begin{tabular}[c]{@{}c@{}}$\sigma_{\theta}$\\ ($^{\circ}$)\end{tabular} & \begin{tabular}[c]{@{}c@{}}$\sigma_{\Theta_{spec}}$\\ ($^{\circ}$)\end{tabular} & \begin{tabular}[c]{@{}c@{}}$\sigma_{\Theta_{*}}$\\ ($^{\circ}$)\end{tabular} & \begin{tabular}[c]{@{}c@{}}$\sigma_{\Theta_{AF}}$\\ (\%)\end{tabular} & \begin{tabular}[c]{@{}c@{}}$\sigma_{PO}$\\ ($^{\circ}$)\end{tabular} & \begin{tabular}[c]{@{}c@{}}$\sigma_{TN}$\\ ($^{\circ}$)\end{tabular} \\
\hline
\hline
2015-05-05 & 0.297                                                          & 0.003                                                                      & 0.098                                                               & 0.02                                                             & 0.01                                                               & 0.068                                                                 & 0.0005                                                                            & 0.016                                                                         & 0.02                                                                  & 0.11                                                             & 0.145                                                             \\
2016-06-15 & 0.065                                                          & 0.001                                                                      & 0.098                                                               & 0.02                                                             & 0.01                                                               & 0.012                                                                 & 0.0001                                                                            & 0.016                                                                         & 0.02                                                                  & 0.11                                                             & 0.060                                                              \\
2017-06-15 & 0.245                                                          & 0.010                                                                      & 0.098                                                               & 0.02                                                             & 0.01                                                               & 0.072                                                                & 0.0015                                                                            & 0.016                                                                         & 0.02                                                                  & 0.11                                                             & 0.040                                                             
\end{tabular}
\\
\raggedright \footnotesize{The table only presents values related to the filters where the lowest uncertainties on each epoch were achieved: 2015-05-05 (H3); 2016-06-15 (BB\_H), and 2017-06-15 (H2), thus referring to the filters also listed on Table \ref{tab_framebyframe}.}
\label{tabuncert}
\end{sidewaystable}






\subsection{Orbital fitting analysis}
\label{sec:orbfit}

\begin{SCfigure*}
\includegraphics[width=0.75\textwidth]{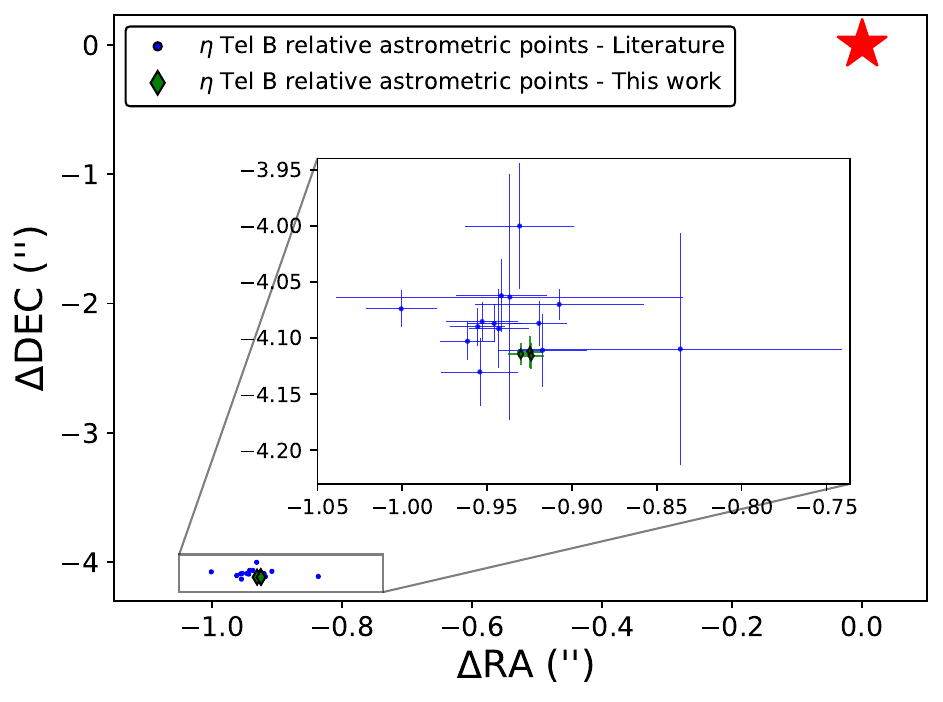}
    \caption{Relative astrometry for $\eta$ Tel B. The red star marks the position of $\eta$ Tel A. The blue-filled circles represent astrometric points extracted from the literature. The green-filled diamonds represent the astrometric points from this work. The astrometric point from epoch 2000.378 was not included due to its high uncertainties.}
    \label{relat_astrom}
\end{SCfigure*}

The system around $\eta$ Tel A was observed with the high-contrast imaging technique for the last two decades. The favorable contrast of the brown dwarf companion and the wide separation between the two objects make the system an ideal target for HCI. The astrometric follow-up of the brown dwarf companion counts 18 epochs of observation spanning almost 20 yr. For the orbit analysis, we included all the astrometrical points presented in the literature, as well as the ones obtained from the new analysis. The complete list is shown in Table~\ref{table_relast} and their positions with respect to the central star are represented in Figure~\ref{relat_astrom}.

\begin{table*}[h!]
\caption{Astrometric positions of $\eta$ Tel B from the literature and our analysis included in the Orvara analysis.}
\begin{center}
\begin{tabular}{llll}
Date (yr) & Separation (\arcsec) & PA ($^{\circ}$) & Ref. \\
\hline
\hline
1998.492  & 4.170$\pm$0.033       & 166.95$\pm$0.36  &    N11\\
2000.307  & 4.107$\pm$0.057       & 166.90$\pm$0.42  &   N11\\
2000.378  & 4.310$\pm$0.270       & 165.80$\pm$6.70  &    G01 \\
2004.329  & 4.189$\pm$0.020       & 167.32$\pm$0.22  &   N11 \\
2004.329  & 4.200$\pm$0.017       & 166.85$\pm$0.22  &   N11\\
2004.329  & 4.199$\pm$0.036       & 167.02$\pm$0.22  &    N11\\
2004.329  & 4.195$\pm$0.017       & 166.97$\pm$0.22  &    N11\\
2006.431  & 4.170$\pm$0.110       & 167.02$\pm$1.40  &   G08 \\
2007.753        & 4.212$\pm$0.033       & 167.42$\pm$0.35  &  N11\\
2008.312   & 4.214$\pm$0.017       & 166.81$\pm$0.22  &    N11\\
2008.599   & 4.195$\pm$0.017       & 166.87$\pm$0.29  &    N11\\
2008.599   & 4.194$\pm$0.016       & 166.20$\pm$0.29  &   N11\\
2009.351        & 4.239$\pm$0.104       & 168.50$\pm$1.30  &   N11 \\
2009.496  & 4.199$\pm$0.031       & 166.99$\pm$0.30  &    N11 \\
2011.576  & 4.170$\pm$0.009       & 167.43$\pm$0.70  &    R13\\
2015.341  & 4.215$\pm$0.004       & 167.33$\pm$0.20  &    $^a$\\
2016.454  & 4.218$\pm$0.004      & 167.26$\pm$0.13  &    $^a$\\
2017.452   & 4.218$\pm$0.004       & 167.35$\pm$0.14   &  $^a$\\
\hline
\end{tabular}
 \end{center}
 \footnotesize \raggedright{N11: \citet{neuhauser2011}; G01: \citet{guenther2001}; G08: \citet{geissler2008}; R13: \citet{rameau2013}\\ $^a$: This work.} \hspace*{-\labelsep}

\label{table_relast}
\end{table*}

We employed the Orvara code \citep[Orbits from Radial Velocity, Absolute, and/or Relative Astrometry;][]{2021AJ....162..186B} to perform the orbital fitting of the system. Orvara accepts input information on the acceleration of the central star from the Hipparcos vs Gaia early data-release 3 \citep[EDR3;][]{gaiadr3} catalogs.

The orbital elements and masses of both the host and the companion were computed by Orvara using a parallel tempered Markov Chain Monte Carlo (MCMC) ensemble sampler, {\tt ptemcee} \citep{vousden2016}, a variant of {\tt emcee} \citep{foremanmackey2013}. Parallel tempering enhances the exploration of complex parameter spaces by simultaneously running multiple chains at different temperatures, allowing for more efficient sampling and improved convergence compared to traditional MCMC methods. In a simulation, different temperatures refer to variations in the parameter that control the acceptance of proposed moves in the MCMC algorithm. Higher temperatures encourage more exploration by accepting moves that might increase energy or objective function values, allowing the algorithm to escape local optima and explore a broader solution space. On the other hand, lower temperatures favor exploitation, focusing on refining solutions and improving the chance of finding the global optimum. Adjusting temperatures during the simulation influences the balance between exploration and exploitation, shaping the algorithm's behavior throughout the MCMC process. The MCMC simulation utilized 10 temperatures, 500 walkers, and 10$^{6}$ steps for each chain. The simulation outputs a point every 1000 steps.

Priors for the MCMC include distributions of the masses of celestial objects, parallax, and proper motion of the system. Additionally, the initial orbital elements distribution for the companions (semi-major axis: \(a\); eccentricity: \(e\); argument of the pericenter: \(\omega\); inclination: \(i\); longitude of the ascending node: \(\Omega\); and longitude at reference epoch: \(l\)) can also be incorporated. We used as priors the common proper motion extracted from \citet{kervellacatalogue,kervella2022}, the parallax from Gaia DR3 \citep{gaiadr3}, the primary mass from \citet{desidera2021,desidera2021catalogue}, and the companion ($\eta$ Tel B) mass with loosened constraints on its uncertainties from \citet{lazzoni2020}. Given the low orbital coverage from relative astrometric points, we just set a mean initial value inspired by the semi-major axis proposed by \citep{neuhauser2011}, leaving other orbital elements to cover wider values in the parameter space, as set by standard values from ORVARA. The specific priors and initial values set are detailed in Table~\ref{priors}. Further orbital constraints and results are analyzed and presented in section \ref{sec:orbitalcutoff}.


\begin{table*}[!htb]
\caption{Priors and initial  and distribution of parameters for $\eta$ Tel system astrometrical fitting$^{a}$.}
\label{priors}
\begin{center}
\begin{tabular}{l|cc}
            Priors & $\eta$ Tel A           & $\eta$ Tel B          \\
                 \hline
Mass \, ($M_{\odot}$) & 2.09$\pm$0.03      & 0.045$\pm$0.014   \\
Parallax (mas)     & 20.603$\pm$0.099          & 20.603$\pm$0.099         \\
RA p.m. (mas/yr)       & 25.689$\pm$0.006        & 25.689$\pm$0.006       \\
DEC p.m. (mas/yr)     & -82.807$\pm$0.006         & -82.807$\pm$0.006         \\
\hline
Initial distribution$^{b}$ & & \\
\hline
\(a\) (au)        & ---        & 220 $^{+300}_{-220}$          \\
$\sqrt{e}*sin(\omega)$     & --- &  0.4$\pm$0.3 \\
$\sqrt{e}*cos(\omega)$      & --- & 0.4$\pm$0.3 \\
\(i\) (radians)     & --- & 1.57$\pm$1.57 \\
$\Omega$ (radians)     & --- & 3.2$\pm$2.2 \\
$l$ (radians)     & --- & 0.8$\pm$0.5 \\
\end{tabular}
\end{center}
\raggedright \footnotesize
$^{a}$ Unphysical values such as negative values for masses or semi-major axis are automatically excluded from the code.
\\
\raggedright \footnotesize $^{b}$ The initial distributions involve a lognormal distribution for the semi-major axis, while all other orbital values follow a normal distribution.
\end{table*}

\section{Discussion}

\subsection{Orbital fitting constraints}
\label{sec:orbitalcutoff}
From the results of the MCMC simulation, we excluded the orbits that may cause instability in the system. In order to prevent disruption of the debris disk by the close-by passage of $\eta$ Tel B, and considering that the masses of the star and companion are well-constrained by the MCMC simulations, we can delimit the possible orbits for $\eta$ Tel B by inspecting the brown dwarf-disk interaction. Before applying any constraint, we obtained as values the mass of $\eta$ Tel A ($m_{*} = 2.09 M_{\odot}$) and mass of $\eta$ Tel B ($m_{comp} = 48.10 \MJup$) from the original ORVARA simulation.

In this context, it is useful to introduce the concept of chaotic zone, a region in the proximity of the orbit of a planet or brown dwarf which is devoid of dust grains, since its gravitational influence sweeps out small dust particles. The chaotic zone depends on the ratio between the mass of the brown dwarf and the mass of the star, on the semi-major axis and eccentricity of the orbit. Given the outer disk's position, it is established that this zone cannot extend below 24 au. As $\eta$ Tel B is positioned outside the disk, we employed the procedure outlined in \cite{Lazzoni2018} to compute the chaotic zone's extension for each orbit during the pericenter passage. Specifically, we utilized equation 10 for cases where the eccentricity is less than a critical value ($e_{crit}$), or equation 12 for cases where it exceeds $e_{crit}$. Here, $e_{crit}$ is determined as 0.21$\mu^{3/7}$, equivalent to 0.022 ($\mu$ is the ratio between the mass of the companion and the mass of the star). In the end, we discarded 326674 orbits from an original number of 500000.

Furthermore, another constraint arises when determining that the pericenter of the brown dwarf (BD) cannot reach the disk. Consequently, considering the semi-major axis of the companion established by the MCMC simulations, ($a_{bd}\footnote[1]{The value of the semi-major axis of 178 au was the one retrieved before the exclusion of orbits. The final value is 218 au, as presented in Table \ref{orvararesults} and Fig. \ref{cornerplot}.} = 178~au$), we can use the following expression to discard highly eccentric orbits: 

\begin{equation}
\centering
a_{bd}(1-e_{max}) > 24 ~au
\end{equation}
where $e_{max}$ is the maximum eccentricity allowed for the BD. Consequently, we obtained a maximum eccentricity of 0.865 and, therefore, we discarded an additional 139 orbits.
Another constraint can be applied if we consider that the $\eta$ Tel system has a wide comoving object, as mentioned previously in \citet{neuhauser2011}. HD 181327 is an F5.5 V star, also a bona-fide member of the $\beta$ Pic moving group, at a separation of $\sim$7 arcmin (20066 au) \citep{holmberg2009,neuhauser2011,gaiadr3}. 
The small proper motion difference in Gaia DR3, corresponding to a velocity difference on the plane of the sky of 370 m/s, is compatible with a bound object. The nominal RV difference derived in \citealp{gaiadr3, 2021A&A...645A..30Z}, is instead larger than the maximum expected one for a bound object ($\sim$400 m/s).  
However, it is well possible that the published RV errors are underestimated for a star with an extremely fast $v \sin i$ such as $\eta$ Tel. There could also be contributions by additional objects, although both the Gaia RUWE (Gaia Renormalized Unit Weight Error) and the analysis of homogeneous RV time series do not indicate the presence of close companions \citep{lagrange2009,grandjean2020}.
Therefore, we consider plausible, although not fully confirmed, that HD 181327 is bound to the $\eta$ Tel system.

\begin{figure}[!htb]
\begin{center}        \includegraphics[width=0.7\textwidth]{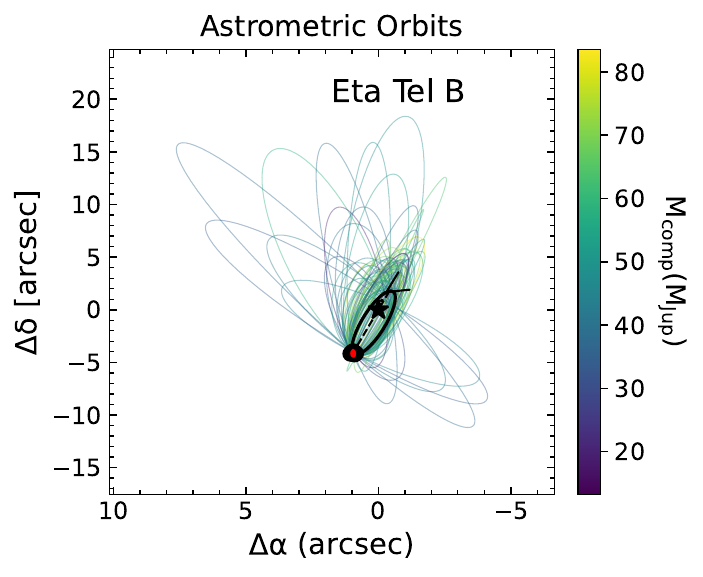}
        \caption{87 randomly selected possible orbits calculated by the Orvara fitting,  after excluding orbits based on constraints described in Section {\color{red}4.1}. The black star indicates the position of $\eta$ Tel A. The best-fit orbit is shown in black.}
    \label{orbits}
\end{center}
\end{figure}

To ensure that HD 181327's presence in the system would not affect the long-term stability of $\eta$ Tel B, we implemented equation 1 of \citep{holman1999}. Therefore, we could exclude orbits where the critical semi-major axis is greater than the periastron of the brown dwarf. To proceed, we used the mass of the perturber as 1.3 $M_{\odot}$ \citep{desidera2021,desidera2021catalogue}. Considering the eccentricities of the binaries as 0, a perturber with a similar mass would only impose constraints with a separation less than 13.25\arcsec (656 au) from $\eta$ Tel A. Otherwise, for HD 181327 to act as a perturber at its separation from the $\eta$ Tel system, the eccentricities between HD 181327 and $\eta$ Tel should be greater than 0.947. Therefore, we choose not to exclude any orbit using the external perturber criteria.
Following, the results of the simulation after the cut-off are shown as the corner plot in Fig.~\ref{cornerplot} and also summarized in Table~\ref{orvararesults}. The values obtained depict a pericenter of 2.9\arcsec (144 au) and an apocenter of 5.9\arcsec (292 au). Despite two decades of observations of $\eta$ Tel B, the coverage only spans a fraction of the wide orbit of the companion, with a total $\Delta$PA of approximately $\sim$2$^{\circ}$. This accounts for less than 1\% of the orbit, assuming a face-on and circular orbit. Consequently, constraining the semi-major axis and eccentricity proves challenging. Nevertheless, the MCMC effectively determines the inclination, yielding an orbit that is nearly edge-on and almost co-planar with the debris disk \citep{2009A&A...493..299S}. 
Conversely, a variety of possible orbits can adequately fit the data, as illustrated in Fig.~\ref{orbits}.

\begin{table*}[!htb]
\caption{Best orbital-fitting parameters for $\eta$ Tel B calculated from the Orvara orbital characterization.}
\label{orvararesults}
\begin{center}
\begin{tabular}{l|l}
Mass ($M_{jup}$)                   & ${48}_{-15}^{+15}$         \\
\(a\) (au)                       & ${218}_{-41}^{+180}$            \\
\(i\) (deg)            & ${81.9}_{-3.5}^{+3.2}$          \\
\(\Omega\) (deg)         & ${174.6}_{-7.1}^{+175}$         \\
Mean longitude (deg)         & ${184}_{-74}^{+164}$           \\
Parallax (mas)               & ${20.60}_{-0.10}^{+0.10}$   \\
Period (yrs)                 & ${2201}_{-592}^{+3224}$         \\
\(\omega\) (deg) & ${159}_{-99}^{+128}$            \\
\(e\)                 & ${0.34}_{-0.23}^{+0.26}$       \\
\(a\) (mas)         & ${4486}_{-845}^{+3704}$        \\
Reference epoch \(T0\)  (JD)                      & ${2740996}_{-107450}^{+488958}$ \\
Mass ratio                   & ${0.0219}_{-0.0068}^{+0.0069}$
\end{tabular}
\end{center}
\end{table*}



\subsection{Possible formation scenarios}
\label{sec:formationscenario}

The coplanarity between $\eta$ Tel B and the debris disk, as described in this work, and the relative spin-alignment between the star and the debris disk \citep{hurt2023}, can offer insights into the formation scenario of the system. Stellar systems form through various mechanisms, typically categorized into three main types: 1) fragmentation of a core or filament, 2) fragmentation of a massive disk, or 3) capture and/or dynamical interactions (for a comprehensive review, including separation and formation time scales, see \citealp{offner2022}). Given the separation of the star and the companion, scenarios involving massive disk fragmentation or capture and/or ejection of $\eta$ Tel B appear more plausible.

Alternatively, core fragmentation, which occurs through direct/turbulent fragmentation (e.g., \citealp{1979ApJ...234..289B, 1997MNRAS.288.1060B}) or rotational fragmentation (e.g., \citealp{larson1972,1994MNRAS.269..837B,1994MNRAS.269L..45B,1994MNRAS.271..999B,1997MNRAS.289..497B}), followed by inward migration, is also a conceivable formation scenario. Binaries formed through direct fragmentation, with well-separated cores, may exhibit uncorrelated angular momenta between the objects \citep{2016ApJ...827L..11O, 2018MNRAS.475.5618B, 2019ApJ...887..232L, 2000MNRAS.314...33B}. In contrast, stars and substellar objects formed through rotational fragmentation within the same plane tend to display preferentially spin-aligned and coplanar systems \citep{2016ApJ...827L..11O, 2018MNRAS.475.5618B}. In such a case, a rotational fragmentation followed by inward migration becomes more probable.

Moreover, the eccentricity of the system plays a crucial role in inferring the formation scenario. If we consider the orbital fitting results where highly eccentric orbits are permitted, the preferred hypothesis for eccentricity enhancement would be recent dynamical interactions.  If so, $\eta$ Tel B did not reach near the debris disk of the star in the last few pericenter approximations. This scenario is highly improbable, given the short best-fitting orbital period in such cases ($\sim{1623}$ years). It is more likely that $\eta$ Tel B has a low eccentric orbit, and the assumptions for excluding highly eccentric orbits may be the most suitable approach. Consequently, the system exhibits quasi-coplanarity and low eccentricity of the companion. Therefore, we tentatively suggest that the preferred formation scenarios for $\eta$ Tel involve either the fragmentation of a massive disk with slow or no inwards migration or rotational fragmentation of a core with fast inwards migration. Long-term RV and astrometry monitoring of the star and the companion, along with multiwavelength observations of the system, could be useful to discard a capture or ejection scenario.


\subsection{The close surroundings of $\eta$ Tel B }
\label{sec:sathunt}

In our Solar System, planets and small-sized bodies are often surrounded by satellites and dust rings or disk-like features (see e.g. \citealp{alibert2005}). For instance, there are approximately 200 natural satellites in the Solar System, most of which orbit giant planets. This raises the possibility of similar objects existing around substellar companions, although such discoveries have not yet been confirmed. Satellites or circumplanetary disks in these environments, if discovered, could provide valuable insights into their formation mechanisms. These mechanisms may include gravitational instability (\citealp{boss1997} and references therein), core-accretion (\citealp{pollack1996} and references therein), or capture and/or orbital crossing, which can lead to satellite companions with specific mass ratios and orbits. For instance, less massive exomoons are likely to form within a circumplanetary disk (CPD), as observed with the Galilean moons (see, e.g., \citealp{canupeward2002}). Conversely, massive companion+satellite candidate systems likely form via orbital crossing+capture \citep{ochiai2014, lazzoni2024}. Furthermore, hydrodynamical simulations have shown that CPDs in the core-accretion scenario are eight times less massive and one order of magnitude hotter than those formed by gravitational instability \citep{szu2017}. Consequently, the characterization of exosatellites can be used to distinguish between these formation scenarios.

\begin{figure*}[!htb]
\begin{center}
    \includegraphics[width=0.9\textwidth]{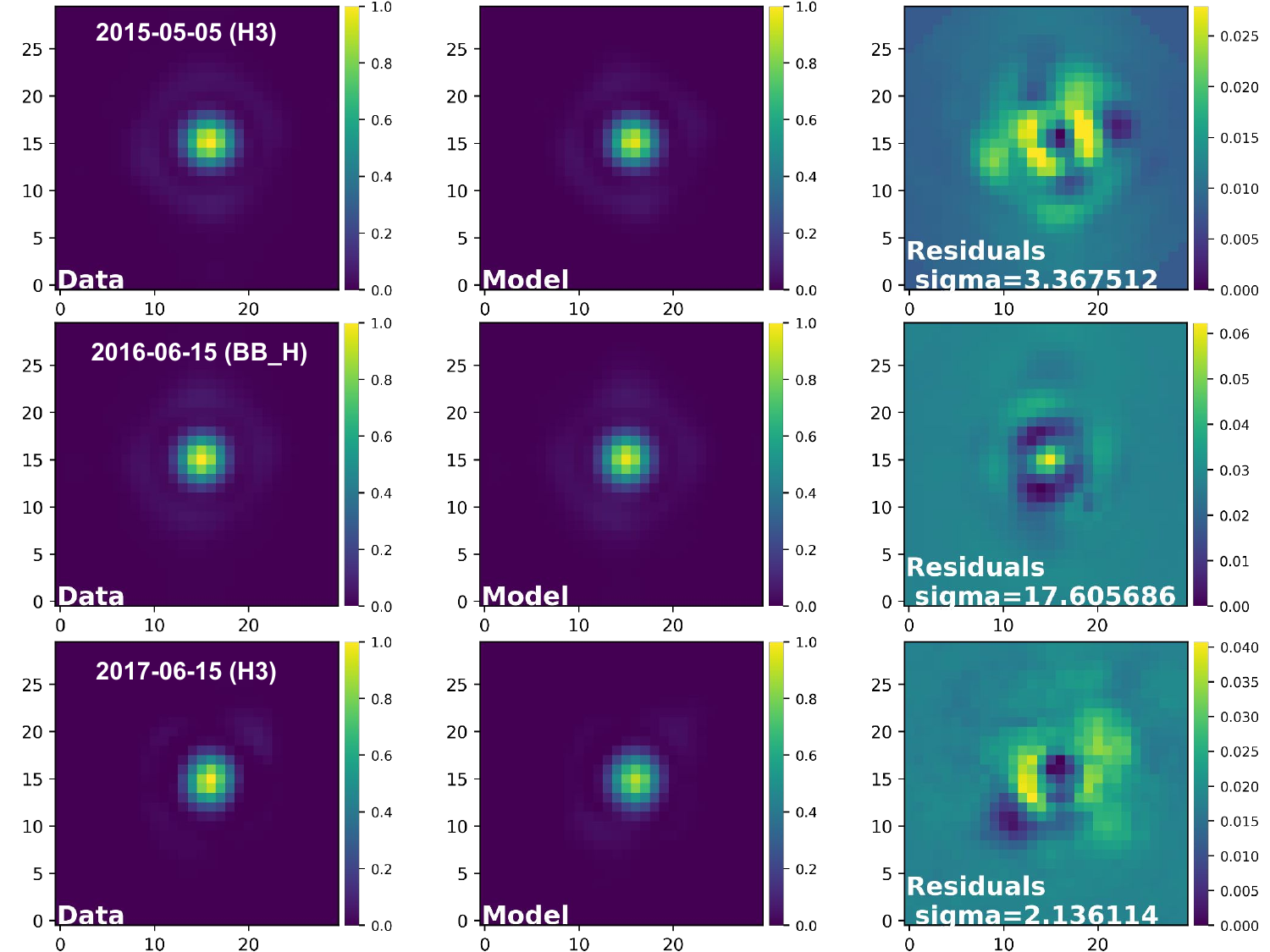}
    \caption{Analysis of the close surroundings of $\eta$ Tel B to look for satellites and/or CPDs. In each row, the first panel shows the ADI image zoomed on $\eta$ Tel B, the second panel shows the model, and the last one the residuals from the subtraction of the two. The three IRDIS epochs are shown. The residuals are expressed in counts. All the panels were normalized to the peak of the central PSF.}
    \label{fig:residuals}
\end{center}
\end{figure*}


To analyze whether a satellite candidate or a CPD is present around the companion, we proceeded as follows: to compensate for the self-subtraction effect induced by post-processing, we forward-modeled the companion per frame based on the observations and the PSF model. With the pre-processed data, the radial distance, position angle, and flux were obtained from the NEGFC approach (refer to Section \ref{sec:astrometry_negfc}). The off-axis PSF extracted before the sequence was positioned and flux-normalized with these parameters in an empty frame, creating what we refer to as ``model''. Following, for each frame, the model was subtracted from the data, producing a residual image. The model and residual cubes were then collapsed using ADI or PCA+ADI. We compare the ADI post-processed data with the collapsed model and residual cubes per night in Fig. \ref{fig:residuals}). The quality of the residuals is strongly related to the number of frames and quality of the night, showing a clearer result for the epoch 2016-06-15, where other epochs are more affected by systematics. Still, the residuals do not show any clear signal of a satellite or other nearby structure, imposing a threshold of detections at the contrast obtained with the NEGFC technique. 

To generate contrast curves around $\eta$ Tel B, we implemented a methodology similar to the one outlined in \citet{lazzoni2020}. The steps can be summarized as follows. Successive annuli, each centered on $\eta$ Tel B and with a width equal to 1 FWHM, were chosen up to the Hill radius of the brown dwarf. The contrast at each radial position is computed as five times the standard deviation inside the annulus divided by the peak of the star ($\eta$ Tel A).

\begin{figure}[!h]
\begin{center}
    \includegraphics[width=0.65\textwidth]{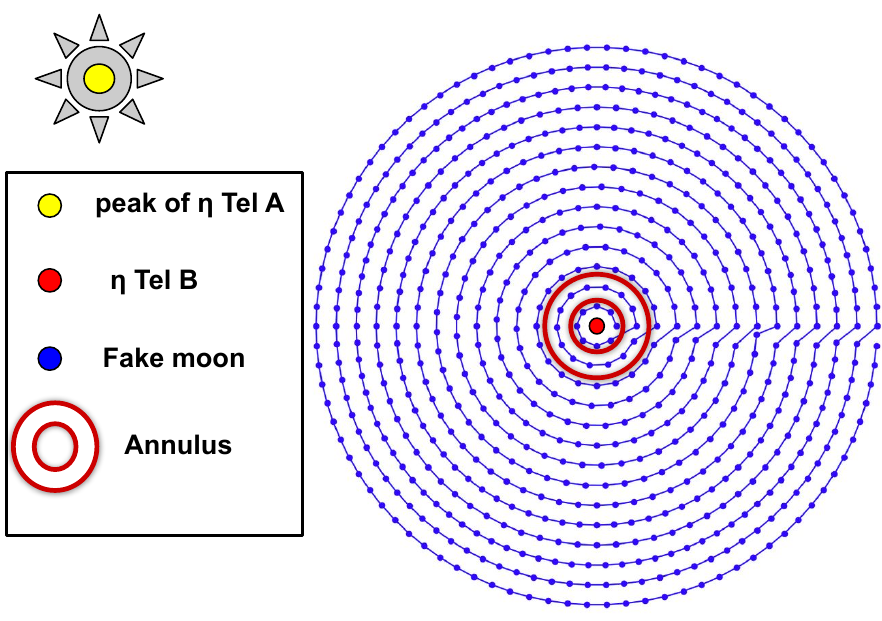}
    \caption{Schematic representation of the placement of putative fake satellites around $\eta$ Tel B to calculate the contrast curves. In this illustration, just one annulus is represented.}
    \label{illustration_cc}
\end{center}
\end{figure}

For each annulus, we injected fake companions at various position angles, and their fluxes were determined after applying the NEGFC and PCA post-processing techniques. The ratio between the retrieved and injected flux provides the throughput value. A mean throughput is then calculated for each annulus and multiplied by the contrast at each separation. As a final step, we adjusted the contrast for small sample statistics, following the discussion presented in \citet{mawet2014}. A schematic representation of the steps used to calculate the contrast curves is illustrated in Fig.~\ref{illustration_cc}.

\begin{figure}[!h]
\begin{center}    
\includegraphics[width=0.75\textwidth]{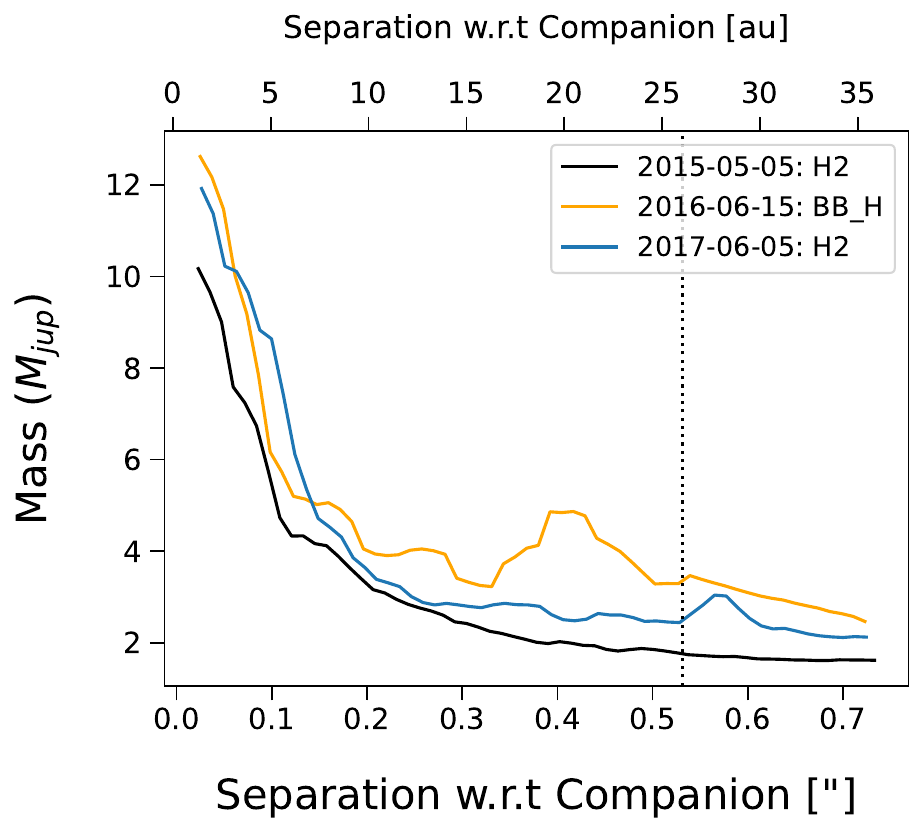}
    \caption{
    Contrast curves with respect to the star $\eta$ Tel A, centered on the companion as depicted in Fig. \ref{illustration_cc}. The three contrast curves, corresponding to different epochs, are showcased.
    The vertical dashed line indicates the Hill radius of $\eta$ Tel B under the assumption of an orbit with an eccentricity of $e$=0.34}
\label{contrast_curves}
\end{center}
\end{figure}

\begin{figure}[htb!]
    \begin{minipage}[t]{1\textwidth} 
        \begin{tabular}[t]{ @{} r @{} }
        \includegraphics[width = 1\linewidth]{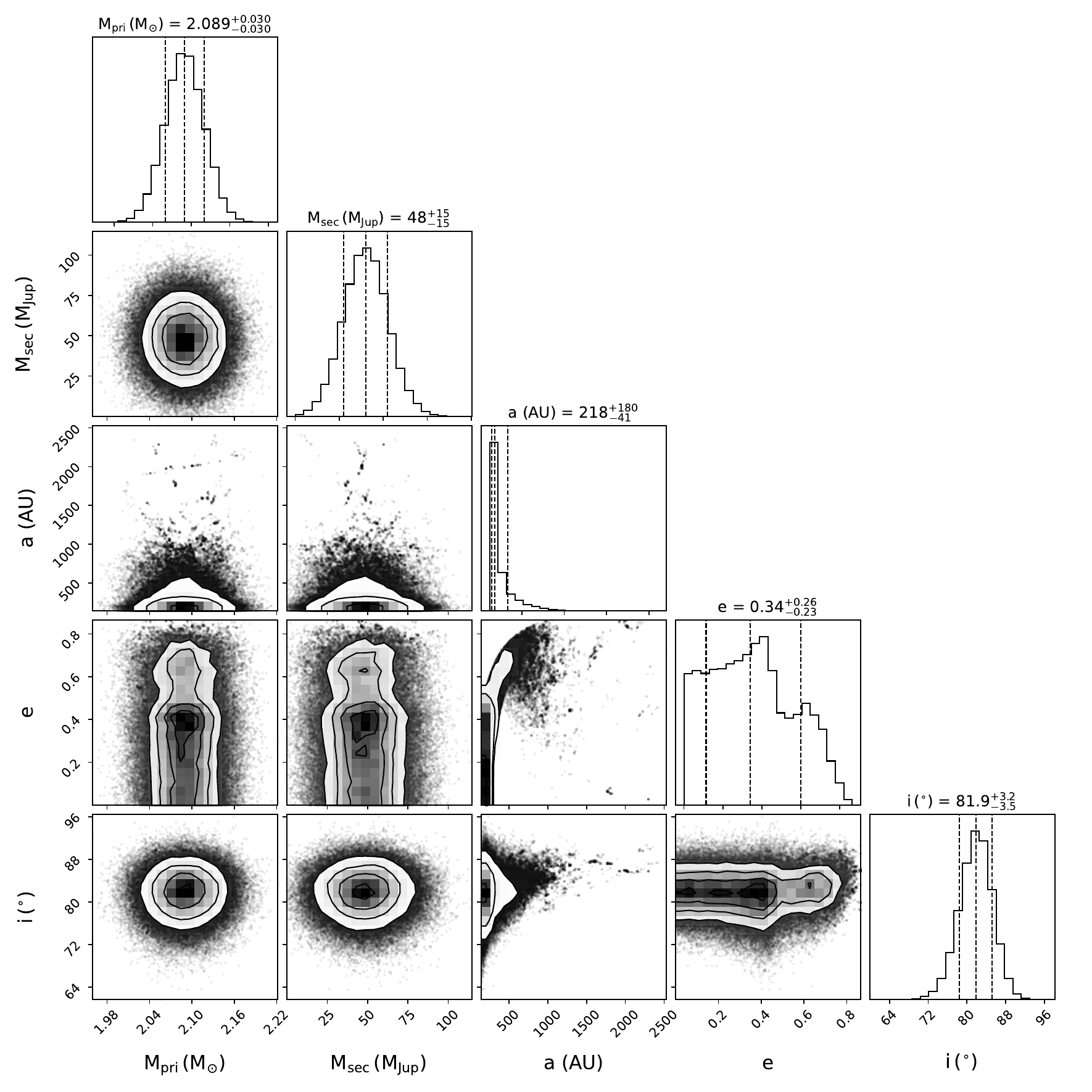} \\

       \end{tabular}
               \caption{Corner plot showing the results of the orbital fitting performed by Orvara, after excluding orbits based on constraints described in section \ref{sec:orbitalcutoff}. The best-fit values are also reported in Table~\ref{orvararesults}.}
    \label{cornerplot}
    \end{minipage}
\end{figure}

Finally, the contrasts were converted into mass constraints using the ATMO 2020 evolutionary models \citep{Phillips}. We utilized an estimated age for the system of 18 Myr \citep{2020A&A...642A.179M} and a distance of 49.5 pc \citep{gaiadr3}. The best contrast curves for each data set are depicted in Fig.~\ref{contrast_curves}. We can discard the detection of satellites around the brown dwarf with masses between 3 and 1.6 \MJup in the range of distances  $[10,33]$ au. Such massive objects, if present, would likely be the result of capture or trapping through tidal interactions \citep{lazzoni2024} or formation in situ via gravitational instability or direct collapsing. However, we cannot exclude the presence of closer-in and/or less massive objects which, for example, could have formed within a CPD via core accretion.
Moreover, we can exclude the presence of an extended CPD from the shape and luminosity of the residuals around the companion.


\section{Conclusions}

\label{sec:sum}
In this paper, we present the most recent photometric and astrometric characterization of $\eta$ Tel B, a brown dwarf situated at an approximate separation of $\sim$4.2\arcsec from its host star. The observations of this system were conducted using SPHERE/IRDIS H2H3 and BB\_H filters, spanning three epochs across three consecutive years (2015-2017).

To robustly establish astrometry and photometry for the sub-stellar companion, we employed the NEGFC customized routine presented in \citep{lazzoni2020}. This approach was applied to each frame of the scientific datacube, as opposed to solely in the post-processed ADI image, as is conventionally practiced. Photometric results were derived by considering the median of each set of parameters per observation. Photometric errors were determined based on the standard deviation of the fluxes obtained per night. Astrometrical results (separation of 4.218$\pm$0.004 arcsecs and position angle of 167.3$\pm$0.2 degrees) were made taking into account not only the NEGFC approach but also incorporated an analysis using 2D Gaussian fitting and position of the peak intensity of the companion. This analysis was made for both the frame-by-frame approach and the median-collapsed ADI image. The final values represent the median of each approach. In addition, for the uncertainties, systematic and statistical uncertainties, akin to the methodology employed by \citet{wertz2017}, were employed. The separation reached depicts a 4-70 times improvement in precision in comparison with previous NACO observations described and observed by \citet{neuhauser2011}, and $\sim$2 times better precision than the L$^{\prime}$ NACO observations from \citet{rameau2013}. 

Furthermore, we conducted a comprehensive orbital characterization by compiling previous astrometric data, the new SPHERE data, and the Hipparcos-Gaia acceleration catalog, resulting in an orbital coverage spanning approximately 19 years. The findings indicate a well-characterized companion mass of 48$\pm$15 \MJup, with the best-fitting orbit demonstrating near-edge-on orientation ($i$=81.9 degrees) and low eccentricity ($e=0.34$) when excluding orbits that can disrupt the debris disk around the star. However, it is essential to note that our observations did not encompass a significant portion of the entire orbit, leading to elevated uncertainties regarding the companion's orbital shape, period, eccentricity, and semi-major axis. We also highlight that due to the low orbital coverage, the orbital fitting presented represents a plausible family of orbits, and the orbital values listed in Table~\ref{orvararesults} must be taken as an example of a possible orbit. Further follow-ups will better constrain the orbit of the BD in the future. Also, based on the high $v sin i$ of $\eta$ Tel A, and the possible uncertainties on its measurements, we consider it plausible that HD 181327, at a separation of 20066 au, is bound to the system. 

A brief discussion about possible formation scenarios has been conducted. The coplanarity between $\eta$ Tel B and its debris disk, along with the relative spin alignment between the star and the debris disk, provides valuable insights into the system's formation. Stellar system formation scenarios were categorized into three main types: fragmentation of a core or filament, fragmentation of a massive disk, or dynamical interactions. Based on the separation of the star and companion, the likelihood of massive disk fragmentation or capture or ejection of $\eta$ Tel B was highlighted as more plausible. Alternatively, the system's low eccentricity also allows for rotational fragmentation from a core scenario, if followed by fast inward migration. In summary, we tentatively propose the preferred formation involves either massive disk fragmentation with slow or no inward migration or rotational fragmentation of a core with faster inward migration, whereas continuous monitoring and multiwavelength observations are crucial for refining these conclusions and further understanding the system's dynamics.

Lastly, a meticulous analysis of the companion's surroundings was undertaken by subtracting the companion's signal using instrument-response models. No clear signal of a substructure or satellite was seen. We conclude the systematics heavily affected the residuals. From the contrast curves generated in the regions surrounding the companion, we can discard satellites down to 3 and 1.6 \MJup in the range of distances [10, 33] au, setting an upper limit on gravitational instability binary pairs or massive objects captured through tidal interactions at wider separations.

Future observations of the system with the next generation of high-contrast imagers mounted on space telescopes, likewise, JWST (James Webb Space Telescope) will further constrain the orbital analysis of the brown dwarf companion and allow for deeper contrast around the central star and the companion. This advancement will enable deeper contrast observations around both the central star and its companion. In the specific case of RV monitoring for detecting objects at closer separations, instruments like CRIRES+ (Cryogenic high-resolution InfraRed Echelle Spectrograph) or HiRISE (High Resolution Imaging Science Experiment),  are suitable. Possible additional companions to the central star or the brown dwarf may be detected in the future. 

\clearpage

\section{Appendix}

\subsection{Cornerplot of the orbital fitting}





  \chapter{Final remarks and conclusions}\label{chap:conclusions}

This thesis's main goal is to highlight the importance of observations regarding fundamental aspects of multiple (sub)stellar formation and evolution. The thesis comprises two comprehensive studies of distinct multiple systems: an eruptive binary system, HBC 494, and a more evolved binary system, $\eta$ Telescopii. Each investigation employs different observational instruments operating at distinct wavelengths. I hope that new readers won't become entangled in the technical details presented in both papers in Chapter \ref{chap:2} and Chapter \ref{chap:3}, as the fundamental concepts are (hopefully) elucidated in the comprehensive introduction (Chapter \ref{ch:intro}). The narrative presented in this thesis unfolds as follows. 

In Chapter~\ref{chap:2}, we refer to the detailed study of HBC 494. HBC 494, as an eruptive YSO system in the Orion Molecular Cloud, was subjected to high-resolution ALMA observations. These observations resolved the system into two components separated by $\sim$ 75 au, HBC 494 N and HBC 494 S, thus offering a comprehensive view of the morphology and dynamics of the system. The FUor-like characteristics of the northern component and EXor-like features in the southern component highlight the diversity within the system. Molecular line observations further revealed bipolar outflows and rotating, infalling envelopes, enriching our understanding of the dynamics in such eruptive binary systems.

Turning our attention to $\eta$ Telescopii (Chapter \ref{chap:3}), a binary system where the companion is a brown dwarf companion, high-contrast imaging campaigns spanning over two decades provided crucial data for orbital and photometric characterization. Detailed analysis of SPHERE/IRDIS coronagraphic observations allowed for precise measurements of the companion's astrometric positions, resulting in the smallest astrometric uncertainties for $\eta$ Tel B to date. The orbital analysis unveiled a low eccentric orbit with a nearly edge-on inclination and a semi-major axis of 214 au. A rather new and promissory analysis using HCI observations was applied to search for satellites or other features around $\eta$ Tel B, failing to detect but useful to provide deeper contrasts for upcoming analysis with different instruments and resolution power.

Together, these studies contribute valuable information to the broader field of multiple stellar formation. The episodic accretion phenomenon observed in HBC 494 showcases the mechanism of episodic accretion within binary systems, while the detailed characterization of $\eta$ Telescopii's binary system enriches our understanding of binary dynamics and high-contrast imaging studies of substellar companions. These findings collectively emphasize the importance of investigating various stellar systems to unravel the intricate processes that govern their formation and evolution.

\section{Specific questions/concepts addressed}

In this work, we address, analyze, and provide insights on the following questions:

\subsection{What is the correlation between stellar multiplicity and accretion processes, and why are binary eruptive systems so infrequently observed?}

These questions remain unanswered, and due to the limited number of known multiple eruptive young stellar object (YSO) systems, expressing the occurrence and significance of these events proves challenging. However, two critical aspects warrant consideration. Firstly, binaries are notably prevalent in the Galaxy, as indicated by various citations. Secondly, certain studies propose that eruptive accretion, occurring on different scales, may be common among low-mass YSOs throughout their evolutionary stages (see, for example, \citealp{cieza2018, hales2020}). Therefore, additional (sub)millimeter observations of eruptive binary systems are imperative to reveal the role of episodic accretion in the evolution of multiple systems. 

To date, only a limited number of eruptive stars have been detected and spectroscopically characterized, with a small fraction of them being part of multiple systems. The findings presented in Chapter \ref{chap:2} suggest that the scarcity of observations may be linked to their resolution power. In the same chapter, attention is drawn to the lack of a precise answer regarding how multiple eruptive YSOs can influence the outbursting and accretion of individual components. The analysis reveals that the disks in multiple eruptive systems are more compact but also more massive than regular YSOs at the same stage/environment. However, the examination of HBC 494 gas on a scale appropriate to the continuum disks remains inconclusive about how the material can flow from one object to the other or remain unaltered by the other component. Therefore, a lingering question that remains and can be addressed with more observations of multiple eruptive YSO systems, using optically thinner tracers, is whether the enhancement of accretion in one disk can trigger effects on another disk, or if they both initiate the FUor/Exor stage simultaneously.

\subsection{How robust astrometric analysis of a system contribute to its characterization}

Despite the revealing nature of HCI observations in uncovering remarkable systems, the observed substellar companions, until now, have commonly shared a characteristic: a large semi-major axis of approximately 10 AU or more, leading to orbits with periods of years. Consequently, continuous HCI astrometric follow-up over the years is essential to refine our understanding of these systems and unravel their dynamics. One example of such an approach and successful results is given for $\eta$ Tel system, in Chapter \ref{chap:3}. The analysis of $\eta$ Tel underscores its significance in achieving a comprehensive understanding of the system when integrated with extensive observations spanning nearly two decades.

However, it is crucial to acknowledge an important aspect. The orbital fitting for widely separated systems lacking thorough orbital coverage from HCI observations is highly sensitive to the precision of astrometric measurements. The initial ``raw'' results obtained for $\eta$ Tel B were deemed physically implausible due to higher eccentricities obtained for the companion, which would cause instability and disruption in the closer-in debris disk. Fortunately, the $\eta$ Tel system's constraints allowed for a more refined assessment of possible orbits in this work. It's worth noting that these priors may not be achievable for all HCI systems. Thus, the study underscores the significance of obtaining more precise astrometric measurements to derive more physically meaningful orbital results.

The importance of robust astrometric measurements for HCI companions becomes even more pronounced, particularly in the case of multiplanetary systems like HR 8799, $\beta$ Pic, or even PDS 70 with its two protoplanets. The impact of dynamical analysis is not limited to orbital properties; it also extends to crucial aspects such as mass retrieval. In subsection \ref{chap:hci-obs}, particularly in the discussion of remarkable systems observed via HCI, it becomes apparent that planetary masses, even when derived from various studies, exhibit considerable uncertainty. The spectrum of masses for each exoplanetary companion discussed in that section spans both dynamical masses and masses calculated from luminosity-age evolutionary models. The latter derivation also highlights the importance of robust photometric retrieval, as was done to $\eta$ Tel B. In my analysis of the companion, however, I only assumed the mass derived from the orbit fitting.

\subsection{How can one deduct the formation scenario for multiple systems?}

I tentatively answer this question in the specific cases of HBC 494 (Chapter \ref{chap:2}) and $\eta$ Tel (Chapter \ref{chap:3}) components. The only way to link formation scenarios from early YSO stages to mature systems is through simulations. While simulations play a pivotal role in this process, observations alone can also offer insights. 

In environments with specific conditions, discerning a formation scenario becomes plausible. In the cases of HBC 494 and $\eta$ Tel components, the proposed scenario suggests rotational disk fragmentation rather than direct collapsing. This conclusion is drawn from the orbital alignment—quasi-coplanarity of HBC 494 N and S inferred from similar inclinations and PAs between the components of the system, and quasi-coplanarity between $\eta$ Tel B and the debris disk inferred from the inclinations. Therefore, observations of multiple systems in different stages may serve as a control test, establishing a connection between (sub)stellar formation and their subsequent evolution. 

Alternative methods of detecting the formation scenario, such as observing regions of protoplanetary disks susceptible to gravitational instability (GI), can also provide useful insights. However, discovering the formation scenario without unveiling the gas/dust dynamics between YSOs (as in the case of HBC 494) or in more evolved systems with (nearly) depleted gas, such as in the case of $\eta$ Tel, remains challenging.




\subsection{Is it possible to search for satellites orbiting substellar companions via HCI?}

The answer to this question is yes. The deep analysis done for $\eta$ Tel B surroundings in Chapter \ref{chap:3} was heavily inspired by the notable satellite candidate unveiled through SPHERE observations around DH Tau B \citep{lazzoni2020}, yet to be confirmed. Although we pushed to the limits of one of the best modern high-contrast imagers (SPHERE) till the moment, the non-detection of a satellite or structure surrounding $\eta$ Tel B does not completely discard the possibility of an orbiting unresolved source. It just provides deeper contrasts showing limitations that HCI observations may overcome to possibly find objects surrounding $\eta$ Tel B. I highlight that no blind detection will necessarily be proven successful, and $\eta$ Tel B's surroundings can be indeed empty. But, as an HCI explorer/researcher, part of the journey is to sail through seas of the unknown.

To tackle the problem for $\eta$ Tel and other systems but with the same instrument, better PSF models need to be created, which can be provided by different observing strategies and cadence of models' creation.  Alternatively, the next generation of ground-based high-contrast imagers, boasting improved resolving power and enhanced Adaptive Optics (AO), or space-based high-contrast imagers like the James Webb Space Telescope, may provide noteworthy increase in sensitivity and/or resolution in the coming years and decades (see \citealp{2023ASPC..534..799C} for an updated review of the next generation instruments dedicated to HCI). Indeed, the future of HCI is promising. 


During the course of my PhD, I, as PI, got three successful observing proposals using SPHERE proposals to detect satellites/structures around 15 promising substellar companions, including $\eta$ Tel. Using the star-hopping (a Reference-star differential imaging observing strategy; \citep{2021A&A...648A..26W}, I plan to observe these substellar companions, obtaining a quasi-simultaneous model of the instrumental PSF. This approach involves alternating between our target and a reference star until the end of the sequence, minimizing turbulence and improving efficiency in retrieving features. Star-hopping enhances data quality by providing multiple PSFs compared to standard IRDIS observations. Additionally, the negative fake companion technique, also used in Chapter \ref{chap:3}, will be applied in post-processing to create models of the companion from the reference star. Subsequently, these models will be subtracted from the data, and a following analysis of the residuals will be performed for further insights. This approach is thus, one example of how HCI can be performed and optimized for the search of companions' companions. 

  \backmatter

  \bibliographystyle{apalike2}
  \bibliography{manuscript}

\end{document}